\documentclass[10pt,a4paper]{iopart}
\usepackage[T1]{fontenc}
\usepackage{amsmath}
\usepackage{amsfonts}
\usepackage{verbatim}
\usepackage{amssymb}
\usepackage{graphicx,bm}
\usepackage{graphicx}
\usepackage{bbm}
\usepackage{amssymb}
\usepackage{subfigure}
\usepackage{caption}
\RequirePackage{latexsym}
\usepackage{hyperref}
\hypersetup{
    colorlinks=true,
    linkcolor=red,
    citecolor=blue,
    filecolor=magenta,      
    urlcolor=cyan,
    pdftitle={A Perturbative Analysis of Interacting Scalar Field Cosmologie},
    pdfpagemode=FullScreen}  
\newcommand{\be}{\begin{equation}}
\newcommand{\ee}{\end{equation}}
\newcommand{\bea}{\begin{eqnarray}}
\newcommand{\eea}{\end{eqnarray}}

\newcommand{\R}{\mathbb{R}}

\newcommand{\bef}{\begin{figure}[ht!]}  \newcommand{\eef}{\end{figure}}
\newcommand{\bec}{\begin{center}}  \newcommand{\eec}{\end{center}}

\usepackage{url} 
\usepackage{pdflscape} 
\usepackage{amsmath}
\usepackage{amssymb} 
\usepackage{epsfig} 
\usepackage{amsfonts}
\usepackage{graphicx}
\usepackage{subfigure}
\usepackage{graphicx}
\usepackage{times,fancyhdr}
\usepackage{enumitem}
\setlist[enumerate,2]{label=\roman*)}

\def\case#1/#2{\textstyle\frac{#1}{#2}}

\newcommand{\ben}{\begin{eqnarray}}
\newcommand{\een}{\end{eqnarray}}
\usepackage{mdframed}
\usepackage{grffile}
\usepackage{amssymb}
\usepackage[T1]{fontenc}

\usepackage{amssymb,amsmath,amsfonts,amsthm,graphicx, graphics, setspace}
\usepackage{bigints}
\usepackage{anysize}
\usepackage{float}
\usepackage{fancyhdr}
\usepackage{subfigure}
\usepackage{pdflscape}
\usepackage{epsfig}
\usepackage{epsf}
\usepackage{epstopdf}
\usepackage{hyperref}

\newtheorem{theorem}{Theorem}[section]
\newtheorem{proposition}{Proposition}
\newtheorem{lemma}{Lemma}[section]
\newtheorem{definition}{Definition}

\providecommand{\U}[1]{\protect\rule{.1in}{.1in}}

\usepackage{tikz,xcolor,hyperref}

\definecolor{lime}{HTML}{A6CE39}
\DeclareRobustCommand{\orcidicon}{%
	\begin{tikzpicture}
	\draw[lime, fill=lime] (0,0) 
	circle [radius=0.16] 
	node[white] {{\fontfamily{qag}\selectfont \tiny ID}};
	\draw[white, fill=white] (-0.0625,0.095) 
	circle [radius=0.007];
	\end{tikzpicture}
	\hspace{-2mm}
}

\foreach \x in {A, ..., Z}{%
	\expandafter\xdef\csname orcid\x\endcsname{\noexpand\href{https://orcid.org/\csname orcidauthor\x\endcsname}{\noexpand\orcidicon}}
}

\begin{document}

\title[A Perturbative Analysis of Interacting Scalar Field Cosmologies]{A Perturbative Analysis of Interacting Scalar Field Cosmologies}

\author{Genly Leon\orcidA{}}
\address{Departamento  de  Matem\'aticas,  Universidad  Cat\'olica  del  Norte, Avda. Angamos  0610,  Casilla  1280  Antofagasta,  Chile.}
\ead{genly.leon@ucn.cl}

\author{Esteban Gonz\'alez\orcidB{}}
\address{Direcci\'on de Investigaci\'on y Postgrado, Universidad de Aconcagua, Pedro de Villagra 2265, Vitacura, 7630367 Santiago, Chile.}
\ead{esteban.gonzalez@uac.cl}

\author{Alfredo D. Millano\orcidC{}}
\address{Departamento  de  Matem\'aticas,  Universidad  Cat\'olica  del  Norte, Avda. Angamos  0610,  Casilla  1280  Antofagasta,  Chile.}
\ead{alfredo.millano@alumnos.ucn.cl@ucn.cl}

\author{Felipe Orlando Franz Silva}
\ead{Felipe.franz.silva@gmail.com}

\begin{abstract}
Scalar field cosmologies with a generalized harmonic potential are investigated in flat and negatively curved Friedmann-Lemaître-Robertson-Walker and Bianchi I metrics. An interaction between the scalar field and matter is considered. Asymptotic methods and averaging theory are used to obtain relevant information about the solution space. In this approach, the Hubble parameter plays the role of a time-dependent perturbation parameter which controls the magnitude of the error between full-system and time-averaged solutions as it decreases. Our approach is used to show that full and time-averaged systems have the same asymptotic behavior. Numerical simulations are presented as evidence of such behavior. Moreover, the asymptotic behavior of the solutions is independent of the coupling function.

\end{abstract}


%\pagestyle{plain}
%\tableofcontents

\maketitle

\section{Introduction}
\noindent 
There are a number of gravitational theories, some of them including scalar fields, that can be studied using local and global variables, providing a qualitative description of the space of solutions. In addition, it is possible  to provide precise schemes to find analytical approximations of the solutions, as well as exact solutions or solutions in quadrature by choosing various approaches, e.g. \cite{Brans:1961sx,Guth:1980zm,Horndeski:1974wa,Copeland:1993jj,Lidsey:1995np,Ibanez:1995zs,Copeland:1997et,Coley:1997nk,Copeland:1998fz,Foster:1998sk,Coley:1999mj,vandenHoogen:1999qq,Albrecht:1999rm,Coley:2000zw,Coley:2000yc,Coley:2003tf,Miritzis:2003ym,Rendall:2004ic,Elizalde:2004mq,Capozziello:2005tf,Curbelo:2005dh,Gonzalez:2005ie,Miritzis:2005hg,Rendall:2005if,Rendall:2005fv,Gonzalez:2006cj,Rendall:2006cq,Hertog:2006rr,Gonzalez:2007ht,Hrycyna:2007gd,Lazkoz:2007mx,Elizalde:2008yf,Dania&Yunelsy,Giambo:2008ck,Leon:2008de,Giambo:2009byn,Leon:2009dt,Leon:2009rc,Leon:2009ce,Leon:2010pu,Basilakos:2011rx,Miritzis:2011zz,Xu:2012jf,Jamil:2012vb,Leon:2012mt,Leon:2013qh,Skugoreva:2013ooa,Fadragas:2013ina,Fadragas:2014mra,Kofinas:2014aka,Tzanni:2014eja,Leon:2014yua,Leon:2014bta,Leon:2014rra,Minazzoli:2014xua,Alho:2014fha,Paliathanasis:2014yfa,DeArcia:2015ztd,Solomon:2015hja,Harko:2015pma,Paliathanasis:2015gga,Leon:2015via,Matsumoto:2015hua,Barrow:2016qkh,Barrow:2016wiy,Cid:2017wtf,Cruz:2017ecg,Paliathanasis:2017ocj,Alhulaimi:2017ocb,Dimakis:2017kwx,Giacomini:2017yuk,Karpathopoulos:2017arc,Matsumoto:2017gnx,VanDenHoogen:2018anx,Leon:2018lnd,Leon:2018skk,DeArcia:2018pjp,Tsamparlis:2018nyo,Paliathanasis:2018vru,Barrow:2018zav,Basilakos:2019dof,Leon:2019mbo,Paliathanasis:2019qch,Leon:2019jnu,Paliathanasis:2019pcl,Quiros:2019ktw,Shahalam:2019jgs,Nojiri:2019riz,Humieja:2019ywy,Giambo:2019ymx}. In particular, relevant information about  the properties of the flow associated with an autonomous system of ordinary differential equations can be obtained by using qualitative techniques of dynamical systems. See textbooks related to qualitative theory of differential equations \cite{Coddington55,Hale69,AP,wiggins,perko,160,Hirsch,165,LaSalle,aulbach} and with some applications in cosmology \cite{TWE,coleybook,Coley:1999uh,bassemah,LeBlanc:1994qm,Heinzle:2009zb}. The tools of averaging theory and qualitative techniques of dynamical systems have been applied successfully in recent years to cosmological models, say in \cite{Alho:2015cza,Alho:2019pku,Paliathanasis:2015cza,Leon:2020ovw,Llibre:2012zz,Zhuravlev_2021,Fajman:2020yjb,Fajman:2021cli,Leon:2021lct,Leon:2021rcx,Leon:2021hxc}. 

In this paper, methods of perturbation theory and averaging theory are applied to differential equations arising from interacting cosmological models. In particular, we will study cosmologies with a scalar field that evolves according to the Klein-Gordon (KG) equation under the influence of a generalized harmonic self-interacting potential. 
Models with and without interaction between the scalar field and the matter (described by an ideal gas with a barotropic equation of state) are investigated. 

That is,
we are interested in the study of models where the matter of the universe is described by a scalar field $ \phi$, which is assumed to be homogeneous, with an energy-momentum tensor given by $ [{T ^ a} _b] = \mathrm{diag} (- \rho_\phi, p_\phi, p_\phi, p_\phi) $, where $ \rho_\phi = \frac{1}{2} \big (\dot \phi ^ 2 + 2 V (\phi) \big) $ and $ p_\phi = \frac {1}{2} \big (\dot \phi ^ 2 - 2 V (\phi) \big) $ are the energy density and isotropic pressure of the scalar field, and $ V (\phi) $ is the self-interacting potential; and by an ideal gas described by the tensor $ [T ^ {\text{matter} \; a} _b] = \mathrm{diag} (- \rho_m, p_m, p_m, p_m) $, where $ \rho_m \geq 0 $ and $ p_m = (\gamma-1) \rho_m $, where $ \gamma \in [0,2] $ is the barotropic index. 

The natural generalization of the models examined in \cite{Leon:2021lct,Leon:2021rcx,Leon:2021hxc} is to consider spatially homogeneous and isotropic matter-scalar field  interactive schemes. 
Interactive matter-scalar field schemes refer to models where the conservation equations have the structure
\begin{equation}
\dot{\rho}_ {m} + 3H \left(\rho_ {m} + p_ {m} \right)= - Q, \quad
\dot {\phi} \left [\ddot {\phi} +3 H \dot {\phi} + V'(\phi) \right] =  Q, \label{interacting-scheme}
\end{equation}%
where a dot means derivative with respect to cosmic time $t$, and comma derivative with respect to $\phi$, $ \rho_m $ is the energy density of matter, $\phi $ is the scalar field, $V(\phi ) $ its potential and $ Q $ is the interaction term  and $H$ stands for the  \emph{ Hubble parameter} $ H=\dot{a}/a$ (which is a general measure of the isotropic rate of spatial expansion) where $a$ denotes the scale factor of the Universe. 

When considering models with interaction, which have different physical implications, different results would be expected from the case without interaction.
An interesting research program is to investigate the dynamics and asymptotic behavior of the solutions of the equations of the gravitational field for various interacting functions $ Q = Q \left (H, \rho_ {m}, \rho_\phi \right) $.
As a first step towards generalization, we investigate interactions of the type $ Q = \lambda/2 \rho_m  \dot {\phi}$. 

Our methodology consists of using perturbation theory, in particular multi-scale methods as well as averaging theory and qualitative analysis to describe oscillating solutions in a wide class of cosmological models going beyond the usual linear stability analysis.  The first sections are devoted to showing that asymptotic methods and the averaging theory are powerful tools for investigating scalar field models, so we will start with examples from low to high complexity. 
The expected results are:
\begin{enumerate}
    \item Obtain relevant information about the solution space of scalar field cosmologies with generalized harmonic potential for the Friedmann-Lemaître-Robertson-Walker (FLRW) metrics, in a vacuum, and in the presence of matter (within minimal or non-minimal interacting schemes) and for the locally and rotationally symmetric (LRS) Bianchi I metric. 
    
    \item Incorporate \emph{asymptotic expansion with multiple timescales, averaging theory, and qualitative analysis of dynamical systems} to describe oscillatory solutions to a wide class of perturbation problems for these models.
    
    \item Build averaged versions of the original systems where oscillations are smoothed out. The analysis can then be reduced to studying the late dynamics of a simpler averaged system where oscillations entering the full system can be controlled through the KG equation.
    
    \item Construct regular equations defined in bounded state spaces that allow giving a global description of the dynamics. In particular, the behavior at early and late-time and the evolution at intermediate stages that may be of physical interest. In addition to proposing suitable differential equations to carry out systematic numerical simulations.
\end{enumerate}

We are particularly interested in the action for a general class of scalar-tensor theories (STT), written in the so-called
Einstein frame (EF), which is given by \cite{Kaloper:1997sh}
\begin{align}&S_{EF}=\int_{M_4} d{ }^4 x \sqrt{|g|}\left\{\frac{1}{2} R-\frac{1}{2} g^{\mu
\nu}\nabla_\mu\phi\nabla_\nu\phi-V(\phi)+\chi(\phi)^{-2}
\mathcal{L}_{\text{matter}}(\mu,\nabla\mu,\chi(\phi)^{-1}g_{\alpha\beta})\right\},\label{eq1}
\end{align}
where $R$ is the curvature scalar, $\phi$ is the
scalar field,   $\nabla_\alpha$
is the covariant derivative, $V(\phi)$ is the quintessence self-interacting potential,
$\chi(\phi)^{-2}$ is the coupling function, $\mathcal{L}_{\text{matter}}$
is the matter Lagrangian, and $\mu$ is a collective name for the
matter degrees of freedom,  repeated indexes mean sum over
them. The energy-momentum tensor of matter is defined by
\begin{equation}T_{\alpha
\beta}=-\frac{2}{\sqrt{|g|}}\frac{\delta}{\delta g^{\alpha
\beta}}\left\{\sqrt{|g|}
 \chi^{-2}(\phi)\mathcal{L}(\mu,\nabla\mu,\chi^{-1}(\phi)g_{\alpha
 \beta})\right\}.\label{Tab}\end{equation}

By considering the conformal transformation $\overline{g}_{\alpha
\beta}=\chi(\phi)^{-1}g_{\alpha \beta}$, defining the Brans-Dicke (BD) coupling ``constant'' $\omega(\chi)$ in such way that
$d\phi=\pm \sqrt{\omega(\chi)+3/2}\chi^{-1} d\chi$ and recalling
$\overline{V}(\chi)=\chi^2 V(\phi(\chi))$, the action (\ref{eq1}) can be
written in the Jordan frame (JF) as  \cite{Coley:2003mj}
\begin{align}& S_{JF}=\int_{M_4} d{ }^4 x \sqrt{|\overline{g}|}\left\{\frac{1}{2}\chi \overline{R}-\frac{1}{2}\frac{\omega(\chi)}{\chi}(\overline{\nabla}\chi)^2-\overline{V}(\chi)
+\mathcal{L}_{\text{matter}}(\mu,\nabla\mu,\overline{g}_{\alpha
\beta})\right\}.\label{eq1JF}
\end{align}
Here the bar is used to denote geometrical objects defined with
respect to the metric $\overline{g}_{\alpha \beta}.$ In the next sections a bar or an over-line will be referring to averaged quantities. 
In the STT given by (\ref{eq1JF}), the energy-momentum of the
matter fields,\begin{equation}\overline{T}_{\alpha
\beta}=-\frac{2}{\sqrt{|\overline{g}|}}\frac{\delta}{\delta \overline{g}^{\alpha
\beta}}\left\{\sqrt{|\overline{g}|}
\mathcal{L}(\mu,\nabla\mu,\overline{g}_{\alpha
 \beta})\right\}, \label{Tabprime}\end{equation} is separately conserved. That is
$\overline{\nabla}^\alpha \overline{T}_{\alpha \beta}=0$.  However, when is written in
the EF (\ref{eq1}), with a matter energy-momentum tensor given by \eqref{Tab}, this is no longer the case (although the
overall energy density is conserved). In fact in the EF we find
that
\begin{equation}
   Q_\beta\equiv\nabla^\alpha T_{\alpha \beta}=-\frac{1}{2}T {\chi(\phi)}^{-1}\frac{\mathrm{d}\chi(\phi)}{\mathrm{d}\phi}\nabla_{\beta}\phi,\quad
 T=T^\alpha_\alpha. 
\end{equation}
By making use of the above ``formal'' conformal equivalence between
the Einstein and Jordan frame we can find, for example, that the
theory formulated in the EF with the coupling function
$\chi(\phi)=\chi_0 \exp((\phi-\phi_0)/\varpi),\; \varpi\equiv
\pm\sqrt{\omega_0+3/2}$ and potential $V(\phi)=\beta
\exp({(\alpha-2){\varpi}/(\phi-\phi_0)})$, corresponds to the
BD theory (BDT) with a power-law potential, i.e.,
$\omega(\chi)=\omega_0$ and $\overline{V}(\chi)=\beta \chi^\alpha.$ Exact
solutions with exponential couplings and exponential potentials
(in the EF) were investigated in \cite{Gonzalez:2006cj}.
Quintessential DE models
\cite{Kolda:1998wq,Sahni:2002kh,Padmanabhan:2002ji}, for instance,
are described by an ordinary scalar field minimally coupled to
gravity. A particular choice of the  scalar field self-interacting
potentials can drive the past and current accelerated expansion.

The natural generalizations to quintessence models evolving
independently from the matter are models that exhibit
non-minimal coupling between both components. 
Several physical theories predict the presence of a scalar
field coupled to matter. For example, in string theory, the dilaton field is generally coupled to matter \cite{Gasperini:2007zz}.
Non-minimally coupling occurs also in STT of gravity
\cite{Fujii:2003pa,Faraoni:2004pi}, in higher order gravity  (HOG) theories
\cite{Capozziello:2007ec} and in models of chameleon gravity
\cite{Waterhouse:2006wv}. Coupled quintessence was investigated
also in
\cite{Amendola:1999er,Tocchini-Valentini:2001wmi,Billyard:2000bh} by
using dynamical systems techniques. The cosmological dynamics of scalar-tensor gravity have been
investigated in \cite{Carloni:2007eu,Tsujikawa:2008uc}.
Phenomenological coupling functions were studied for instance in
\cite{Boehmer:2008av} which can describe either the decay of dark
matter into radiation, the decay of the curvaton field into
radiation or the decay of dark matter into dark energy \cite{Boehmer:2008av}.  In the reference \cite{Tsujikawa:2008uc}, the
authors constructed a family of viable scalar-tensor models of dark
energy, which includes pure $F(R)$ theories and quintessence. 
There is the possibility of a universal coupling of dark energy to
all sorts of matter, including baryons, but excluding radiation
\cite{Chimento:2003iea}. 

The strength of the coupling between the perfect fluid and the scalar field is 
$Q=\frac{1}{2}(4-3\gamma)\rho_m\dot\phi
\frac{\mathrm{d}\ln \chi(\phi)}{\mathrm{d}\phi}$, where $ \chi(\phi)$ is an input function. In reference  \cite{Billyard:2000bh}  the interaction terms (in
the flat FLRW geometry) $Q=\alpha\dot\phi\rho_m$ and
$Q=\alpha\rho_m H$ were investigated, here $\alpha$ is a constant, $\phi$ is the
scalar field, $\rho$ is the energy density of background matter
and $H$ is the Hubble parameter. The first choice
corresponds to an exponential coupling function $\chi(\phi)=\chi_0
\exp\left(2 \alpha \phi/(4-3\gamma)\right).$ The second case
corresponds to the choice $\chi=\chi_0 a^{-2\alpha/(4-3\gamma)}$
(and then, $\rho\propto a^{\alpha-3\gamma}$).

Here, some perturbation problems in scalar field cosmologies in a vacuum and including matter will be studied. Relevant information about the solution's space for scalar field cosmologies in FLRW and Bianchi I metrics is expected to be obtained using qualitative techniques, asymptotic methods, and averaging theory. In this regard, this paper is a continuation of \cite{Leon:2020ovw,Leon:2020pfy}. There, some well-known results were reviewed and new theorems in the context of scalar field cosmologies with arbitrary potential (and with an arbitrary coupling to matter) were proved. In particular,   cosine-like corrections with small phase were incorporated to the harmonic potential for FLRW metric and Bianchi I metrics inspired in \cite{Sharma:2018vnv}.  Following this line, we select a self-interacting potential 
\begin{equation}
    V(\phi)= \frac{\phi ^2}{2} + f\left[1- \cos \left(\frac{\phi }{f}\right)\right] = \frac{(f+1) \phi ^2}{2 f}+\mathcal{O}\left(\phi ^3\right), \; f> 0, \label{pot1}
\end{equation}
and the coupling function 
\begin{equation}
 \label{coupling}
 \chi(\phi)=\chi_0 e^{\frac{\lambda \phi}{4-3\gamma}}, \; \lambda\; \text{is a constant and}\;  \gamma \neq \frac{4}{3}.
\end{equation}

We must emphasize that there is a close relationship between the KG equation and that of a harmonic oscillator with non-linear damping, where the damping depends on time through the coupling of the Einstein equations with the KG equation through the   Hubble parameter  $ H $. Motivated by the works \cite{Leon:2020ovw,Leon:2020pfy} and based on the previous analogy, an amplitude-phase transformation $(\dot \phi, \phi )\rightarrow (r, \varphi)$ (chapter 11 of \cite{Verhulst}; p~ 22, 24-27, 42, 54, 361 of \cite{SandersEtAl2010}), which is defined as
\begin{equation}\label{E: amplitude-phase}
\dot{\phi}(t)= r(t) \cos (t-\varphi(t)), \quad \phi(t)  = r(t) \sin (t-\varphi(t)),
\end{equation}
such that
\begin{equation}\label{eq_64}
r(t)=\sqrt{\dot \phi(t)^2+{\phi}(t)^2}, \quad \varphi(t) =t-\arctan\left(\frac{\phi
   (t)}{\dot \phi(t)}\right), 
\end{equation}
will be used \cite {Verhulst}; which allows obtaining new equations which will be averaged with respect to time to obtain new systems. With this approach, the oscillations present in the non-linear systems, which enter/modify the dynamics through the KG equation, can be controlled and smoothed as long as the Hubble parameter $ H $, which acts as a time-dependent perturbation parameter, decreases monotonically. We will use the methods of the averaging theory of systems of nonlinear differential equations to prove that the original time-dependent systems and their corresponding averaged versions have the same late dynamics. Therefore, to determine the future asymptotic behavior, the simpler averaged systems are investigated. Numerical simulations will be carried out to show the oscillatory behavior of the solutions. This simulations will also show how the averaged solutions behave as compared to the original ones. These results will allow to make conjectures about the dynamics of the universe at local or cosmological scales, and will establish demonstration schemes to prove them.

The paper is organized as follows. In section \ref{section_2} we discuss some asymptotic expansion techniques, in particular the two-timing method. In section \ref{SECT:II} we present a review on averaging techniques, with special emphasis on applications in cosmology. In section \ref{section_5} some applications of perturbation and averaging methods in cosmology are presented. In particular, in section  \ref{SECT:4.5} is studied a scalar field with generalized harmonic potential  \eqref{pot1} non-minimally coupled to matter with coupling \eqref{coupling}. Sections \ref{SECTION_4.5} and \ref{Sect:2.7.3} are devoted to the minimally coupled  and vacuum cases, respectively. We are focused on studying the imprint of coupling function, as well as the influence of the metric on the dynamics of the averaged problem. In section \ref{Numerical} we present numerical simulations as evidence that the solutions of the full system for each model follow the track of the solutions of their corresponding averaged version when $H$ is monotonically decreasing. Section \ref{Sect:6} is devoted to results and conclusions.

\section{Perturbation problems}
\label{section_2}

Perturbation problems focus on the study of the phase portrait of the differential system
\begin{equation}
\dot x=X(x;\varepsilon), \quad x\in \mathbb{R}^k, \quad \varepsilon \sim 0,
\end{equation}
near the zero of $X(x;0)$ \cite{Verhulst,SandersEtAl2010,dumortier,fenichel,Fusco,Berglund,holmes,Kevorkian1}. In general,  perturbation problems are expressed in Fenichel's normal form, i. e., given $(x,y)\in \mathbb{R}^{n+m}$ and $f, g$ smooth functions, the equations can be written as
\begin{equation}
\label{eq:1.1}
\dot x=f(x,y; \varepsilon), \quad \dot y=\varepsilon g(x,y; \varepsilon), \quad x=x(t), \quad y=y(t).
\end{equation}
The system \eqref{eq:1.1} is called ``fast system'', unlike the system
\begin{equation}
\label{eq:1.2}
\varepsilon  x'=f(x,y; \varepsilon), \quad y'=g(x,y; \varepsilon), \quad x=x(\tau), \quad y=y(\tau),
\end{equation}
obtained after the re-scaling $\tau=\varepsilon t$, that is called the ``slow system''.  Notice that for $\varepsilon>0$, the phase portraits of \eqref{eq:1.1} and \eqref{eq:1.2} coincide.  However, this two problems manifestly depend on two scales: (i) the problem in terms of the ``slow time'' variable, whose solution is analogous to the outer solution in a boundary layer problem; and (ii) the fast system, a change of scale on the system which describes the rapid evolution that occurs in shorter times, analogous to the inner solution of a boundary layer problem.
The solution of each subsystem will be sought in the form of a regular perturbation expansion. For singularly perturbed problems the subsystems will have
simpler structures than the full problem, allowing the characterization of the slow and fast dynamics in terms of a reduced phase line or phase plane dynamics.

For $\varepsilon >0$, let $\mathcal{S}$ denotes the singular points of \eqref{eq:1.1}. Equations \eqref{eq:1.2} define  a dynamical system on $\mathcal{S}$ called the reduced problem. The implicit equation
$f(x,y; 0)= 0$ is called the slow manifold or ``slow solution curve''. 
Very often  the solution is pushed out of the slow manifold  at which point the solution is no longer described by the dynamics of the slow system; all out
the slow manifold in the phase plane is part of the fast problem. 
Combining the results of these two limiting problems, some information of the dynamics for small values of $\varepsilon$ is obtained. This technique is used to construct uniformly valid approximations of the solutions of perturbation problems using as seed solutions those which satisfy the original equations in the limit of $\varepsilon\rightarrow 0$.  
One approach used to construct
that asymptotic expansions is to introduce the two time scales $t_1 = t$ and $t_2 = \varepsilon t$.
For this reason, the method is sometimes called two-timing, and $t_1$ is said to
be the fast time scale and $t_2$ the slow scale.  The list of possible
scales includes the following \cite{holmes}:
\begin{enumerate}
\item 
Several time scales like
$t_1 = t/\varepsilon, \quad t_2 = t, \quad t_3 = t \varepsilon, \quad t_4 = t \varepsilon ^2\ldots$ may be needed. 
\item More complex dependence on $\varepsilon$, for example, $t_1 =
t \left(1 + \omega_1 \varepsilon +\omega_2 \varepsilon^2 + \ldots \right)$ and $t_2 = t \varepsilon $  where the $\omega_n$ are determined while solving the problem
(Poincarè-Lindstedt's method).

\item The correct scaling may not be immediately apparent, and one starts off
with something like $t_1 = t \varepsilon^\alpha$ and $t_2 = t \varepsilon^\beta$, where $\alpha < \beta$.

\item Nonlinear time dependence, for example, one may have to assume $t_1 =
f(t,\varepsilon)$ and $t_2 = \varepsilon t_1$, where the function $f(t,\varepsilon)$ is determined from the
problem. 
\end{enumerate}

Perturbations methods and averaging methods were used, for example, in \cite{Rendall:2006cq}, in investigations of the oscillating behavior in scalar field cosmologies with harmonic potential using amplitude-phase variables of the form \eqref{E: amplitude-phase} (chapter 11 of \cite{Verhulst}; p~ 22, 24-27, 42, 54, 361 of \cite{SandersEtAl2010}). In \cite{Alho:2015cza}, these techniques were used to prove statements about how the relationship between the Equation of State (EoS) of the fluid and the monomial exponent of the scalar field affects the asymptotic source dominance and asymptotic late time behavior. Slow-fast methods were used for example in GUP theories, say in \cite{Paliathanasis:2015cza}.
In \cite{Leon:2020ovw} averaging  over an angle $\varphi$ by using an amplitude-angle
transformation (p~358 \cite{SandersEtAl2010}) of the form $\dot\phi(t) = r(t) \sin \varphi(t)$ and $\phi (t) = r(t) \cos \varphi(t)$ was used to study oscillations of the scalar field driven by generalized harmonic potentials. In the reference, \cite{Llibre:2012zz} was applied the averaging theory of first-order to study the periodic orbits of Hamiltonian systems describing a universe filled with a scalar field. There were provided sufficient conditions on the parameters of these cosmological models which guarantee that at any
positive or negative Hamiltonian level, the Hamiltonian system has periodic orbits. Additionally, it was shown the non-integrability of these cosmological systems in the sense
of Liouville-Arnold, proving that there cannot exist any second first integral of class
$C^1$. These techniques can be applied to Hamiltonian systems with an arbitrary number of degrees of freedom. 

In reference, \cite{Zhuravlev_2021} the method of multiple scales was applied to the analysis of cosmological dynamics. This method was used to construct solutions to the governing equations of the Universe filled with a scalar field in the Friedman-Lemaître-Robertson-Walker (FLRW) metric. A general scheme is described for choosing small dimensionless parameters of the expansion of model functions and applying the multiple scales method to the cosmological equations for two different types of a small parameter, a small field value, and a small slow-roll parameter.

In general, the regular asymptotic expansion fails in presence of resonant (secular) terms. One alternative is to use Poincarè-Lindstedt's method. This method would determine solutions of perturbed oscillators by suppressing resonant forcing terms that would yield spurious secular terms in the asymptotic expansions. The $t_1$ and $t_2$ time variables are introduced to keep a well ordered expansion, 
where $t_1$ is the regular (or ``fast'') time variable and $t_2$ is a new variable describing the ``slow-time'' dependence of the solution. The idea is to use any freedom that is in the  $t$-dependence of $t_1$ and $t_2$ to minimize the approximation's error, and whenever is possible to remove unbounded or secular terms. 
To our knowledge, Poincarè-Lindstedt's method has not been implemented yet in the cosmological setup. However, basic examples of oscillators show that by implementing a time-averaged version of the model instead of multiple scales, the issue of secular terms is overcome; getting the same accuracy as in the two-timing method. \footnote{We elaborate more on averaging techniques in subsection
\ref{SECT:II}.} Alternatively, the method of
multiple time scales makes a less restrictive assumption on the form of the solution
than those employed by Poincarè-Lindstedt's method. It assumes that the solution can
be expressed as a function of multiple (just two for our purposes) time variables, which are introduced to keep a well-ordered expansion,
\begin{equation}
x(t) = X(t, \tau),   
\end{equation}
where $t$ is the regular (or ``fast'') time variable and $\tau$ is a new variable describing the ``slow-time'' dependence of the solution. As commented before, the idea is to use any freedom that is in the  $\tau$-dependence to minimize the approximation's error, and whenever is possible to remove unbounded or secular terms. Some examples to illustrate the use of perturbation methods are the following.

\subsection{Example 1}  Considering the following initial value problem with $t>0$
\begin{align}
\displaystyle{\frac{d^2y}{d{t}^2}=-\frac{1}{(1+\varepsilon y)^2}}, \quad y(0)=0,\quad  y'(0)=1. \label{example1}
\end{align}
Assuming the solution has an asymptotic expansion of the form
\begin{equation}
y(t) \sim y_0(t) + {\varepsilon}  y_1(t) +\ldots,
\end{equation}
and considering a very small $z$, $(1 + z)^{-2} \sim 1-2 z$, the original problem becomes
\begin{equation}
y''_0(t) + {\varepsilon} y_1''(t)+\ldots=-\frac{1}{[1+\varepsilon (y_0 (t)+\ldots)]^2} \sim -1+2\varepsilon y_0 (t) + \ldots,
\end{equation}
with initial conditions  
\begin{center}
$y_0(0)+ {\varepsilon} y_1(0)+\ldots=0$, and $y'_0(0)+{\varepsilon} y'_1(0)+\ldots=1$
\end{center}
Collecting terms the following systems are obtained:

To order $\mathcal{O}(1)$:  $ {y_0}''(t) =-1, \quad y_0'(0)=1, \quad  y_0(0)=0$
has solution   $y_0(t)=-\frac{1}{2}{t}^2 + t$.

To order  $\mathcal{O}(\varepsilon)$:   $y_1''(t) =2y_0(t), \quad y_1'(0)=0,\quad y_1(0)=0$
has solution $y_1(t)=\frac{1}{3}{t}^3 -\frac{1}{12}{t}^4$.

Finally, the solution is given by 
\begin{equation}
    y(t) \sim t\left(1-\frac{1}{2}t\right) + \frac{1}{3} \varepsilon {t}^3 \left( 1-\frac{1}{4}t\right). \label{asympt_expasion}
\end{equation}   

\begin{figure}[ht!]
    \centering
    \includegraphics[width=0.9\textwidth]{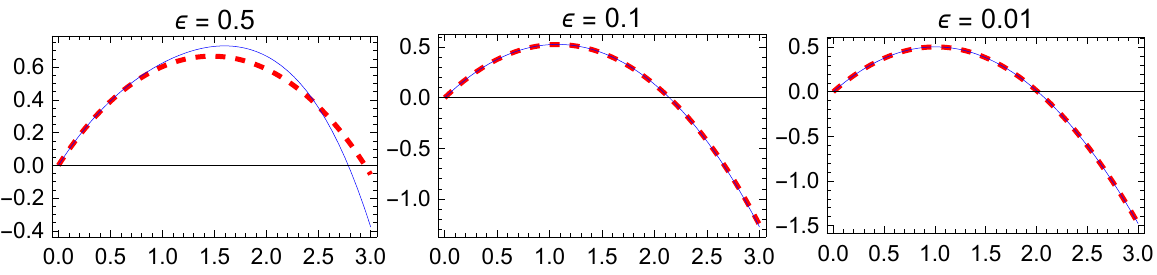}
    \caption{Numerical solution of \eqref{example1} (solid line) vs asymptotic expansion \eqref{asympt_expasion} (dashed line).}
    \label{fig:Example_1}
\end{figure}

\noindent This example illustrate how the regular asymptotic expansion method works. As shown in Figure \ref{fig:Example_1} as $\epsilon$ becomes small the numerical solution of \eqref{example1} (solid line) and the asymptotic expansion \eqref{asympt_expasion} coincide. 

The next example shows the failure of the regular asymptotic expansion due to the appearance of spurious secular terms in the asymptotic expansions.

\subsection{Example 2}
\begin{figure}[ht!]
    \centering
    \includegraphics[width=0.9\textwidth]{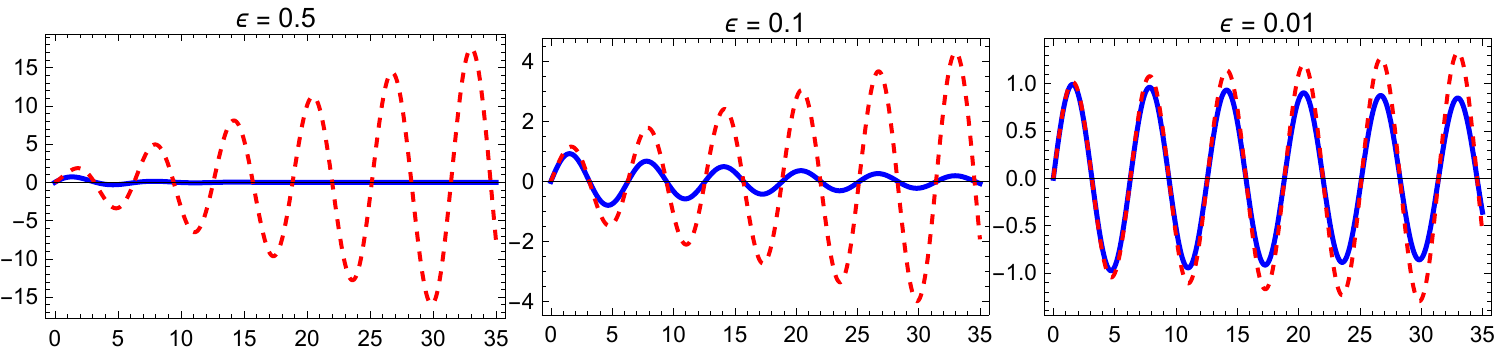}
    \caption{Exact solution \eqref{exact2} of \eqref{oscillator_1} (thick blue line) vs asymptotic expansion \eqref{expansion_0} (thick red line).}
    \label{Regular_Expansion2}
\end{figure}

Considering the classical example \cite{holmes}, given by the ordinary differential equation 
\begin{equation}
\label{oscillator_1}
y''+\varepsilon y'+y=0, \quad t>0, \quad y(0)=0, \quad  y'(0)=1. 
\end{equation}
Equation \eqref{oscillator_1} admits an exact solution of the form 
\begin{equation}
y(t)= \frac{2 e^{-\frac{t \varepsilon }{2}} \sin
   \left(\frac{1}{2} t \sqrt{4-\varepsilon ^2}\right)}{\sqrt{4-\varepsilon
   ^2}}. \label{exact2}
\end{equation}
Using regular asymptotic expansions to solve \eqref{oscillator_1} would yield spurious secular terms, for instance, the solution by  regular expansion is
\begin{equation}\label{expansion_0}
    x(t, \varepsilon)= \sin (t) - \varepsilon t \sin (t) + \mathcal{O}(\varepsilon^2),
\end{equation}
notice that the ``next to leading term'' $\varepsilon t \sin (t)$ is dominant on scales $\varepsilon t = \mathcal{O}(1)$. Therefore,  it becomes larger than the zeroth-order terms as the time increases as shown in Figure \ref{Regular_Expansion2}. 

\begin{figure}[ht!]
    \centering
    \includegraphics[width=0.9\textwidth]{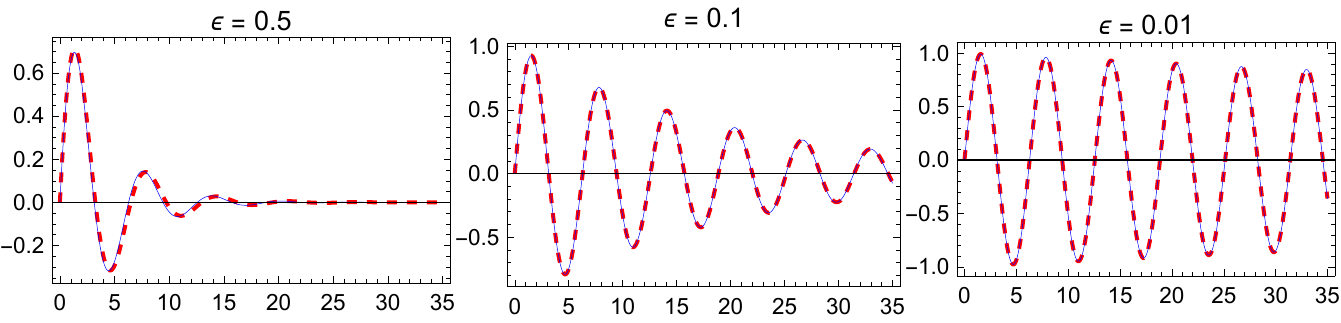}
    \caption{Exact solution \eqref{exact2} of equation \eqref{oscillator_1} (thin blue line) vs two-timing expansion \eqref{expansion_0} (thick dashed red line).}
    \label{Regular_Expansion2b}
\end{figure}

Observe that  solution \eqref{exact2} has an oscillatory component running on the scale of order $\mathcal{O}(1)$, as well as a slow variation of order $\mathcal{O}(\varepsilon^{-1})$.
Therefore, two time scales  $t, \tau=\varepsilon t$ are introduced and treated as independent variables. 
Using the chain rule
\begin{equation}
\frac{d f}{d t}=\frac{\partial f}{\partial t}+\varepsilon \frac{\partial f}{\partial \tau}, \quad \frac{d^2 f}{d t^2}=\frac{\partial^2 f}{\partial t^2}+2 \varepsilon  \frac{\partial^2 f}{\partial t \partial \tau} +\varepsilon ^2 \frac{\partial^2 f}{\partial \tau^2}
\end{equation}
the initial value problem of a scalar differential equation 
\begin{align}
& y_{tt} + 2 \varepsilon y_t y_\tau +\varepsilon^2 y_{\tau \tau}+\varepsilon (y_t+ \varepsilon y_{t \tau })+y=0, \\
& y=0, \quad y_t +\varepsilon y_{\tau }=1 \; \text{for}\; t=0=\tau,  
\end{align}
is obtained, where the subscripts $y_t, \quad y_\tau, \ldots$, denote the partial derivatives. 
Now, using a series expansion of the form 
\begin{equation}
y\sim y_0(t,\tau)+\varepsilon y_1(t,\tau) + \ldots, 
\end{equation}
the following equation
\begin{equation}
y_{0 tt}+y_0 +\varepsilon \left(y_{1 tt}+y_1 +2 y_{0 t \tau}+y_{0 t}\right)+ \mathcal{O}(\varepsilon^2)=0, 
\end{equation}
is obtained. 
Collecting terms of order $1$  and $\varepsilon$  leads to 

 $\mathcal{O}(1)$:  \begin{equation}
                         y_{0 tt}+y_0=0,
                       \end{equation}
with a general solution 
\begin{equation}
y_0(t,\tau )= A(\tau ) \sin (t)+B(\tau ) \cos (t),
\end{equation}
and 

$\mathcal{O}(\varepsilon)$:
\begin{align}
 y_{1 tt}+y_1 &=-\left(2 y_{0 t \tau}+y_{0 t}\right) \nonumber\\
& = \sin (t) \left(2 B'(\tau )+B(\tau )\right)-\cos (t) \left(2 A'(\tau
   )+A(\tau )\right).
\end{align}
Then,  secular terms $\propto \varepsilon t \sin(t)$, $\propto \varepsilon t \cos(t)$ are removed by setting 
\begin{align} 
\left(2 B'(\tau )+B(\tau )\right)=0, \quad  \left(2 A'(\tau )+A(\tau )\right)=0.
\end{align}
After imposing the initial conditions, it follows that
$B(\tau)=0$ and $A(\tau)=e^{-\frac{\tau}{2}}$ and the solution 
\begin{equation}
y\sim e^{-\frac{\tau}{2}}\sin(t)=e^{-\frac{\varepsilon t}{2}}\sin(t),
\end{equation}
valid up to the first order of $\varepsilon$ is obtained, which gives a good approximation to the solution of the problem. 
Indeed, the previous approximation holds up to $\varepsilon t = \mathcal{O}(1)$, that is, it holds for $0 \leq \varepsilon t \leq T$ where $T$ is fixed. Therefore, this procedure alleviates the failure of the regular asymptotic expansion \eqref{expansion_0} that yielded spurious secular terms  $\propto \varepsilon t \sin(t)$ in the asymptotic expansion. A comparison between figures \ref{Regular_Expansion2} and \ref{Regular_Expansion2b} illustrates the benefit of two-timing procedure over the regular asymptotic expansion when secular terms appears.

\subsection{Example 3}
 The so-called induced gravity model has the action  \cite{Kamenshchik:2013dga,Andrianov:2011fg}
\begin{align}
S_{IG}=\int\sqrt{- {g}}\left( \frac{\sigma^2}{8 \omega_0}  {R}-\frac{1}{2} {g}^{\mu\nu} {\partial}_{\mu}\sigma {\partial}_{\nu}\sigma-\frac{\gamma ^2 U_0 \sigma^2}{4-6 \gamma^2}\right),
\end{align} where $\omega_0>0$ and $\gamma\geq 0$. A massless scalar field is added to the action in \cite{Cid:2015pja} of the form
	\begin{align}
S_{IG\phi}=S_{IG}+\int\sqrt{- {g}}\left(-\frac{1}{2} {g}^{\mu\nu} {\partial}_{\mu}\phi {\partial}_{\nu}\phi\right).
\end{align}
The equation of motion for a massless scalar field is given by 
\begin{equation}
\ddot\phi +3\frac{\dot a}{a}\dot\phi=0,
\end{equation}
and admits the solution $\dot\phi=\varepsilon a^{-3}$, where $\varepsilon$ is an integration constant. 	Using the parametrization 	 \cite{Kamenshchik:2013dga}
	\begin{subequations}
	\label{eqC2}
	\begin{align}
	a=\sigma^{-1} \exp(u+v),\\
	\sigma=\exp(A(u-v)),
	\end{align}
	\end{subequations} with $A=\sqrt{\frac{3}{2}}\gamma,$ 
the Raychaudhuri equation and the equation of motion for $\sigma$ lead to 
\begin{subequations}
\begin{align}
& \frac{3 \left(\sqrt{6} \gamma +2\right) \left(3 \gamma ^2-2\right) \varepsilon ^2 \exp \left(2 \sqrt{6} \gamma  (u-v)-6 u-6 v\right)}{\gamma^2}\nonumber \\
& +12  
   \left(\sqrt{6} \gamma -6\right) {\dot u}^2+  \left(\sqrt{6} \gamma +6\right) U_0-24   \ddot u=0, \label{eq:27a}\\
	&-\frac{3 \left(\sqrt{6} \gamma -2\right) \left(3 \gamma ^2-2\right) \varepsilon ^2
   \exp \left(2 \sqrt{6} \gamma  (u-v)-6 u-6 v\right)}{\gamma ^2}\nonumber \\
	&-12 \left(\sqrt{6} \gamma +6\right) {\dot v}^2+\left(6-\sqrt{6} \gamma
   \right) U_0-24
   \ddot v=0.
\end{align}
\end{subequations}
where the Friedmann equation
\begin{align}
\label{eq:28}
&\dot u \dot v= \frac{1}{12} \left(\frac{\left(2-3
   \gamma ^2\right) \varepsilon^2 \exp
   \left(2 \sqrt{6} \gamma 
   (u-v)-6 u-6
   v\right)}{\gamma
   ^2}+U_0\right),
\end{align}
is used to eliminate the mixed terms $\propto \dot u \dot v$.
	
Using series expansion of the form  
\begin{equation}
u\sim u_0(t,\tau)+\varepsilon u_1(t,\tau) + \mathcal{O}(\varepsilon^2), \quad v\sim v_0(t,\tau)+\varepsilon v_1(t,\tau) + \mathcal{O}(\varepsilon^2),
\end{equation}
where the time variables $t, \tau=\varepsilon t$, are introduced and treated as independent variables.
Collecting terms of order $\varepsilon$ (see reference \cite{Cid:2015pja}) the following problems are found:
\\
$\mathcal{O}(1)$:
\begin{equation}
\left\{\begin{array}{c}
12 \left(\sqrt{6} \gamma -6\right) u_{0 t}(t,\tau )^2-24 u_{0 t t}(t,\tau )+\left(\sqrt{6} \gamma +6\right) U_0=0\\\\
-12 \left(\sqrt{6} \gamma   +6\right) v_{0 t}(t,\tau )^2-24 v_{0 t t}(t,\tau )+\left(6-\sqrt{6} \gamma \right) U_0=0\\\\
	u_{0 t}(t,\tau ) v_{0 t}(t,\tau )-\frac{U_0}{12}=0\\\\
	\text{(see, e.g., similar equations (28) in \cite{Kamenshchik:2013dga}, (2.24) in \cite{Andrianov:2011fg})}
\end{array}\right.,
\end{equation} 
\\
$\mathcal{O}(\varepsilon)$: \begin{equation}
\left\{
\begin{array}{c}
-2 u_{0 t \tau}(t,\tau )+\left(\sqrt{6} \gamma -6\right) u_{0 t}(t,\tau ) \left(u_{0 \tau}(t,\tau )+u_{1 t}(t,\tau
   )\right)-u_{1 t t}(t,\tau )=0\\\\
(2 v_{0 t \tau}(t,\tau )+\left(\sqrt{6} \gamma +6\right) v_{0 t}(t,\tau ) \left(v_{0 \tau}(t,\tau )+v_{1 t}(t,\tau )\right)+v_{1 t t}(t,\tau )=0\\\\
	v_{0 t}(t,\tau ) \left(u_{0 \tau}(t,\tau )+u_{1 t}(t,\tau
   )\right)+u_{0 t}(t,\tau ) \left(v_{0 \tau}(t,\tau )+v_{1 t}(t,\tau )\right)=0
\end{array} \right..
\end{equation}
Solving 
up to order $\mathcal{O}(1)$, the following systems are obtained
	\begin{subequations}
	\label{EQ29}
			\begin{equation}
			u_0(t, \tau)=\left\{\begin{array}{cc}
			c_2(\tau)-\frac{2 \ln \left(\cosh \left(\Delta\right)\right)}{\sqrt{6} \gamma -6}, & \gamma^2<6   \\
	c_2(\tau)-\frac{2 \ln \left(\cos \left(\Delta\right)\right)}{\sqrt{6} \gamma -6}, & \gamma^2\geq 6
	\end{array}
	\right., \quad 
				v_0(t,\tau)= \left\{ \begin{array}{cc} c_3(\tau)+ \frac{2 \ln \left(\sinh \left(\Delta\right)\right)}{\sqrt{6} \gamma
   +6}, & \gamma^2<6 \\  c_3(\tau)+ \frac{2 \ln \left(\sin \left(\Delta\right)\right)}{\sqrt{6} \gamma
   +6}, & \gamma^2 \geq 6
	\end{array}
	\right.,
				\end{equation}
	\end{subequations}
where $c_1(\tau), c_2(\tau)$ and $c_3(\tau)$ are integration functions, and
\begin{equation}
\Delta:=\Delta(t,\tau)=\frac{\sqrt{|\gamma ^2-6|} \sqrt{U_0} \left(24 c_1(\tau )+t\right)}{2 \sqrt{2}}.
\end{equation}
Substituting \eqref{EQ29} into the equations at order $\mathcal{O}(\varepsilon)$, the following is obtained \begin{small}
\begin{align}
&u_{1 t t}=
 \left\{
\begin{array}{cc}
  -\frac{\sqrt{U_0} \left(12 \sqrt{U_0} \left(\gamma ^2-6\right) c_1'(\tau )+\left(\sqrt{3} \gamma -3 \sqrt{2}\right) \sqrt{6-\gamma ^2} \tanh (\Delta ) \left(c_2'(\tau )+u_{1 t}\right)\right)}{\sqrt{6} \gamma -6}, & \gamma^2<6 \\
	\frac{\sqrt{U_0} \left(\left(\sqrt{3} \gamma -3 \sqrt{2}\right) \sqrt{\gamma ^2-6} \tan (\Delta ) \left(c_2'(\tau )+u_{1 t}\right)-12 \sqrt{U_0} \left(\gamma ^2-6\right) c_1'(\tau
   )\right)}{\sqrt{6} \gamma -6}, & \gamma^2 \geq 6 \\
\end{array}
\right., \label{EQv1}\end{align}
\begin{align}
& v_{1 t t}=
 \left\{
\begin{array}{cc}
-\frac{\sqrt{U_0} \text{csch}^3(\Delta ) \left(4 \left(\sqrt{6} \gamma -6\right) \sqrt{6-\gamma ^2} \cosh ^3(\Delta ) \left(u_{1 t}+c_2'(\tau )\right)+12 \sqrt{2} \left(\gamma ^2-6\right)
   \sqrt{U_0} (5 \sinh (\Delta )+\sinh (3 \Delta )) c_1'(\tau )\right)}{8 \left(\sqrt{3} \gamma +3 \sqrt{2}\right)}, & \gamma^2<6 \\
 \frac{\sqrt{U_0} \csc ^3(\Delta ) \left(4 \left(\sqrt{6} \gamma -6\right) \sqrt{\gamma ^2-6} \cos ^3(\Delta ) \left(u_{1 t}+c_2'(\tau )\right)+12 \sqrt{2} \left(\gamma ^2-6\right) \sqrt{U_0} (5
   \sin (\Delta )+\sin (3 \Delta )) c_1'(\tau )\right)}{8 \left(\sqrt{3} \gamma +3 \sqrt{2}\right)}, & \gamma^2 \geq 6 \\
\end{array}
\right.,\end{align}
\begin{align}
&v_{1 t}=
 \left\{
\begin{array}{cc}
  \frac{\left(\sqrt{6} \gamma -6\right) c_2'(\tau ) \coth ^2(\Delta )+\sqrt{6} \gamma  u_{1 t} \coth ^2(\Delta )-6 u_{1 t} \coth ^2(\Delta )-24 \sqrt{2} \sqrt{U_0} \sqrt{6-\gamma ^2}
   c_1'(\tau ) \coth (\Delta )-\sqrt{6} \gamma  c_3'(\tau )-6 c_3'(\tau )}{\sqrt{6} \gamma +6}, & \gamma^2 < 6 \\
	-\frac{\left(\sqrt{6} \gamma -6\right) c_2'(\tau ) \cot ^2(\Delta )+\sqrt{6} \gamma  u_{1 t} \cot ^2(\Delta )-6 u_{1 t} \cot ^2(\Delta )+24 \sqrt{2} \sqrt{U_0} \sqrt{\gamma ^2-6}
   c_1'(\tau ) \cot (\Delta )+\sqrt{6} \gamma  c_3'(\tau )+6 c_3'(\tau )}{\sqrt{6} \gamma +6}, & \gamma^2 \geq 6\\
\end{array}
\right..
\end{align}
\end{small}
The Integration of \eqref{EQv1} leads to
\begin{small}
\begin{equation}
u_1= c_4(\tau ) +  \left\{
\begin{array}{cc}
  \left(-t-24 c_1(\tau )\right) c_2'(\tau )+\frac{2 \tanh (\Delta ) \left(\left(2 \sqrt{3} \gamma -6 \sqrt{2}\right) c_2(\tau )-6 \sqrt{2} U_0 \left(\gamma ^2-6\right) \left(t+24 c_1(\tau )\right) c_1'(\tau )+\left(\sqrt{3}
   \gamma -3 \sqrt{2}\right) c_2'(\tau )\right)}{\sqrt{U_0} \left(\sqrt{6} \gamma -6\right) \sqrt{6-\gamma ^2}}, & \gamma^2<6 \\
	\left(-t-24 c_1(\tau )\right) c_2'(\tau )+\frac{2 \tan (\Delta ) \left(\left(2 \sqrt{3} \gamma -6 \sqrt{2}\right) c_2(\tau )-6 \sqrt{2} U_0 \left(\gamma ^2-6\right) \left(t+24 c_1(\tau )\right) c_1'(\tau )+\left(\sqrt{3}
   \gamma -3 \sqrt{2}\right) c_2'(\tau )\right)}{\sqrt{U_0} \left(\sqrt{6} \gamma -6\right) \sqrt{\gamma ^2-6}}, & \gamma^2\geq 6\\
\end{array}
 \right..
\end{equation}
\end{small}
Avoiding the two secular terms $\propto t$, conditions $c_1'(\tau)=c_2'(\tau)=0$ are imposed, i.e.,  $c_1$ and $c_2$ are constants. 
Hence,
\begin{equation}
u_1= c_4(\tau ) +  \left\{
\begin{array}{cc}
  \frac{2 \sqrt{2} c_2 \tanh (\Delta )}{\sqrt{U_0 \left(6-\gamma ^2\right)}}, & \gamma^2<6 \\
	\frac{2 \sqrt{2} c_2 \tan (\Delta )}{\sqrt{U_0 \left(\gamma ^2-6\right)}}, & \gamma^2 \geq 6 \\
\end{array}
 \right.,
\end{equation}
where $\Delta:=\Delta(t)=\frac{(24 c_1+t) \sqrt{U_0}\sqrt{|\gamma ^2-6|}}{4 \sqrt{3}}$.
Then,
\begin{align}
&v_{1 t t}=\left\{
\begin{array}{cc}
  \frac{\left(6-\sqrt{6} \gamma \right) \sqrt{U_0 \left(6-\gamma ^2\right)} c_2\coth (\Delta ) \text{csch}^2(\Delta )}{\sqrt{2} \left(\sqrt{6} \gamma +6\right)}, & \gamma^2<6  \\
	\frac{\left(\sqrt{6} \gamma -6\right) \sqrt{U_0 \left(\gamma ^2-6\right)} c_2 \cot (\Delta ) \csc ^2(\Delta )}{\sqrt{2} \left(\sqrt{6} \gamma +6\right)}, & \gamma^2 \geq 6 \\
\end{array}\right.,\\
& v_{1 t}=\left\{
\begin{array}{cc}
  \frac{\left(\sqrt{6} \gamma -6\right) c_2 \text{csch}^2(\Delta )}{\sqrt{6} \gamma +6}-c_3'(\tau ), & \gamma^2<6 \\
	-\frac{\left(\sqrt{6} \gamma -6\right) c_2 \csc ^2(\Delta )}{\sqrt{6} \gamma +6}-c_3'(\tau ), & \gamma^2 \geq 6 \\
\end{array}\right..
\end{align}
Solving the second equation the following is obtained
\begin{equation}
v_1= \left\{
\begin{array}{cc}
 -\frac{4 \left(\sqrt{3} \gamma -3 \sqrt{2}\right) c_2 \coth \left(\Delta\right)}{\left(\sqrt{6} \gamma +6\right) \sqrt{U_0 \left(6-\gamma
   ^2\right)}}-t c_3'(\tau )+c_5(\tau ), & \gamma^2 <6  \\
 \frac{4 \left(\sqrt{3} \gamma -3 \sqrt{2}\right) c_2 \cot \left(\Delta\right)}{\left(\sqrt{6} \gamma +6\right) \sqrt{U_0 \left(\gamma
   ^2-6\right)}}-t c_3'(\tau )+c_5(\tau ), & \gamma^2 \geq 6\\
\end{array}\right.,
\end{equation}
such that both differential equations for $v_1$ are identically satisfied. To avoid the  secular terms $\propto t$, the condition $c_3'(\tau)=0$ is imposed, i.e.,  $c_3$ is a constant. For simplicity, we set $c_4=c_5=0.$  
Therefore, it follows that
	\begin{subequations}
\label{seed_solution_first_order}
\begin{align}
  & u(t;\varepsilon)=    c_2 - \left\{\begin{array}{cc}
			\frac{2 \ln \left(\cosh \left(\Delta\right)\right)}{\sqrt{6} \gamma -6}, & \gamma^2<6   \\
	\frac{2 \ln \left(\cos \left(\Delta\right)\right)}{\sqrt{6} \gamma -6}, & \gamma^2\geq 6
	\end{array}
	\right. +\varepsilon  \left\{
\begin{array}{cc}
  \frac{2 \sqrt{2} c_2 \tanh (\Delta )}{\sqrt{U_0 \left(6-\gamma ^2\right)}}, & \gamma^2<6 \\
	\frac{2 \sqrt{2} c_2 \tan (\Delta )}{\sqrt{U_0 \left(\gamma ^2-6\right)}}, & \gamma^2 \geq 6 \\
\end{array}
 \right.+O(\varepsilon^2),\\
	& v(t;\varepsilon)=c_3+\left\{ \begin{array}{cc} 
	\frac{2 \ln \left(\sinh \left(\Delta\right)\right)}{\sqrt{6} \gamma
   +6}, & \gamma^2<6 \\ \frac{2 \ln \left(\sin \left(\Delta\right)\right)}{\sqrt{6} \gamma
   +6}, & \gamma^2 \geq 6
	\end{array}
	\right. +\varepsilon \left\{
\begin{array}{cc}
 -\frac{4 \left(\sqrt{3} \gamma -3 \sqrt{2}\right) c_2 \coth \left(\Delta\right)}{\left(\sqrt{6} \gamma +6\right) \sqrt{U_0 \left(6-\gamma
   ^2\right)}}, & \gamma^2 <6  \\
 \frac{4 \left(\sqrt{3} \gamma -3 \sqrt{2}\right) c_2 \cot \left(\Delta\right)}{\left(\sqrt{6} \gamma +6\right) \sqrt{U_0 \left(\gamma
   ^2-6\right)}}, & \gamma^2 \geq 6\\
\end{array}\right.+O(\varepsilon^2).
\end{align}
\end{subequations} 
The relative errors in the approximation of \eqref{seed_solution_first_order} by $u=u(t;0), v=v(t;0)$ are
\begin{subequations}
\label{relative-errors}
\begin{align}
& E_r(u):=\frac{u(t;\varepsilon)-u(t;0)}{u(t;\varepsilon)}=\frac{\varepsilon 
   \left\{
\begin{array}{cc}
  \frac{2 \sqrt{2} c_2 \tanh (\Delta )}{\sqrt{U_0 \left(6-\gamma ^2\right)}}, & \gamma^2<6 \\
	\frac{2 \sqrt{2} c_2 \tan (\Delta )}{\sqrt{U_0 \left(\gamma ^2-6\right)}}, & \gamma^2 \geq 6 \\
\end{array}
 \right.}{c_2 - \left\{\begin{array}{cc}
			\frac{2 \ln \left(\cosh \left(\Delta\right)\right)}{\sqrt{6} \gamma -6}, & \gamma^2<6   \\
	\frac{2 \ln \left(\cos \left(\Delta\right)\right)}{\sqrt{6} \gamma -6}, & \gamma^2\geq 6  .
	\end{array}
	\right.}+O\left(\varepsilon ^2\right),\end{align}
	\begin{align}
& E_r(v):=\frac{v(t;\varepsilon)-v(t;0)}{v(t;\varepsilon)}=\frac{\varepsilon \left\{
\begin{array}{cc}
 -\frac{4 \left(\sqrt{3} \gamma -3 \sqrt{2}\right) c_2 \coth \left(\Delta\right)}{\left(\sqrt{6} \gamma +6\right) \sqrt{U_0 \left(6-\gamma
   ^2\right)}}, & \gamma^2 <6  \\
 \frac{4 \left(\sqrt{3} \gamma -3 \sqrt{2}\right) c_2 \cot \left(\Delta\right)}{\left(\sqrt{6} \gamma +6\right) \sqrt{U_0 \left(\gamma
   ^2-6\right)}}, & \gamma^2 \geq 6\\
\end{array}\right.}{c_3+\left\{ \begin{array}{cc} 
	\frac{2 \ln \left(\sinh \left(\Delta\right)\right)}{\sqrt{6} \gamma
   +6}, & \gamma^2<6 \\ \frac{2 \ln \left(\sin \left(\Delta\right)\right)}{\sqrt{6} \gamma
   +6}, & \gamma^2 \geq 6
	\end{array}
	\right.}+O\left(\varepsilon ^2\right).
\end{align}
\end{subequations}
Taking the limit $t\rightarrow +\infty$ it follows that the above relative errors tend to zero. Thus, the linear terms in $\varepsilon$ in the equation \eqref{seed_solution_first_order} can be made a small percent of the contribution of the zeroth-solutions by taking $\tau$ large enough. Henceforth, this shows that the behavior of the solutions for the induced gravity model does not change abruptly when a massless scalar field $\phi$ with a small kinetic term is added to the setup.

\section{Review on averaging techniques}
\label{SECT:II}

The averaging methods applied extensively in \cite{Alho:2015cza,Alho:2019pku,Paliathanasis:2015cza,Leon:2020ovw,Llibre:2012zz,Fajman:2020yjb,Fajman:2021cli,Leon:2021lct,Leon:2021rcx,Leon:2021hxc} to single field scalar field cosmologies are extended to scalar field cosmologies of two fields in \cite{Chakraborty:2021vcr}. New dynamic variables and dimensionless time variables were adopted, which have not been used to analyze these cosmological dynamics. The main difficulties that arise when using standard dynamical systems approaches are due to the oscillations that enter the nonlinear system through the KG equations. This motivates the analysis of the oscillations using averaging techniques.
  
The theory of averaging studies initial value problems of the general form
\begin{align*}
\dot{\mathbf x} = \mathbf{f}(\mathbf x, t, \varepsilon), \quad \mathbf x(0)=\mathbf{x}_0,
\end{align*}
with $\mathbf x, \mathbf f(\mathbf x, t, \varepsilon)\in\R^n$, where $\varepsilon$ plays the role of a, usually small, perturbation parameter. Typically one would then perform a Taylor expansion of $\mathbf f$ in $\varepsilon$ around $\varepsilon=0$. For the simplest form of averaging, \emph{periodic averaging}, the zeroth order term usually vanishes, and one is typically looking at problems of the standard form
\begin{align}\label{E: standard form}
\dot{\mathbf x} = \varepsilon\,\mathbf f^1(\mathbf x, t) + \varepsilon^2\,\mathbf f^{[2]}(\mathbf x,t,\varepsilon),
\quad
\mathbf x(0) = \mathbf{x}_0,
\end{align}
with $\mathbf f^1$ and $\mathbf f^{[2]}$ $T$-periodic in $t$. The exponents correspond to the respective perturbative order, and the square bracket marks the remainder of the series (Notation~1.5.2, p~13 \cite{SandersEtAl2010}).

To first order, the theory is then concerned with the question to what degree solutions of~\eqref{E: standard form} can be approximated by the solutions of an associated \emph{averaged system}
\begin{align}\label{E: averaged system}
\dot{\mathbf{y}} &= \varepsilon\,\overline{\mathbf f}^1(\mathbf{y}), \quad \mathbf{y}(0)=\mathbf{x}_0,
\end{align}
with
\begin{align}\label{E: f bar}
\overline{\mathbf f}^1(\mathbf{y}) &= \frac{1}{T}\int_0^T\mathbf f^1(\mathbf{y}, s)\,\mathrm ds.
\end{align}

Take the following two definitions from \cite{SandersEtAl2010}: 
\begin{definition}[p 31 ~\cite{SandersEtAl2010}]\label{D: D}
$D\subset\R^n$ is a connected, bounded open set (with compact closure) containing the initial value $\mathbf{x}_0$, and constants $L>0, \varepsilon_0>0$, such that the solutions $\mathbf x(t,\varepsilon)$ and $\mathbf{y}(t,\varepsilon)$ with $0\leq\varepsilon\leq\varepsilon_0$ remain in $D$ for $0\leq t\leq L/\varepsilon$.
\end{definition}

\begin{definition}[Definition 4.2.4 of \cite{SandersEtAl2010}]\label{D: KBM}
Consider the vector field $\mathbf f(\mathbf x, t)$ with $\mathbf f:\R^n\times\R\to\R^n$. Let $\mathbf f$ be Lipschitz continuous in $\mathbf x$ on $D\subset\R^n,t\geq0$. Let further $\mathbf f$ be continuous in $t$ and $\mathbf x$ on $\R^+\times D$. If the average
\begin{align}
\label{avrgd}
\overline{\mathbf f}(\mathbf x) &= \lim_{T\to\infty} \frac{1}{T}\int^T_0\mathbf f(\mathbf x, s)\mathrm ds,
\end{align}
exists and the limit is uniform in $\mathbf x$ on compact subsets of $D$, then $\mathbf f$ is called a \textbf{KBM-vector field} (from the initials Krylov, Bogoliubov and Mitropolsky). If the vector field $\mathbf f(\mathbf x, t)$ contains parameters, we assume that the parameters and the initial conditions are independent of $\varepsilon$ and that the limit is uniform in the parameters.
\end{definition}

The basic result is given by the following theorem:

 \begin{lemma}[Theorem 11.1 of \cite{Verhulst}]
 Let be the $n$- dimensional system \eqref{E: standard form}. 
 Supposing that $\mathbf{f}^1(t,\mathbf{x})$ is $T$-periodic in $t$, with $T>0$ a constant independent of $\varepsilon$. Performing the averaging process \eqref{E: f bar}
 where $y$ is considered as a parameter that is kept constant during integration.  Let be the associated  initial value problem
 \begin{equation}
 \dot{\mathbf{y}} = \varepsilon \overline{\mathbf{f}}^1(\mathbf{y}), \;  \mathbf{y}(0)=\mathbf{x}_0.
 \end{equation}
 Then, we have $\mathbf{y}(t)=\mathbf{x}(t)+ \mathcal{O}(\varepsilon)$ on  the time scale $1/\varepsilon$, under fairly general conditions:

 \begin{enumerate}
 \item The vector functions  $\mathbf f^1$ and $\mathbf f^{[2]}$ are continuously differentiable in a bounded $n$-dimensional domain $D$, with $\mathbf{x}_0$ an interior point, on the time scale $1/\varepsilon$.
 \item $\mathbf{y}(t)$ remains interior to the domain $D$ on the time scale $1/\varepsilon$ to avoid boundary effects. 
 \end{enumerate} 
 \end{lemma}
 
Similar results is:
\begin{lemma}[Theorem 2.8.1, p 31 \cite{SandersEtAl2010}]\label{L: averaging}
Let $\mathbf f^1$ be Lipschitz continuous, let $\mathbf f^{[2]}$ be continuous, and let $\varepsilon_0, D, L$ be as in {Definition}~\ref{D: D}. Then there exists a constant $c>0$ such that
\begin{align*}
||\mathbf x(t,\varepsilon)-\mathbf{y}(t,\varepsilon)|| &< c\varepsilon
\end{align*}
for $0\leq\varepsilon\leq\varepsilon_0$ and $0\leq t\leq L/\varepsilon$, and where $||\,.\, ||$ denotes the norm $||\mathbf u ||:=\sum_{i=1}^n|u_i|$ for $\mathbf u\in\R^n$.
\end{lemma}

Now, supposing that the slowly varying system $\dot{\mathbf{x}}= \varepsilon \mathbf{f}^1(t, \mathbf{x})$ is such that $\mathbf{f}^1(t, \mathbf{x})$ is not periodic, nor a finite sum of periodic vector fields as before, we have the following result:  
 \begin{lemma}[Theorem 11.3 of \cite{Verhulst}]
 Let be the $n$- dimensional system \eqref{E: standard form}. 
 Supposing that $\mathbf{f}^1(t, \mathbf{x})$ can be averaged over $t$ in the sense that the limit \eqref{avrgd}
 exists.  Let be the associated initial value problem
 \begin{equation}
 \dot{\mathbf{y}}= \varepsilon \overline{\mathbf{f}}^1(\mathbf{y}), \;  \mathbf{y}(0)=\mathbf{x}_0,
 \end{equation} where $\mathbf{y}$ is again considered a parameter that is kept constant during
 integration.
 Then, we have 
 \begin{equation}
 \mathbf{y}(t) = \mathbf{x}(t) + \mathcal{O}(\delta(\varepsilon)),
 \end{equation} on the timescale $1/\varepsilon$ under fairly general
 conditions:
 \begin{enumerate}
 \item The vector functions $\mathbf{f}^1$ and $\mathbf{f}^{[2]}$ are continuously differentiable in a bounded
 $n$-dimensional domain $D$ with $\mathbf{x}_0$  an interior point on the timescale $1/\varepsilon$.
 \item $\mathbf{y}(t)$ remains interior to the domain $D$ on the timescale $1/\varepsilon$ to avoid boundary
 effects.
 \end{enumerate}
 For the error $\delta(\varepsilon)$, we have the explicit estimate
 \begin{equation}
 \delta(\varepsilon)= \sup_{\mathbf{x}\in D} \sup_{0\leq \varepsilon t \leq C} \varepsilon \Bigg\|\int_{0}^{t}
 (\mathbf{f}^1(s,\mathbf{x}) - \overline{\mathbf{f}}^1(\mathbf{x}))ds\Bigg\|,
 \end{equation}
 with $C$ a constant independent of $\varepsilon$.
 \end{lemma}
In other words, the error made when approximating the entire system \eqref{E: standard form} by the averaged system \eqref{E: averaged system} will be of the order $\varepsilon$ on timescales of the order $ \varepsilon^{-1}$. When the solutions of the complete or averaged system are attracted by an asymptotically stable critical point, the approximation domain can be extended to all times (see chapter 5 of \cite{SandersEtAl2010}). 
For instance:
\begin{lemma}[Theorem~5.5.1 by Eckhaus/Sanchez-Palencia of p~101 \cite{SandersEtAl2010}]\label{L: Eckhaus}
Consider the initial value problem
\begin{align*}
\dot{\mathbf x} &= \varepsilon\,\mathbf f^1(\mathbf x,t), \quad \mathbf x(0)=\mathbf{x}_0,
\end{align*}
with $\mathbf{x}_0, \mathbf x\in D\subset\R^n$. Suppose $\mathbf f^1$ is a KBM-vector field ({Definition}~\ref{D: KBM}) producing the averaged equation
\begin{align*}
\dot{\mathbf{y}} &= \varepsilon\,\overline{\mathbf f}^1(\mathbf{y}), \quad \mathbf{y}(0)=\mathbf{x}_0,
\end{align*}
where $\mathbf{y}=0$ is an asymptotically stable critical point in the linear approximation, $\overline{\mathbf f}^1$ is continuously differentiable with respect to $\mathbf{y}$ in $D$ and has a domain of attraction $D^o\subset D$. Then for any compact $K\subset D^o$ there exists a $\delta(\varepsilon)>0$ such that for all $\mathbf{x}_0\in K$
\begin{align*}
\mathbf x(t)-\mathbf{y}(t) &= \mathcal O\big(\delta(\varepsilon)\big), \quad 0\leq t<\infty,
\end{align*}
with $\delta(\varepsilon)= o(1)$ in the general case and $\mathcal O\big(\varepsilon\big)$ in the periodic case. 
\end{lemma}
 For periodic solutions we have the following: 
 \begin{lemma}[Theorem 11.4 of \cite{Verhulst}]
 $\dot{\mathbf{x}}= \varepsilon \mathbf{f}(t, \mathbf{x})$ is such that $\mathbf{f}(t, \mathbf{x})$ is $T$-periodic and
 the averaged equations \begin{equation}
 \dot{\mathbf{y}}= \varepsilon \overline{\mathbf{f}}(\mathbf{y}),
 \end{equation} with \begin{equation}
 \overline{\mathbf{f}}(\mathbf{y}) =  \frac{1}{T} \int_{0}^{T} \mathbf{f}(t, \mathbf{y})dt,
 \end{equation} where $\mathbf{y}_0$ is a stationary solution (equilibrium point) of the averaged
 equation $\overline{\mathbf{f}}(\mathbf{y}_0)=0$. If
 \begin{enumerate}
 \item $\mathbf{f}(t, \mathbf{x})$ is a smooth vector field,
 \item  for the Jacobian in $\mathbf{y}_0$ we have
 \begin{equation}
 \Bigg| \frac{\partial \overline{\mathbf{f}}}{\partial \mathbf{y}}\Big|_{\mathbf{y}=\mathbf{y}_0}\Bigg| \neq 0,
 \end{equation}
 \end{enumerate}
 then a $T$-periodic solution of the equation $\dot{x} = \varepsilon \mathbf{f}(t, \mathbf{x})$  exists in an $\varepsilon$-
 neighborhood of $\mathbf{x} = \mathbf{y}_0$.  We can establish the stability of the periodic solution as it matches
 exactly the stability of the stationary solution of the averaged equation. This reduces the stability problem of the periodic solution to determine the eigenvalues
 of a matrix.
 \end{lemma}
To summarize, methods from the theory of averaging nonlinear dynamical systems allow us to prove that time-dependent systems and their corresponding time-averaged versions have the same late-time dynamics. Therefore, simple time-averaged systems determine the future asymptotic behavior.
    
\subsection{Example 4: Harmonic oscillator}
\label{harmonic-oscillator}

Giving a differential equation 
$\dot x= f(x,t,\varepsilon)$  with $f$ periodic in $t$. An approximation scheme that can be used consists of solving the problem for $\varepsilon=0$ (unperturbed problem). Then, use this approximated unperturbed solution to formulate variational equations in standard form which can be averaged. 

Take the simple equation 
\begin{equation} \label{HarmonicOscillator}
\ddot \phi +\phi = \varepsilon (-2 \dot \phi), 
\end{equation}
with $\phi(0), \dot\phi(0)$  given. The unperturbed problem: 
\begin{equation}
\ddot \phi +\phi = 0, 
\end{equation}
have as solution
\begin{equation}
\dot\phi(t)= r_0 \cos (t-\varphi_0), \quad \phi(t)= r_0 \sin (t-\varphi_0),
\end{equation}
where $r_0$ and $\varphi_0$ are constants depending on the initial conditions. 
Using the amplitude-phase variables defined as \eqref{E: amplitude-phase}
with inverse transformation \eqref{eq_64}. Then, under the coordinate transformation $(\dot \phi, \phi )\rightarrow (r, \varphi)$, equation \eqref{HarmonicOscillator} leads to
\begin{equation}
\dot r= -2 r \varepsilon  \cos ^2(t-\varphi), \quad \dot\varphi = -  \varepsilon \sin (2 (t-\varphi )).
\end{equation}
These equations mean that $r$ and $\varphi$ are varying slowly with time,
and the system is in the form $\dot y= \varepsilon f(y)$.  The idea is to consider only the nonzero average of the right-hand-sides, keeping $r$ and $\varphi$
fixed, and leave out the terms with average zero ignoring the slow-varying dependence of $r$ and $\varphi$ on $t$ in the averaging process. 
Now,  replacing $r, \varphi$ by their averaged approximations $\overline{r}, \overline{\varphi}$, is obtained
\begin{align}
\label{EQ:49}
&\dot {\overline{r}}= -\varepsilon \frac{1}{2 \pi} \int_{0}^{2\pi} 2 r  \cos ^2(t-\overline{\varphi}) dt = - \varepsilon \overline{r}, \quad \dot{\overline{\varphi}}=  -\varepsilon \frac{1}{2 \pi} \int_{0}^{2\pi} \sin (2 (t-\overline{\varphi} )) dt = 0, 
\end{align}
where, by {Lemma}~\ref{L: averaging}, we know that the error between $[r, \varphi]^{\mathrm T}$ and $[\overline r, \overline\varphi]^{\mathrm T}$ will be of order $\varepsilon$ on timescales of order $\varepsilon^{-1}$. 

Solving \eqref{EQ:49} with initial conditions $\overline{r}(0)=r_0$ and $\overline{\varphi}(0)= \varphi_0$, the approximation takes the form  
\begin{equation}
\overline{\phi}= r_0 e^{-\varepsilon t} \sin (t-\varphi_0),
\end{equation}
which coincides with the result that would be obtained using the two-timing expansion procedure. These two procedures alleviate the failure of the  regular asymptotic expansion that would yield spurious secular terms in the asymptotic expansions, say, on the regular asymptotic expansion \eqref{expansion_0},
the ``next to leading term'' $\varepsilon t \sin t$ is dominant on scales $\varepsilon t = \mathcal{O}(1)$. 

These techniques can be extended to homogeneous cosmologies when $H$, the Hubble parameter, is considered as a time-dependent perturbation parameter. Examples are the model in \cite{Fajman:2020yjb} for  LRS Bianchi III Einstein-KG system. This system is analogous to a harmonic oscillator with nonlinear damping, and where the time dependence of the latter is governed by the coupling of the Einstein equations with the KG equation \begin{equation}
\ddot \phi + \phi = H[-3\dot\phi], \label{E: KG}
\end{equation} via $H$. 
In \cite{Fajman:2020yjb} the state vector  $\mathbf x=[\Sigma_+,\Omega,\varphi]^{\mathrm T}$, $\Omega=r^2/(6H^2)$, and $(r, \varphi)$ defined by \eqref{eq_64} is introduced, and the system takes the form
\begin{align}\label{E: quasi-standard form}
\begin{bmatrix} \dot H \\ \dot{\mathbf x} \end{bmatrix} &=
H\,\mathbf F^1(\mathbf x, t) + H^2\,\mathbf F^{[2]}(\mathbf x, t) =
H \begin{bmatrix} 0 \\ \mathbf f^1(\mathbf x, t) \end{bmatrix} + H^2 \begin{bmatrix} f^{[2]}(\mathbf x, t) \\ \mathbf 0 \end{bmatrix},
\end{align}
where $\mathbf f^1, f^{[2]}$ are independent of $H$. One can see that~\eqref{E: quasi-standard form} is resembling the standard form~\eqref{E: standard form} with $H(t)$ playing the role of the perturbation parameter $\varepsilon$. The resulting system was studied in \cite{Fajman:2020yjb} using averaging tools.  

Let $\overline{\mathbf{y}}(t)$ denote the solution of the corresponding averaged system. Then from {Lemma}~\ref{L: averaging} one knows that $\mathbf y(t) - \overline{\mathbf{y}}(t) = \mathcal O(H_*)$ on time scales of $\mathcal O(H_*^{-1})$, where $H_*$ is the value of $H$ at a large truncation time $t^*$, $H(t^*)$. Furthermore, one have a case of averaging with attraction and, one can extend the validity of this error estimate for all times for the $\mathbf{x}$-components. 
In  {\cite{Fajman:2021cli}}, a more general result was proved, where the long-term behavior of solutions of a general class of systems in standard form  \eqref{E: quasi-standard form} was studied; where $H>0$ is strictly decreasing in $t$ and $\lim_{t\rightarrow \infty}H(t)=0$. 
Theorem by {\cite{Fajman:2021cli}}, gives local-in-time asymptotics for system \eqref{E: quasi-standard form}. Let the norm $\|\cdot\|$ denotes the standard discrete $\ell^1$- norm $\|\mathbf{u}\| := \sum_i^n |u_i|$ for $\mathbf{u}\in \mathbb{R}^n$. Let also $L_{\mathbf{x}, t}^\infty$ denotes the standard $L^{\infty}$ space in both $t$ and $\mathbf{x}$ variables with  norm defined as  $\|f\|_{L_{\mathbf{x}, t}^\infty}:= \sup_{\mathbf{x}, t}|f(\mathbf{x}, t)|.$

\begin{theorem}[Theorem 3.1 of \textcolor{red}{\cite{Fajman:2021cli}}]
\label{localintime}
Suppose  $H(t)>0$ is strictly decreasing in $t$ and $\lim_{t\rightarrow \infty} H(t)=0.$ Fix any $\varepsilon>0$ with $\varepsilon<H(0)$ and define $t_*>0$ such that $\varepsilon=H(t_*).$ Suppose that $\|\mathbf{f}^1\|_{L_{\mathbf{x}, t}^\infty},\quad \|f^{[2]}\|_{L_{\mathbf{x}, t}^\infty}<\infty$ and that $\mathbf{f}^1(\mathbf{x}, t)$ is Lipschitz continuous and $f^{[2]}$ is continuous with respect to $x$ for all $t\geq t_*.$ Also, assume that $\mathbf{f}^1$ and $f^{[2]}$ are $T$-periodic for some $T>0.$ Then for all $t>t_*$ with $t=t_*+\mathcal{O}\Big(H(t_*)^{-\delta}\Big)$ for any given $\delta \in (0,1)$ we have $$\mathbf{x}(t)-\mathbf{y}(t)=\mathcal{O}\Big(H(t_*)^{\min\{1,2-2\delta\}}\Big),$$  where $\mathbf{x}$ is the solution of system \eqref{E: quasi-standard form} with initial condition $\mathbf{x}(0)=\mathbf{x}_0$ and $\mathbf{y}(t)$ is the solution of the time-averaged system 
\begin{equation*}
    \dot{\mathbf{y}}=H(t_*)\overline{\mathbf{f}}^1(\mathbf{y}),\quad \text{for} \quad t>t_*,
\end{equation*} 
with initial condition $\mathbf{y}(t_*)=\mathbf{x}(t_*)$ where the time-averaged vector $\overline{\mathbf{f}}^1$ is defined as 
\begin{equation*}
   \overline{\mathbf{f}}^1(\mathbf{y})=\frac{1}{T}\int_{t_*}^{t_*+T}\mathbf{f}^1(\mathbf{y},s)ds.
\end{equation*}
\end{theorem}
In references \cite{Leon:2021lct,Leon:2021rcx}, systems which are not in the standard form \eqref{E: quasi-standard form}, but can be expressed as a series with center in $H=0$ according to the equation
\begin{align}
\label{nonstandtard}
\begin{bmatrix}
       \dot{H} \\
        \dot{\mathbf{x}}
\end{bmatrix}= & \begin{bmatrix}
    0 \\
       \mathbf{f}^0 (\mathbf{x}, t)
 \end{bmatrix}+ H \begin{bmatrix}
       0 \\
       \mathbf{f}^1 (\mathbf{x}, t)
 \end{bmatrix}  + H^2 \begin{bmatrix}
       f^{[2]} (\mathbf{x}, t)  \\
       \mathbf{0}
\end{bmatrix}+ \mathcal{O}(H^3),
  \end{align}
 were studied. These systems depend on a parameter $\omega$ which is a free frequency that can be tuned to make $\mathbf{f}^0 (\mathbf{x}, t)= \mathbf{0}$. Therefore,  systems  can be expressed in the standard form \eqref{E: quasi-standard form}. 
The examples worked in reference \cite{Leon:2021lct} correspond to  generalized scalar-field cosmologies with matter in LRS Bianchi III and open FLRW model with generalized harmonic potential 
\begin{equation}
\label{pot_v2}
    V(\phi)=\mu ^2 \phi ^2 + f^2 \left(\omega ^2-2 \mu ^2\right) \left(1-\cos \left(\frac{\phi
   }{f}\right)\right). 
\end{equation}
The asymptotic features of potential \eqref{pot_v2} are the following.  Near the global minimum $\phi=0$, we have 
$V(\phi) \sim \frac{\omega ^2 \phi ^2}{2}+\mathcal{O}\left(\phi ^3\right), \quad \text{as} \; \phi\rightarrow 0$. 
That is, $\omega^2$ can be related to the mass of the scalar field near its global minimum.  As $\phi\rightarrow \pm \infty$ the cosine- correction is bounded, then  $V(\phi) \sim \mu ^2 \phi ^2+\mathcal{O}\left(1\right)$.  This makes it suitable to describe oscillatory behavior in cosmology. 

The state vector is $\mathbf{x}= \left(\Omega, \Sigma, \Omega_k, \Phi \right)^T$, the system can be symbolically written  as a Taylor series of the form \eqref{nonstandtard}. The term 
$\mathbf{f}^0 (t, \mathbf{x})$ in  expression \eqref{nonstandtard} is eliminated imposing the condition $b \mu ^3+2 f \mu ^2-f \omega ^2=0$,  which defines an angular frequency $\omega \in\mathbb{R}$. Then, order zero terms in the series expansion around $H=0$ are eliminated  assuming $\omega ^2>2 \mu ^2$ and setting $f=\frac{b \mu ^3}{\omega ^2-2 \mu ^2}$, which is equivalent to tune $\omega$. 
In Theorem 2 of \cite{Leon:2021lct} it was proved that if $\overline{\Omega}, \overline{\Sigma}, \overline{\Omega}_k,  \overline{\Phi}$ and $H$ are the solutions of averaged  equations. Then, there exist continuously differentiable functions $g_1, g_2, g_3$ and $g_4$,  such that   $\Omega, \Sigma, \Omega_k$ and $\Phi$  are locally given by {\cite{Alho:2015cza,Alho:2019pku}} 
\begin{align}
&\mathbf{x}_0:=\left(\Omega_{0}, \Sigma_{0}, \Omega_{k0}, \Phi_{0}\right)^T  \mapsto \mathbf{x}:=\left(\Omega, \Sigma, \Omega_k, \Phi\right)^T, \nonumber \\
& \mathbf{x}=\mathbf{x}_0 + H \mathbf{g}(H, \mathbf{x}_0,t), \quad \mathbf{g}(H, \mathbf{x}_0,t)= \begin{bmatrix}
    g_1(H , \Omega_{0}, \Sigma_{0}, \Omega_{k0}, \Phi_{0}, t)\\
    g_2(H , \Omega_{0}, \Sigma_{0}, \Omega_{k0}, \Phi_{0}, t)\\
    g_3(H , \Omega_{0}, \Sigma_{0}, \Omega_{k0}, \Phi_{0}, t)\\
    g_4(H , \Omega_{0}, \Sigma_{0}, \Omega_{k0}, \Phi_{0}, t)\\ 
\end{bmatrix}, \label{AppBIIIquasilinear211}
\end{align} where $\Omega_{0}, \Sigma_{0}, \Omega_{k0}, \Phi_0$ are order zero approximations of them as $H\rightarrow 0$. Then,  functions $\Omega_{0}, \Sigma_{0}, \Omega_{k0}, \Phi_0$ and averaged solution $\overline{\Omega},  \overline{\Sigma}, \overline{\Omega}_k, \overline{\Phi}$  have the same limit as $t\rightarrow \infty$. 
Setting $\Sigma=\Sigma_0=0$ are derived the analogous results for the negatively curved FLRW model. Theorem 3 of \cite{Leon:2021lct} shows that the late time attractors of the full system and averaged system for Bianchi III line element are the same.  The results from the linear stability analysis combined with Theorem 2 of  \cite{Leon:2021lct} (for $\Sigma=0$, open FLRW) lead to Theorem 4 in \cite{Leon:2021lct}, which shows that the late time attractors of the full system and the averaged system are the same.  
The examples worked in reference \cite{Leon:2021rcx} corresponds to generalized scalar-field cosmologies with the matter in LRS Bianchi I and flat  FLRW model. 
Denoting $\mathbf{x}= \left(\Omega, \Sigma, \Phi \right)^T$ and  using the condition $b \mu ^3+2 f \mu ^2-f \omega ^2=0$, to obtain a system can be expressed in the standard form \eqref{E: quasi-standard form}. 
Proceeding in analogous way as in references \cite{Alho:2015cza,Alho:2019pku} but for 3 dimensional systems  instead of a 1-dimensional one, it was implemented a local nonlinear transformation 
\begin{align}
&\mathbf{x}_0:=\left(\Omega_{0}, \Sigma_{0},  \Phi_{0}\right)^T  \mapsto \mathbf{x}:=\left(\Omega, \Sigma,  \Phi\right)^T, \nonumber \\
 &   \mathbf{x}= \mathbf{x}_0 + H \mathbf{g}(H, \mathbf{x}_0,t), \; \mathbf{g}(H, \mathbf{x}_0,t)= \begin{bmatrix}
    g_1(H , \Omega_{0}, \Sigma_{0}, \Phi_{0}, t)\\
    g_2(H , \Omega_{0}, \Sigma_{0}, \Phi_{0}, t)\\
    g_3(H , \Omega_{0}, \Sigma_{0}, \Phi_{0}, t)
 \end{bmatrix}. \label{EQT55} 
\end{align}
Theorem 1 of \cite{Leon:2021rcx}, states that, given  the functions 
$\overline{\Omega}, \overline{\Sigma}, \overline{\Phi}$  and $H$, be defined as solutions of averaged equations. Then, there exist continuously differentiable functions $g_1, g_2$ and $g_3$ such that  $\Omega, \Sigma, \Phi$ are locally given by  \eqref{EQT55} where $\Omega_{0}, \Sigma_{0}, \Phi_{0}$ are zero order approximations of $\Omega, \Sigma, \Phi$ as $H\rightarrow 0$. Then, functions $\Omega_{0}, \Sigma_{0}, \Phi_0$ and averaged solution $\overline{\Omega},  \overline{\Sigma}, \overline{\Phi}$  have the same limit as $t\rightarrow \infty$.
Setting $\Sigma=\Sigma_0=0$  analogous results for flat FLRW model are derived. 
Results from the linear stability analysis which are combined with Theorem 1 of \cite{Leon:2021rcx},  lead to Theorem 2 of \cite{Leon:2021rcx}, where the late-time attractors of the full system and time-averaged system for LRS Bianchi I line element are proved to be the same.  For flat FLRW metric,  Theorem 3 of \cite{Leon:2021rcx} shows that the late-time attractors of the full system and averaged system with $\Sigma=0$   are the same too. 
The core of these examples is to show how methods from the theory of averaging in nonlinear dynamical systems can be used to prove that time-dependent systems and their corresponding time-averaged versions have the same late-time dynamics. Therefore, the simplest time-averaged system determines the future asymptotic behavior. Depending on the values of free parameters,  we can find the late-time attractors of physical interests. With this approach, the oscillations entering the system through the  KG equation can be controlled and smoothed out as the Hubble parameter $H$ - acting as time-dependent perturbation parameter - tends monotonically to zero. In other words, these results show that one can ``average out'' the oscillations arising due to the harmonic functions, thus simplifying the problem.

\section{Perturbation and averaging methods applied to interacting scalar field cosmology}
\label{section_5}

It is worth noticing that when Hubble-normalized quantities are used more often the evolution equation for $H$, which is given by the Raychaudhuri equation, decouples.  The asymptotic of the remaining reduced system is then typically given by the equilibrium points and often it can be determined by a dynamical system analysis \cite{Coley:1999uh,Coley:2003mj,Wainwrightellis1997}. In particular, this is always the case for a scalar field with exponential potential. This is due to the fact the exponential potential has symmetry such that its derivative is also an exponential function. For other potentials that do not satisfy the above symmetry, like the harmonic potential $V(\phi)= \mu^2 \phi^2+ \text{cosine corrections}$, the Raychaudhuri equation fails to decouple \cite{Alho:2014fha}. Hubble-normalized equations often are very difficult to be analyzed using the standard dynamical systems approach due to oscillations entering the system via the KG equation \cite{Fajman:2020yjb,Leon:2021lct,Leon:2021rcx}. 

The preliminary analysis of oscillations in scalar-field cosmologies with generalized harmonic potentials of type $V(\phi)= \mu^2 \phi^2 + \text{cosine corrections}$ is extended here using averaging techniques similar to those used in \cite{Fajman:2020yjb,Leon:2021lct,Leon:2021rcx} for a family of generalized harmonic potentials when $H$ monotonically tends to zero. In this approach, the Hubble scalar plays a role of a time-dependent perturbation parameter which controls the magnitude of the error between full-system and time-averaged solutions. These oscillations can be viewed as perturbations that can be smoothed out with the benefit that the averaged Raychaudhuri equation decouples in the averaged system. In the end, the analysis of the system is reduced to the study of corresponding averaged equations. 
 
In this section, we investigate a cosmological model obtained by varying the action \eqref{eq1} for FLRW and Bianchi I geometries. An auxiliary function is used to include them, defined by
\begin{align}\label{G0a}
&G_0(a)=\left\{ \begin{array}{cc}
-3\frac{k}{a^2}, k=0, \pm 1, & \text{spatial curvature of FLRW metrics}\\
\frac{\sigma_0^2}{a^6}, & \text{anisotropies of Bianchi I metric}
\end{array} 
\right..
\end{align} 
We assume that the energy-momentum tensor \eqref{Tab} is in the
form of a perfect fluid
$$T^\alpha_\beta=\text{diag} \left(-\rho_m,p_m,p_m,p_m\right),$$ where $\rho_m$ and
$p_m$ are respectively the isotropic energy density and the
isotropic pressure (consistently with FLRW metric, pressure is
necessarily isotropic \cite{Trodden:2004st}). For simplicity we
will assume a barotropic EoS $p_m=(\gamma-1)\rho_m.$
Also we consider a quintessence scalar field, $\phi,$ interacting
in the action with the perfect fluid. In this case, the equations for FLRW and Bianchi
I metrics are \cite{Fadragas:2014mra,Gonzalez:2007ht}:
\begin{subequations}
	\label{Non_min}
	\begin{align}
	&\ddot\phi+3 H \dot \phi +\frac{d V(\phi)}{d\phi}=\frac{1}{2}(4-3\gamma)\rho_m \frac{d\ln\chi(\phi)}{d \phi},\label{Fried1b}\\
	&\dot{\rho_m}+3\gamma H\rho_m=-\frac{1}{2}(4-3\gamma)\rho_m  \dot{\phi}\frac{d\ln\chi(\phi)}{d \phi},\label{consb}\\
	&\dot a = a H, \\
	& \dot{H}=-\frac{1}{2}\left(\gamma \rho_m+{\dot \phi}^2\right)+\frac{1}{6}a G_0'(a),\label{Rachd}\\
	& 3H^2=\rho_m+\frac{1}{2}\dot\phi^2+V(\phi)+G_0(a),\label{Fried2b}
	\end{align}
\end{subequations}
where $a(t)$ denotes the scale factor of the Universe, 
 $H=\frac{\dot{a}}{a}$ denotes the Hubble parameter, a dot accounts for the derivative with respect to $t$, $\phi$ is the scalar field, $V(\phi)$ the scalar field self-interacting potential  which is assumed to be of class $C^2$, $\chi(\phi)^{-2}$ is the coupling function,
  $ \rho_m$ corresponds to the energy density of matter  with  EoS parameter  $w_m = \frac{p_m}{\rho_m}: = \gamma -1 $,  where $0\leq \gamma \leq 2$ denotes the barotropic index. 
The integration of  \eqref{consb} leads to
\begin{equation}
\rho_m=\frac{\rho_{m,0}}{a^{3\gamma}}\chi(\phi(a))^{-2+\frac{3\gamma}{2}}. \label{int-matter}
\end{equation} 
As in \cite{Gonzalez:2006cj}, here the baryons (a subdominant
component at present, but important in the past of the cosmic
evolution) are included in the background of dark matter. We assume a generalized harmonic potential  \eqref{pot1} non-minimally coupled to matter with coupling \eqref{coupling}.

\begin{figure}[ht!]
\begin{center}
	 \includegraphics[scale=0.8]{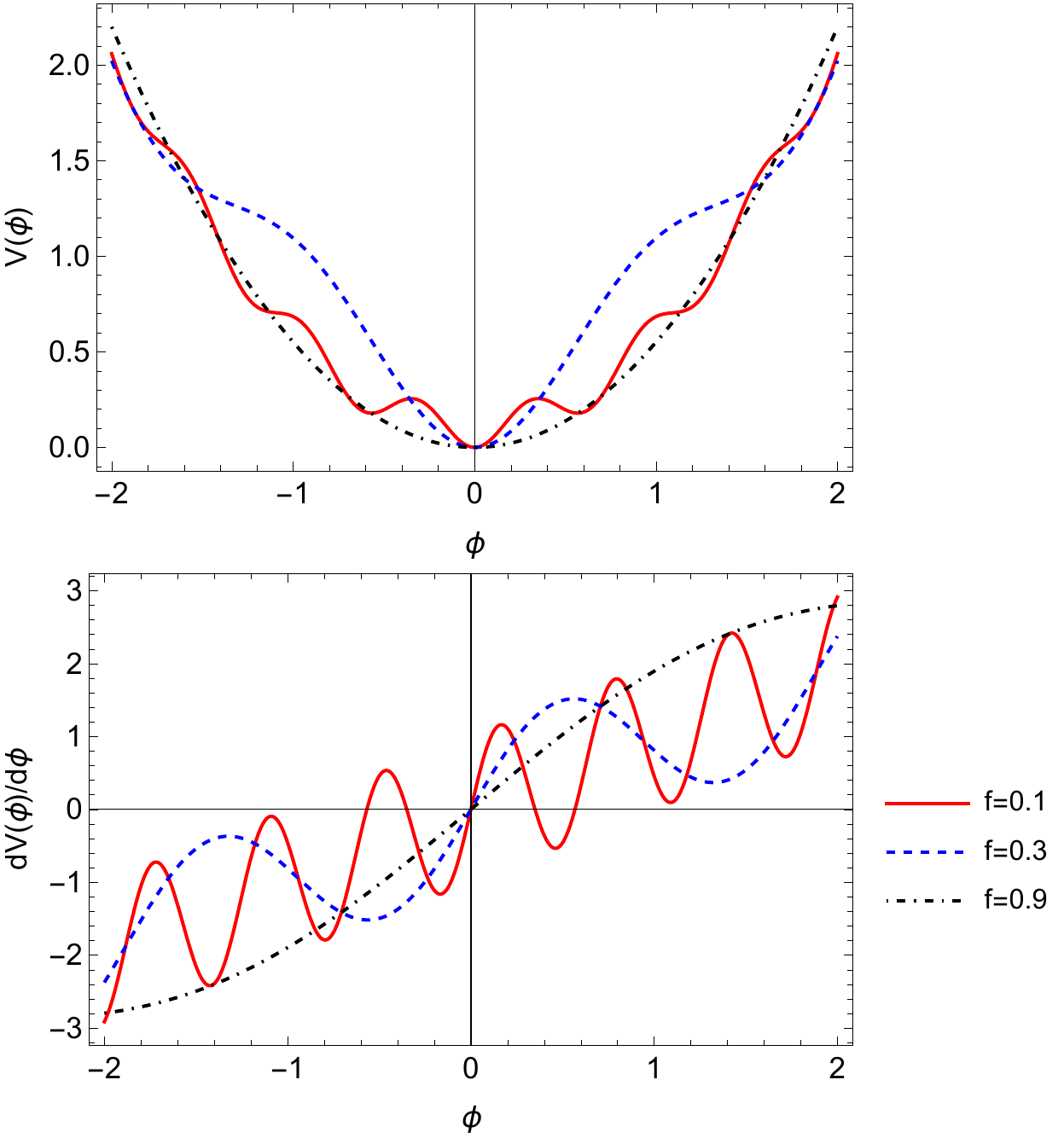}
	\caption{\label{FIG3af}  Generalized harmonic potential $V(\phi)= -f \cos \left(\frac{\phi }{f}\right)+f+\frac{\phi ^2}{2}$ and   its derivative for $f=0.1, 0.3, 0.9$.}
	\end{center}
\end{figure}

Potential \eqref{pot1} belongs to the class of potentials studied by  \cite{Rendall:2006cq}. 
In the Fig. \ref{FIG3af}, it is presented this the generalized harmonic potential and its derivative for $f=0.1$, $f=0.3$ and $f=10$. 
In first case the potential has three local minimums and two local maximums. In other two cases the origin is the unique stationary point and the global minimum of the potential. 
Harmonic potentials plus cosine corrections were introduced in the context of inflation in loop-quantum cosmology in \cite{Sharma:2018vnv}. 
In  \cite{Leon:2020pfy}, some theorems related to the asymptotic behavior of a very general cosmological model given by system \eqref{Non_min} were presented.  Using the Hubble-normalized formulation for a scalar field non-minimally coupled to matter  with generalized harmonic potential \eqref{pot1}  and with coupling function  \eqref{coupling} where  $\lambda$ is a constant and  $0\leq \gamma \leq 2, \quad \gamma \neq \frac{4}{3}$   the late time attractors corresponding to the non zero local minimums of the potential for FLRW metrics and for the Bianchi I metric were found.  These equilibrium points are related to de Sitter solutions. The global minimum of $V(\phi)$ at $\phi=0$  is unstable to curvature perturbations for $\gamma>\frac{2}{3}$ in the  case of a negatively curved FLRW model. This confirms the result in \cite{Giambo:2019ymx}, that in a non-degenerated minimum with zero critical value, the curvature will eventually dominate both the perfect fluid and the scalar field densities on the late evolution of the universe for $\gamma> 2/3$. For the Bianchi I model the global minimum $V(0)=0$ is unstable to shear perturbations. Equations for a scalar field cosmology minimally coupled to matter for FLRW metrics and for Bianchi I metrics are obtained by setting $\chi(\phi)\equiv1$ in \eqref{Non_min} with $G_0(a)$ given by \eqref{G0a} \cite{Chimento:1995da,Coley:1999uh}. Equation \eqref{int-matter} reduces to  $\rho_m=\frac{\rho_{m,0}}{a^{3\gamma}}$. The field equations of a scalar field with self-interacting potential $V(\phi)$ in vacuum for flat FLRW metric are obtained by setting $\chi(\phi)\equiv1, \rho_m=0$ in \eqref{Non_min} with $G_0(a)=0$. 
In \cite{Leon:2020ovw},  a local dynamical systems analysis for arbitrary $V(\phi)$ and $\chi(\phi)$ using Hubble normalized equations was provided. The analysis relies on two arbitrary functions $f(\lambda)$ and $g(\lambda)$ which encode a potential and a coupling function through a quadrature. Afterward,  a global dynamical systems formulation using the Alho \& Uggla's approach \cite{Alho:2014fha}  was implemented. The equilibrium points that represent some solutions of cosmological interest were obtained. In particular, several scaling solutions are found, as well as stiff solutions, and a solution dominated by the effective energy density of the geometric term $G_0(a)$, a quintessence scalar field dominated solution, the vacuum de Sitter solution associated to the minimum of the potential and a non-interacting matter-dominated solution. All of which reveals a very rich cosmological behavior.

\subsection{Scalar field with generalized harmonic potential non-minimally coupled to matter.}
\label{SECT:4.5}

In this section the averaging methods are applied for FLRW  and Bianchi I metrics for the generalized harmonic potential \eqref{pot1}
 coupled to matter with coupling function \eqref{coupling}.
In the following sections the FLRW and Bianchi I models will be studied separately.

\subsubsection{FLRW metric}
In this case the field equations are:  
\begin{subequations}
	\label{Non_minProb2FLRW}
	\begin{align}
	&\ddot\phi+3 H \dot \phi +\phi + \sin\left( \frac{\phi}{f}\right)=\frac{\lambda}{2}\rho_m ,\\
	&\dot{\rho_m}+3\gamma H\rho_m=-\frac{\lambda}{2}\rho_m  {\dot\phi},\\
	&\dot a = a H, \\
	& \dot{H}=-\frac{1}{2}\left(\gamma \rho_m+{\dot \phi}^2\right)+\frac{k}{a^2},\\
	& 3H^2=\rho_m+\frac{1}{2}\dot\phi^2+\frac{\phi ^2}{2}+f\left[1- \cos \left(\frac{\phi }{f}\right)\right]-\frac{3 k}{a^2}.
	\end{align}
\end{subequations}
Using the amplitude-phase variables \eqref{E: amplitude-phase}
with inverse transformation \eqref{eq_64},
it follows 
\begin{align}
    & \dot r= \frac{\dot \phi}{r}\left[\ddot \phi +\phi\right] = \frac{\dot \phi}{r}\left[ -3 H \dot \phi- \sin\left( \frac{\phi}{f}\right)+\frac{\lambda}{2}\rho_m\right]\nonumber \\
    & = -3 H  r \cos^2 (t-\varphi) - \cos (t-\varphi)\sin\left( \frac{r \sin (t-\varphi)}{f}\right)+\frac{\lambda}{2}\rho_m \cos (t-\varphi
   ),
\end{align}
and
\begin{align}
   & \dot \varphi=\frac{\phi }{r^2} \left[\ddot \phi +\phi\right]= \frac{\phi }{r^2} \left[ -3 H \dot \phi- \sin\left( \frac{\phi}{f}\right)+\frac{\lambda}{2}\rho_m\right]\nonumber \\
   & =-3 {H}  \sin (t-\varphi) \cos (t-\varphi
   ) -\frac{\sin (t-\varphi) \sin \left(\frac{r \sin (t-\varphi)}{f}\right)}{r} +\frac{\lambda}{2}\rho_m \frac{\sin (t-\varphi)}{r}.
\end{align}
Defining 
\begin{align}
\Omega= \frac{r^2}{6 H^2}, \quad 
 \Omega_m= \frac{\rho_m}{3 H^2}, \quad 
 \Omega_k=-\frac{k}{a^2 H^2},  \label{135}
\end{align}
such that  
\begin{equation}
    f \cos \left(\frac{\sqrt{6} \sqrt{\Omega } {H}  \sin (t-\varphi)}{f}\right)=f-3 {H} ^2 (1-\Omega -\Omega_{k}-\Omega_{m}), \label{163}
\end{equation}
the following dynamical system is obtained
\begin{small}
\begin{equation}
\label{XXEQ3.19XX}
\left\{
    \begin{array}{c}
 \dot{{H}}= -\frac{1}{2} {H} ^2 \left(3 \gamma  \Omega_m+6 \Omega  \cos ^2(t-\varphi)+2 \Omega_k\right) \\
 \dot{\Omega}=\frac{1}{2} {H}  \Big(2 \Omega  (3 \gamma  \Omega_m+3 (\Omega -1) \cos (2 (t-\varphi))+3 \Omega +2 \Omega_k-3)\\
  +\sqrt{6} \lambda  \sqrt{\Omega } \Omega_m \cos (t-\varphi)\Big) -\frac{\sqrt{\frac{2}{3}} \sqrt{\Omega } \cos (t-\varphi) \sin \left(\frac{\sqrt{6} \sqrt{\Omega } {H}  \sin (t-\varphi)}{f}\right)}{{H} } \\
 \dot{\Omega}_m= \frac{1}{2} \Omega_m {H}  \left(6 \gamma  (\Omega_m-1)-\sqrt{6} \lambda  \sqrt{\Omega } \cos (t-\varphi)+6 \Omega  \cos (2 (t-\varphi))+6 \Omega +4 \Omega_k\right) \\
  \dot{\Omega}_k= \Omega_k {H}  \left(3 \gamma  \Omega_m+6 \Omega  \cos ^2(t-\varphi)+2 \Omega_k-2\right) \\
 \dot{\varphi}=\frac{1}{4} {H}  \left(\frac{\sqrt{6} \lambda  \Omega_m \sin (t-\varphi)}{\sqrt{\Omega }}-6 \sin (2 (t-\varphi))\right)-\frac{\sin (t-\varphi) \sin \left(\frac{\sqrt{6} \sqrt{\Omega } {H}  \sin (t-\varphi
   )}{f}\right)}{\sqrt{6} \sqrt{\Omega } {H} } \\
\end{array}
    \right.. 
\end{equation}
\end{small}

\begin{figure}[ht!]
    \centering
    \includegraphics[scale=0.9]{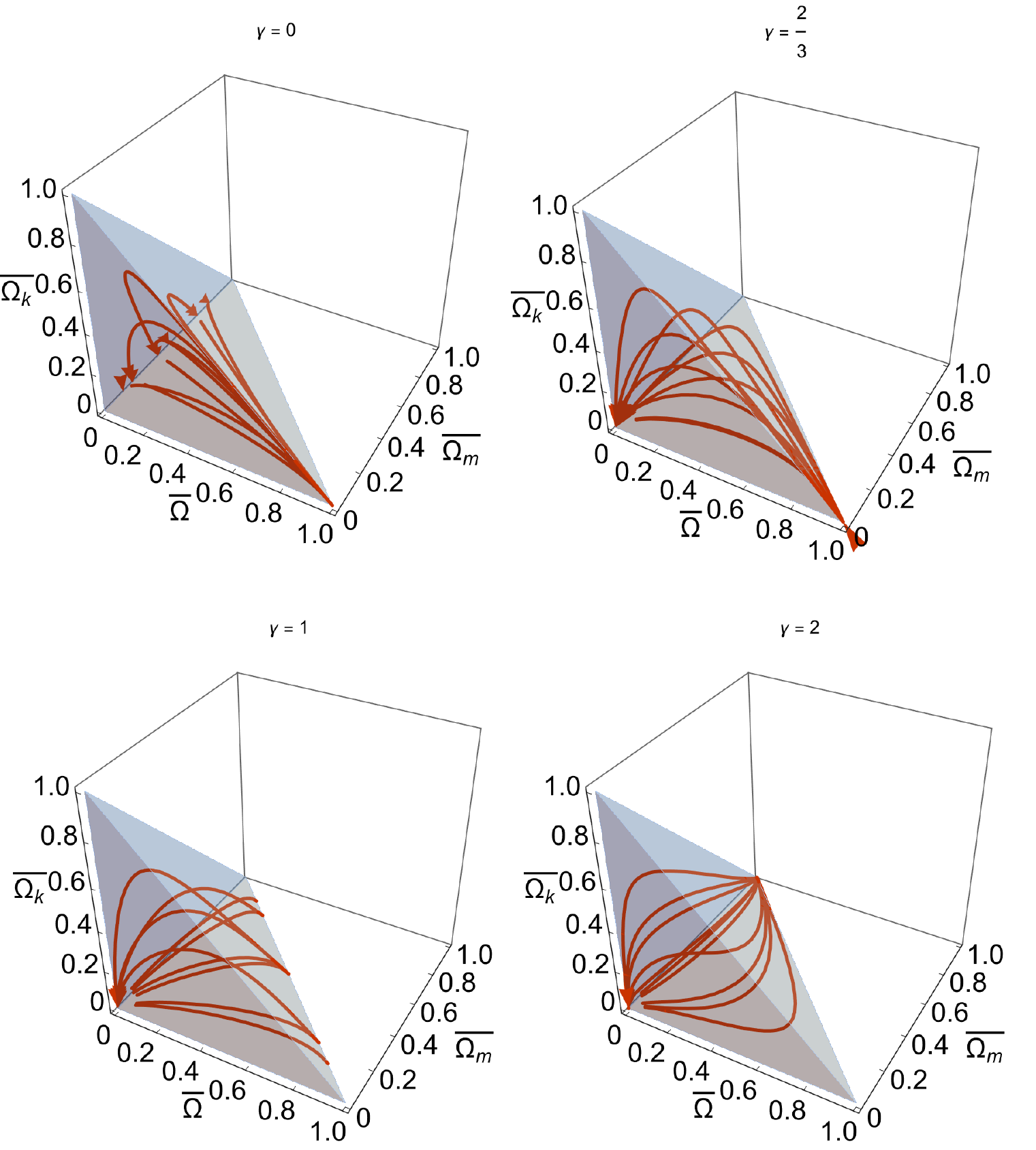}
    \caption{Phase portrait of the system \eqref{sispa} for $\gamma =0, 1, 2/3, 2$.}
    \label{Fig5}
\end{figure}

For the problem \eqref{XXEQ3.19XX}, using the techniques of section \ref{harmonic-oscillator},
 we obtain the averaged system
\begin{equation}\label{EQsEps}
\left\{
\begin{array}{c}
 \dot{H}=-\frac{1}{2} {H} ^2 (3 \gamma  \overline{\Omega}_m+3 \overline{\Omega} +2 \overline{\Omega}_k)\\
 \dot{\overline{\Omega}}= \overline{\Omega}  {H}  (3 \gamma  \overline{\Omega}_m+3 \overline{\Omega} +2 \overline{\Omega}_k-3)\\
  \dot{\overline{\Omega}}_m=\overline{\Omega}_m {H}  (3 \gamma  (\overline{\Omega}_m-1)+3 \overline{\Omega} +2 \overline{\Omega}_k)\\
  \dot{\overline{\Omega}}_k=\overline{\Omega}_k {H}  (3 \gamma  \overline{\Omega}_m+3 \overline{\Omega} +2 \overline{\Omega}_k-2)\\
 \dot{\overline{\varphi}}= -\frac{1}{2f}\\
\end{array}
\right.,
\end{equation}
where the angular equation is decoupled. 
Defining the new temporary variable $ \tau = \ln a$, the following guiding system is obtained:
\begin{subequations}\label{sispa}
\begin{align}
& \partial_{\tau}{\overline{\Omega}}= \overline{\Omega}    (3 \gamma  \overline{\Omega}_m+3 \overline{\Omega} +2 \overline{\Omega}_k-3), \\
& \partial_{\tau}{\overline{\Omega}}_m=\overline{\Omega}_m   (3 \gamma  (\overline{\Omega}_m-1)+3 \overline{\Omega} +2 \overline{\Omega}_k), \\
& \partial_{\tau}{\overline{\Omega}}_k=\overline{\Omega}_k  (3 \gamma  \overline{\Omega}_m+3 \overline{\Omega} +2 \overline{\Omega}_k-2).
\end{align}
\end{subequations}
The equilibrium points for system \eqref{sispa} are $ P_1 = (0,1,0) $, $ P_2 (0,0,1) $, $ P_3 = (0,0,0 ) $ and $ P_4 = (1,0,0)$. By evaluating the linearization matrix of system \eqref{sispa} on each of the equilibrium points and calculating its
eigenvalues, we obtain the stability of each point depending on $\gamma$, this results are summarized in the table \ref{PCsispa}. Furthermore, in Fig. \ref{Fig5}   is shown that the origin is a sink as indicated in Table \ref{PCsispa}. 

\begin{table}[t]
\begin{center}
\begin{tabular}{|c|c|c|c|}
\hline
Label & $({\overline{\Omega}},{\overline{\Omega}}_m, {\overline{\Omega}}_k)$ &  Eigenvalues  & Stability \\ \hline 
$P_1$ &  $(0, 1,0)$ & $\lbrace 3(\gamma-1), 3\gamma, 3\gamma-2 \rbrace$ & nonhyperbolic for  $\gamma=0,2/3,1$ \\
&&& Saddle for $0<\gamma <2/3$ or $2/3<\gamma <1$\\
&&& Source for $1<\gamma \leq 2$\\ \hline 
$P_2$ & $(0,0,1)$ & $\lbrace 2, -1, 2-3 \gamma  \rbrace$& Saddle for  $\gamma \neq 2/3$\\
&&& nonhyperbolic for  $\gamma =2/3$\\ \hline
$P_3$ & $(0,0,0)$ & $\lbrace -3,-2,-3 \gamma \rbrace$ & Sink for $0<\gamma\leq 2$ \\
&&& nonhyperbolic for $\gamma =0$\\ \hline
$P_4$ & $(1,0,0)$ & $\lbrace 3,1,-3 (\gamma -1) \rbrace$ & Source for  $0\leq\gamma<1$\\
&&& Saddle for $1<\gamma\leq 2$\\
&&& nonhyperbolic for $\gamma=1$ \\ \hline
\end{tabular}
\end{center}\caption{\label{PCsispa} Stability criteria for the equilibrium points of the system \eqref{sispa}. }
\end{table}

\subsubsection{Bianchi I metric}
In this case, the field equations are:  
\begin{subequations}
	\label{Non_minProb2BI}
	\begin{align}
	&\ddot\phi+3 H \dot \phi +\phi + \sin\left( \frac{\phi}{f}\right)=\frac{\lambda}{2}\rho_m ,\\
	&\dot{\rho_m}+3\gamma H\rho_m=-\frac{\lambda}{2}\rho_m  {\dot\phi},\\
	&\dot a = a H, \\
	& \dot{H}=-\frac{1}{2}\left(\gamma \rho_m+{\dot \phi}^2\right)-\frac{\sigma_0^2}{a^6},\\
	& 3H^2=\rho_m+\frac{1}{2}\dot\phi^2+\frac{\phi ^2}{2}+f\left[1- \cos \left(\frac{\phi }{f}\right)\right]+\frac{\sigma_0^2}{a^6}.
	\end{align}
\end{subequations}
Using the amplitude- phase transformation \eqref{E: amplitude-phase} with 
\eqref{eq_64}, 
and defining  \begin{align} \label{Bianchi_I_vars}
\Omega= \frac{r^2}{6 H^2}, \quad 
 \Omega_m= \frac{\rho_m}{3 H^2},  \quad \Sigma=\frac{\sigma_0}{a^3 H},
\end{align}
such that
\begin{equation}
    f \cos \left(\frac{\sqrt{6} \sqrt{\Omega } {H}  \sin (t-\varphi)}{f}\right)=f-{H} ^2 \left(3 (1-\Omega -\Omega_{m})-\Sigma ^2\right), \label{168}
\end{equation}
the following dynamical system is obtained
\begin{equation}
\label{YYEQ3.29YY}
\left\{
\begin{array}{c}
\dot{{H}}= -\frac{1}{2} {H} ^2 \left(3 \gamma  \Omega_m+2 \Sigma ^2+6 \Omega  \cos ^2(t-\varphi )\right) \\
\dot{\Omega}= \frac{1}{2} {H}  \Big(2 \Omega  \left(3 \gamma  \Omega_m+2 \Sigma ^2+3 (\Omega -1) \cos (2 (t-\varphi ))+3 \Omega -3\right)\\
 +\sqrt{6} \lambda  \sqrt{\Omega } \Omega_m \cos (t-\varphi
   )\Big) -\frac{\sqrt{\frac{2}{3}} \sqrt{\Omega } \cos (t-\varphi ) \sin \left(\frac{\sqrt{6} \sqrt{\Omega } {H}  \sin (t-\varphi )}{f}\right)}{{H} } \\
\dot{\Omega}_m= \frac{1}{2} \Omega_m {H}  \left(6 \gamma  (\Omega_m-1)+4 \Sigma ^2-\sqrt{6} \lambda  \sqrt{\Omega } \cos (t-\varphi )+6 \Omega  \cos (2 (t-\varphi ))+6 \Omega \right) \\
\dot{\Sigma}= \frac{1}{2} \Sigma  {H}  \left(3 \gamma  \Omega_m+2 \Sigma ^2+6 \Omega  \cos ^2(t-\varphi )-6\right) \\
 \dot{\varphi}= \frac{1}{4} {H}  \left(\frac{\sqrt{6} \lambda  \Omega_m \sin (t-\varphi )}{\sqrt{\Omega }}-6 \sin (2 (t-\varphi ))\right)
 -\frac{\sin (t-\varphi ) \sin \left(\frac{\sqrt{6} \sqrt{\Omega } {H}  \sin (t-\varphi
   )}{f}\right)}{\sqrt{6} \sqrt{\Omega } {H} } \\
\end{array}
\right..
\end{equation}

For the problem \eqref{YYEQ3.29YY}, using the techniques of section \ref{harmonic-oscillator}, we obtain the averaged system
\begin{equation}
\label{BIYEQ3.29Y}
\left\{
\begin{array}{c}
 \dot{{H}}=-\frac{1}{2} {H} ^2 \left(3 (\gamma  \overline{\Omega}_{m}+\Omega )+2 \overline{\Sigma}^2\right) \\
 \dot{\overline{\Omega}}=\overline{\Omega}  {H}  \left(3 (\gamma  \overline{\Omega}_{m}+\Omega -1)+2 \overline{\Sigma}^2\right) \\
\dot{\overline{\Omega}}_m= \overline{\Omega}_{m} {H}  \left(3 \gamma  (\overline{\Omega}_{m}-1)+2 \overline{\Sigma}^2+3 \overline{\Omega} \right) \\
\dot{\overline{\Sigma}}= \frac{1}{2} \overline{\Sigma}{H}  \left(3 (\gamma  \overline{\Omega}_{m}+\overline{\Omega} -2)+2 \overline{\Sigma} ^2\right) \\
\dot{\overline{\varphi}}=  -\frac{1}{2f} \\
\end{array}
\right..
\end{equation}    
where the angular equation is decoupled. Introducing the new variable  $\tau= \ln a$, the following guiding system is obtained: 
\begin{subequations}\label{Sis3.30}
\begin{align}
& \partial_{\tau}{\overline{\Omega}}=\overline{\Omega}   \left(3 (\gamma  \overline{\Omega}_{m}+\Omega -1)+2 \overline{\Sigma}^2\right), \\
& \partial_{\tau}{\overline{\Omega}}_m= \overline{\Omega}_{m}  \left(3 \gamma  (\overline{\Omega}_{m}-1)+2 \overline{\Sigma}^2+3 \overline{\Omega} \right), \\
& \partial_{\tau}{\overline{\Sigma}}= \frac{1}{2} \overline{\Sigma}  \left(3 (\gamma  \overline{\Omega}_{m}+\overline{\Omega} -2)+2 \overline{\Sigma} ^2\right).
\end{align}
\end{subequations}

\begin{figure}[ht!]
    \centering
    \includegraphics[scale=0.9]{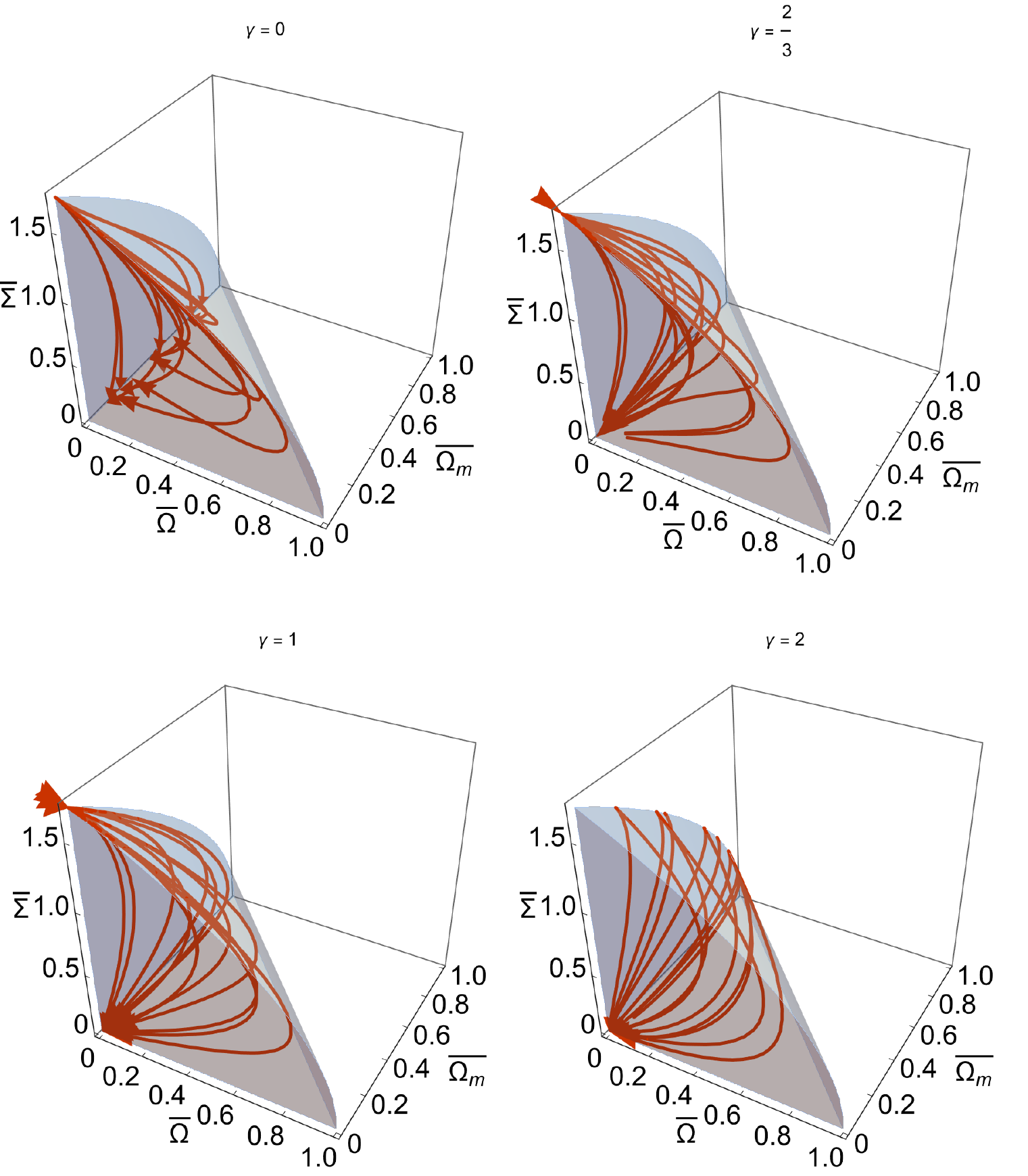}
    \caption{\label{Fig6} Phase portrait of the system \eqref{Sis3.30} for $\gamma =0, 1, 2/3, 2$.}
   \end{figure}

Observe that the system \eqref{Sis3.30} is invariant under the change of coordinates $\Sigma \rightarrow -\Sigma$, therefore it can be investigated in only one part of the phase portrait. 
\\
The equilibrium points of the system  \eqref{Sis3.30} are $P_1=(0,1,0)$, $P_2=(1,0,0)$, $P_3=(0,0,-\sqrt{3})$, $P_4=(0,0,\sqrt{3})$ and $P_5=(0,0,0)$. The stability criteria for each of them is summarized in Table \ref{PCsis3.31}.
\begin{table}[t]
\begin{center}
\begin{tabular}{|c|c|c|c|}
\hline
Label & $({\overline{\Omega}},{\overline{\Omega}}_m,\Sigma)$ & Eigenvalues & Stability \\ \hline 
$P_1$ & $(0,1,0)$ & $\lbrace \frac{3 (\gamma -2)}{2},3 (\gamma -1),3 \gamma \rbrace$ &  saddle for $0<\gamma<1$ or $1<\gamma<2$\\
&&& nonhyperbolic for $\gamma=0,1,2$\\ \hline
$P_2$ & $(1,0,0)$& $\lbrace 3,-\frac{3}{2},-3 (\gamma -1) \rbrace$& saddle for $0\leq\gamma<1 $ or $1<\gamma\leq 2$ \\
&&& nonhyperbolic saddle for  $\gamma=1$\\ \hline
$P_3$ & $(0,0,-\sqrt{3})$& $\lbrace 6,3,-3 (\gamma -2) \rbrace$&  source for $0\leq \gamma <2 $ \\
&&& nonhyperbolic for $\gamma=2$\\\hline
$P_4$ & $(0,0,\sqrt{3})$&$\lbrace 6,3,-3 (\gamma -2) \rbrace$& source for $0\leq \gamma <2 $ \\
&&& nonhyperbolic for $\gamma=2$\\\hline
$P_5$ & $(0,0,0)$& $\lbrace -3,-3,-3 \gamma \rbrace$& sink for $0<\gamma\leq 2$\\
&&& nonhyperbolic for $\gamma =0$\\\hline
\end{tabular}
\end{center}\caption{\label{PCsis3.31} Stability criteria for the equilibrium points of the system   \eqref{Sis3.30}.}
\end{table}
In figure \ref{Fig6}, it can be corroborated that the origin is a sink as it was indicated in table \ref{PCsis3.31}. 

\subsection{Scalar field with generalized harmonic potential minimally coupled to matter.}
\label{SECTION_4.5}

In this section, a scalar field cosmology is investigated in the presence of matter for FLRW metrics and Bianchi I metrics. The averaging methods are applied for a generalized harmonic potential of the type \eqref{pot1}. In every case, the stability criteria of their equilibrium points are obtained. 

\subsubsection{FLRW metric}
For the minimally coupled case of the FLRW metric, the field equations are given by setting $\lambda=0$ in \eqref{Non_minProb2FLRW}.
Using the amplitude-phase variables \eqref{E: amplitude-phase}
with \eqref{eq_64}
and defining  \eqref{135}, which satisfy \eqref{163},
we obtain the following dynamical system: 
\begin{equation}
\label{EQ3.38}
\left\{
\begin{array}{c}
 \dot{{H}}=-\frac{1}{2} {H} ^2 \left(3 \gamma  \Omega_m+6 \Omega  \cos ^2(t-\varphi)+2 \Omega_k\right)\\
  \dot{\Omega}=\Omega  {H}  (3 \gamma  \Omega_m+3 (\Omega -1) \cos (2 (t-\varphi))+3 \Omega +2
   \Omega_k-3)  \\
    -\frac{\sqrt{\frac{2}{3}} \sqrt{\Omega } \cos (t-\varphi) \sin \left( \frac{\sqrt{6} \sqrt{\Omega } {H}  \sin (t-\varphi
   )}{f}\right)}{{H} }\\
 \dot{\Omega}_m= \Omega_m {H}  \left(3 \gamma  (\Omega_m-1)+6 \Omega  \cos ^2(t-\varphi)+2 \Omega_k\right)\\
 \dot{\Omega}_k=\Omega_k {H}  \left(3 \gamma  \Omega_m+6 \Omega  \cos ^2(t-\varphi)+2 \Omega_k-2\right)\\
   \dot\varphi=-\frac{3}{2}
   {H}  \sin (2 (t-\varphi))-\frac{\sin (t-\varphi) \sin \left(\frac{\sqrt{6} \sqrt{\Omega } {H}  \sin (t-\varphi
   )}{f}\right)}{\sqrt{6} \sqrt{\Omega } {H} }
\end{array}\right..
\end{equation}
For the problem \eqref{EQ3.38},  the corresponding averaged system is again \eqref{EQsEps}. Introducing the time variable $\tau= \ln a$, we obtain once again the guiding system \eqref{sispa}. Therefore, we find the same equilibrium points  $P_1=(0,1,0)$, $P_2(0,0,1)$, $P_3=(0,0,0)$ and $P_4=(1,0,0)$. Their stability conditions are summarized in Table \ref{PCsispa}.  Then, the asymptotic behavior of the model on average is independent of the coupling function. Although, obviously, non-averaged systems have different dynamics.

\subsubsection{Bianchi I metric}
For the minimally coupled case of the Bianchi I metric the field equations are obtained from \eqref{Non_minProb2BI} by setting $\lambda=0$. 
Using the amplitude- phase transformation \eqref{E: amplitude-phase} with 
\eqref{eq_64}, 
and defining  \eqref{Bianchi_I_vars}, which satisfies \eqref{168},
it is derived the dynamical system:  
\begin{equation}
\label{EQ3.46}
\left\{
\begin{array}{c}
\dot{{H}}= -\frac{1}{2} {H} ^2 \left(3 \gamma  \Omega_{m}+2 \Sigma ^2+6 \Omega  \cos ^2(t-\varphi)\right)\\
\dot{\Omega}= \Omega  {H}  \left(3 (\gamma  \Omega_{m}+\Omega -1)+2 \Sigma ^2+3 (\Omega -1) \cos (2 (t-\varphi))\right) \nonumber \\
 -\frac{\sqrt{\frac{2}{3}} \sqrt{\Omega } \cos (t-\varphi) \sin \left( \frac{\sqrt{6} \sqrt{\Omega } {H}  \sin (t-\varphi )}{f}\right)}{{H} }
\\
\dot{\Omega}_m= \Omega_{m} {H}  \left(3 \gamma  (\Omega_{m}-1)+2 \Sigma ^2+6 \Omega  \cos ^2(t-\varphi)\right) \\
 \dot{\Sigma}= \frac{1}{2} \Sigma  {H}  \left(3 \gamma  \Omega_{m}+2 \Sigma ^2+6 \Omega  \cos ^2(t-\varphi)-6\right) \\
\dot{\varphi}= -\frac{3}{2} {H}  \sin (2 (t-\varphi))-\frac{\sin (t-\varphi) \sin \left( \frac{\sqrt{6} \sqrt{\Omega } {H}  \sin (t-\varphi )}{f}\right)}{\sqrt{6} \sqrt{\Omega } {H} }
\end{array}\right..
\end{equation}
For the problem \eqref{EQ3.46}, the corresponding averaged system is again \eqref{BIYEQ3.29Y}.    
Introducing the time variable $\tau= \ln a$, we obtain again the guiding system \eqref{Sis3.30}.  Therefore, the equilibrium points are the same: $P_1=(0,1,0)$, $P_2=(1,0,0)$, $P_3=(0,0,-\sqrt{3})$, $P_4=(0,0,\sqrt{3})$ and $P_5=(0,0,0)$. The stability criteria of the equilibrium points for system  \eqref{Sis3.30} are summarized in table  \ref{PCsis3.31}. Then, the asymptotic behavior of the model on average is independent of the coupling function. Although, obviously, non-averaged systems have different dynamics.

\subsection{A  scalar field in vacuum with generalized harmonic potential.}
\label{Sect:2.7.3} 
In this section, the perturbation methods are applied for analyzing the dynamics of a scalar field in a vacuum with generalized harmonic potential \eqref{pot1}.
The amplitude-phase variables \eqref{E: amplitude-phase}
produce the system:
\begin{subequations}
\begin{align}
   & \dot r= -\cos (t-\varphi) \sin \left(\frac{r \sin (t-\varphi)}{f}\right)-3 r {H}  \cos ^2(t-\varphi),\\
   & \dot\varphi=-\frac{\sin (t-\varphi) \sin \left(\frac{r \sin (t-\varphi)}{f}\right)}{r}-3 {H}  \sin (t-\varphi) \cos (t-\varphi
   ), \end{align}
   with restriction
   \begin{align}
   & f \cos \left(\frac{r \sin (t-\varphi )}{f}\right)-f-\frac{r^2}{2}+3 {H} ^2=0.
\end{align}
\end{subequations}
Defining the transformation $r \rightarrow \Omega= \frac{r^2}{6 H^2}$, it follows: 
\begin{equation}
\label{P({H})} 
\left\{
\begin{array}{cc}
&\dot {H}=-3 \Omega  {H} ^2 \cos ^2(t-\varphi ) \\
& \dot \Omega=6 (\Omega -1) \Omega  {H}  \cos ^2(t-\varphi )-\frac{\sqrt{\frac{2}{3}} \sqrt{\Omega } \cos (t-\varphi ) \sin \left(\frac{\sqrt{6} \sqrt{\Omega } {H}  \sin (t-\varphi
   )}{f}\right)}{{H} }\\
  &\dot \varphi =-\frac{\sin (t-\varphi ) \sin \left(\frac{\sqrt{6} \sqrt{\Omega } {H}  \sin (t-\varphi )}{f}\right)}{\sqrt{6} \sqrt{\Omega } {H} }-\frac{3}{2} {H}  \sin (2 (t-\varphi )
    \end{array}\right.,
\end{equation}
where 
\begin{equation}
\label{Hamiltonian2}
f \cos \left(\frac{\sqrt{6} \sqrt{\Omega } {H}  \sin (t-\varphi)}{f}\right)-f-3 (\Omega -1) {H} ^2=0.
\end{equation}

\begin{proposition}
\label{Prop4}
System \eqref{P({H})} admits the approximated solution as $H\rightarrow 0$: 
\begin{subequations}  
\label{solnoe}
\begin{align}
& \Omega_0(t)= c_2 \left(\cos \left(\frac{2 \left(2 c_1+t\right)}{\sqrt{\frac{f}{f+1}}}\right)+2
   f+1\right), \label{eq98a}\\
& \varphi_0(t)= t-\arctan\left(\sqrt{\frac{f}{f+1}} \tan \left(\frac{2 c_1+t}{\sqrt{\frac{f}{f+1}}}\right)\right), \label{98b}\end{align}
\end{subequations}
where $c_1$ and $c_2$ are integration constants. 
\end{proposition}
\textbf{Proof}. The sketch  of the  proof is given in \ref{App4}. 

From \eqref{eq_64}, using the approximation $\varphi\approx \varphi_0$, given by  \eqref{98b}, and restricting the domain where the $\arctan(x)$ is a one-to-one function, we have for large $t$ (and as $H\rightarrow 0$), 
\begin{equation}
  \dot{\phi} \approx  \phi  \underbrace{\left(\sqrt{\frac{f+1}{f}} \cot\left(\frac{2 c_1+t}{\sqrt{\frac{f}{f+1}}}\right)\right)}_{\Phi(t)}. \label{eq99}
\end{equation}
Substituting \eqref{eq99} in \eqref{135} and using $H(t)=\dot{a}(t)/a(t)$, $H(t)=r(t)/\sqrt{6 \Omega(t)}$, $r(t)= \sqrt{(1 + \Phi(t)^2)\phi(t)}$, where $\Phi(t)$ is defined in \eqref{eq99}, and using the approximation $\Omega\approx \Omega_0$ given by \eqref{eq98a},  we have  for large $t$ (and as $H\rightarrow 0$), 
\begin{equation} 
\dot{a} \approx a \phi  \frac{\csc \left(\frac{\sqrt{f+1} (2
  c_1+t)}{\sqrt{f}}\right)}{2 \sqrt{3} \sqrt{c_2}
   \sqrt{f}}. \label{eq100}
\end{equation}
Solving the system \eqref{eq99}-\eqref{eq100} we obtain 
\begin{align}
  \phi (t)= c_3 \sin \left(\sqrt{\frac{1}{f}+1}
   (t+2 c_1)\right), \quad  a(t)= c_4 e^{\frac{c_3 (t+2 c_1)}{2
   \sqrt{3} \sqrt{c_2 f}}}.
\end{align}
That is, asymptotically we have a de Sitter solution with ``small'' $H \approx \frac{c_3}{2 \sqrt{3} \sqrt{c_2 f}}$.

Now, continuing with the applications of the perturbation theory tools it is proved the following: 
\begin{proposition}
\label{Prop5}
System \eqref{P({H})} admits the expansion 
\begin{subequations}
\label{eqs_204}
\begin{align}
 &\Omega \equiv \Omega (t)= \Omega_0(t)+{H}(t)\Omega_1(t) + \mathcal{O}({H}^2), \\
 & \varphi \equiv  \varphi (t)=  \varphi_0(t)+{H}(t)\varphi_1(t) + \mathcal{O}({H}^2),
\end{align}
\end{subequations}
where 
$\Omega_0(t)$ and $\varphi_0(t)$ are the solutions \eqref{solnoe} of the unperturbed problem  $P(0)$,  
\begin{align}
\label{eq_208}
\varphi_1(t)=\frac{1}{2} \left(\frac{2 c_3-3 f (2 f+1)}{\cos \left(2 \sqrt{\frac{1}{f}+1} \left(2 c_1+t\right)\right)+2 f+1}+3 f\right),  
\end{align}
and $\Omega_1$ is given in quadratures
\begin{align}
&\Omega_{1}(t)= \exp \left(-\int_1^t \frac{\sin (2 (s-\varphi_0(s)))}{f} \, ds\right) \left(\int_1^t \frac{g(s_1) \exp
   \left(\int_1^{s_1} \frac{\sin (2 (s-\varphi_0(s_1)))}{f} \, ds\right)}{f} \, ds_1+c_1\right), \label{eq_206}
\end{align}
where 
\begin{align}
& g(t)=c_3 \left[\cos \left(\frac{2 \left(2 c_1+t\right)}{\sqrt{\frac{f}{f+1}}}\right)+2 f+1\right] \times \nonumber \\
& \left[\frac{\left(3 f \cos \left(\frac{2 \left(2
   c_1+t\right)}{\sqrt{\frac{f}{f+1}}}\right)+2 c_2\right) \cos \left(2 \arctan\left(\sqrt{\frac{f}{f+1}} \tan \left(\frac{2
   c_1+t}{\sqrt{\frac{f}{f+1}}}\right)\right)\right)}{\cos \left(\frac{2 \left(2 c_1+t\right)}{\sqrt{\frac{f}{f+1}}}\right)+2 f+1} \right. \nonumber \\
   & \left. +\frac{f (f+1) \left(c_3
   \cos \left(\frac{2 \left(2 c_1+t\right)}{\sqrt{\frac{f}{f+1}}}\right)+2 c_3 f+c_3-1\right)}{f \tan ^2\left(\frac{2
   c_1+t}{\sqrt{\frac{f}{f+1}}}\right)+f+1}\right], \label{eq_207}
\end{align}
where $c_1$, $c_2$ and $c_3$ are integration constants. 
\end{proposition}
\textbf{Proof}. The sketch  of the  proof is given in \ref{App5}.

The system \eqref{P({H})} can be expressed as 
\begin{align}
\label{AAsistemapertubado1a}
\frac{dY}{d\eta}= {H} G(Y, t, {H}), \quad\frac{dt}{d\eta}={H},
\end{align}
where 
\begin{equation}
G(Y, t, {H})= \begin{bmatrix}
       -3 \Omega  {H} ^2 \cos ^2(t-\varphi)\\
       6 (\Omega -1) \Omega  {H}  \cos ^2(t-\varphi )-\frac{\sqrt{\frac{2}{3}} \sqrt{\Omega } \cos (t-\varphi) \sin \left(\frac{\sqrt{6} \sqrt{\Omega } {H}  \sin (t-\varphi
   )}{f}\right)}{{H} }\\
  -\frac{3}{2} {H}  \sin (2
   (t-\varphi))-\frac{\sin (t-\varphi) \sin \left(\frac{\sqrt{6} \sqrt{\Omega } {H}  \sin (t-\varphi
   )}{f}\right)}{\sqrt{6} \sqrt{\Omega } {H} }
    \end{bmatrix},
\end{equation}
where $Y$ denotes the phase vector  $\left(H, \Omega, \varphi\right)^T$. 

For the problem \eqref{P({H})} the following averaged system is deduced: 
\begin{equation}
\label{Ypromedio2Y}
\left\{
\begin{array}{cc}
&\dot {H}=-\frac{3 \overline{\Omega}  {H} ^2}{2}\\
& \dot{\overline{\Omega}}=-3 (1-\overline{\Omega}) \overline{\Omega}  {H}\\
  &\dot{\overline{\varphi}} =-\frac{1}{2 f}
    \end{array}\right.,
\end{equation}
where the angular equation is decoupled. 
Introducing the new variable $\tau= \ln a$, the following guiding equation is obtained 
\begin{equation}
   \partial_{\tau} {\overline{\Omega}}=-3 (1-\overline{\Omega})\overline{\Omega}, 
\end{equation}
for which $\overline{\Omega}=0$ is a sink and $\overline{\Omega}=1$ is a source. 

Starting with the averaged equations \eqref{Ypromedio2Y}, it is proved that ${\Omega}, {\varphi}$  evolve at first order according to the averaged equations for $\overline{\Omega}, \overline{\varphi}$.
\begin{proposition} 
\label{Prop_6}
Given $(H, \Omega,\varphi, t)$ solutions of \eqref{AAsistemapertubado1a}, there exists a transformation 
\begin{subequations}
\label{YquasilinearY}
\begin{align}
   & t= t_0+{H} \alpha_1(t_0, {\varphi}_0),\\
   & \Omega=\Omega_0+{H} \left[\alpha_2(t_0,  {\varphi}_0)-\frac{\eta}{f}\sin (2 (t_0-\varphi_0)) \Omega_0 \right],\\
   & \varphi=\varphi_0+{H} \left[\alpha_3(t_0,  {\varphi}_0) +  \frac{\eta}{2 f}\cos\left(2(t_0-\varphi_0)\right) \right],
\end{align}
\end{subequations}
where $\alpha_i(t_0,  {\varphi}_0), i=1,2,3$ are differentiable, 
such  that the functions  ${t}_0, {\Omega}_0, {\varphi}_0$  have the same asymptotic of the averaged solutions $\overline{t}, \overline{\Omega}, \overline{\varphi}$ of  \eqref{Ypromedio2Y}
as $H\rightarrow 0$ and $\eta\rightarrow \infty$.
\end{proposition}
\textbf{Proof}. The sketch  of the  proof is given in \ref{App6}.

\section{Numerical simulations}
\label{Numerical}
In this section, we present the numerical results obtained from the integration of the full system and its corresponding averaged version of the scalar field with a generalized harmonic potential model in the non-minimally coupled, for FLRW and Bianchi I metrics, and vacuum cases, as evidence that the full and averaged systems have the same dynamics when $H\rightarrow 0$. To that end, we elaborated an algorithm in the programming language \textit{Python}, where the systems of differential equations were numerically integrated using the \textit{solve\_ivp} code provided by the \textit{Scipy} open-source \textit{Python}-based ecosystem. As an integration method, we use \textit{Radau}, which is an implicit Runge-Kutta method of the Radau IIa family of order 5, with relative and absolute tolerances of $10^{-3}$ and $10^{-6}$, respectively. In the numerical integration, we use as a time variable $\tau$, which is related to the cosmic time $t$ through the expression $dt/d\tau=1/H$, in an integration range of $-40\leq\tau\leq 3$ for the full systems and $-40\leq\tau\leq 40$ for the averaged system, all of them partitioned in 20000 and 60000 data points for the non-minimal coupling and vacuum cases, respectively. Furthermore, the full and time-averaged systems were solved for a value of $\gamma$ equal to $0$ (CC), $2/3$, $1$ (dust) and $2$ (stiff fluid); all of them for a value of $f=0.1$, $0.3$ and $0.9$, for the non-minimally coupling case with a value of $\lambda=0.1$. The vacuum case was integrated only for the same values of $f$ as the non-minimal coupling case. It is worth noticing that in the case of the scalar field with generalized harmonic potential minimally coupled to matter model ($\lambda=0$), for FLRW and  Bianchi I metrics, the numerical results are very similar to their respective non-minimally coupling cases ($\lambda\neq 0$). Observe that the interaction appears in the equations explicitly in  the form $\lambda \sin (t-\varphi ) , \;    \lambda \cos (t-\varphi )$, which are zero in average. 

\subsection{Scalar field with generalized harmonic potential non-minimally coupled to matter.}
\subsubsection{FLRW metric}
\begin{figure}[ht!]
\centering
\begin{minipage}{.48\textwidth}
  \centering
    \subfigure[\label{FLRWNonminimallyCCf013Dm} Projections in the space $(\Omega_{m},H,\Omega)$. The surface is given by the constraint $\Omega=1-\Omega_{m}$.]{\includegraphics[scale = 0.25]{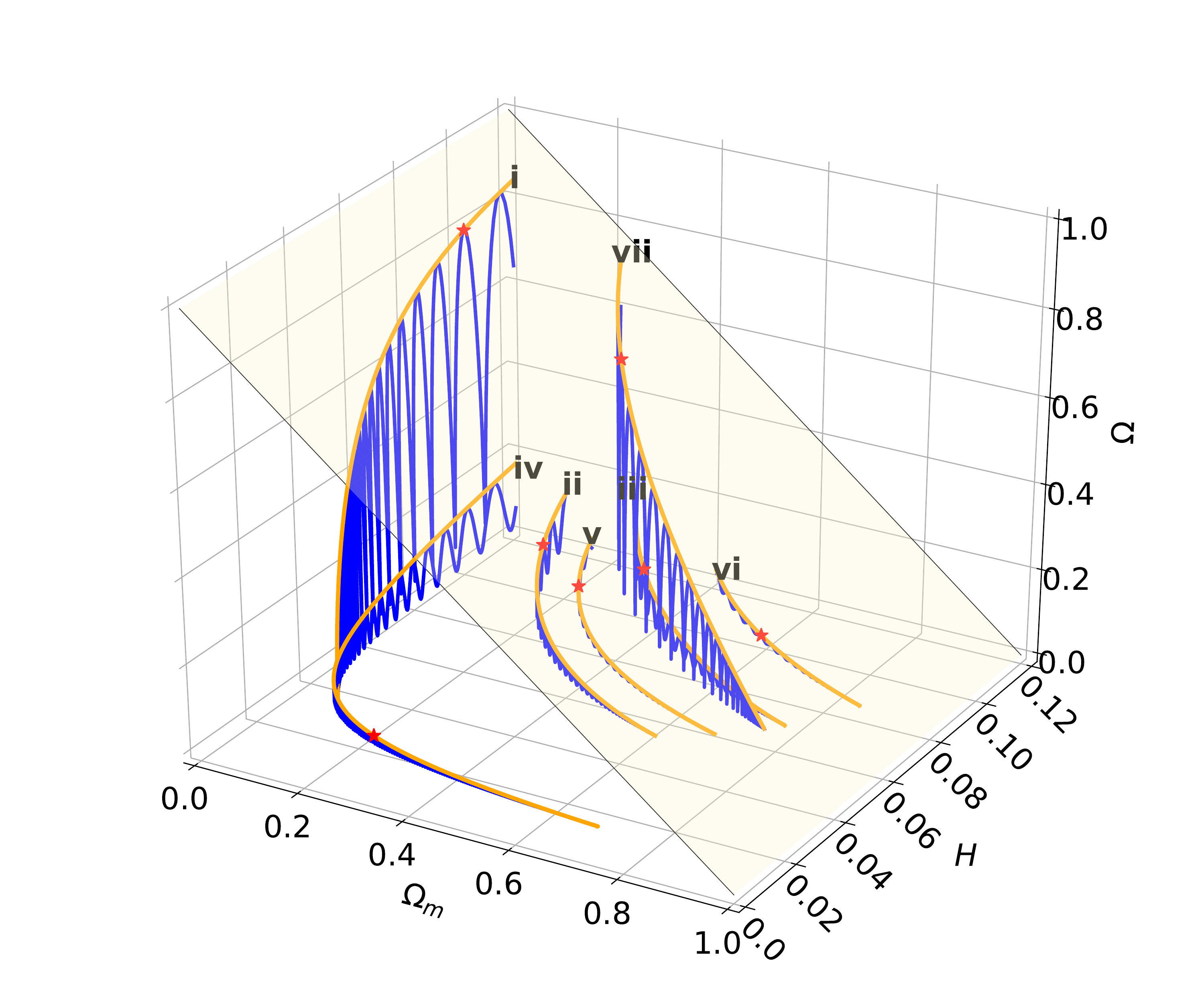}}
    \subfigure[\label{FLRWNonminimallyCCf013Dk} Projections in the space $(\Omega_{k},H,\Omega)$. The surface is given by the constraint $\Omega=1-\Omega_{k}$.]{\includegraphics[scale = 0.25]{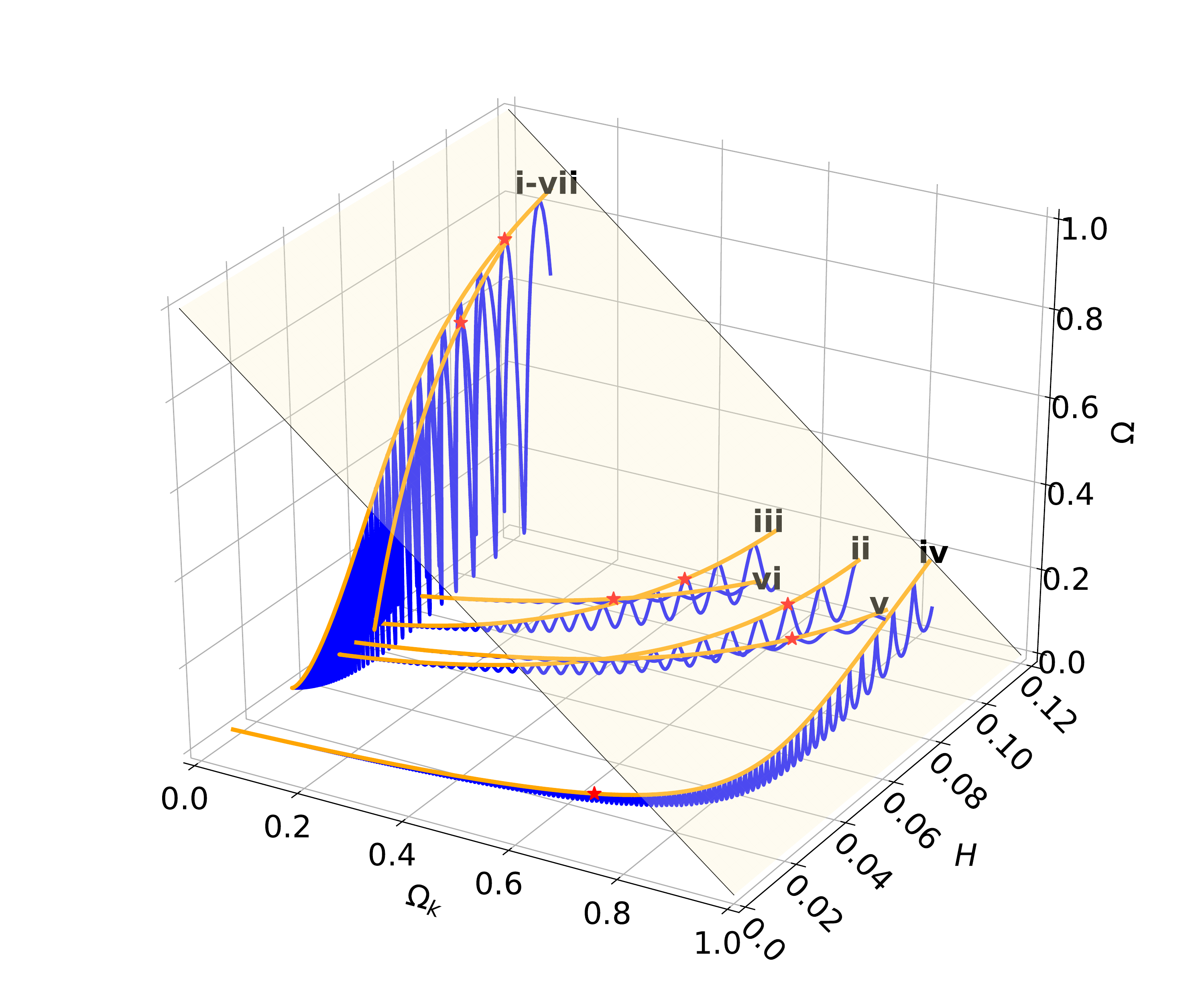}}
    \subfigure[\label{FLRWNonminimallyCCf013Dp} Projections in the space $(\Omega_{m},\Omega_{k},\Omega)$.]{\includegraphics[scale = 0.25]{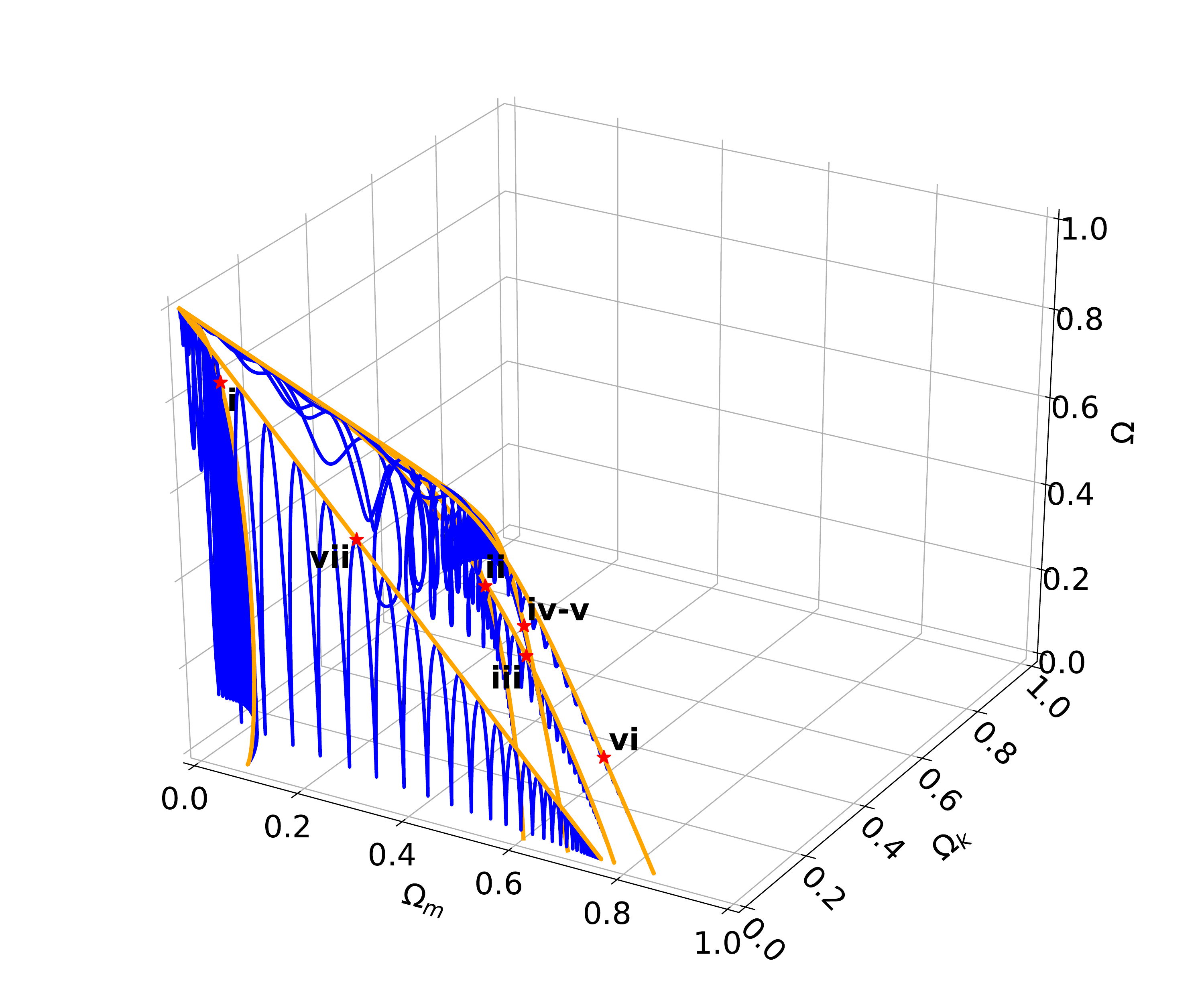}}
  \captionof{figure}{Some solutions of the full system \eqref{XXEQ3.19XX} (blue) and time-averaged system \eqref{EQsEps} (orange) for a scalar field with generalized harmonic potential non-minimally coupled to matter in the FLRW metric when $\lambda = 0.1$, $f=0.1$ and $\gamma=0$. We have used for both systems the initial data sets presented in Table \ref{tab:FLRW}.}
  \label{fig:FLRWNonminimallyCCf01}
\end{minipage}%
\hspace{.02\textwidth}
\begin{minipage}{.48\textwidth}
  \centering
     \subfigure[\label{FLRWNonminimallyBiff013Dm} Projections in the space $(\Omega_{m},H,\Omega)$. The surface is given by the constraint $\Omega=1-\Omega_{m}$.]{\includegraphics[scale = 0.25]{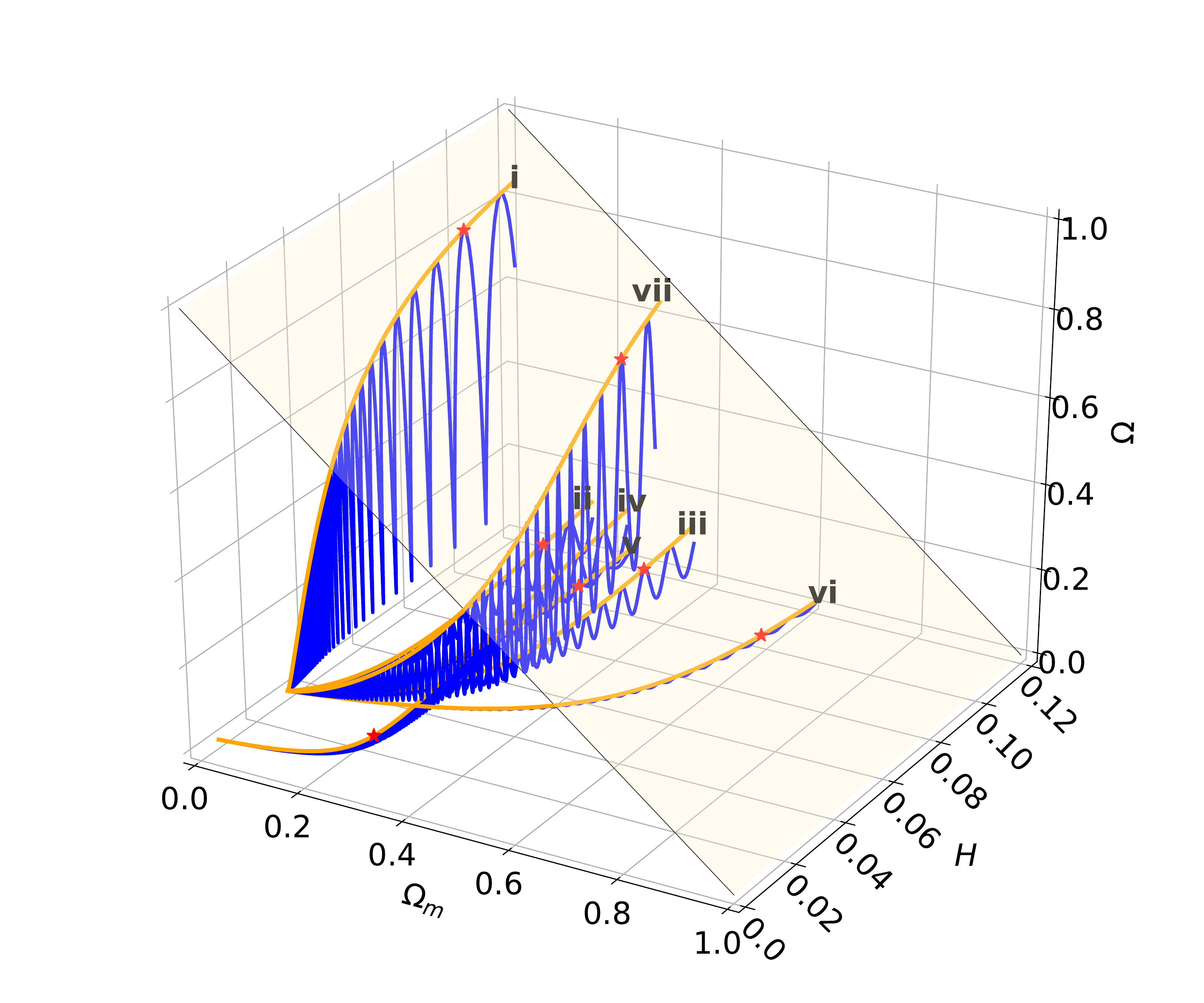}}
    \subfigure[\label{FLRWNonminimallyBiff013Dk} Projections in the space $(\Omega_{k},H,\Omega)$. The surface is given by the constraint $\Omega=1-\Omega_{k}$.]{\includegraphics[scale = 0.25]{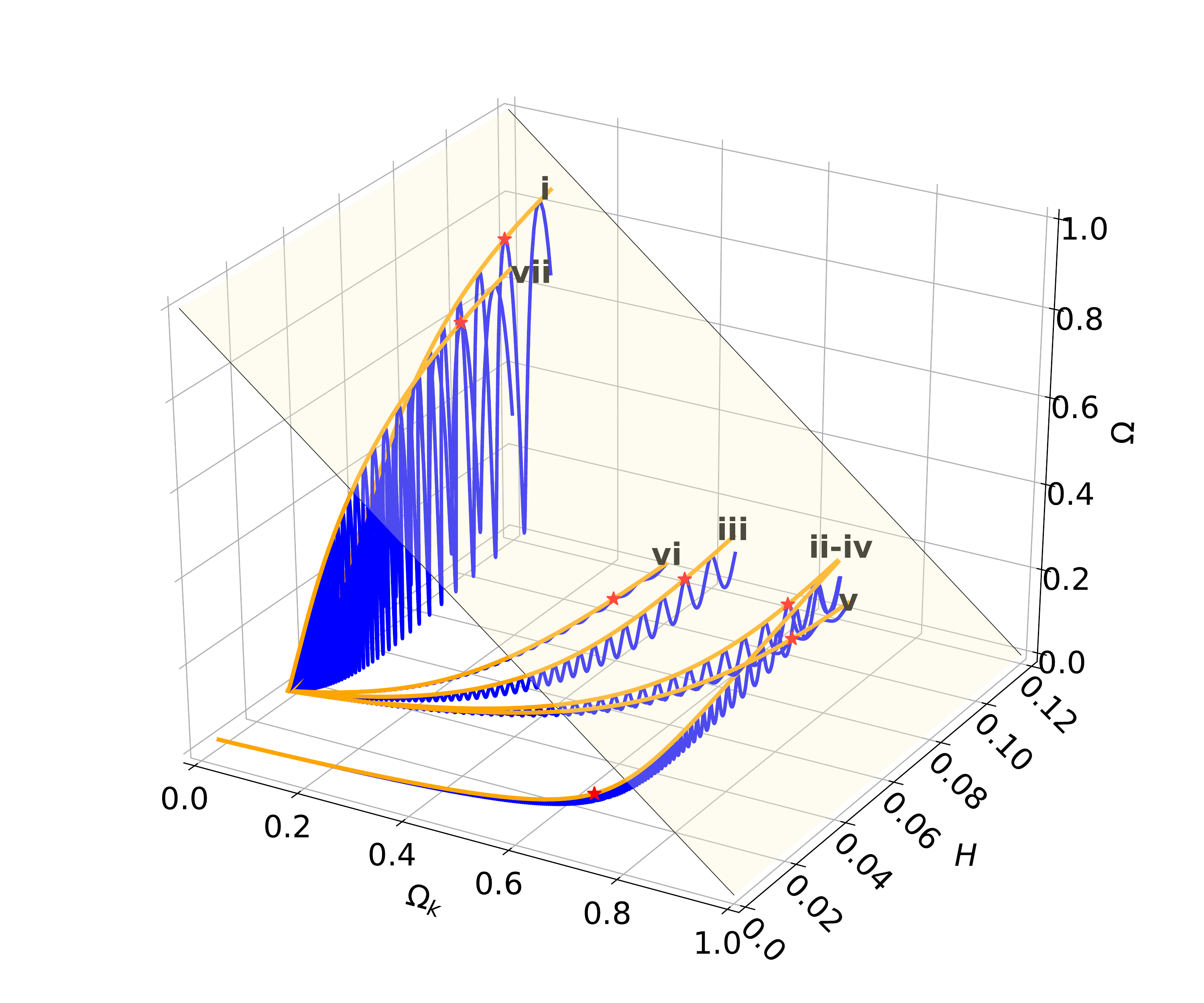}}
    \subfigure[\label{FLRWNonminimallyBiff013Dp} Projections in the space $(\Omega_{m},\Omega_{k},\Omega)$.]{\includegraphics[scale = 0.25]{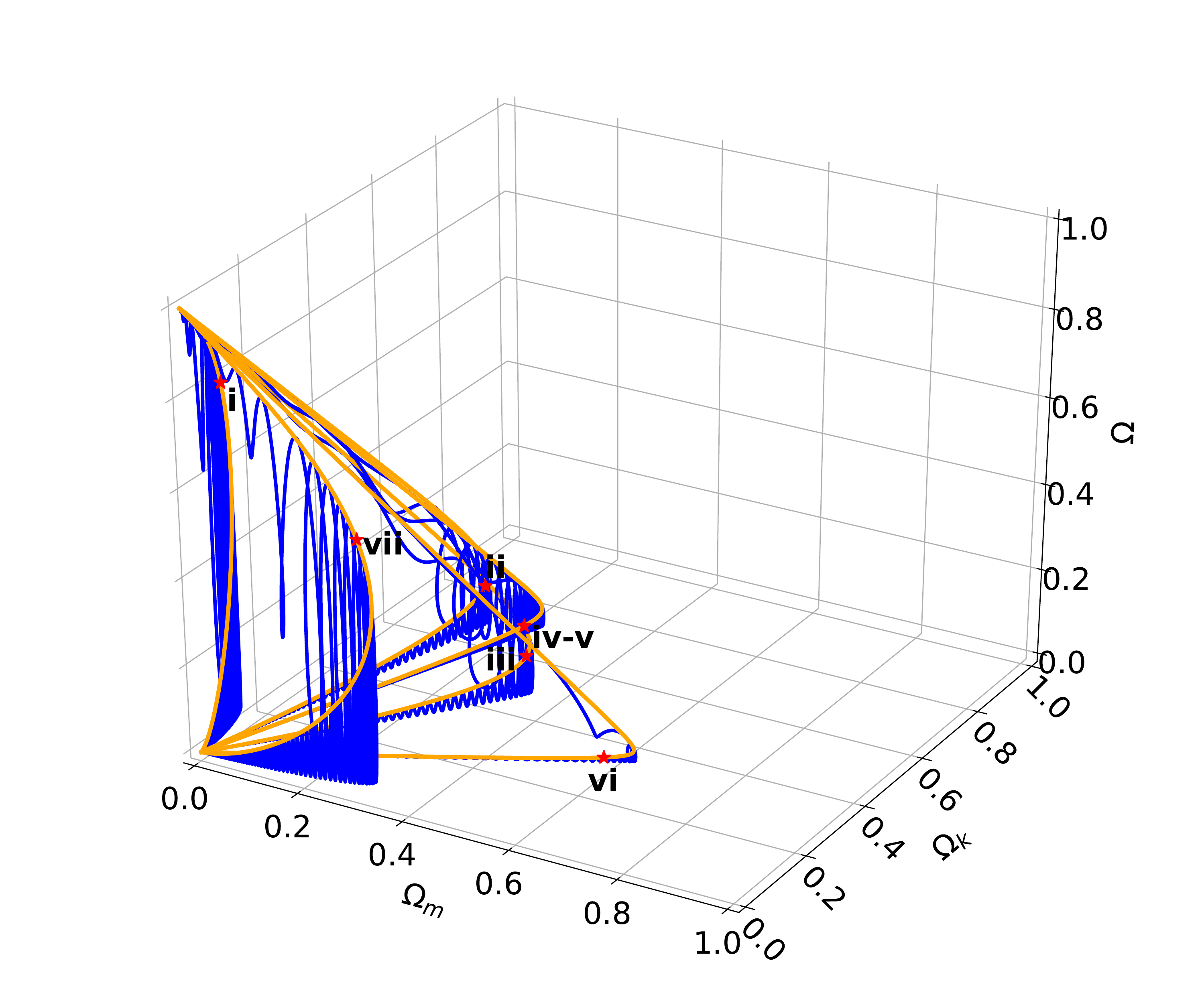}}
  \caption{Some solutions of the full system \eqref{XXEQ3.19XX} (blue) and time-averaged system \eqref{EQsEps} (orange) for a scalar field with generalized harmonic potential non-minimally coupled to matter in the FLRW metric when $\lambda = 0.1$, $f=0.1$ and $\gamma=\frac{2}{3}$. We have used for both systems the initial data sets presented in Table \ref{tab:FLRW}.  }
  \label{fig:FLRWNonminimallyBiff01}
\end{minipage}
\end{figure}

\begin{figure}[ht!]
\centering
\begin{minipage}{.48\textwidth}
  \centering
    \subfigure[\label{FLRWNonminimallyDustf013Dm} Projections in the space $(\Omega_{m},H,\Omega)$. The surface is given by the constraint $\Omega=1-\Omega_{m}$.]{\includegraphics[scale = 0.25]{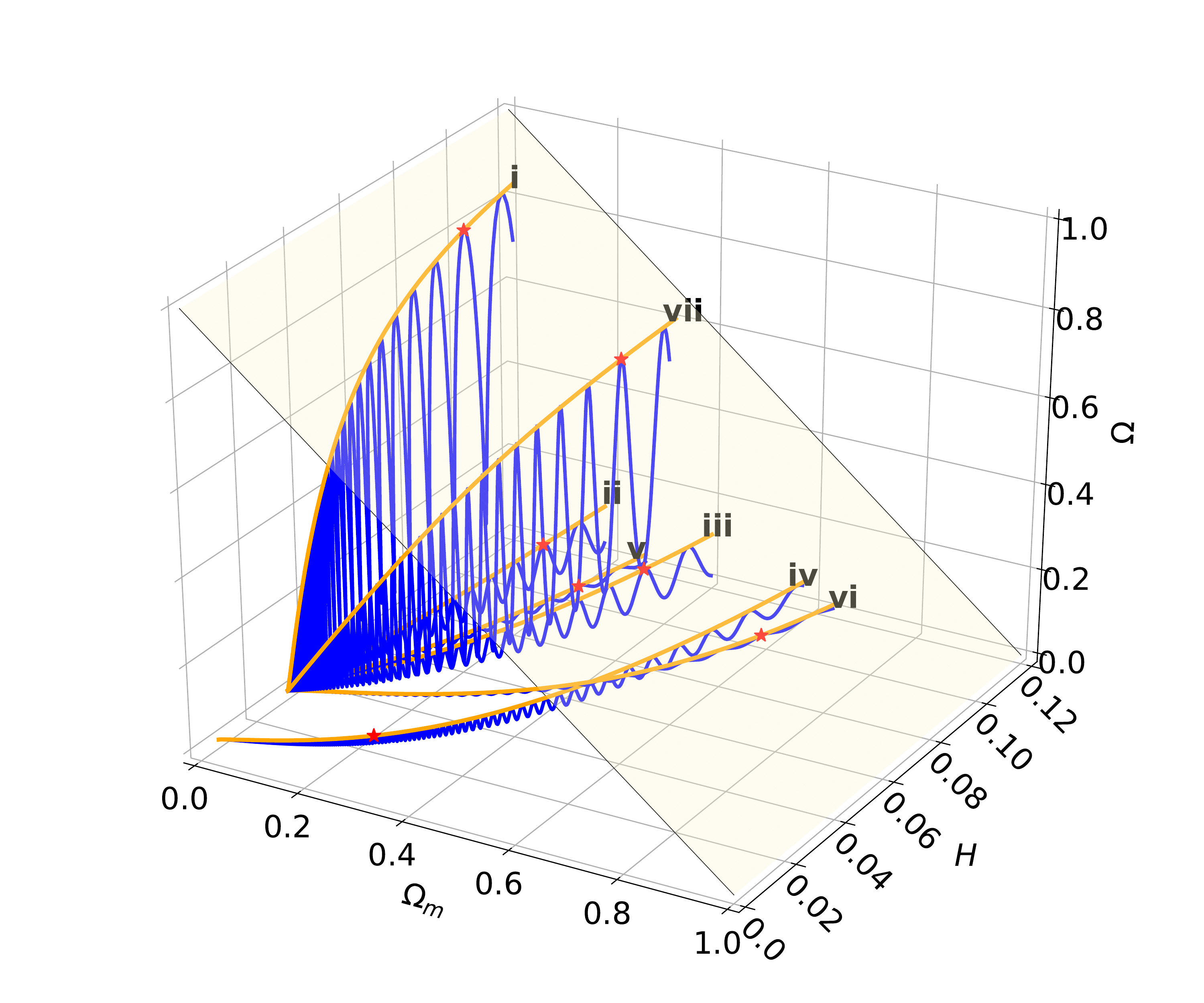}}
    \subfigure[\label{FLRWNonminimallyDustf013Dk} Projections in the space $(\Omega_{k},H,\Omega)$. The surface is given by the constraint $\Omega=1-\Omega_{k}$.]{\includegraphics[scale = 0.25]{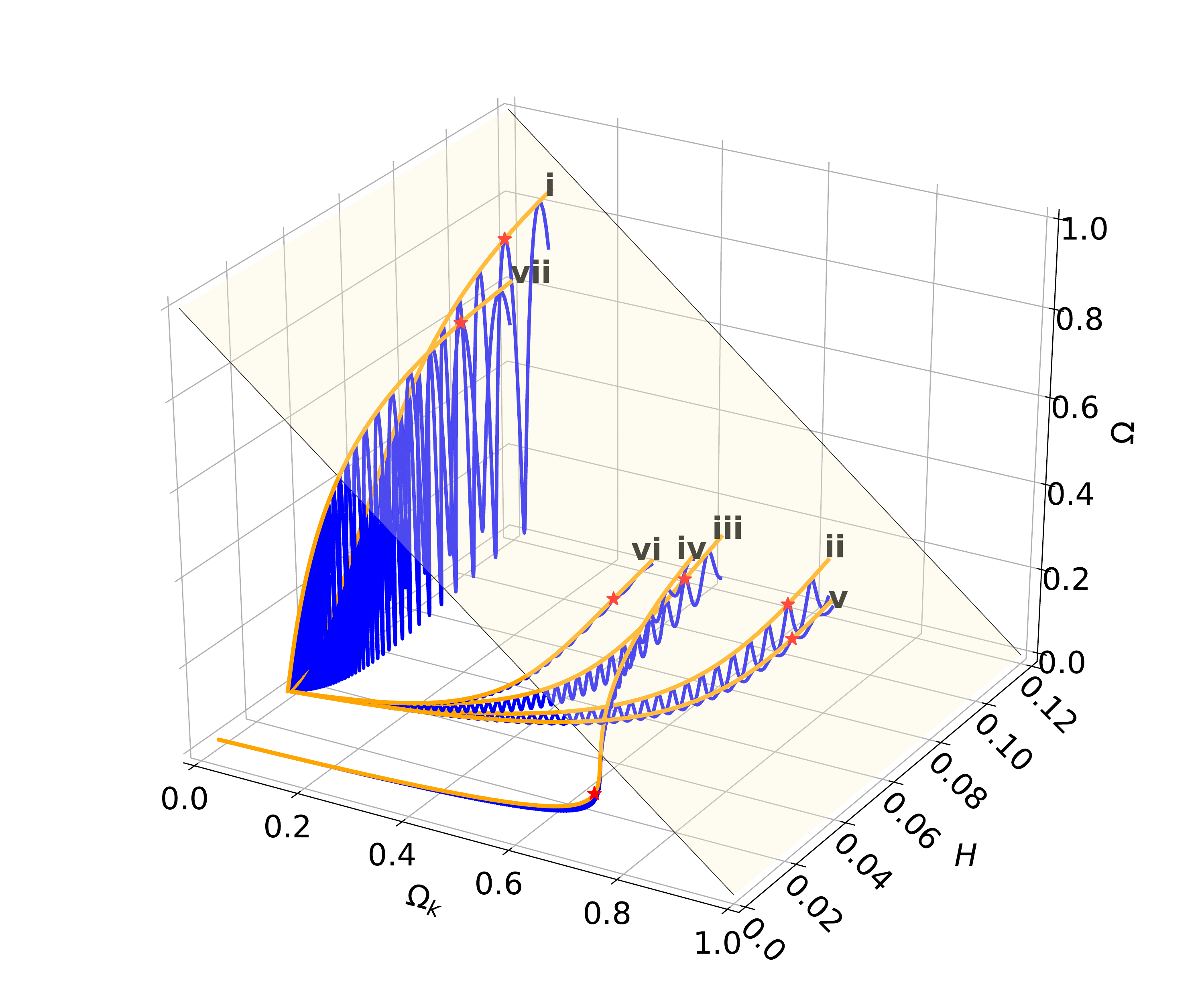}}
    \subfigure[\label{FLRWNonminimallyDustf013Dp} Projections in the space $(\Omega_{m},\Omega_{k},\Omega)$.]{\includegraphics[scale = 0.25]{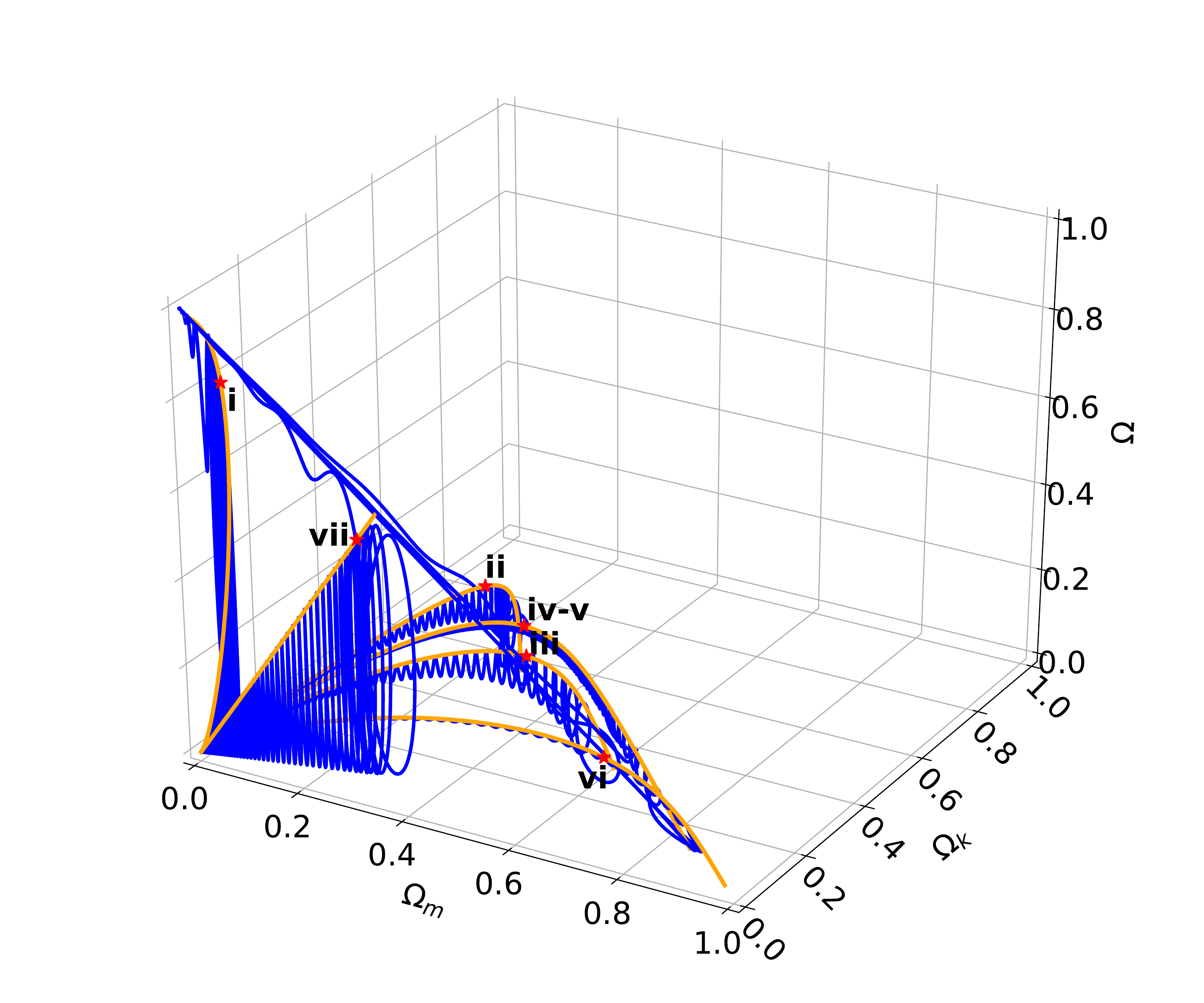}}
  \captionof{figure}{Some solutions of the full system \eqref{XXEQ3.19XX} (blue) and time-averaged system \eqref{EQsEps} (orange) for a scalar field with generalized harmonic potential non-minimally coupled to matter in the FLRW metric when $\lambda = 0.1$, $f=0.1$ and $\gamma=1$. We have used for both systems the initial data sets presented in Table \ref{tab:FLRW}.}
  \label{fig:FLRWNonminimallyDustf01}
\end{minipage}%
\hspace{.02\textwidth}
\begin{minipage}{.48\textwidth}
  \centering
     \subfigure[\label{FLRWNonminimallyStifff013Dm} Projections in the space $(\Omega_{m},H,\Omega)$. The surface is given by the constraint $\Omega=1-\Omega_{m}$.]{\includegraphics[scale = 0.25]{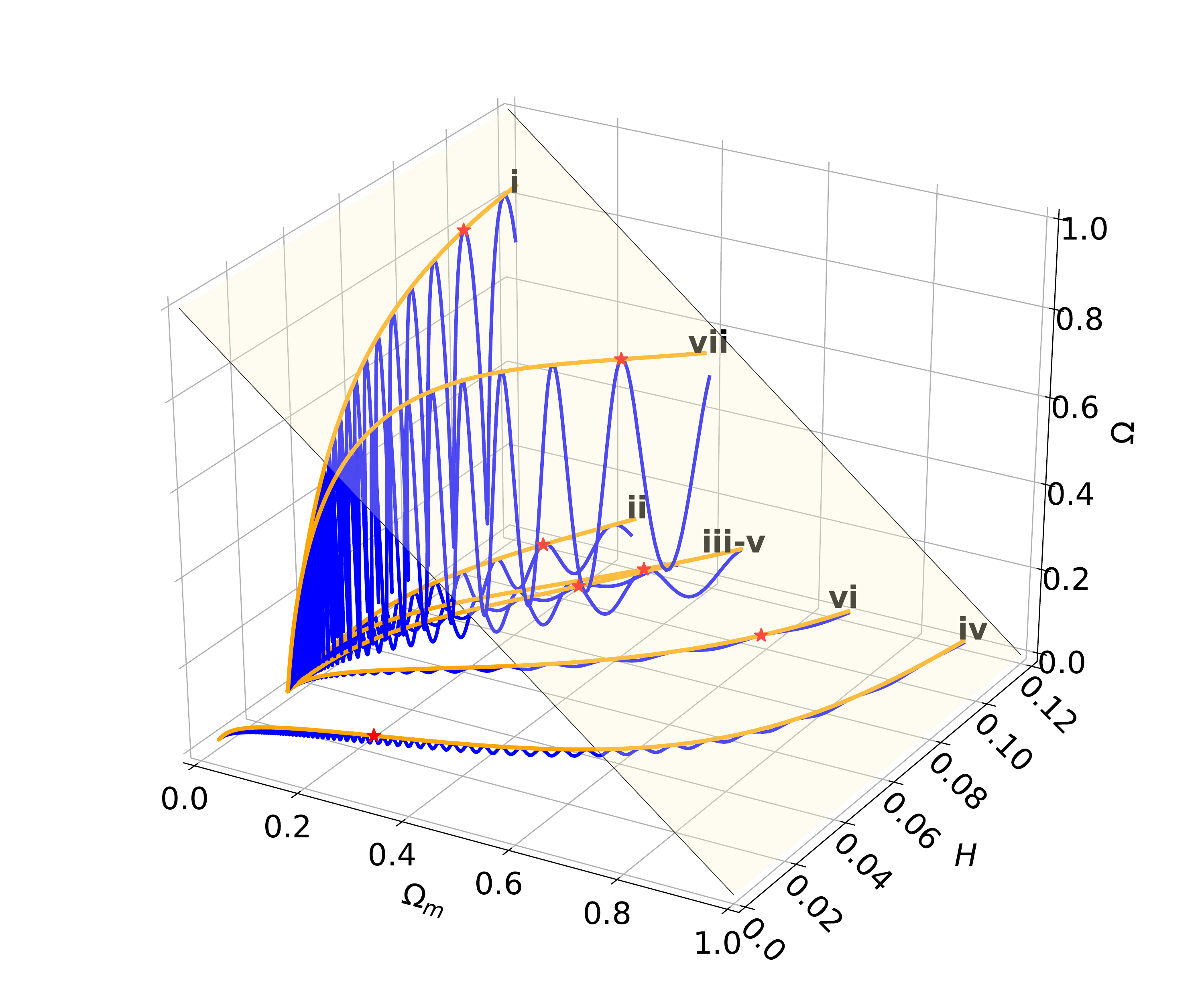}}
    \subfigure[\label{FLRWNonminimallyStifff013Dk} Projections in the space $(\Omega_{k},H,\Omega)$. The surface is given by the constraint $\Omega=1-\Omega_{k}$.]{\includegraphics[scale = 0.25]{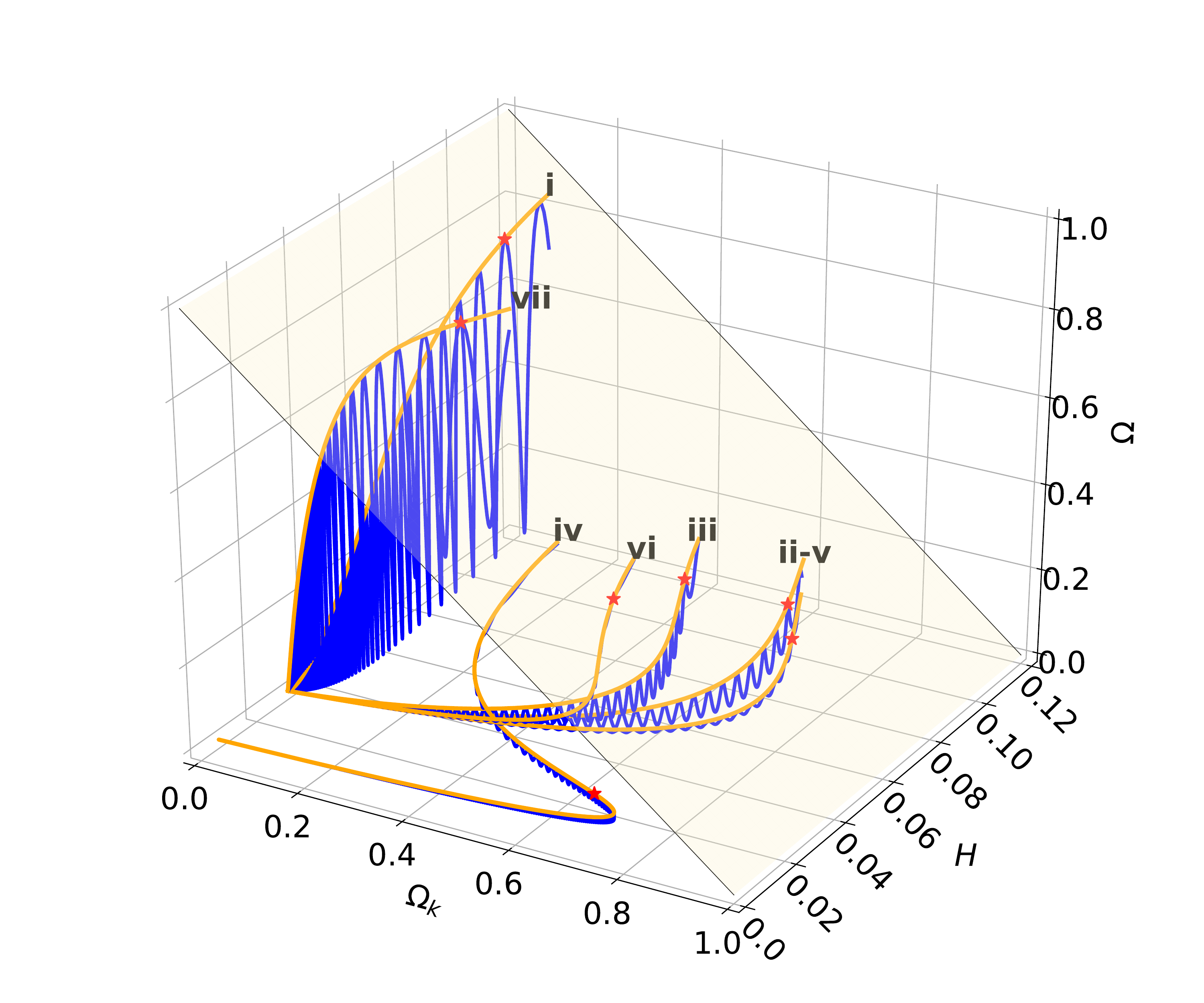}}
    \subfigure[\label{FLRWNonminimallyStifff013Dp} Projections in the space $(\Omega_{m},\Omega_{k},\Omega)$.]{\includegraphics[scale = 0.25]{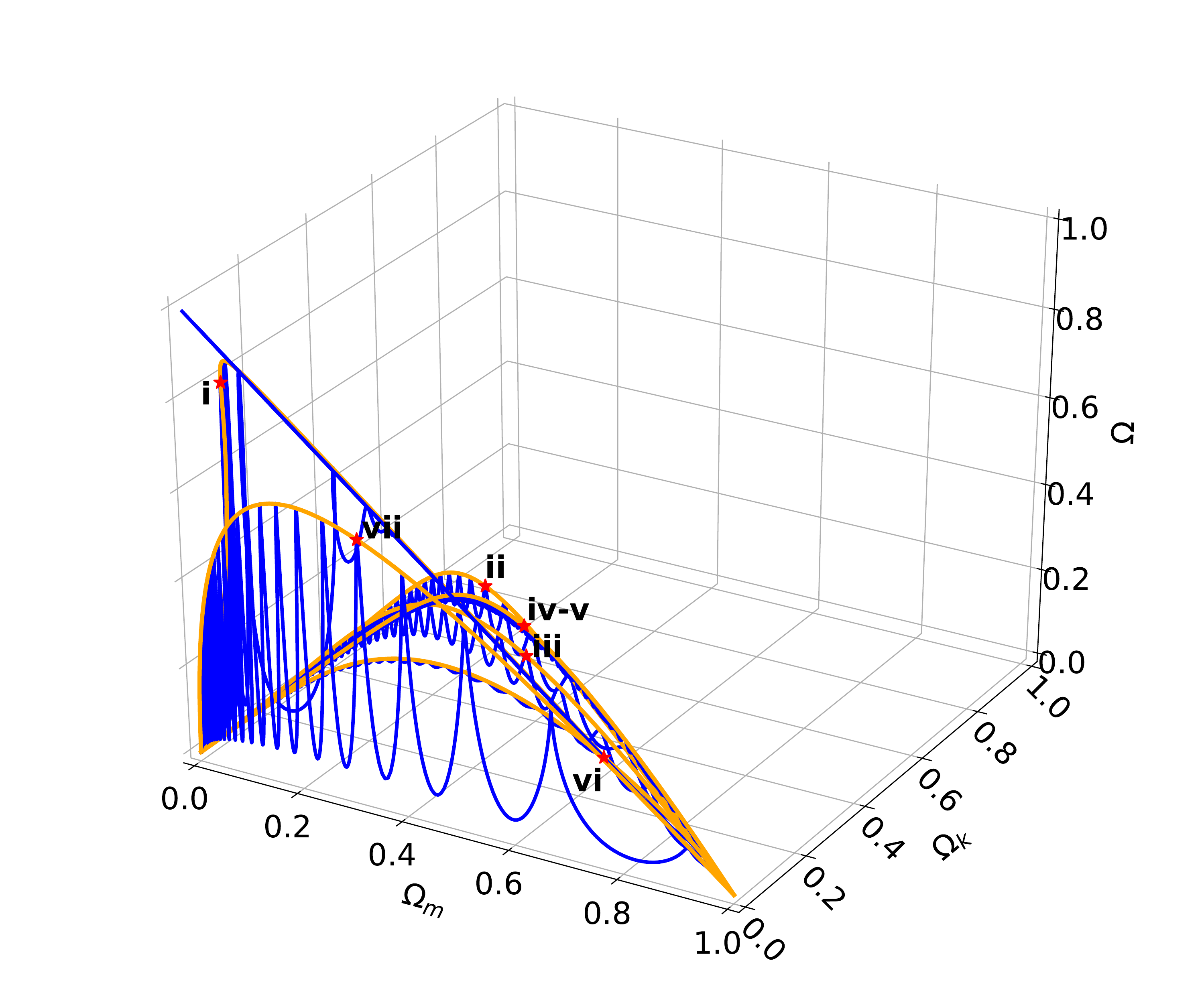}}
  \caption{Some solutions of the full system \eqref{XXEQ3.19XX} (blue) and time-averaged system \eqref{EQsEps} (orange) for a scalar field with generalized harmonic potential non-minimally coupled to matter in the FLRW metric when $\lambda = 0.1$, $f=0.1$ and $\gamma=2$. We have used for both systems the initial data sets presented in Table \ref{tab:FLRW}.}
  \label{fig:FLRWNonminimallyStifff01}
\end{minipage}
\end{figure}

\begin{figure}[ht!]
\centering
\begin{minipage}{.48\textwidth}
  \centering
    \subfigure[\label{FLRWNonminimallyCCf033Dm} Projections in the space $(\Omega_{m},H,\Omega)$. The surface is given by the constraint $\Omega=1-\Omega_{m}$.]{\includegraphics[scale = 0.25]{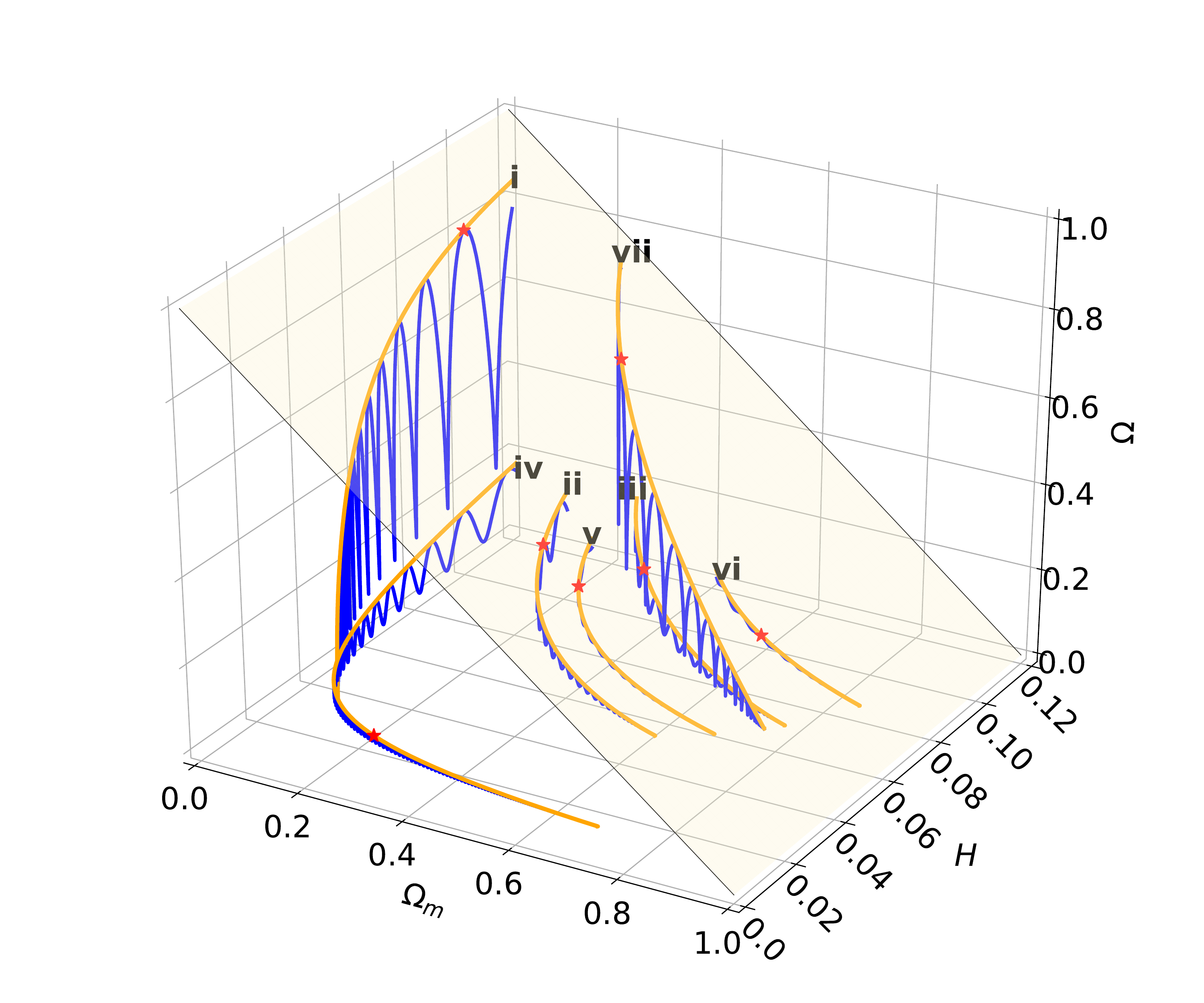}}
    \subfigure[\label{FLRWNonminimallyCCf033Dk} Projections in the space $(\Omega_{k},H,\Omega)$. The surface is given by the constraint $\Omega=1-\Omega_{k}$.]{\includegraphics[scale = 0.25]{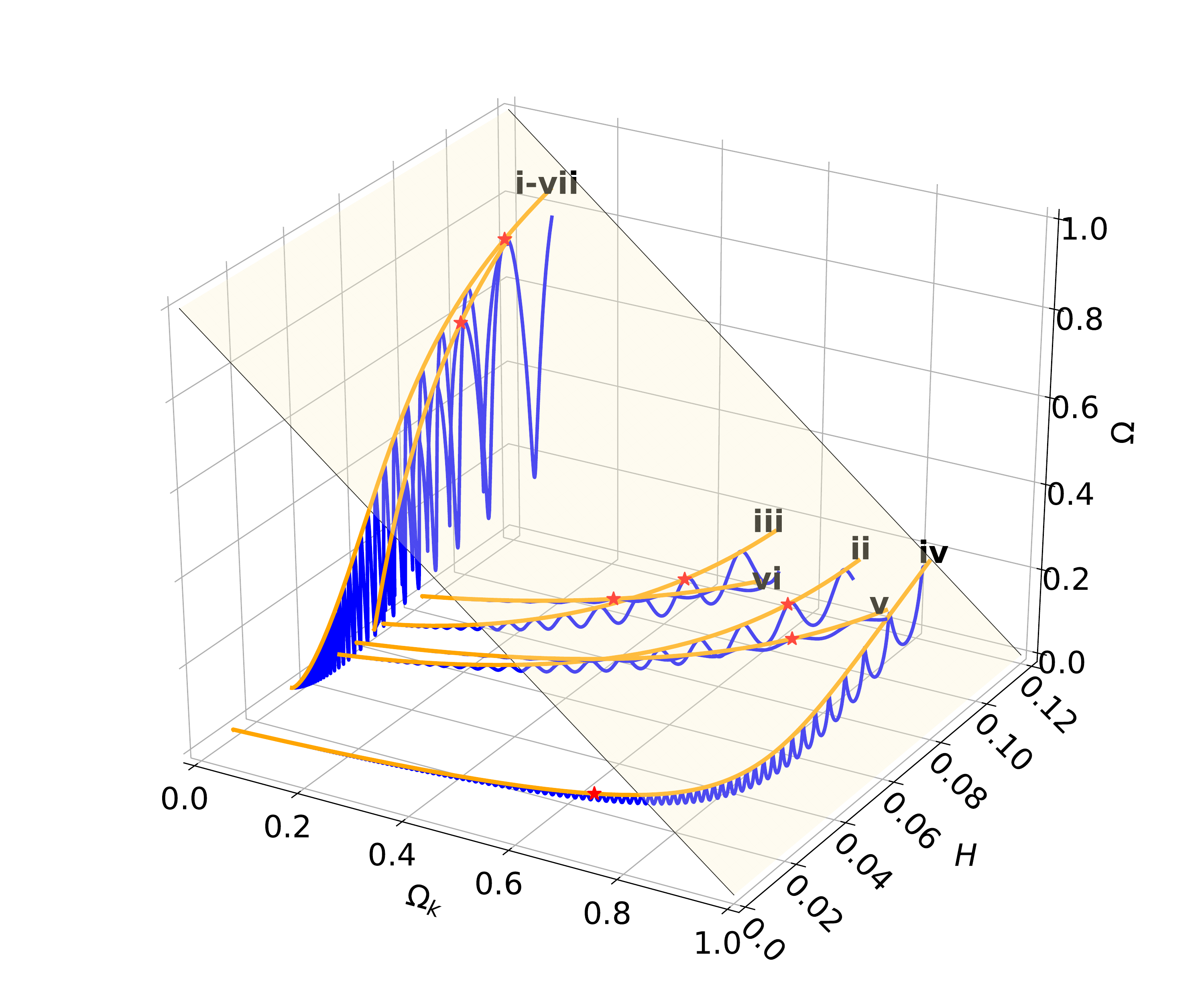}}
    \subfigure[\label{FLRWNonminimallyCCf033Dp} Projections in the space $(\Omega_{m},\Omega_{k},\Omega)$.]{\includegraphics[scale = 0.25]{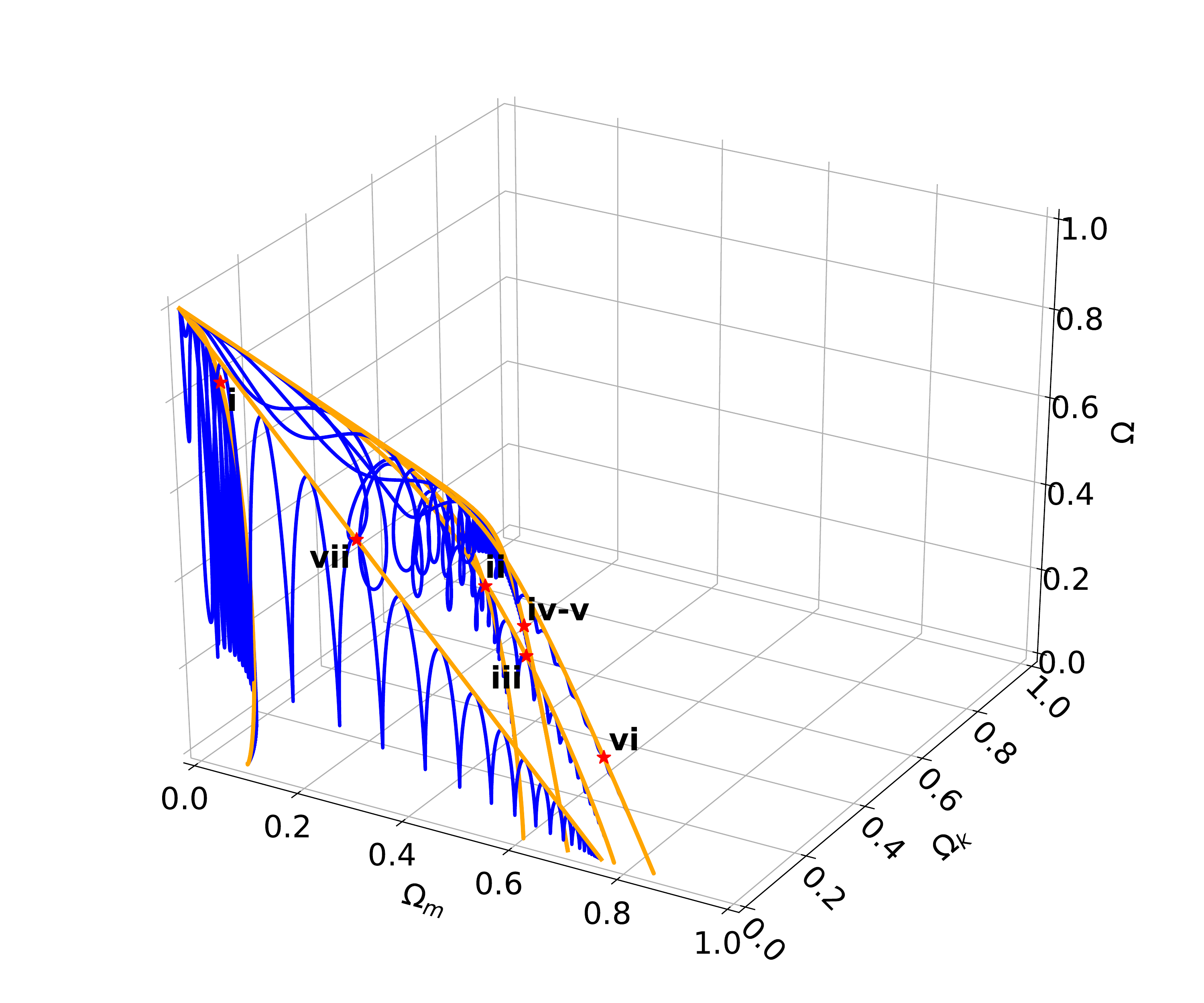}}
  \captionof{figure}{Some solutions of the full system \eqref{XXEQ3.19XX} (blue) and time-averaged system \eqref{EQsEps} (orange) for a scalar field with generalized harmonic potential non-minimally coupled to matter in the FLRW metric when $\lambda = 0.1$, $f=0.3$ and $\gamma=0$. We have used for both systems the initial data sets presented in Table \ref{tab:FLRW}.}
  \label{fig:FLRWNonminimallyCCf03}
\end{minipage}%
\hspace{.02\textwidth}
\begin{minipage}{.48\textwidth}
  \centering
     \subfigure[\label{FLRWNonminimallyBiff033Dm} Projections in the space $(\Omega_{m},H,\Omega)$. The surface is given by the constraint $\Omega=1-\Omega_{m}$.]{\includegraphics[scale = 0.25]{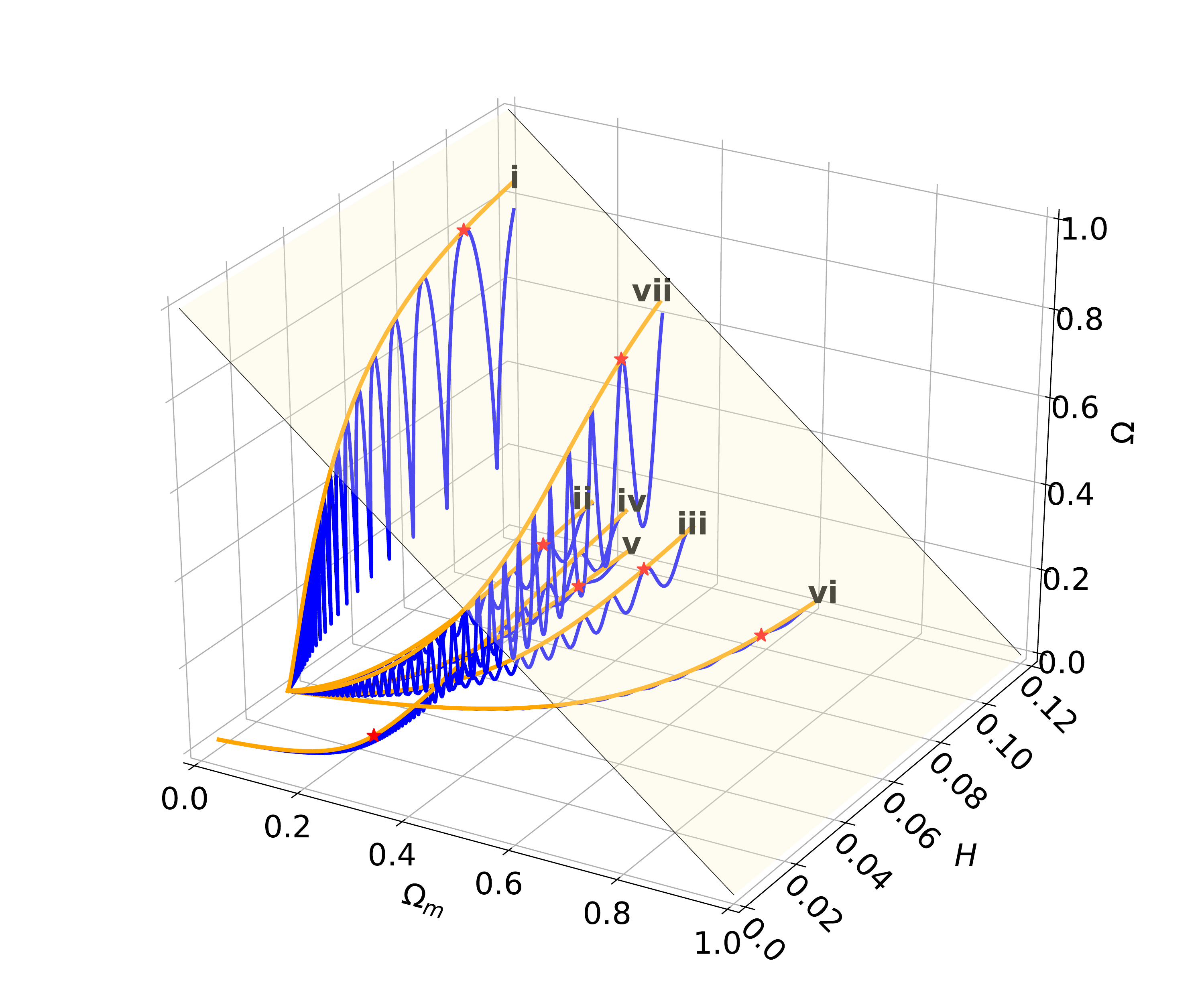}}
    \subfigure[\label{FLRWNonminimallyBiff033Dk} Projections in the space $(\Omega_{k},H,\Omega)$. The surface is given by the constraint $\Omega=1-\Omega_{k}$.]{\includegraphics[scale = 0.25]{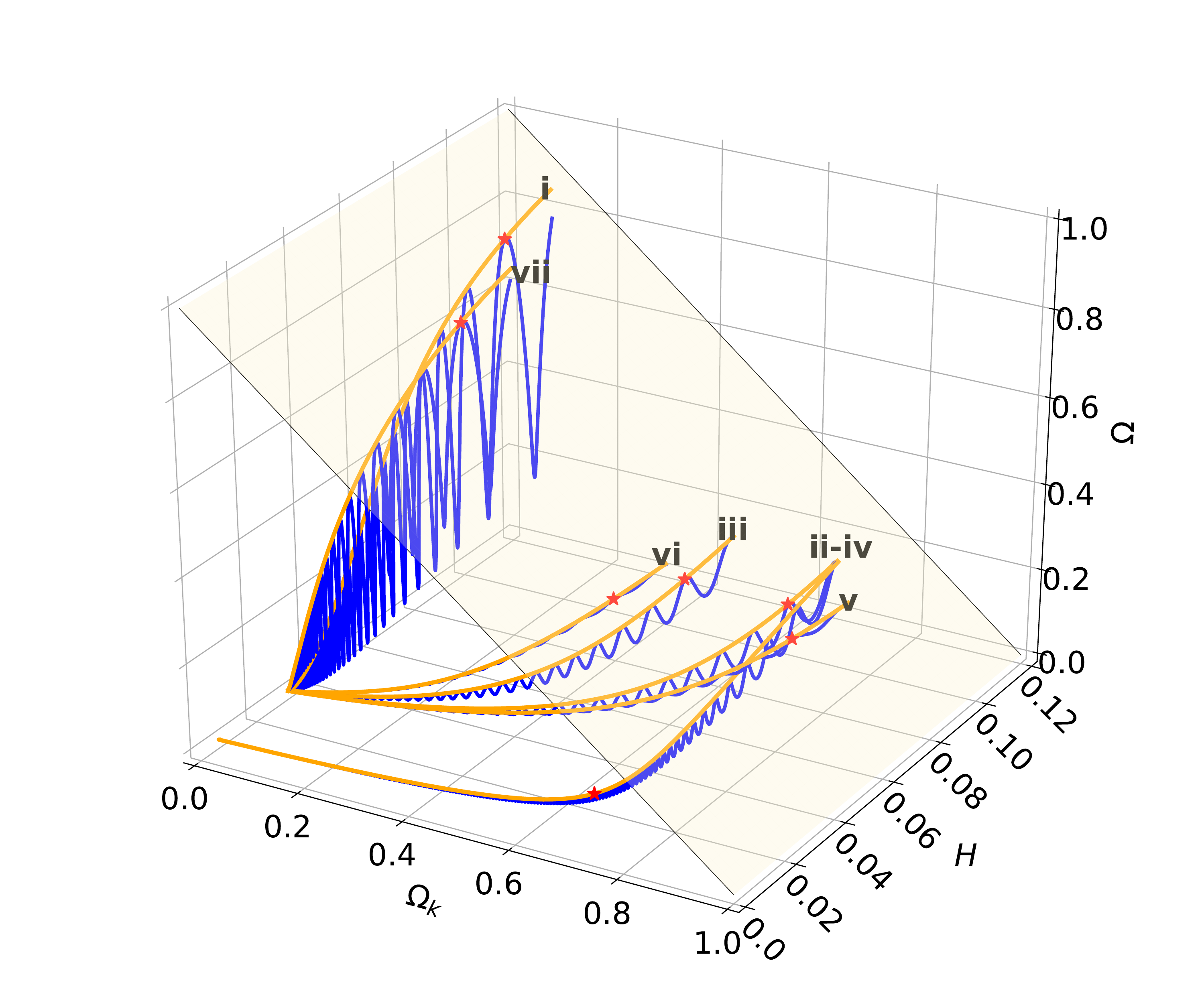}}
    \subfigure[\label{FLRWNonminimallyBiff033Dp} Projections in the space $(\Omega_{m},\Omega_{k},\Omega)$.]{\includegraphics[scale = 0.25]{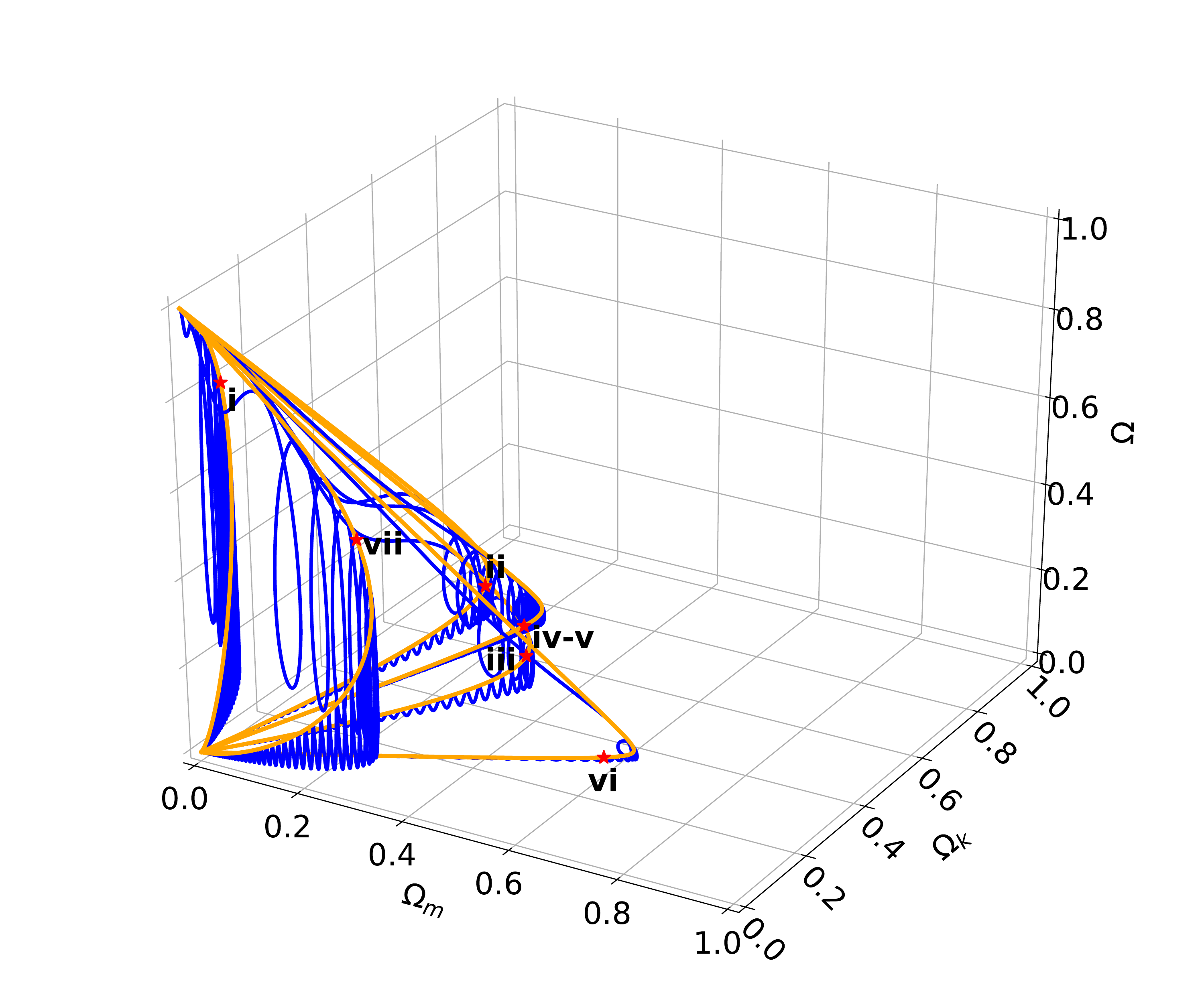}}
  \caption{Some solutions of the full system \eqref{XXEQ3.19XX} (blue) and time-averaged system \eqref{EQsEps} (orange) for a scalar field with generalized harmonic potential non-minimally coupled to matter in the FLRW metric when $\lambda = 0.1$, $f=0.3$ and $\gamma=\frac{2}{3}$. We have used for both systems the initial data sets presented in Table \ref{tab:FLRW}.}
  \label{fig:FLRWNonminimallyBiff03}
\end{minipage}
\end{figure}

\begin{figure}[ht!]
\centering
\begin{minipage}{.48\textwidth}
  \centering
    \subfigure[\label{FLRWNonminimallyDustf033Dm} Projections in the space $(\Omega_{m},H,\Omega)$. The surface is given by the constraint $\Omega=1-\Omega_{m}$.]{\includegraphics[scale = 0.25]{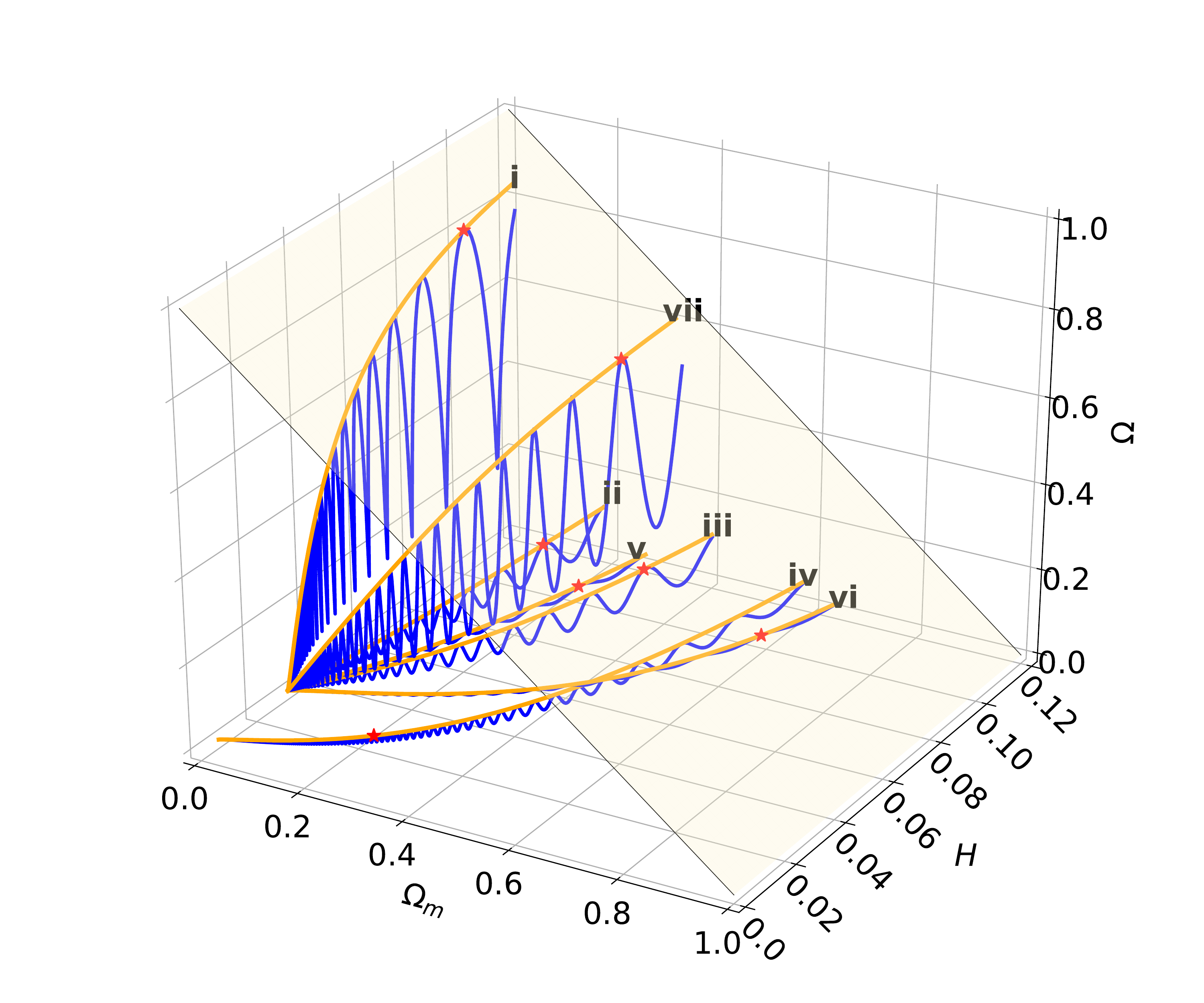}}
    \subfigure[\label{FLRWNonminimallyDustf033Dk} Projections in the space $(\Omega_{k},H,\Omega)$. The surface is given by the constraint $\Omega=1-\Omega_{k}$.]{\includegraphics[scale = 0.25]{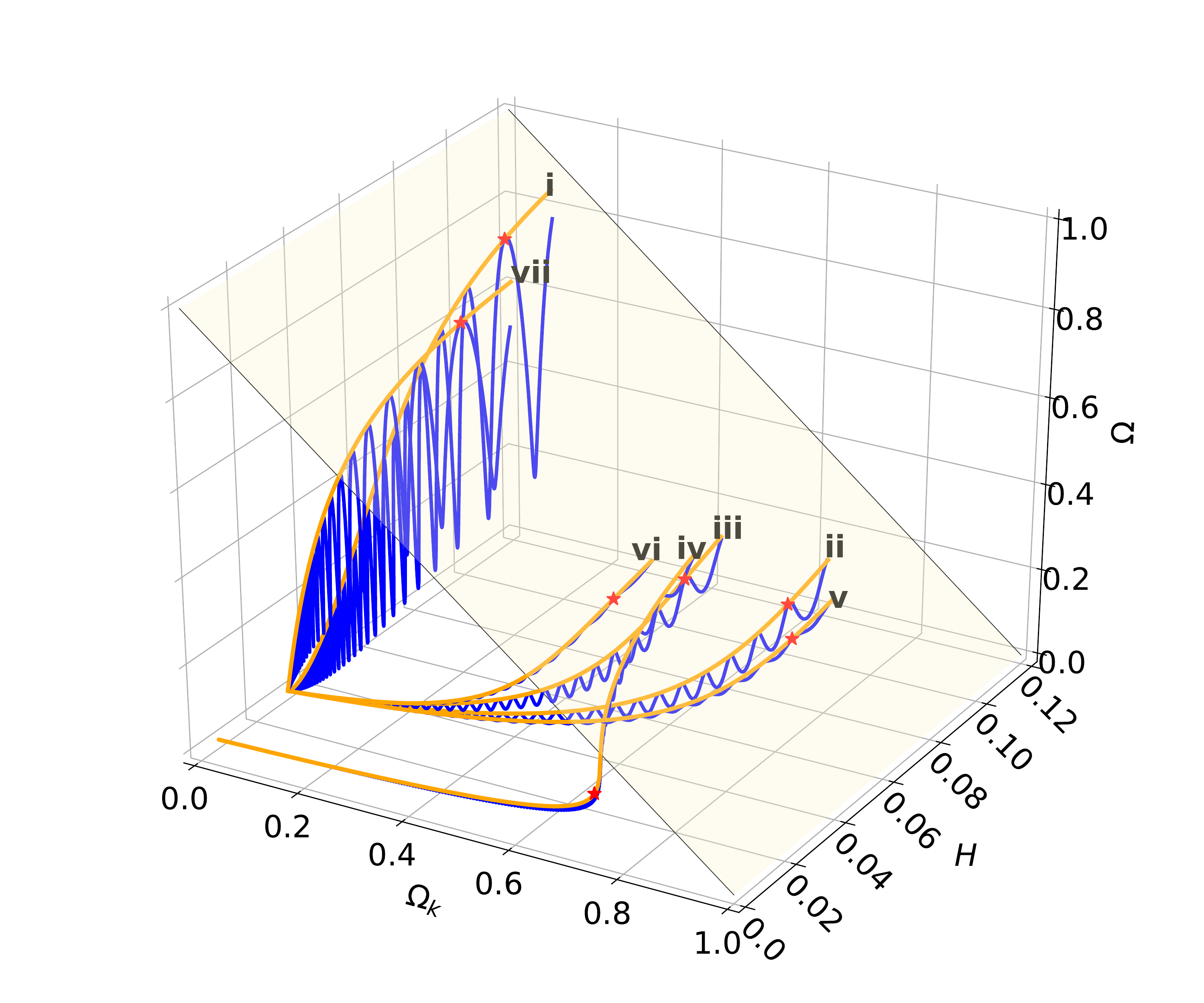}}
    \subfigure[\label{FLRWNonminimallyDustf033Dp} Projections in the space $(\Omega_{m},\Omega_{k},\Omega)$.]{\includegraphics[scale = 0.25]{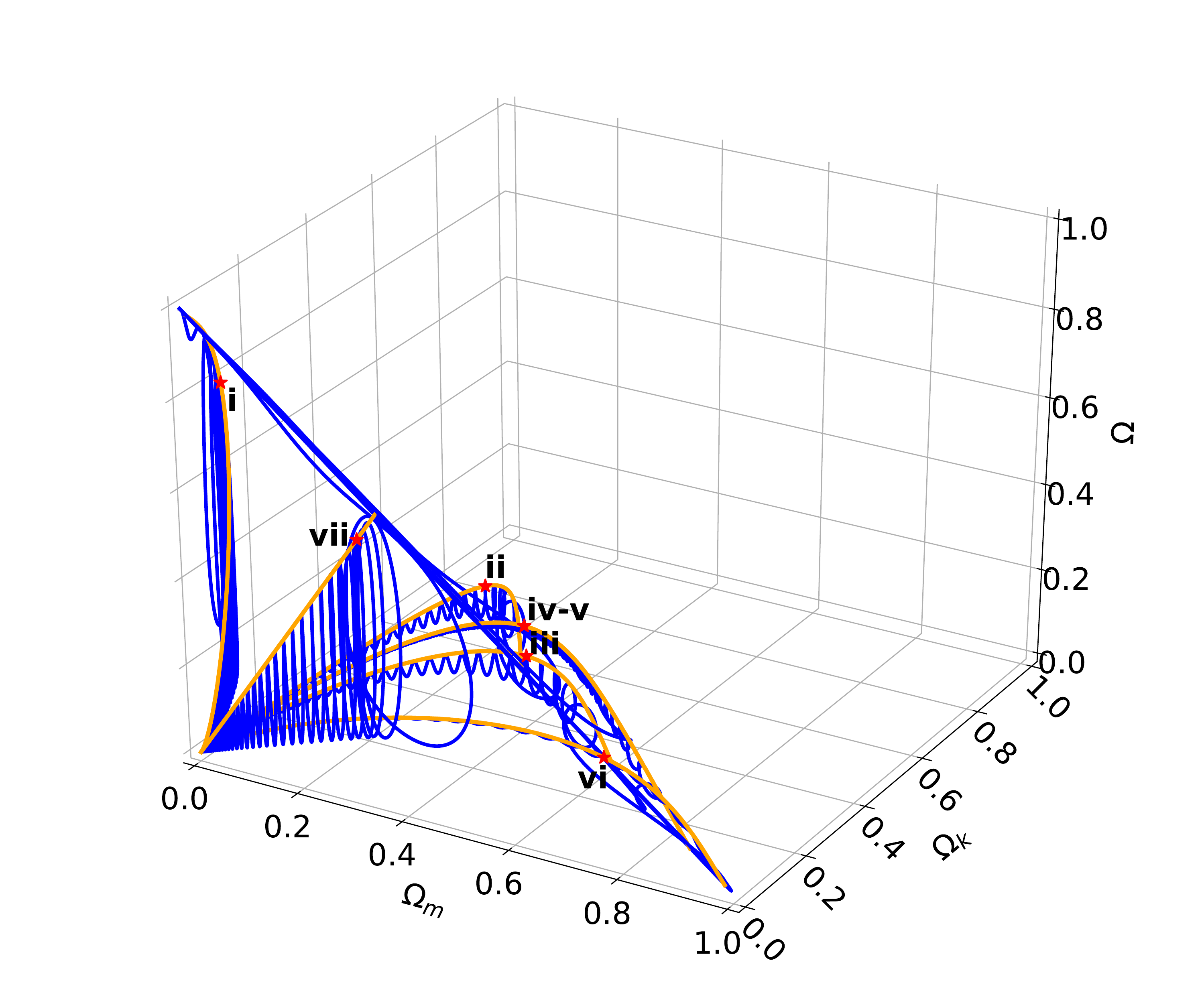}}
  \captionof{figure}{Some solutions of the full system \eqref{XXEQ3.19XX} (blue) and time-averaged system \eqref{EQsEps} (orange) for a scalar field with generalized harmonic potential non-minimally coupled to matter in the FLRW metric when $\lambda = 0.1$, $f=0.3$ and $\gamma=1$. We have used for both systems the initial data sets presented in Table \ref{tab:FLRW}.}
  \label{fig:FLRWNonminimallyDustf03}
\end{minipage}%
\hspace{.02\textwidth}
\begin{minipage}{.48\textwidth}
  \centering
     \subfigure[\label{FLRWNonminimallyStifff033Dm} Projections in the space $(\Omega_{m},H,\Omega)$. The surface is given by the constraint $\Omega=1-\Omega_{m}$.]{\includegraphics[scale = 0.25]{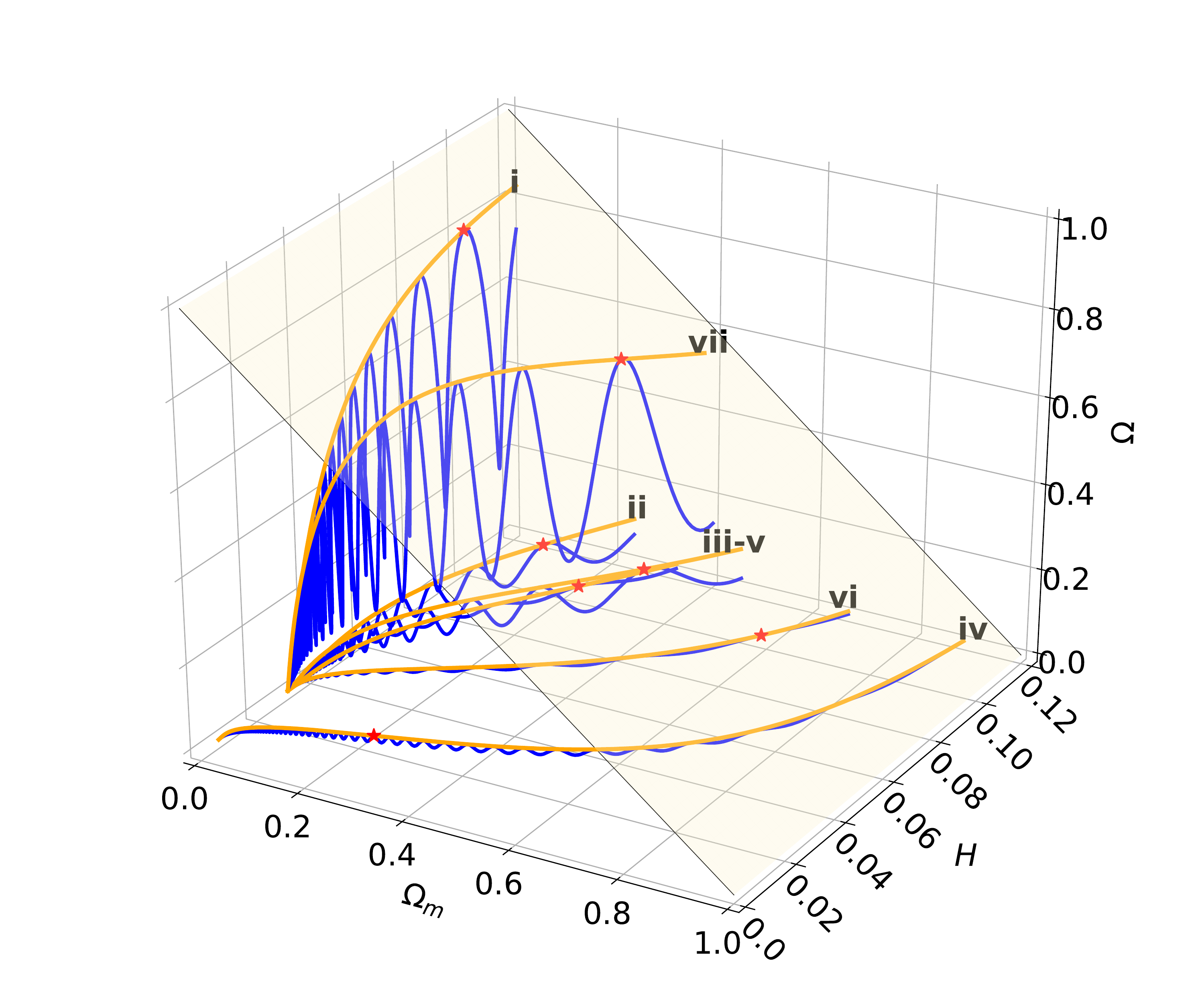}}
    \subfigure[\label{FLRWNonminimallyStifff033Dk} Projections in the space $(\Omega_{k},H,\Omega)$. The surface is given by the constraint $\Omega=1-\Omega_{k}$.]{\includegraphics[scale = 0.25]{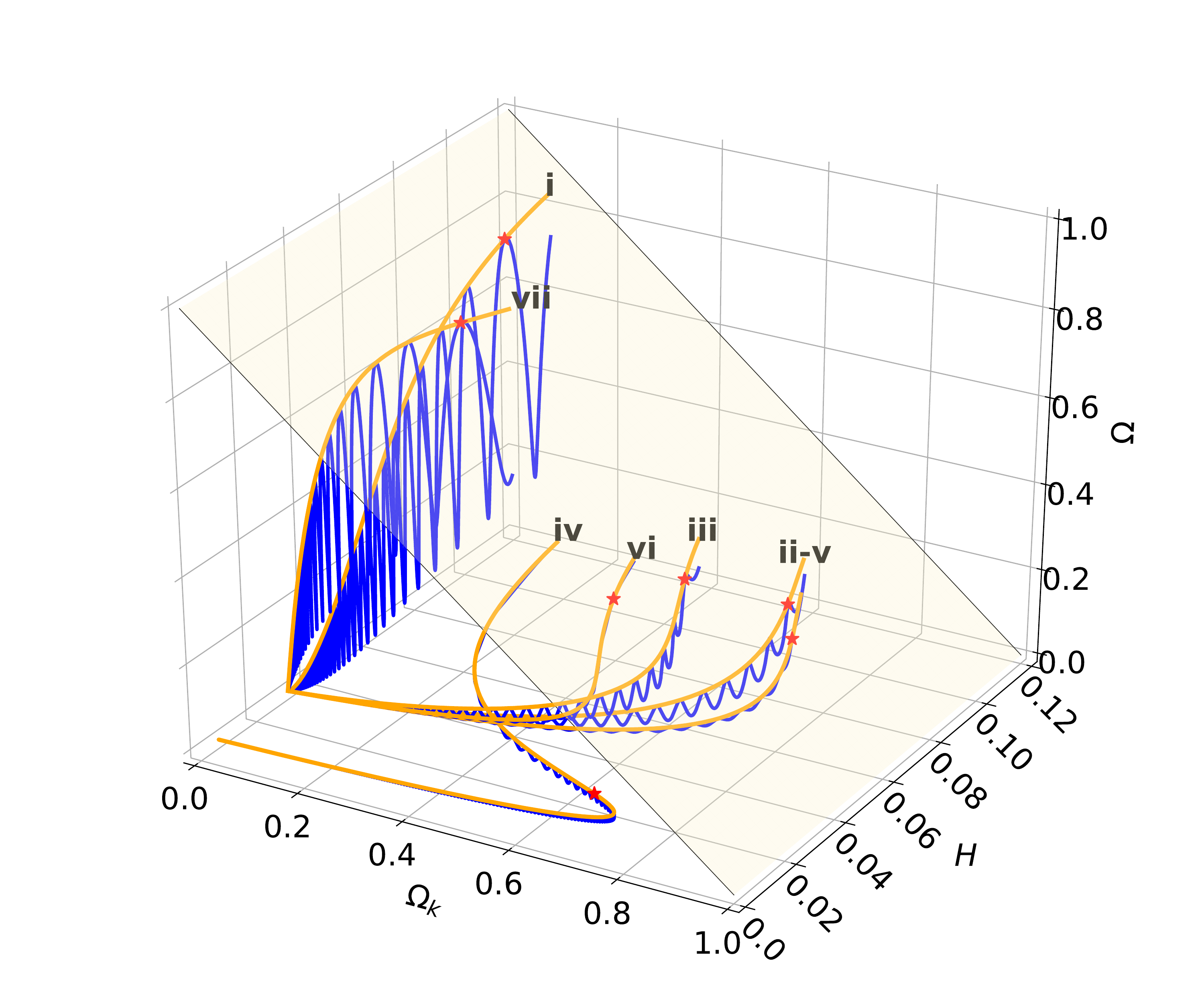}}
    \subfigure[\label{FLRWNonminimallyStifff033Dp} Projections in the space $(\Omega_{m},\Omega_{k},\Omega)$.]{\includegraphics[scale = 0.25]{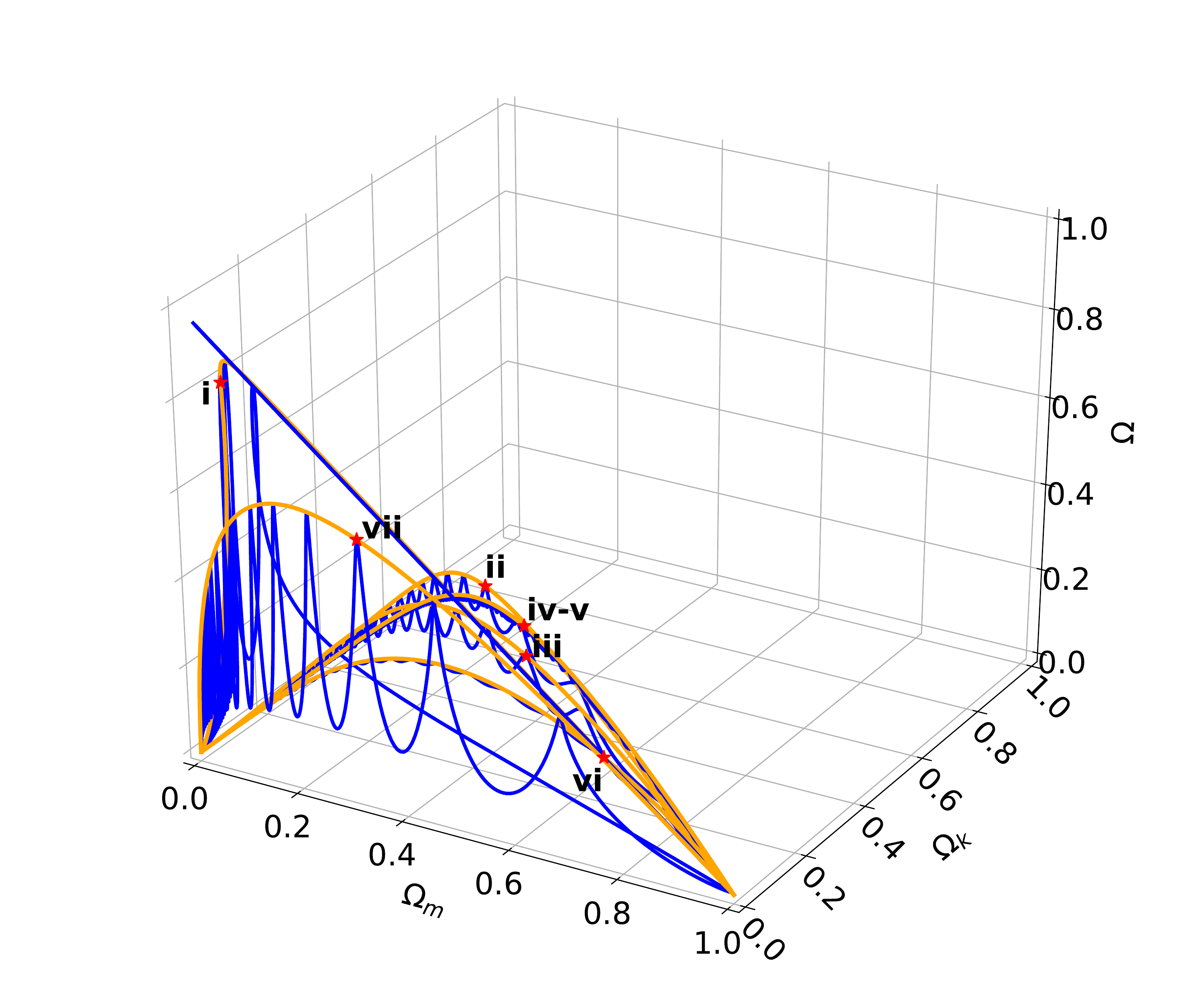}}
  \caption{Some solutions of the full system \eqref{XXEQ3.19XX} (blue) and time-averaged system \eqref{EQsEps} (orange) for a scalar field with generalized harmonic potential non-minimally coupled to matter in the FLRW metric when $\lambda = 0.1$, $f=0.3$ and $\gamma=2$. We have used for both systems the initial data sets presented in Table \ref{tab:FLRW}.}
  \label{fig:FLRWNonminimallyStifff03}
\end{minipage}
\end{figure}

\begin{figure}[ht!]
\centering
\begin{minipage}{.48\textwidth}
  \centering
    \subfigure[\label{FLRWNonminimallyCCf093Dm} Projections in the space $(\Omega_{m},H,\Omega)$. The surface is given by the constraint $\Omega=1-\Omega_{m}$.]{\includegraphics[scale = 0.25]{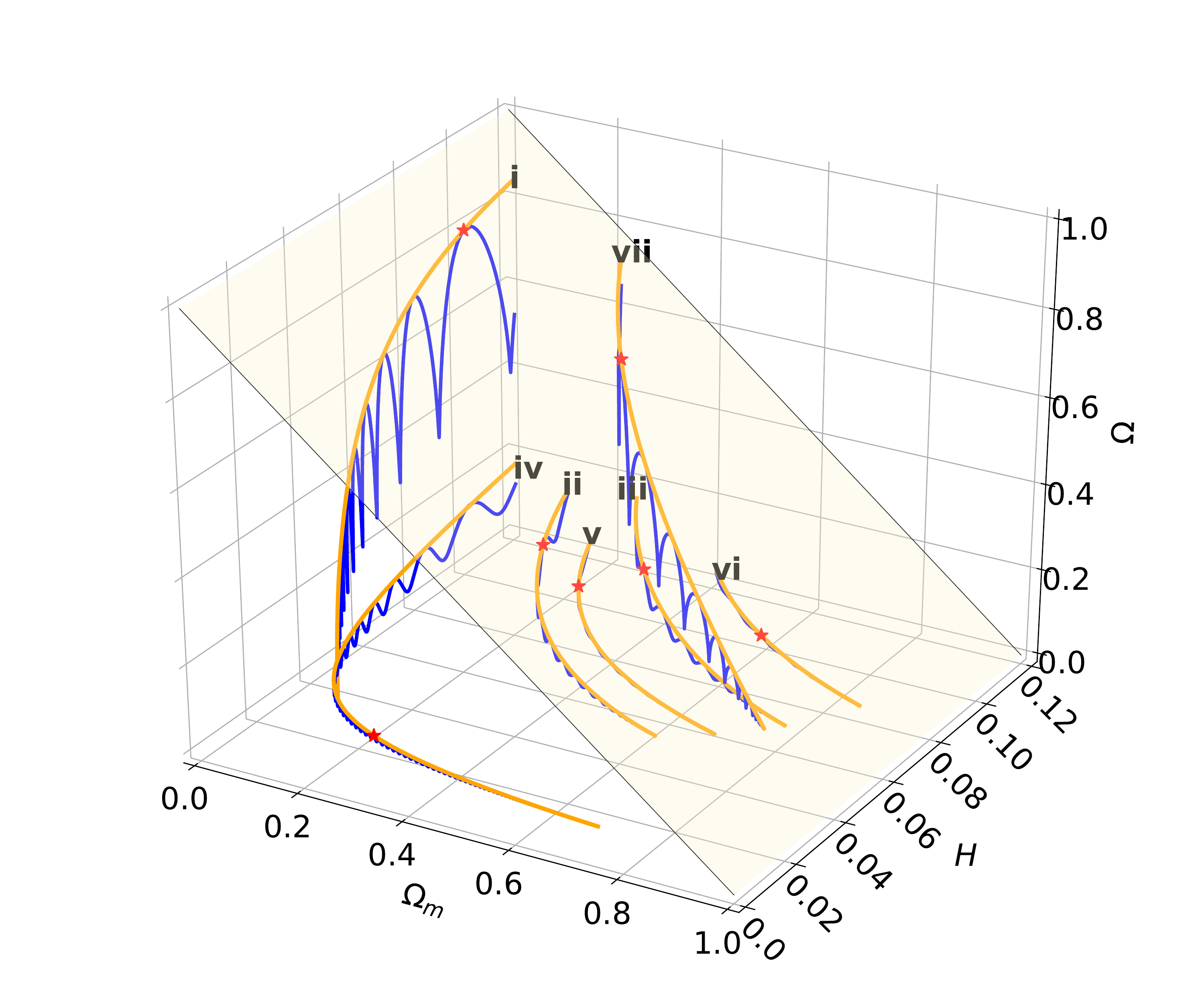}}
    \subfigure[\label{FLRWNonminimallyCCf093Dk} Projections in the space $(\Omega_{k},H,\Omega)$. The surface is given by the constraint $\Omega=1-\Omega_{k}$.]{\includegraphics[scale = 0.25]{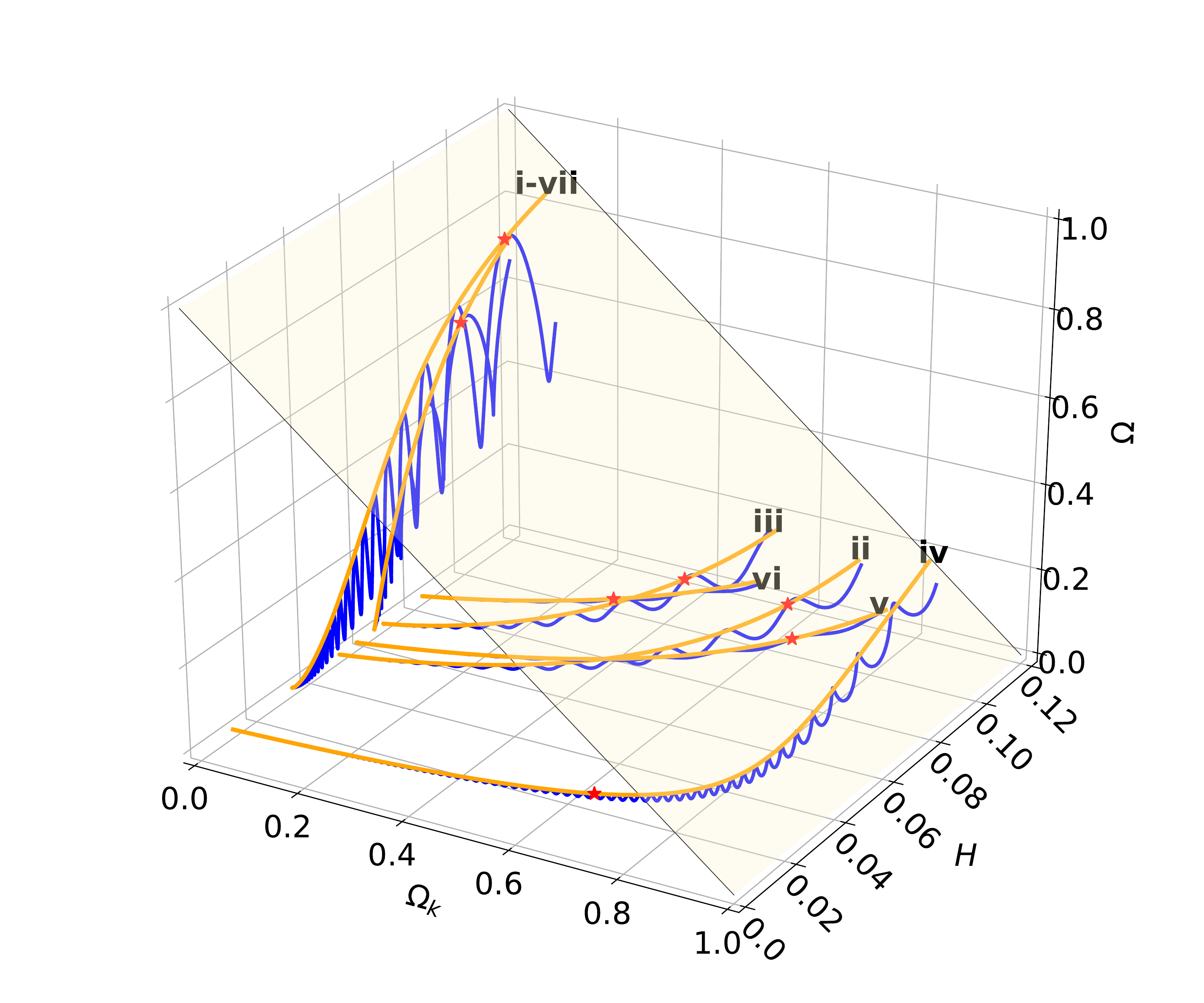}}
    \subfigure[\label{FLRWNonminimallyCCf093Dp} Projections in the space $(\Omega_{m},\Omega_{k},\Omega)$.]{\includegraphics[scale = 0.25]{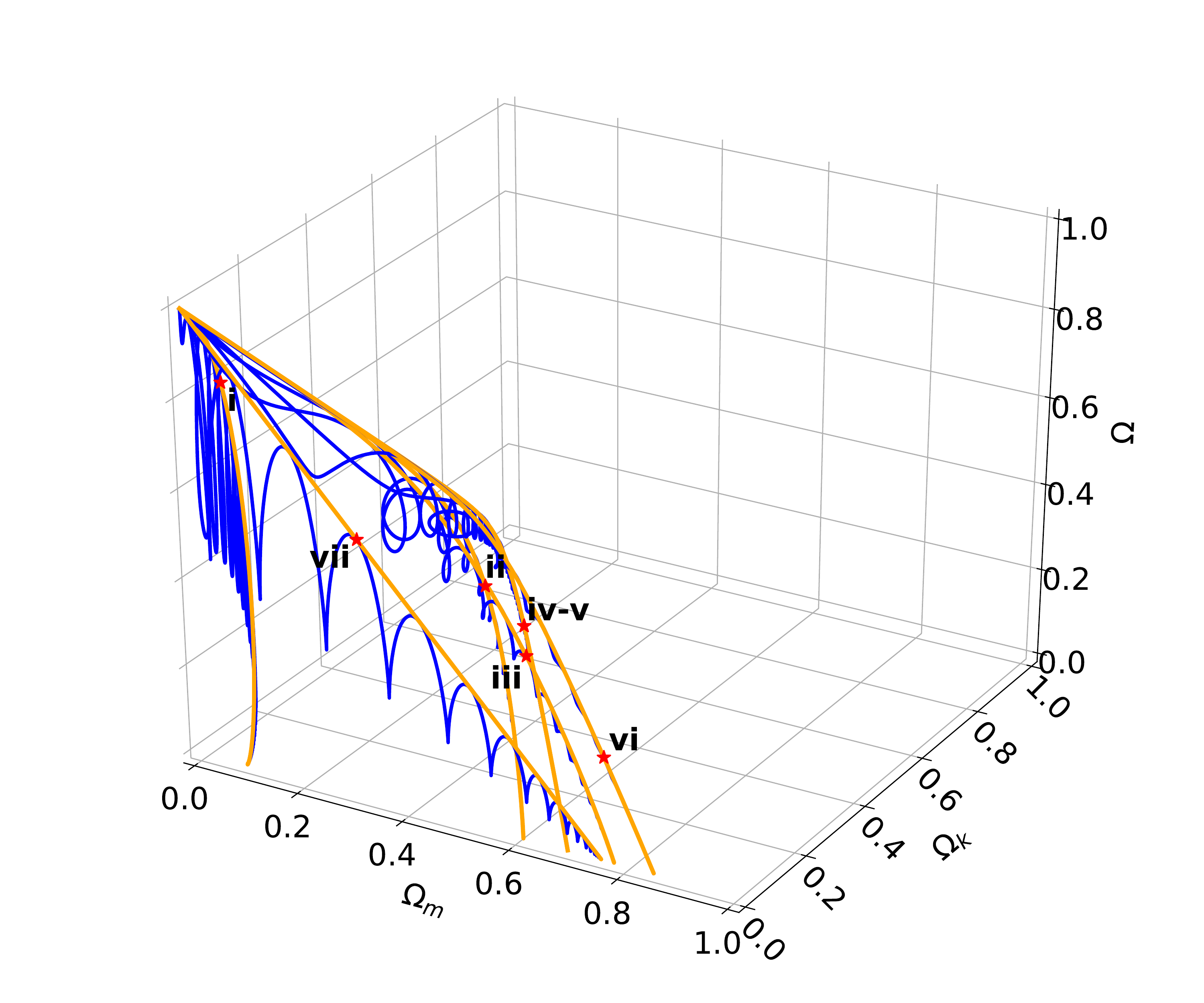}}
  \captionof{figure}{Some solutions of the full system \eqref{XXEQ3.19XX} (blue) and time-averaged system \eqref{EQsEps} (orange) for a scalar field with generalized harmonic potential non-minimally coupled to matter in the FLRW metric when $\lambda = 0.1$, $f=0.9$ and $\gamma=0$. We have used for both systems the initial data sets presented in Table \ref{tab:FLRW}.}
  \label{fig:FLRWNonminimallyCCf09}
\end{minipage}%
\hspace{.02\textwidth}
\begin{minipage}{.48\textwidth}
  \centering
     \subfigure[\label{FLRWNonminimallyBiff093Dm} Projections in the space $(\Omega_{m},H,\Omega)$. The surface is given by the constraint $\Omega=1-\Omega_{m}$.]{\includegraphics[scale = 0.25]{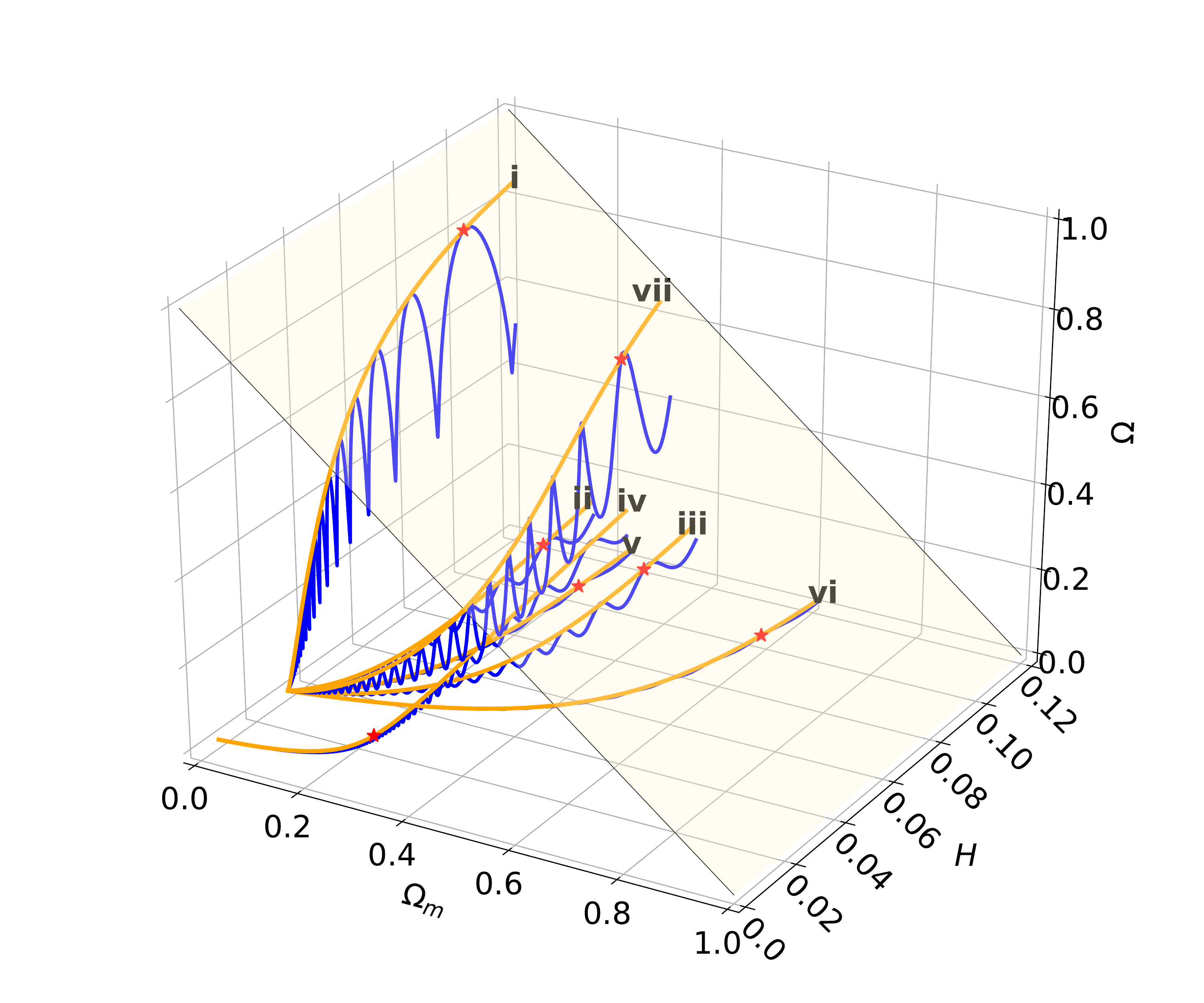}}
    \subfigure[\label{FLRWNonminimallyBiff093Dk} Projections in the space $(\Omega_{k},H,\Omega)$. The surface is given by the constraint $\Omega=1-\Omega_{k}$.]{\includegraphics[scale = 0.25]{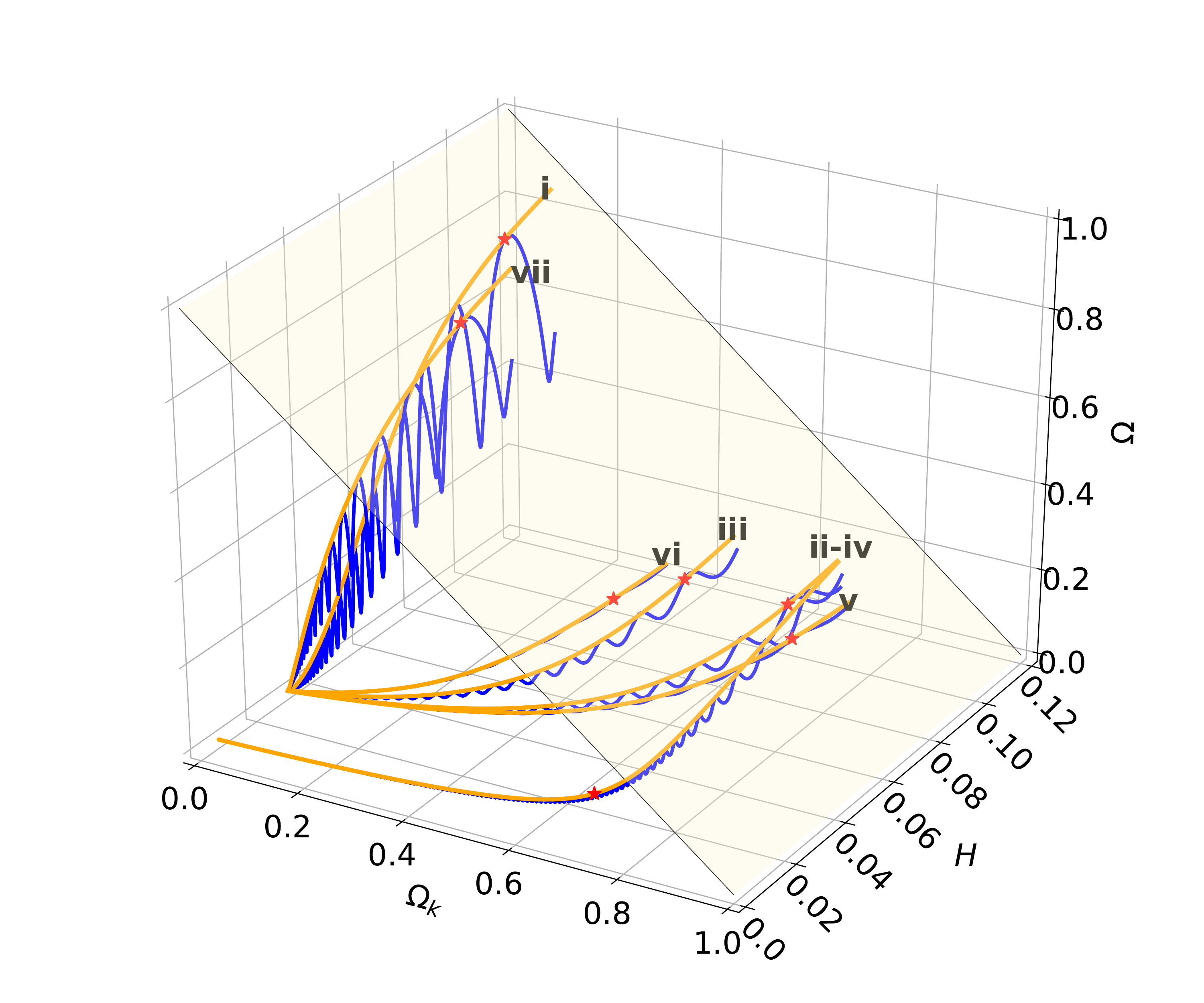}}
    \subfigure[\label{FLRWNonminimallyBiff093Dp} Projections in the space $(\Omega_{m},\Omega_{k},\Omega)$.]{\includegraphics[scale = 0.25]{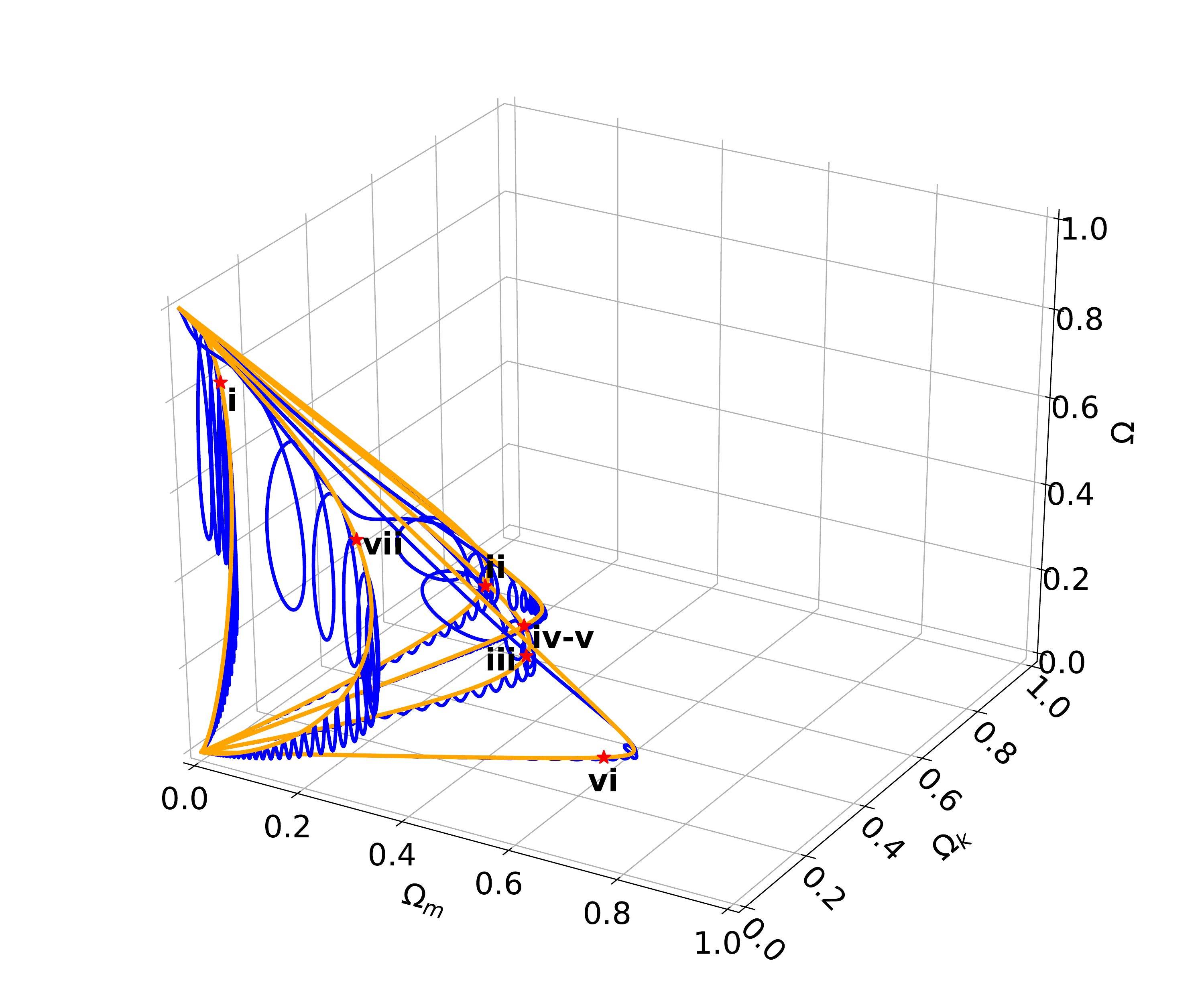}}
  \caption{Some solutions of the full system \eqref{XXEQ3.19XX} (blue) and time-averaged system \eqref{EQsEps} (orange) for a scalar field with generalized harmonic potential non-minimally coupled to matter in the FLRW metric when $\lambda = 0.1$, $f=0.9$ and $\gamma=\frac{2}{3}$. We have used for both systems the initial data sets presented in Table \ref{tab:FLRW}.}
  \label{fig:FLRWNonminimallyBiff09}
\end{minipage}
\end{figure}

\begin{figure}[ht!]
\centering
\begin{minipage}{.48\textwidth}
  \centering
    \subfigure[\label{FLRWNonminimallyDustf093Dm} Projections in the space $(\Omega_{m},H,\Omega)$. The surface is given by the constraint $\Omega=1-\Omega_{m}$.]{\includegraphics[scale = 0.25]{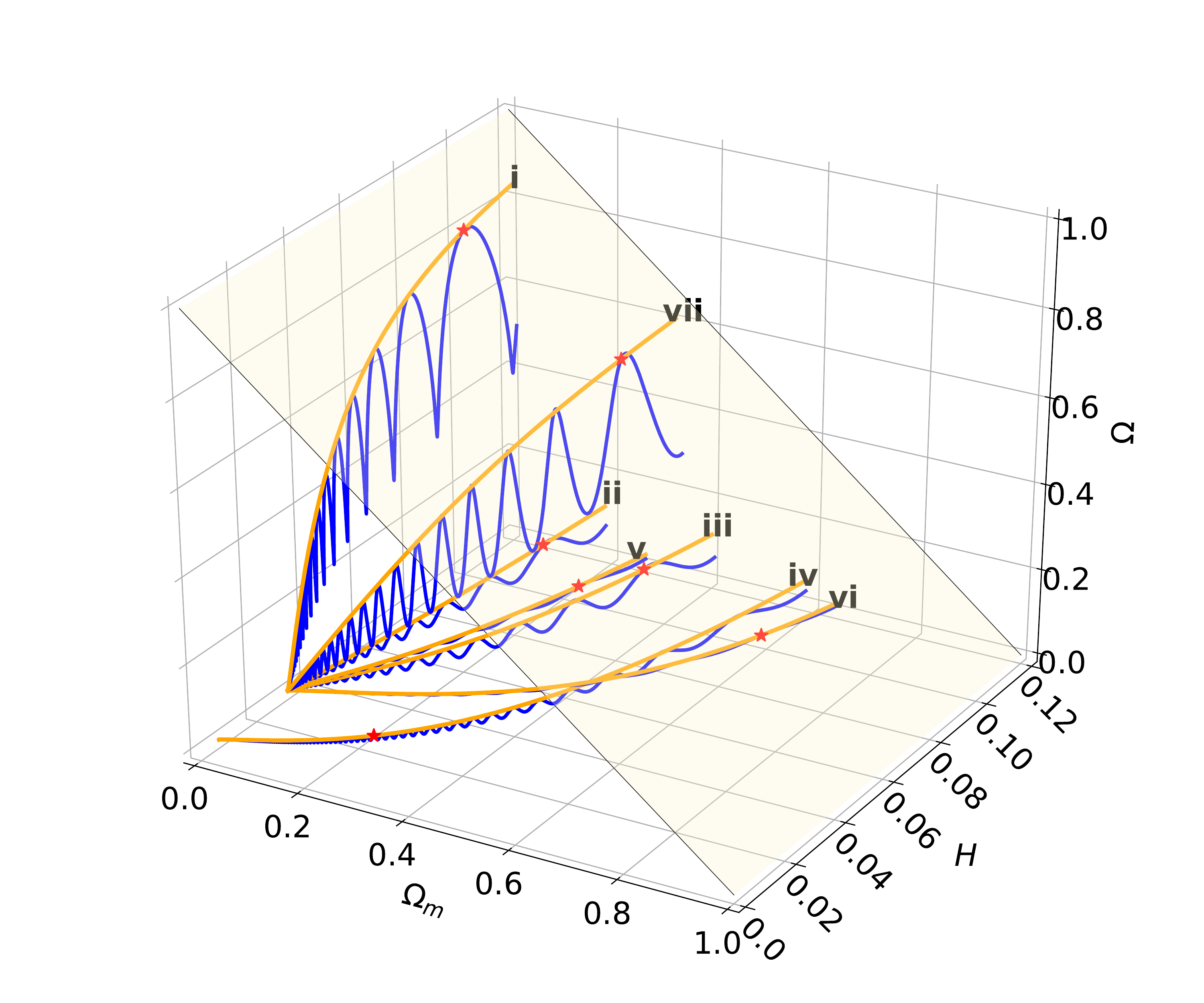}}
    \subfigure[\label{FLRWNonminimallyDustf093Dk} Projections in the space $(\Omega_{k},H,\Omega)$. The surface is given by the constraint $\Omega=1-\Omega_{k}$.]{\includegraphics[scale = 0.25]{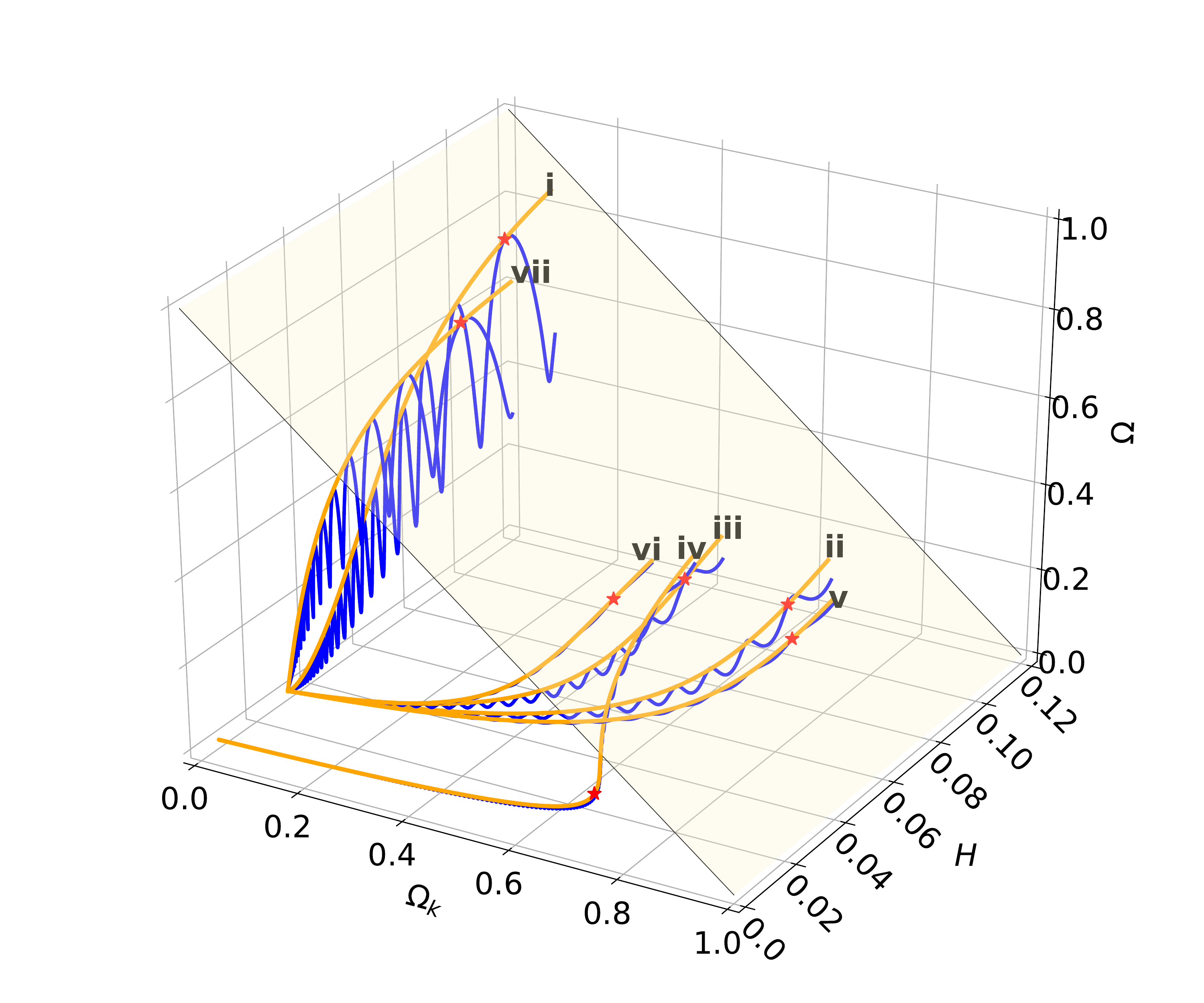}}
    \subfigure[\label{FLRWNonminimallyDustf093Dp} Projections in the space $(\Omega_{m},\Omega_{k},\Omega)$.]{\includegraphics[scale = 0.25]{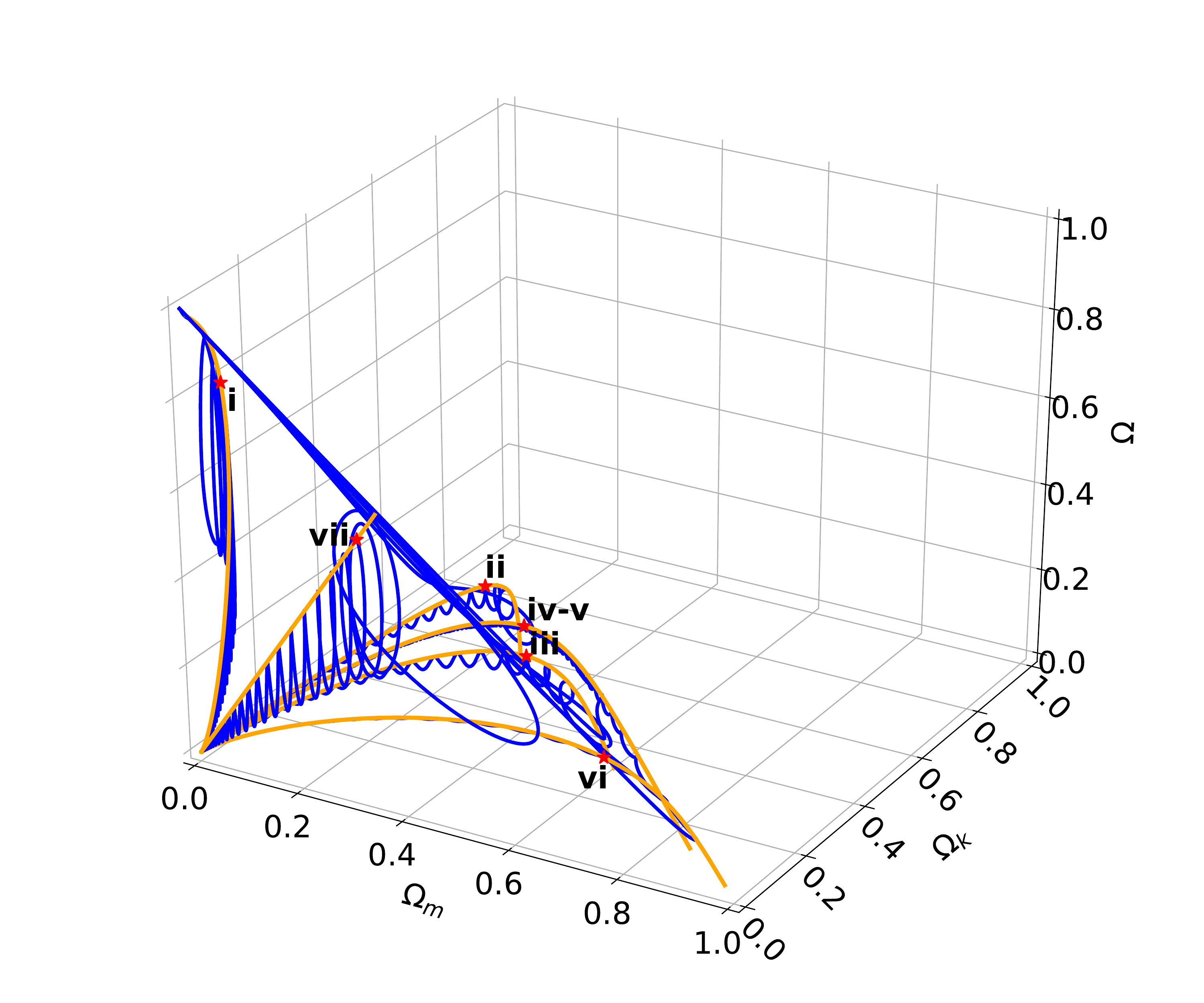}}
  \captionof{figure}{Some solutions of the full system \eqref{XXEQ3.19XX} (blue) and time-averaged system \eqref{EQsEps} (orange) for a scalar field with generalized harmonic potential non-minimally coupled to matter in the FLRW metric when $\lambda = 0.1$, $f=0.9$ and $\gamma=1$. We have used for both systems the initial data sets presented in Table \ref{tab:FLRW}.}
  \label{fig:FLRWNonminimallyDustf09}
\end{minipage}%
\hspace{.02\textwidth}
\begin{minipage}{.48\textwidth}
  \centering
     \subfigure[\label{FLRWNonminimallyStifff093Dm} Projections in the space $(\Omega_{m},H,\Omega)$. The surface is given by the constraint $\Omega=1-\Omega_{m}$.]{\includegraphics[scale = 0.25]{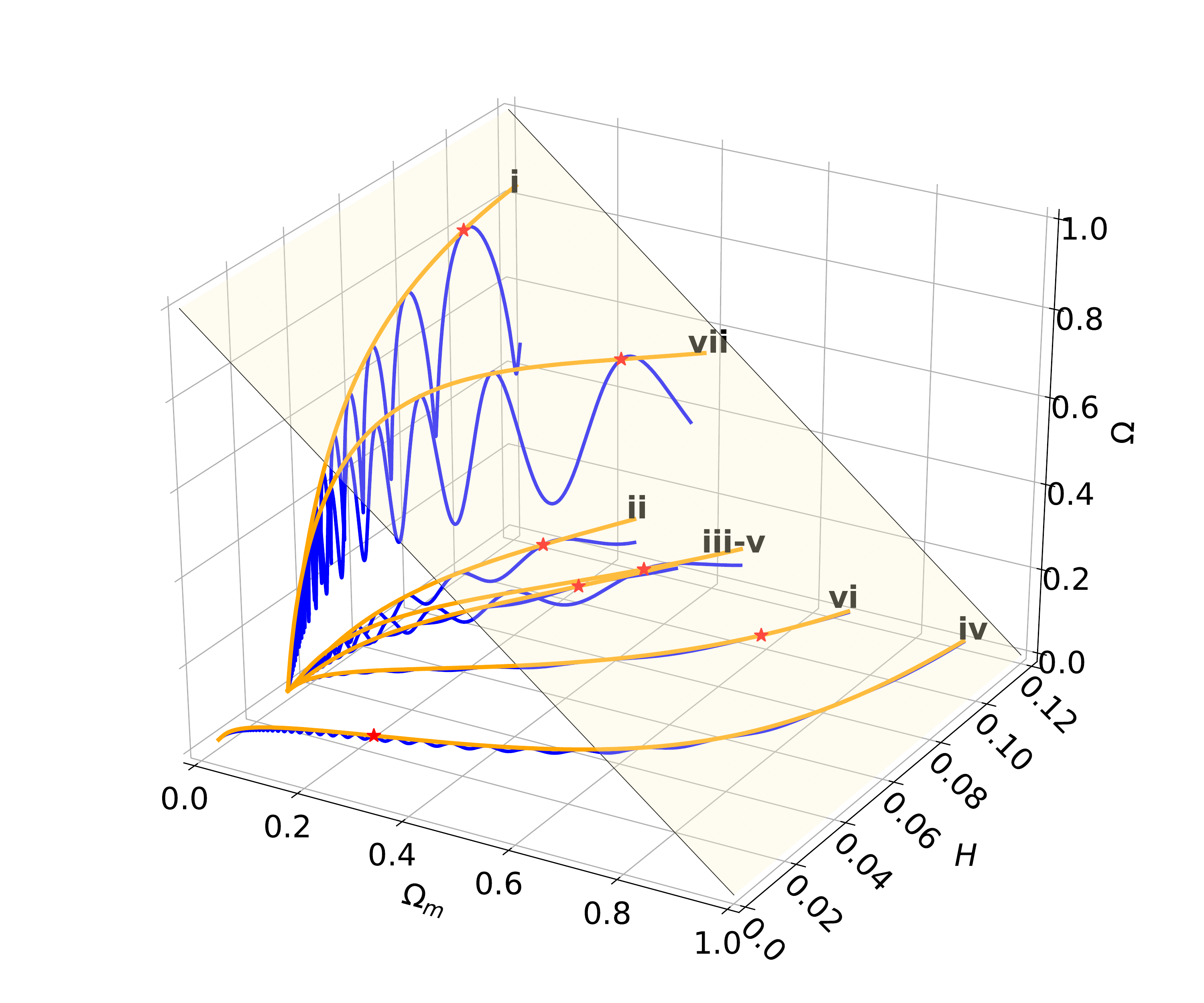}}
    \subfigure[\label{FLRWNonminimallyStifff093Dk} Projections in the space $(\Omega_{k},H,\Omega)$. The surface is given by the constraint $\Omega=1-\Omega_{k}$.]{\includegraphics[scale = 0.25]{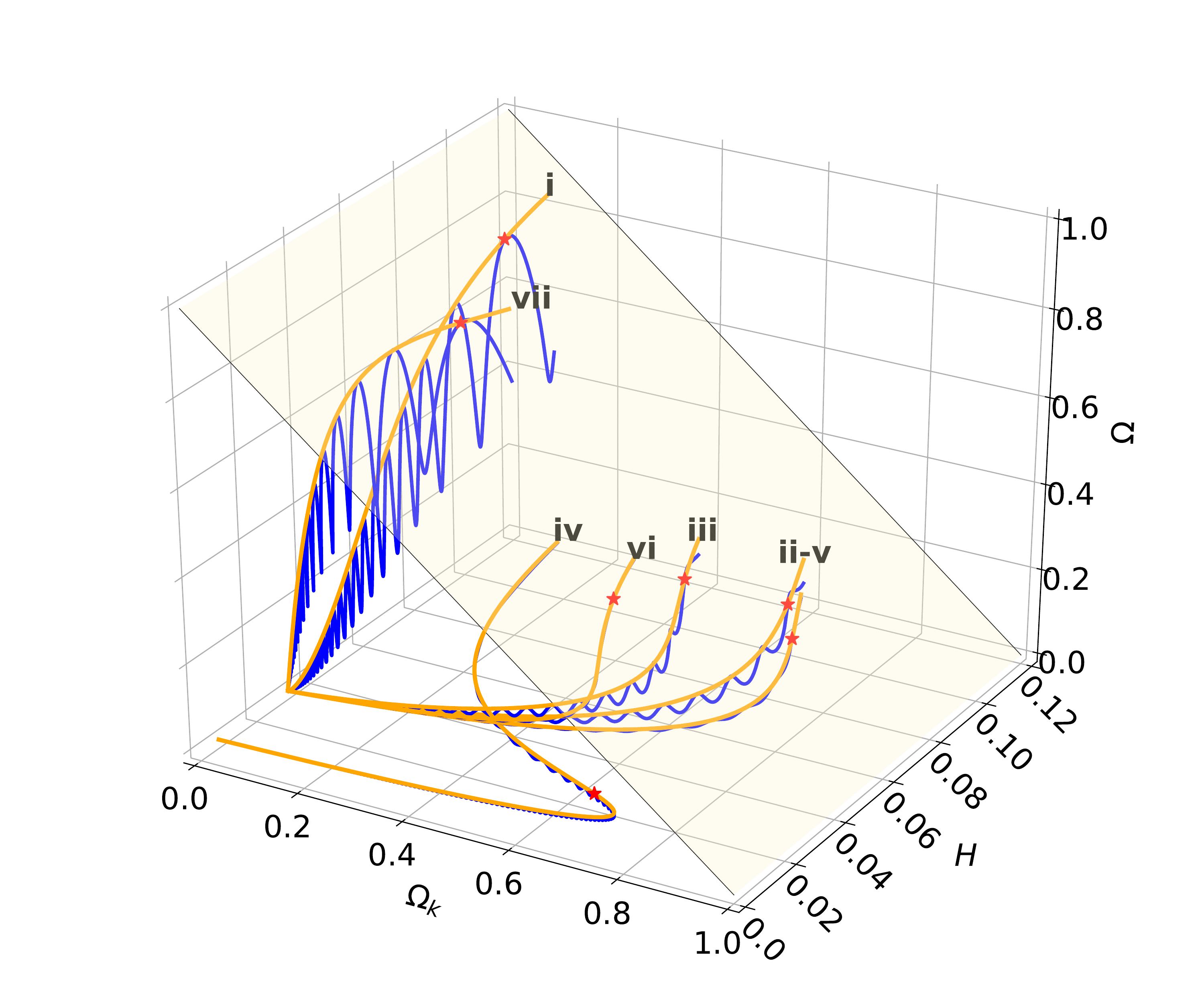}}
    \subfigure[\label{FLRWNonminimallyStifff093Dp} Projections in the space $(\Omega_{m},\Omega_{k},\Omega)$.]{\includegraphics[scale = 0.25]{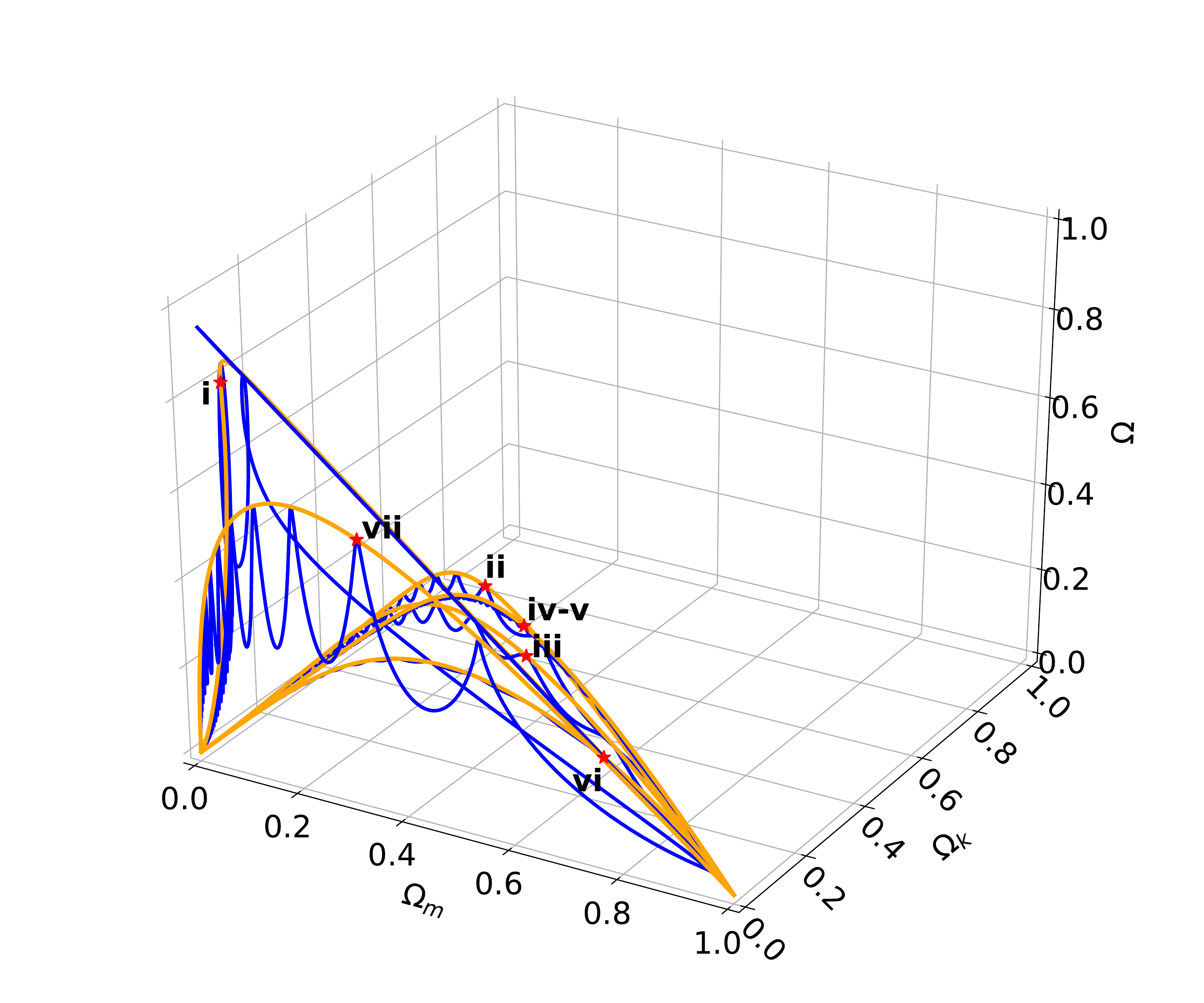}}
  \caption{Some solutions of the full system \eqref{XXEQ3.19XX} (blue) and time-averaged system \eqref{EQsEps} (orange) for a scalar field with generalized harmonic potential non-minimally coupled to matter in the FLRW metric when $\lambda = 0.1$, $f=0.9$ and $\gamma=2$. We have used for both systems the initial data sets presented in Table \ref{tab:FLRW}.}
  \label{fig:FLRWNonminimallyStifff09}
\end{minipage}
\end{figure}

In Figures \ref{fig:FLRWNonminimallyCCf01}, \ref{fig:FLRWNonminimallyBiff01}, \ref{fig:FLRWNonminimallyDustf01}, \ref{fig:FLRWNonminimallyStifff01}, \ref{fig:FLRWNonminimallyCCf03}, \ref{fig:FLRWNonminimallyBiff03}, \ref{fig:FLRWNonminimallyDustf03}, \ref{fig:FLRWNonminimallyStifff03}, \ref{fig:FLRWNonminimallyCCf09}, \ref{fig:FLRWNonminimallyBiff09}, \ref{fig:FLRWNonminimallyDustf09} and \ref{fig:FLRWNonminimallyStifff09} we present the numerical results obtained from the integration of the full system \eqref{XXEQ3.19XX} (blue lines) and time-averaged system \eqref{EQsEps} (orange lines) for the non-minimally coupled case in the FLRW metric, using for both systems the seven initial data set presented in the Table \ref{tab:FLRW}. 

In Figures \ref{fig:FLRWNonminimallyCCf01}, \ref{fig:FLRWNonminimallyCCf03} and \ref{fig:FLRWNonminimallyCCf09} we depict the results obtained for $\gamma=0$ when $f=0.1$, $0.3$ and $0.9$, respectively. Figures \ref{FLRWNonminimallyCCf013Dm}, \ref{FLRWNonminimallyCCf033Dm} and \ref{FLRWNonminimallyCCf093Dm} shows the projections in the space $(\Omega_{m},H,\Omega)$, Figures \ref{FLRWNonminimallyCCf013Dk}, \ref{FLRWNonminimallyCCf033Dk} and \ref{FLRWNonminimallyCCf093Dk} shows the projections in the space $(\Omega_{k},H,\Omega)$, and Figures \ref{FLRWNonminimallyCCf013Dp}, \ref{FLRWNonminimallyCCf033Dp} and \ref{FLRWNonminimallyCCf093Dp} shows the projections in the space $(\Omega_{m}, \Omega_{k}, \Omega)$. 

\begin{table}[ht!]
    \centering
       \begin{tabular}{ccccccc}
        \hline
        Sol. & $H(0)$ & $\Omega(0)$ & $\Omega_{m}(0)$ & $\Omega_{k}(0)$ & $\varphi(0)$ & $t(0)$ \\
        \hline
        i & $0.1$ & $0.8$ & $0.01$ & $0.09$ & $0$ & $0$ \\
        ii & $0.1$ & $0.1$ & $0.16$ & $0.64$ & $0$ & $0$ \\
        iii & $0.1$ & $0.1$ & $0.36$ & $0.44$ & $0$ & $0$ \\
        iv & $0.02$ & $0.02$ & $0.2304$ & $0.6496$ & $0$ & $0$ \\
        v & $0.1$ & $0.02$ & $0.2304$ & $0.6496$ & $0$ & $0$ \\
        vi & $0.1$ & $0.01$ & $0.59$ & $0.3$ & $0$ & $0$ \\
        vii & $0.1$ & $0.584$ & $0.315$ & $0.001$ & $0$ & $0$ \\
        \hline
    \end{tabular}
     \caption{Seven initial data set for the simulations of the full and time averaged system for the FLRW metric in the non-minimal and minimal coupling cases. The initial conditions were chosen in order to fulfill the inequality $\Omega(0)+\Omega_{m}(0)+\Omega_{k}(0)<1$.}
    \label{tab:FLRW}
\end{table}

In Figures \ref{fig:FLRWNonminimallyBiff01}, \ref{fig:FLRWNonminimallyBiff03} and \ref{fig:FLRWNonminimallyBiff09} we depict the results obtained for $\gamma=2/3$ when $f=0.1$, $0.3$ and $0.9$, respectively. Figures \ref{FLRWNonminimallyBiff013Dm}, \ref{FLRWNonminimallyBiff033Dm} and \ref{FLRWNonminimallyBiff093Dm} shows the projections in the space $(\Omega_{m},H,\Omega)$, Figures \ref{FLRWNonminimallyBiff013Dk}, \ref{FLRWNonminimallyBiff033Dk} and \ref{FLRWNonminimallyBiff093Dk} shows the projections in the space $(\Omega_{k},H,\Omega)$, and Figures \ref{FLRWNonminimallyBiff013Dp}, \ref{FLRWNonminimallyBiff033Dp} and \ref{FLRWNonminimallyBiff093Dp} shows the projections in the space $(\Omega_{m}, \Omega_{k}, \Omega)$. 

In Figures \ref{fig:FLRWNonminimallyDustf01}, \ref{fig:FLRWNonminimallyDustf03} and \ref{fig:FLRWNonminimallyDustf09} we depict the results obtained for $\gamma=1$ when $f=0.1$, $0.3$ and $0.9$, respectively. Figures \ref{FLRWNonminimallyDustf013Dm}, \ref{FLRWNonminimallyDustf033Dm} and \ref{FLRWNonminimallyDustf093Dm} shows the projections in the space $(\Omega_{m},H,\Omega)$, Figures \ref{FLRWNonminimallyDustf013Dk}, \ref{FLRWNonminimallyDustf033Dk} and \ref{FLRWNonminimallyDustf093Dk} shows the projections in the space $(\Omega_{k},H,\Omega)$, and Figures \ref{FLRWNonminimallyDustf013Dp}, \ref{FLRWNonminimallyDustf033Dp} and \ref{FLRWNonminimallyDustf093Dp} shows the projections in the space $(\Omega_{m}, \Omega_{k}, \Omega)$.

In Figures \ref{fig:FLRWNonminimallyStifff01}, \ref{fig:FLRWNonminimallyStifff03} and \ref{fig:FLRWNonminimallyStifff09} we depict the results obtained for $\gamma=2$ when $f=0.1$, $0.3$ and $0.9$, respectively. Figures \ref{FLRWNonminimallyStifff013Dm}, \ref{FLRWNonminimallyStifff033Dm} and \ref{FLRWNonminimallyStifff093Dm} shows the projections in the space $(\Omega_{m},H,\Omega)$, Figures \ref{FLRWNonminimallyStifff013Dk}, \ref{FLRWNonminimallyStifff033Dk} and \ref{FLRWNonminimallyStifff093Dk} shows the projections in the space $(\Omega_{k},H,\Omega)$, and Figures \ref{FLRWNonminimallyStifff013Dp}, \ref{FLRWNonminimallyStifff033Dp} and \ref{FLRWNonminimallyStifff093Dp} shows the projections in the space $(\Omega_{m}, \Omega_{k}, \Omega)$.

These figures are evidence that the solutions of the full system (blue lines), obtained for a scalar field with generalized harmonic potential non-minimally coupled to matter in the FLRW metric, follow the track of the solutions of the averaged system (orange lines), therefore, have the same asymptotic behavior. Furthermore, we can see that the amplitude of oscillations decreases when the value of $f$ increases.

\subsubsection{Bianchi I metric}
\begin{figure}[ht!]
\centering
\begin{minipage}{.48\textwidth}
  \centering
    \subfigure[\label{BianchiINonminimallyCCf013Dm} Projections in the space $(\Omega_{m},H,\Omega)$. The surface is given by the constraint $\Omega=1-\Omega_{m}$.]{\includegraphics[scale = 0.25]{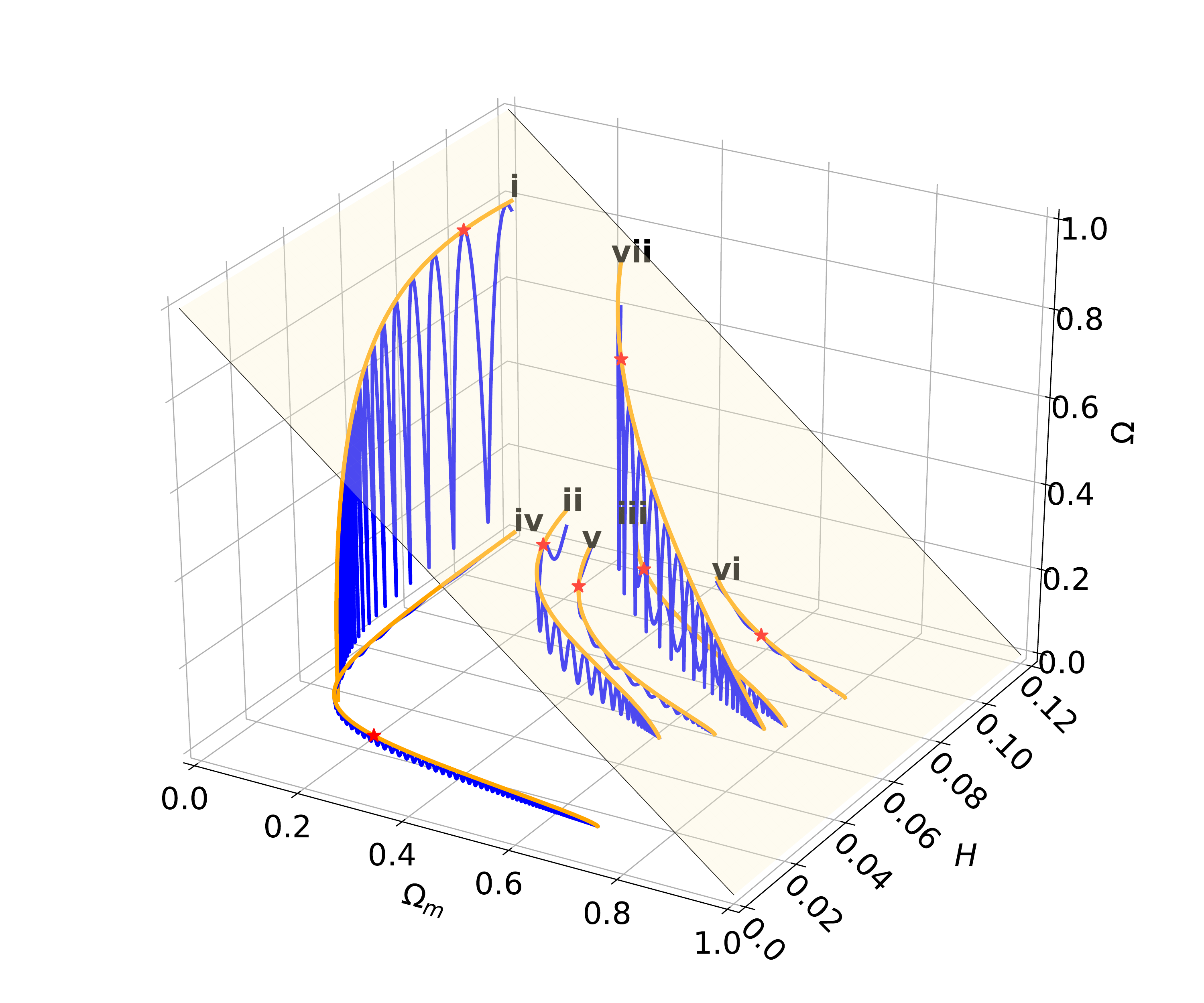}}
    \subfigure[\label{BianchiINonminimallyCCf013DS} Projections in the space $(\Sigma,H,\Omega)$. The surface is given by the constraint $\Omega=1-\Sigma^{2}/3$.]{\includegraphics[scale = 0.25]{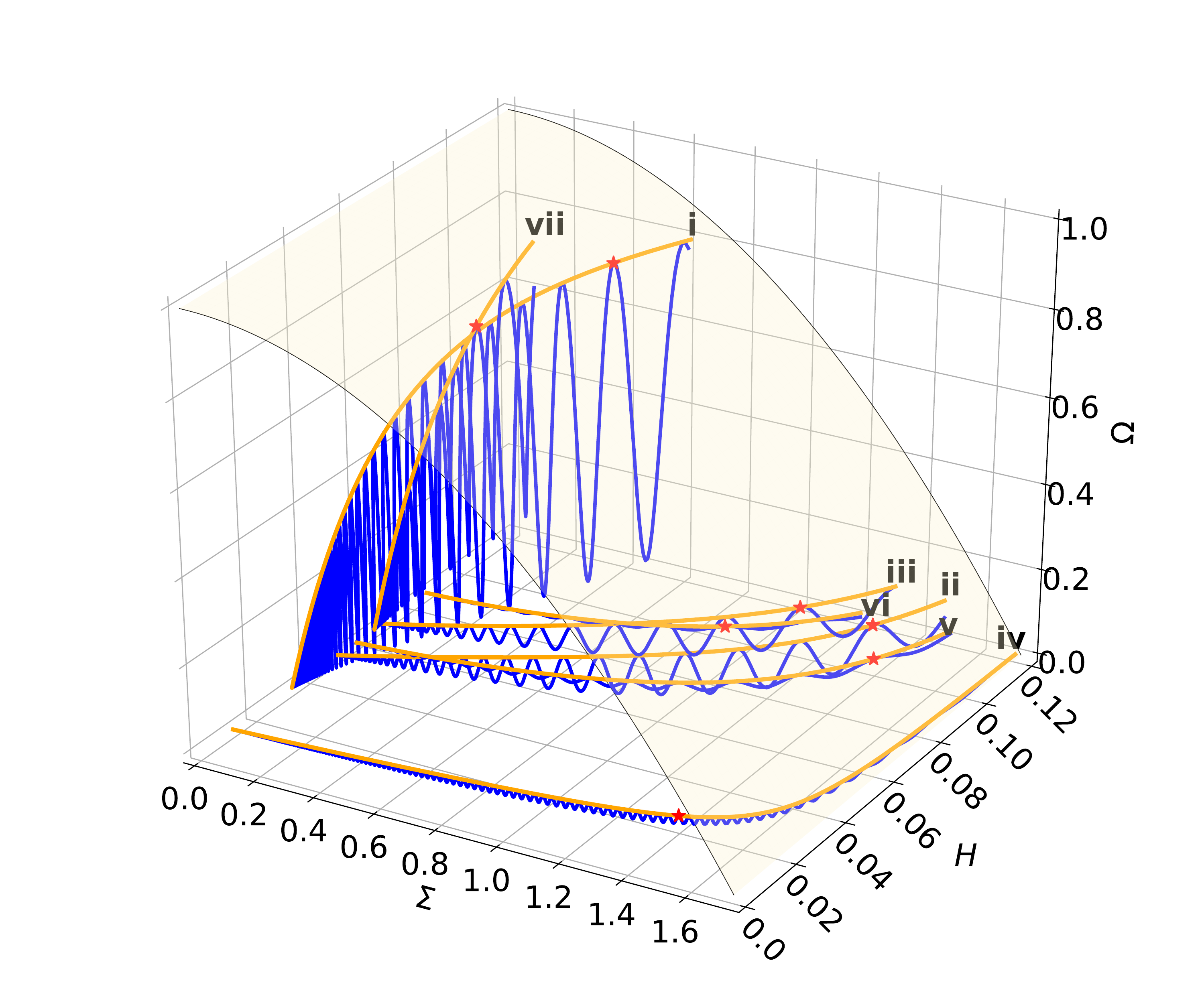}}
    \subfigure[\label{BianchiINonminimallyCCf013Dp} Projections in the space $(\Omega_{m},\Sigma,\Omega)$.]{\includegraphics[scale = 0.25]{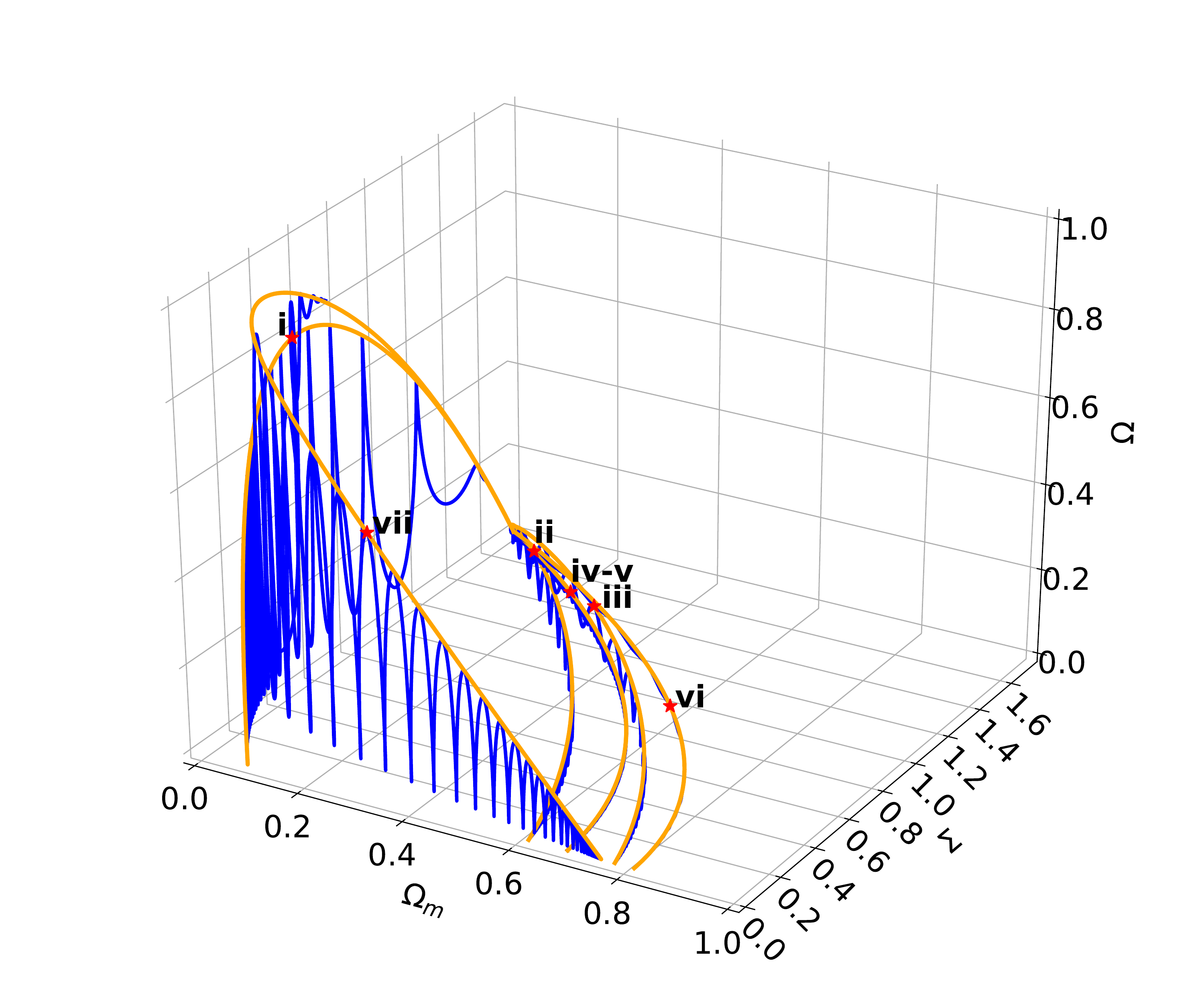}}
  \captionof{figure}{Some solutions of the full system \eqref{YYEQ3.29YY} (blue) and time-averaged system \eqref{BIYEQ3.29Y} (orange) for a scalar field with generalized harmonic potential non-minimally coupled to matter in the Bianchi I metric when $\lambda = 0.1$, $f=0.1$ and $\gamma=0$. We have used for both systems the initial data sets presented in Table \ref{tab:BianchiI}.}
  \label{fig:BianchiINonminimallyCCf01}
\end{minipage}%
\hspace{.02\textwidth}
\begin{minipage}{.48\textwidth}
  \centering
     \subfigure[\label{BianchiINonminimallyBiff013Dm} Projections in the space $(\Omega_{m},H,\Omega)$. The surface is given by the constraint $\Omega=1-\Omega_{m}$.]{\includegraphics[scale = 0.25]{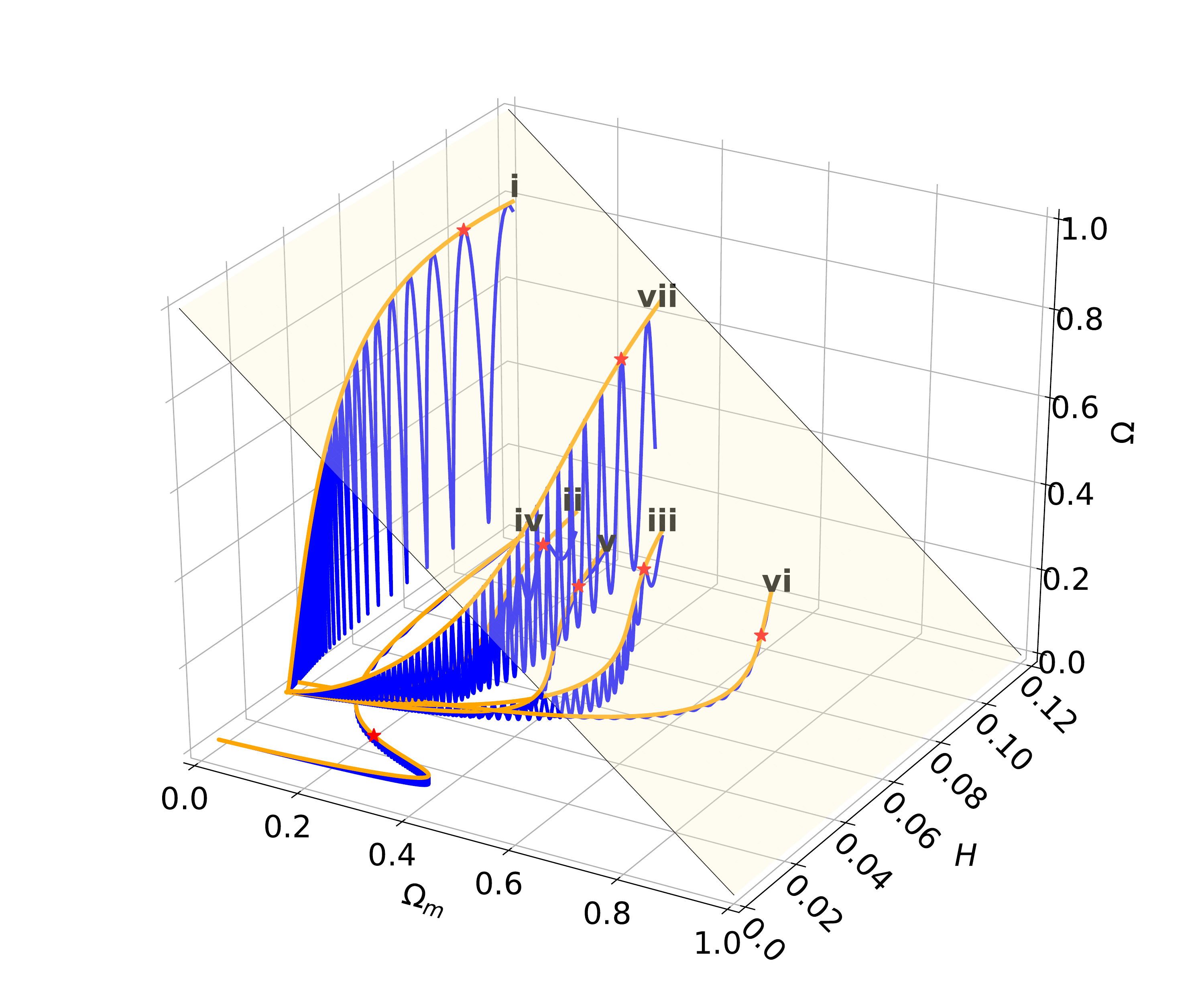}}
    \subfigure[\label{BianchiINonminimallyBiff013DS} Projections in the space $(\Sigma,H,\Omega)$. The surface is given by the constraint $\Omega=1-\Sigma^{2}/3$.]{\includegraphics[scale = 0.25]{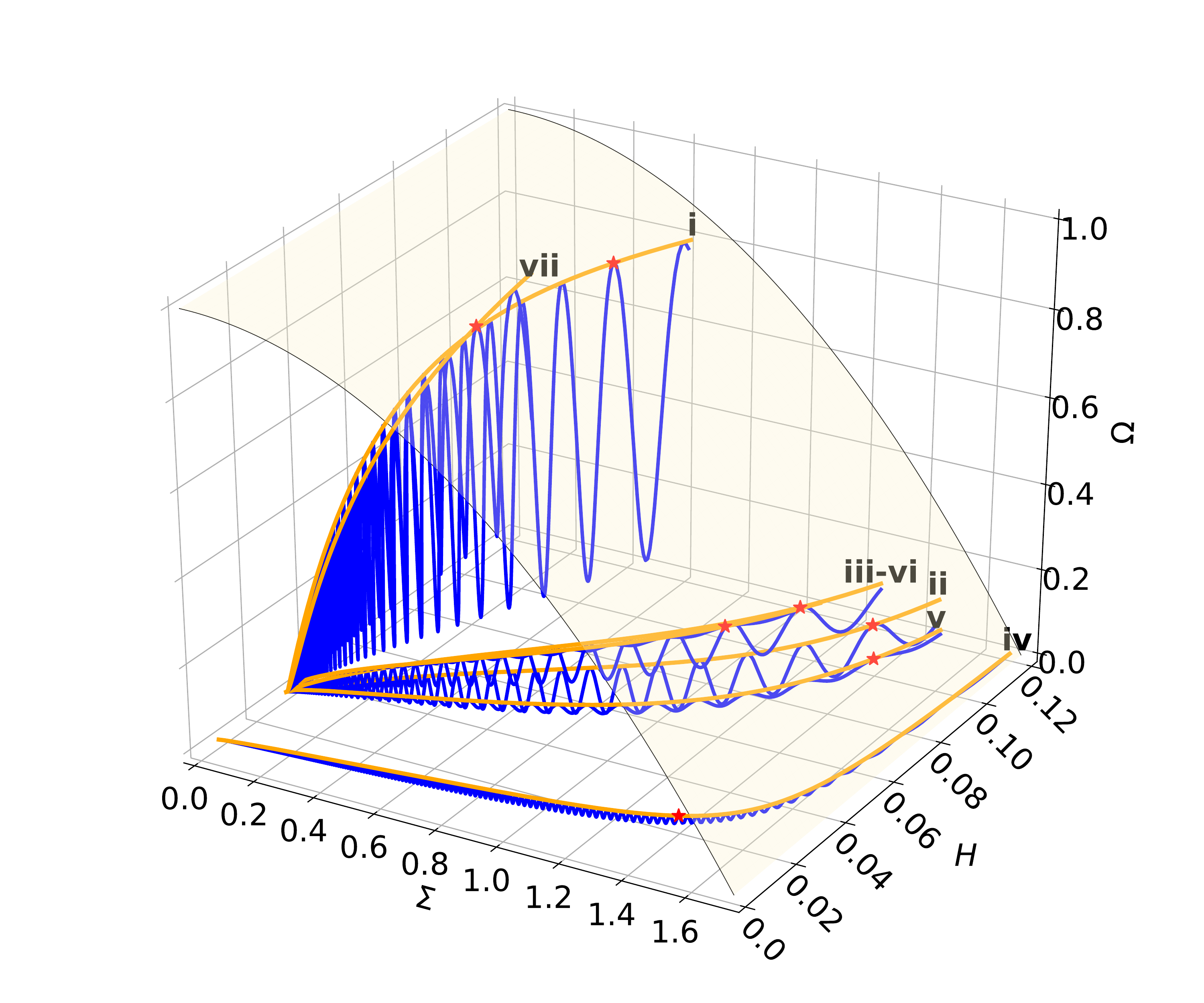}}
    \subfigure[\label{BianchiINonminimallyBiff013Dp} Projections in the space $(\Omega_{m},\Sigma,\Omega)$.]{\includegraphics[scale = 0.25]{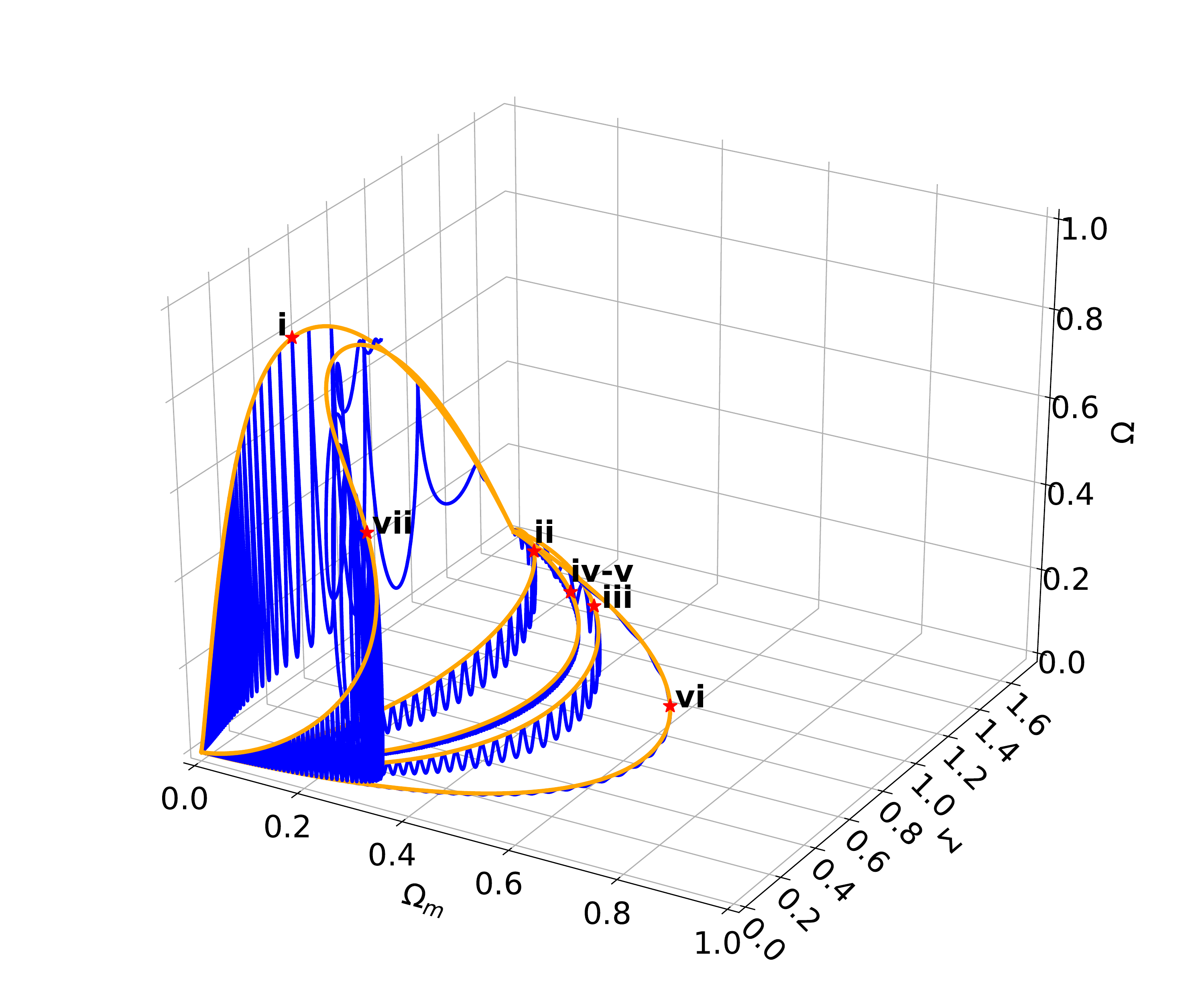}}
  \caption{Some solutions of the full system \eqref{YYEQ3.29YY} (blue) and time-averaged system \eqref{BIYEQ3.29Y} (orange) for a scalar field with generalized harmonic potential non-minimally coupled to matter in the Bianchi I metric when $\lambda = 0.1$, $f=0.1$ and $\gamma=\frac{2}{3}$. We have used for both systems the initial data sets presented in Table \ref{tab:BianchiI}.}
  \label{fig:BianchiINonminimallyBiff01}
\end{minipage}
\end{figure}

\begin{figure}[ht!]
\centering
\begin{minipage}{.48\textwidth}
  \centering
    \subfigure[\label{BianchiINonminimallyDustf013Dm} Projections in the space $(\Omega_{m},H,\Omega)$. The surface is given by the constraint $\Omega=1-\Omega_{m}$.]{\includegraphics[scale = 0.25]{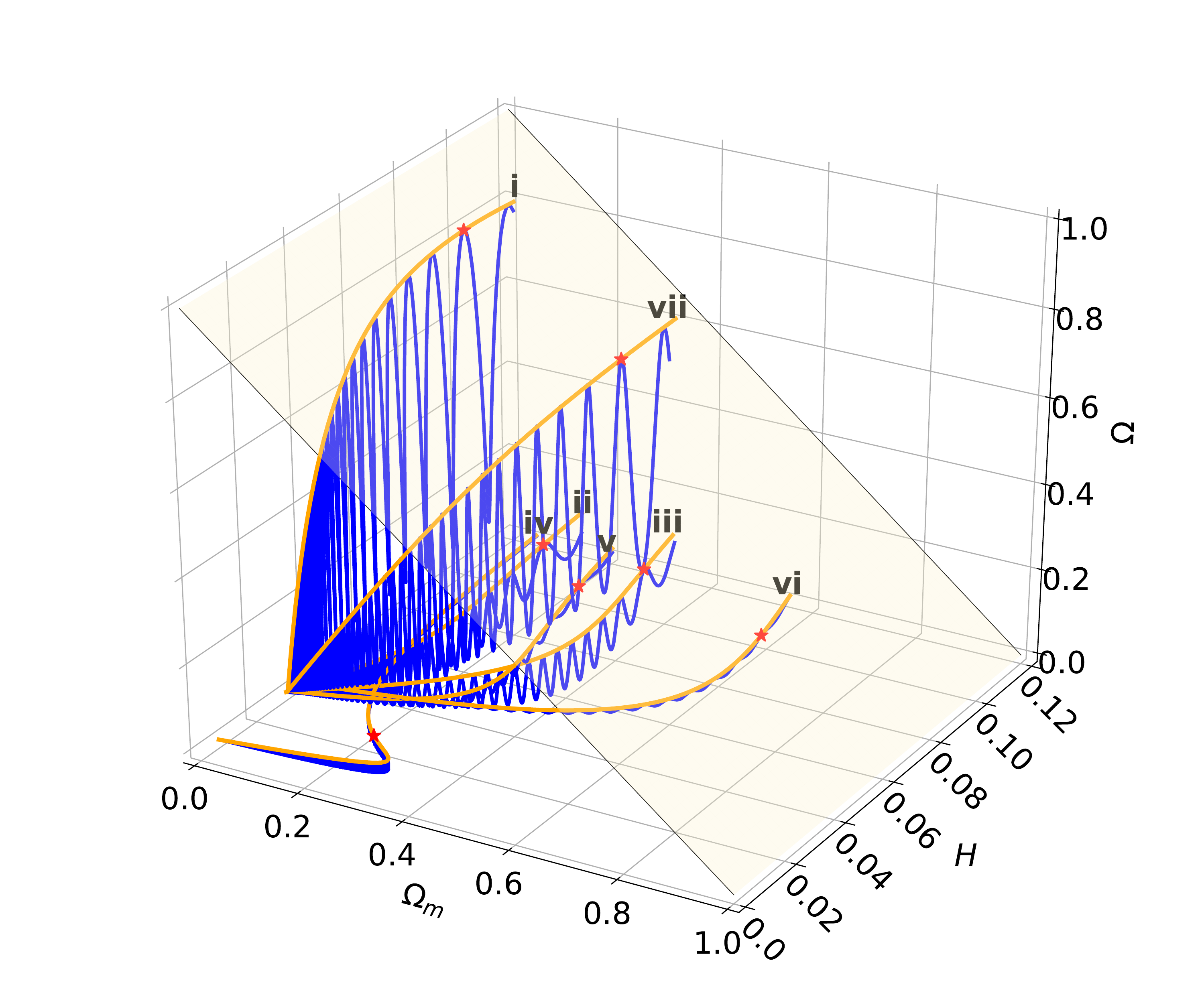}}
    \subfigure[\label{BianchiINonminimallyDustf013DS} Projections in the space $(\Sigma,H,\Omega)$. The surface is given by the constraint $\Omega=1-\Sigma^{2}/3$.]{\includegraphics[scale = 0.25]{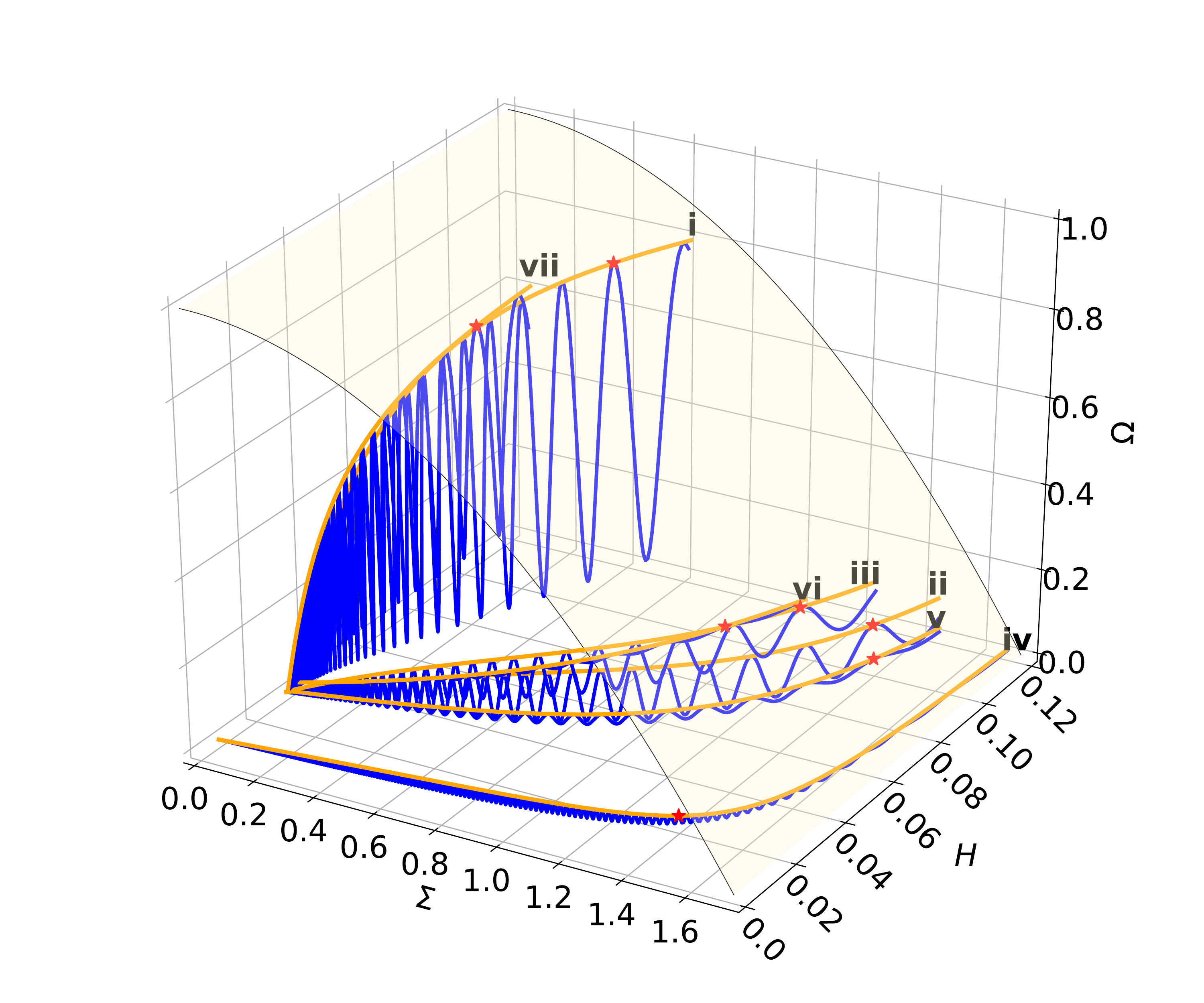}}
    \subfigure[\label{BianchiINonminimallyDustf013Dp} Projections in the space $(\Omega_{m},\Sigma,\Omega)$.]{\includegraphics[scale = 0.25]{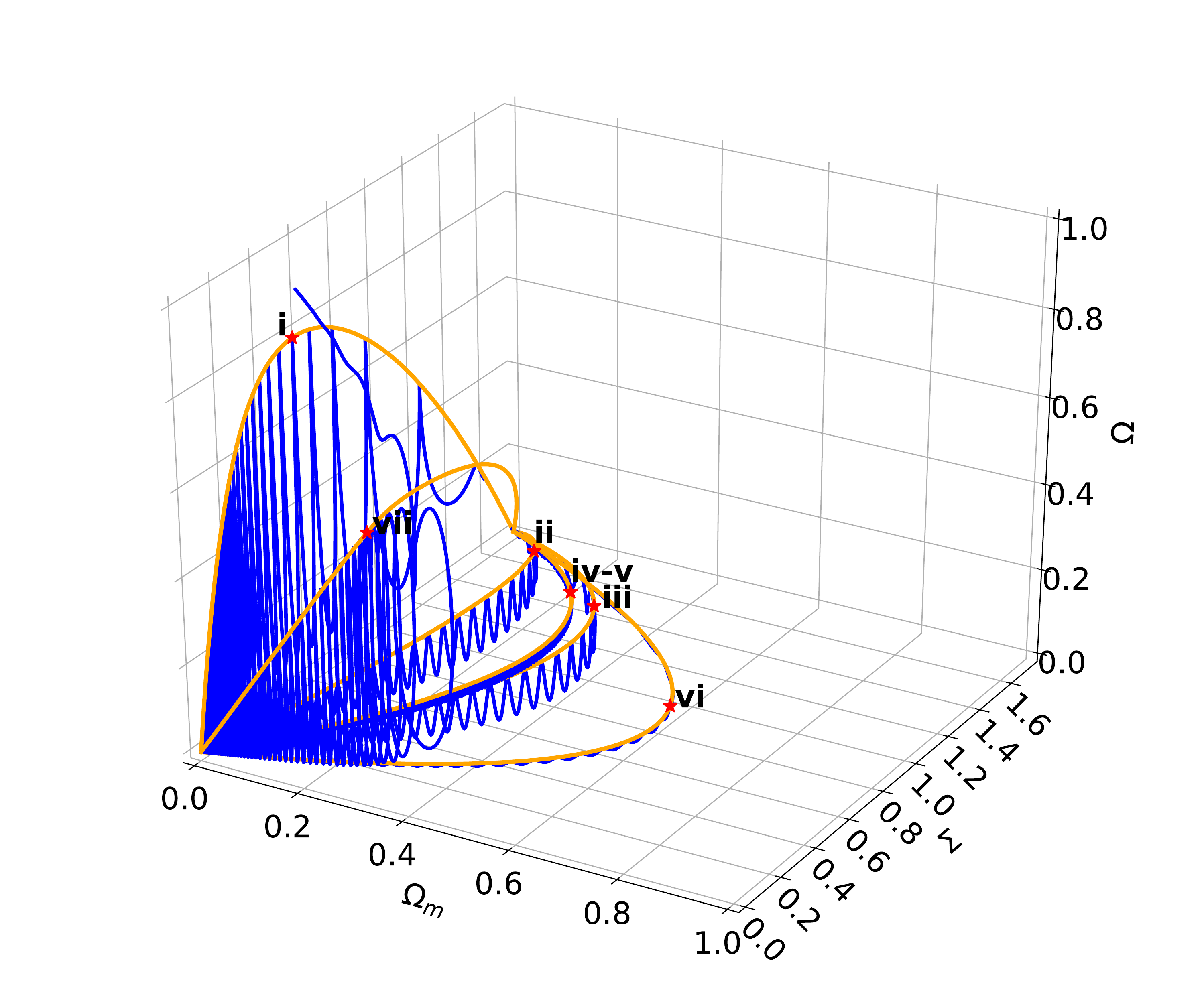}}
  \captionof{figure}{Some solutions of the full system \eqref{YYEQ3.29YY} (blue) and time-averaged system \eqref{BIYEQ3.29Y} (orange) for a scalar field with generalized harmonic potential non-minimally coupled to matter in the Bianchi I metric when $\lambda = 0.1$, $f=0.1$ and $\gamma=1$. We have used for both systems the initial data sets presented in Table \ref{tab:BianchiI}.}
  \label{fig:BianchiINonminimallyDustf01}
\end{minipage}%
\hspace{.02\textwidth}
\begin{minipage}{.48\textwidth}
  \centering
     \subfigure[\label{BianchiINonminimallyStifff013Dm} Projections in the space $(\Omega_{m},H,\Omega)$. The surface is given by the constraint $\Omega=1-\Omega_{m}$.]{\includegraphics[scale = 0.25]{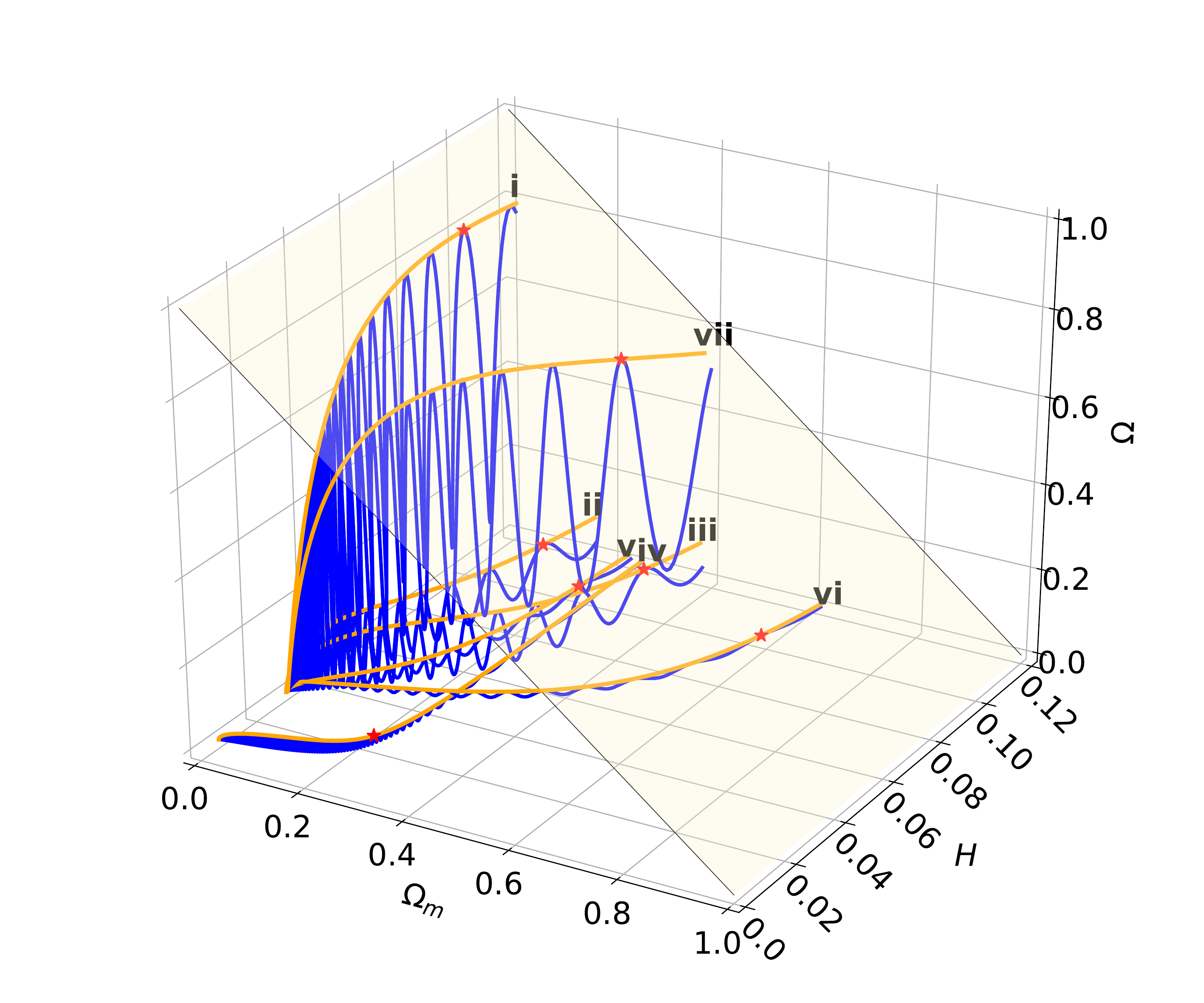}}
    \subfigure[\label{BianchiINonminimallyStifff013DS} Projections in the space $(\Sigma,H,\Omega)$. The surface is given by the constraint $\Omega=1-\Sigma^{2}/3$.]{\includegraphics[scale = 0.25]{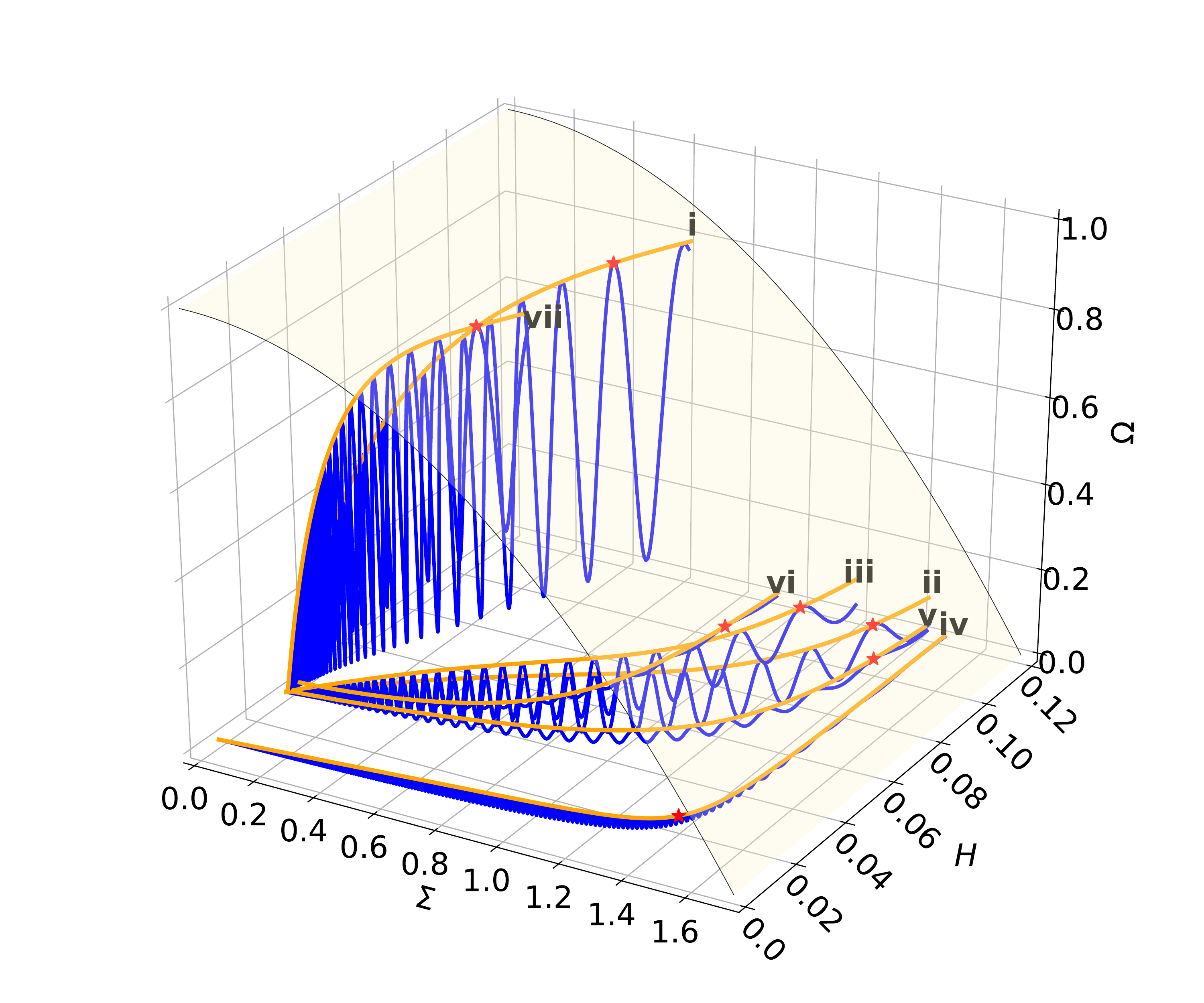}}
    \subfigure[\label{BianchiINonminimallyStifff013Dp} Projections in the space $(\Omega_{m},\Sigma,\Omega)$.]{\includegraphics[scale = 0.25]{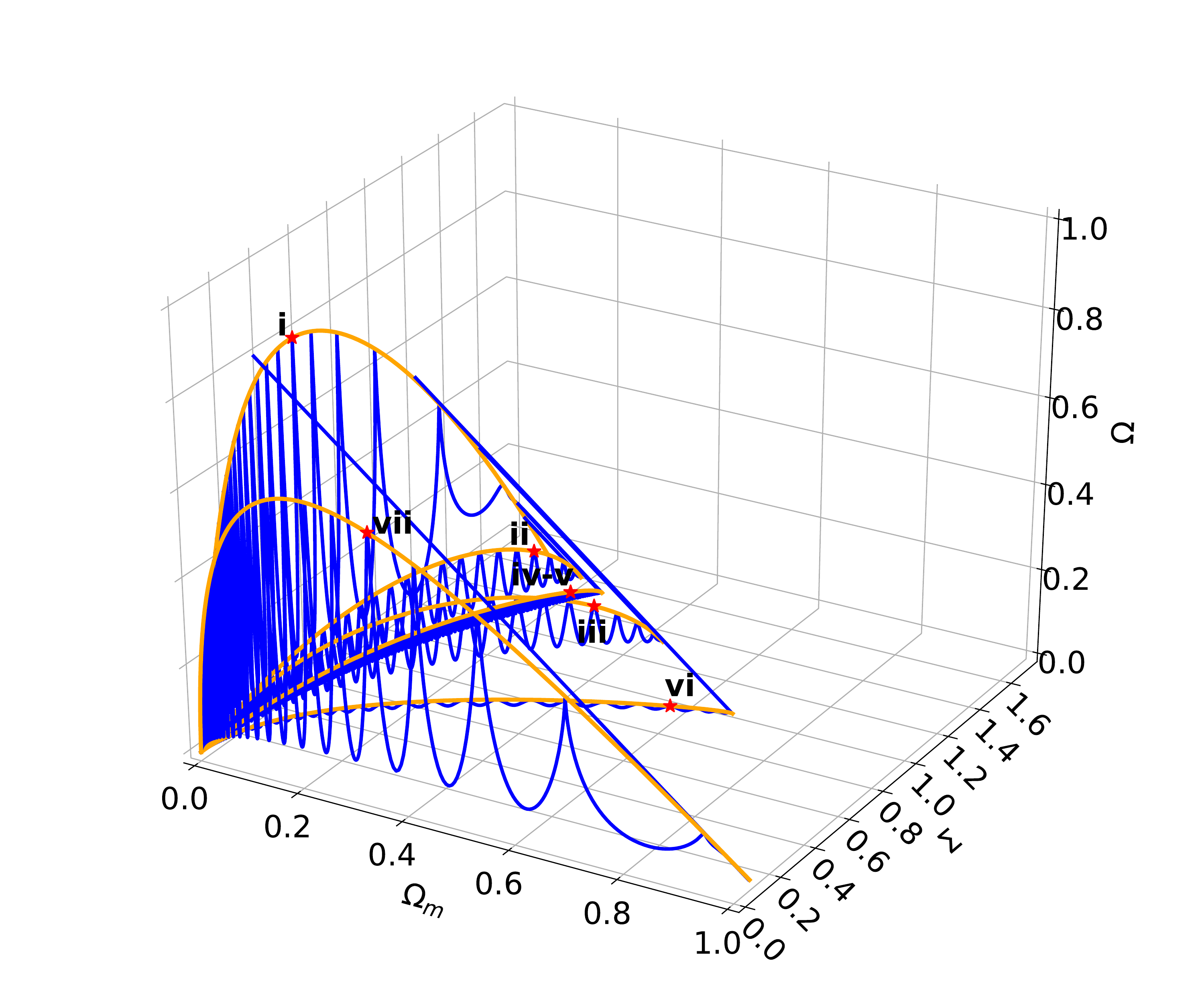}}
  \caption{Some solutions of the full system \eqref{YYEQ3.29YY} (blue) and time-averaged system \eqref{BIYEQ3.29Y} (orange) for a scalar field with generalized harmonic potential non-minimally coupled to matter in the Bianchi I metric when $\lambda = 0.1$, $f=0.1$ and $\gamma=2$. We have used for both systems the initial data sets presented in Table \ref{tab:BianchiI}.}
  \label{fig:BianchiINonminimallyStifff01}
\end{minipage}
\end{figure}

\begin{figure}[ht!]
\centering
\begin{minipage}{.48\textwidth}
  \centering
    \subfigure[\label{BianchiINonminimallyCCf033Dm} Projections in the space $(\Omega_{m},H,\Omega)$. The surface is given by the constraint $\Omega=1-\Omega_{m}$.]{\includegraphics[scale = 0.25]{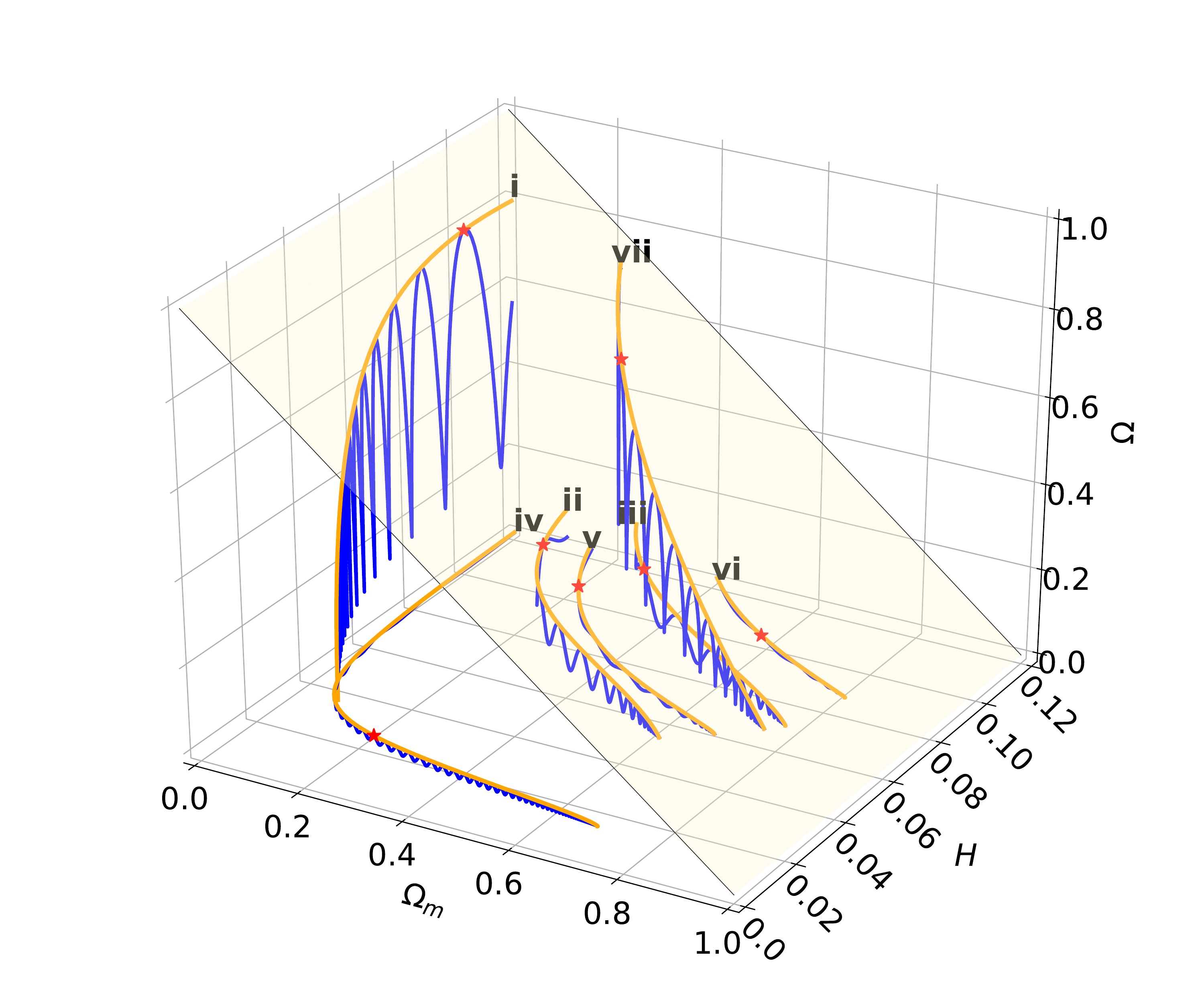}}
    \subfigure[\label{BianchiINonminimallyCCf033DS} Projections in the space $(\Sigma,H,\Omega)$. The surface is given by the constraint $\Omega=1-\Sigma^{2}/3$.]{\includegraphics[scale = 0.25]{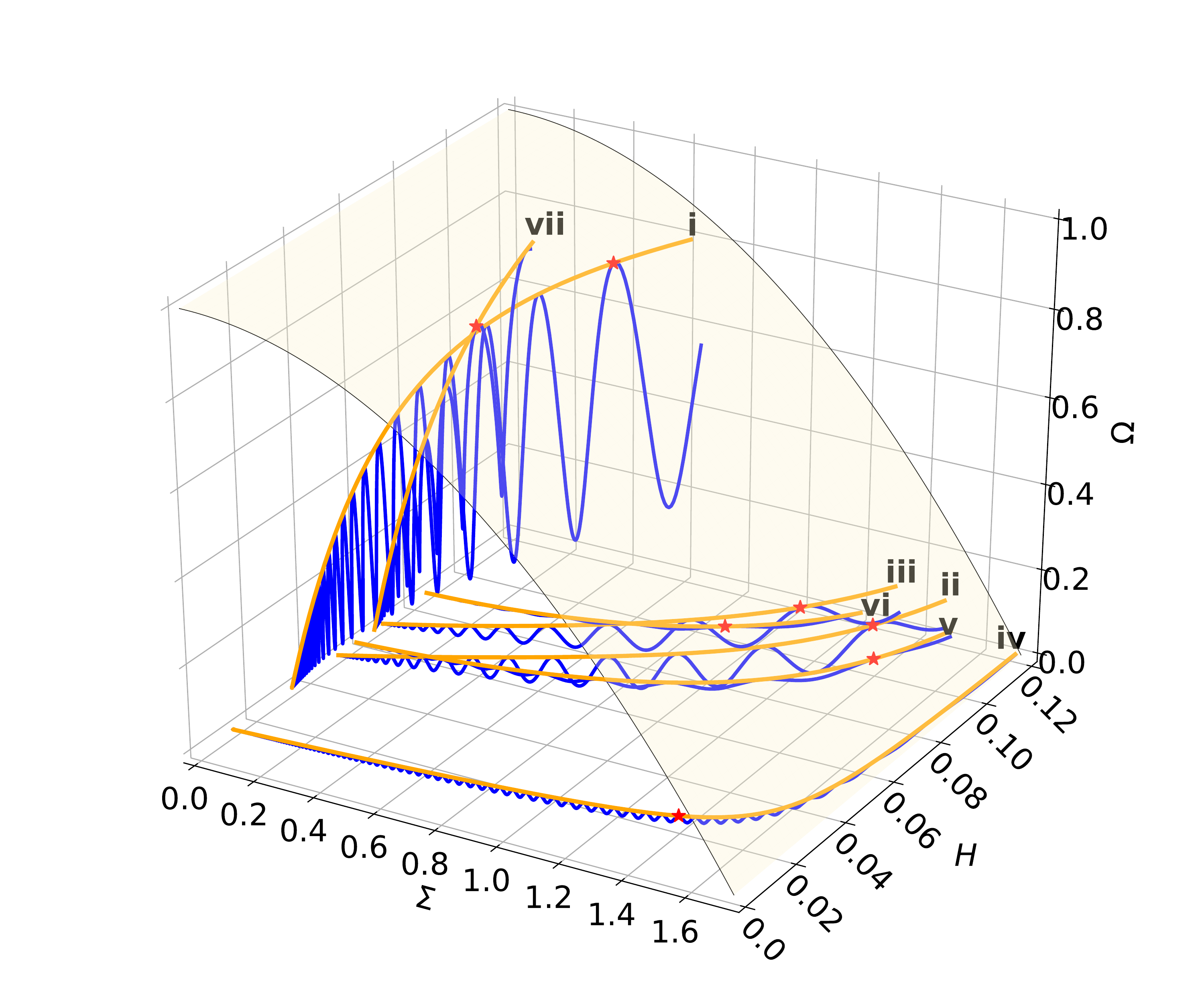}}
    \subfigure[\label{BianchiINonminimallyCCf033Dp} Projections in the space $(\Omega_{m},\Sigma,\Omega)$.]{\includegraphics[scale = 0.25]{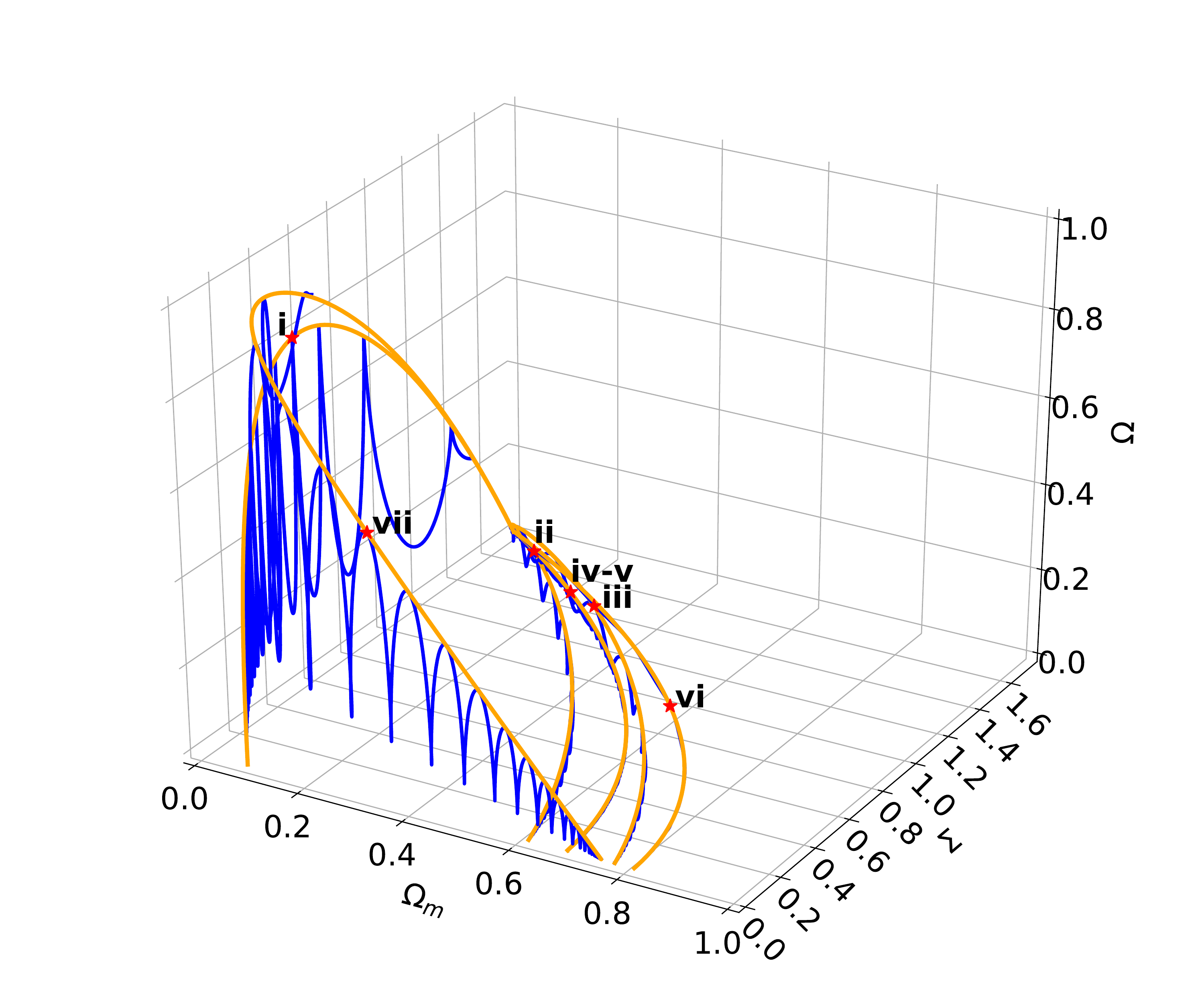}}
  \captionof{figure}{Some solutions of the full system \eqref{YYEQ3.29YY} (blue) and time-averaged system \eqref{BIYEQ3.29Y} (orange) for a scalar field with generalized harmonic potential non-minimally coupled to matter in the Bianchi I metric when $\lambda = 0.1$, $f=0.3$ and $\gamma=0$. We have used for both systems the initial data sets presented in Table \ref{tab:BianchiI}.}
  \label{fig:BianchiINonminimallyCCf03}
\end{minipage}%
\hspace{.02\textwidth}
\begin{minipage}{.48\textwidth}
  \centering
     \subfigure[\label{BianchiINonminimallyBiff033Dm} Projections in the space $(\Omega_{m},H,\Omega)$. The surface is given by the constraint $\Omega=1-\Omega_{m}$.]{\includegraphics[scale = 0.25]{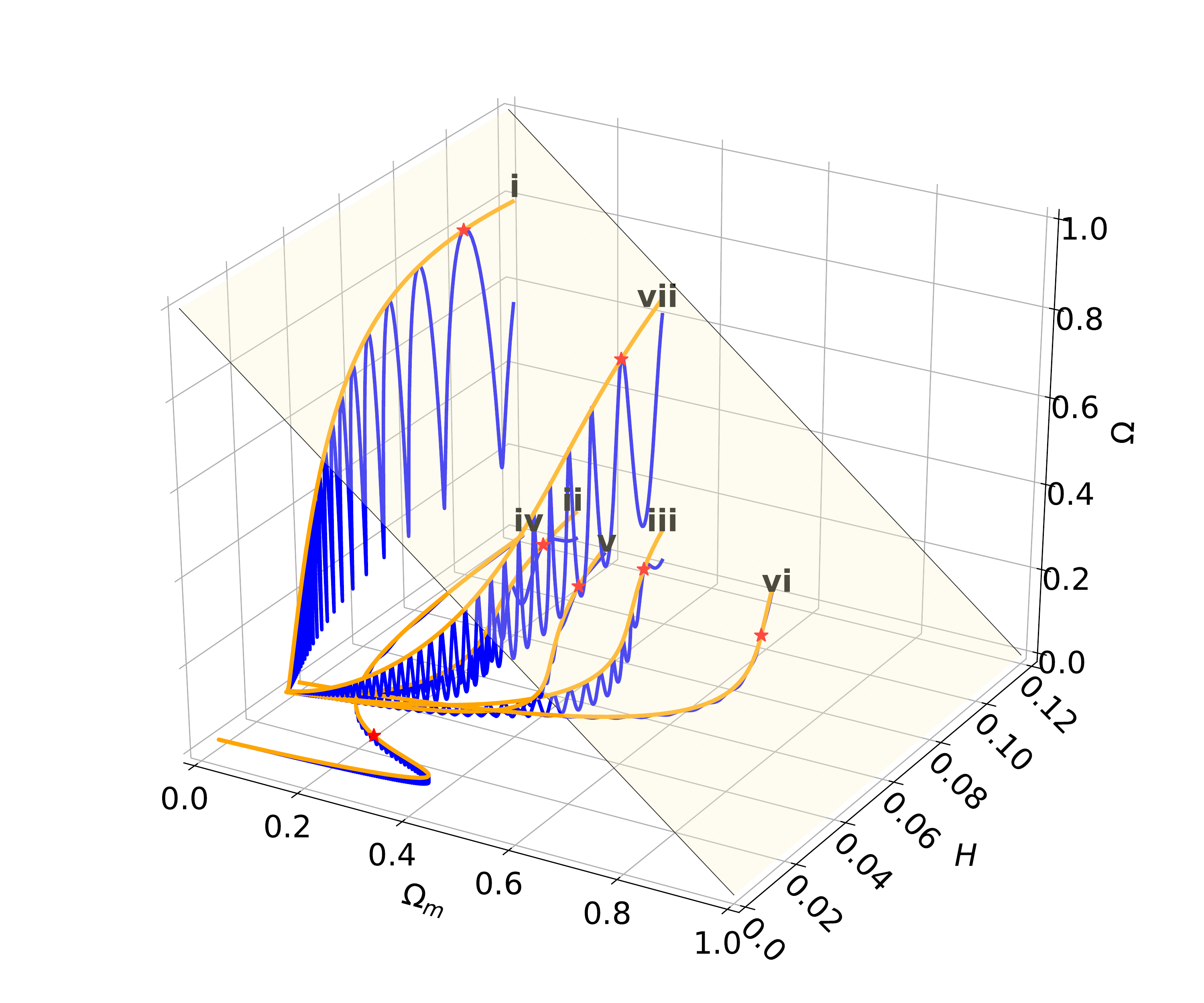}}
    \subfigure[\label{BianchiINonminimallyBiff033DS} Projections in the space $(\Sigma,H,\Omega)$. The surface is given by the constraint $\Omega=1-\Sigma^{2}/3$.]{\includegraphics[scale = 0.25]{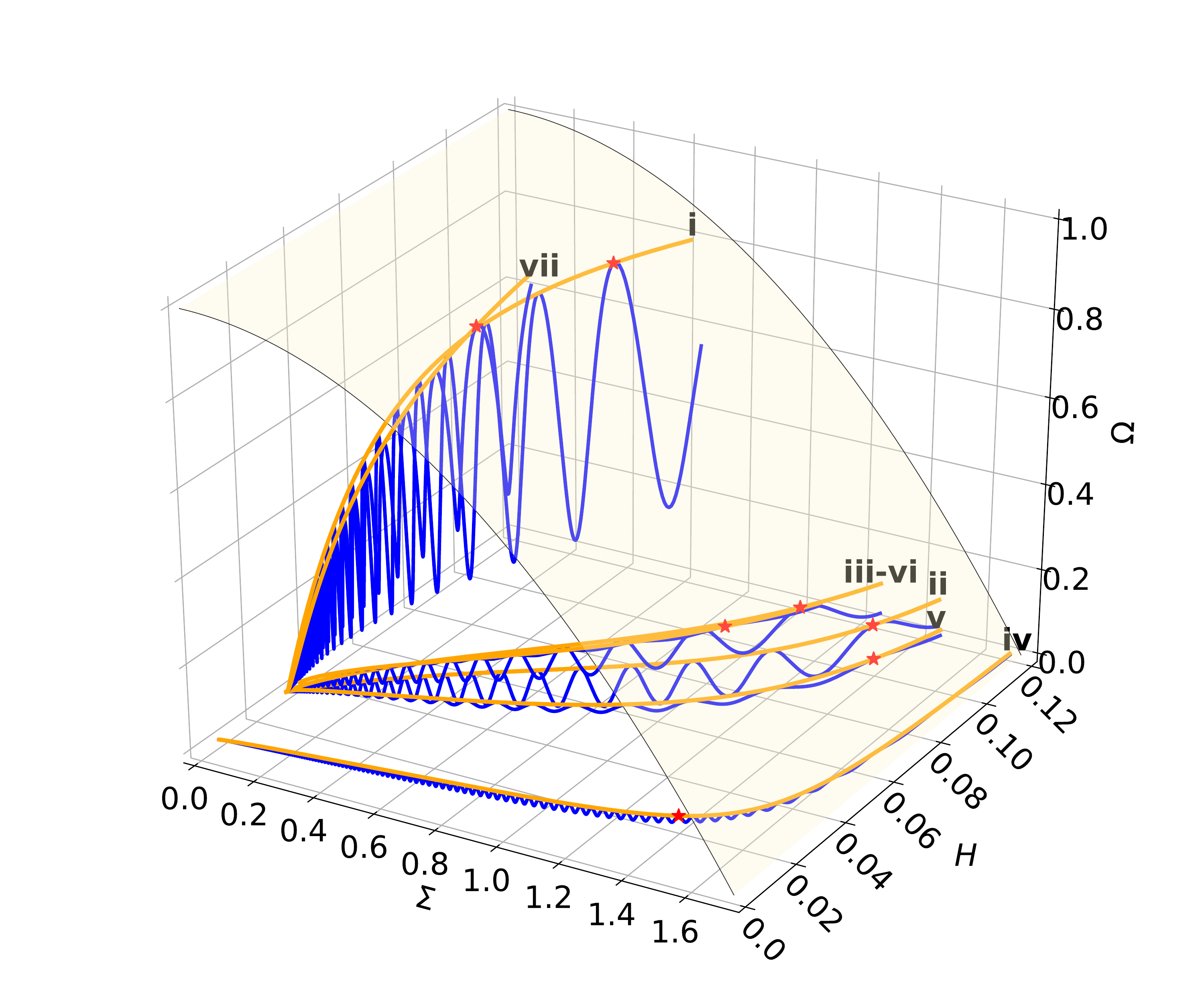}}
    \subfigure[\label{BianchiINonminimallyBiff033Dp} Projections in the space $(\Omega_{m},\Sigma,\Omega)$.]{\includegraphics[scale = 0.25]{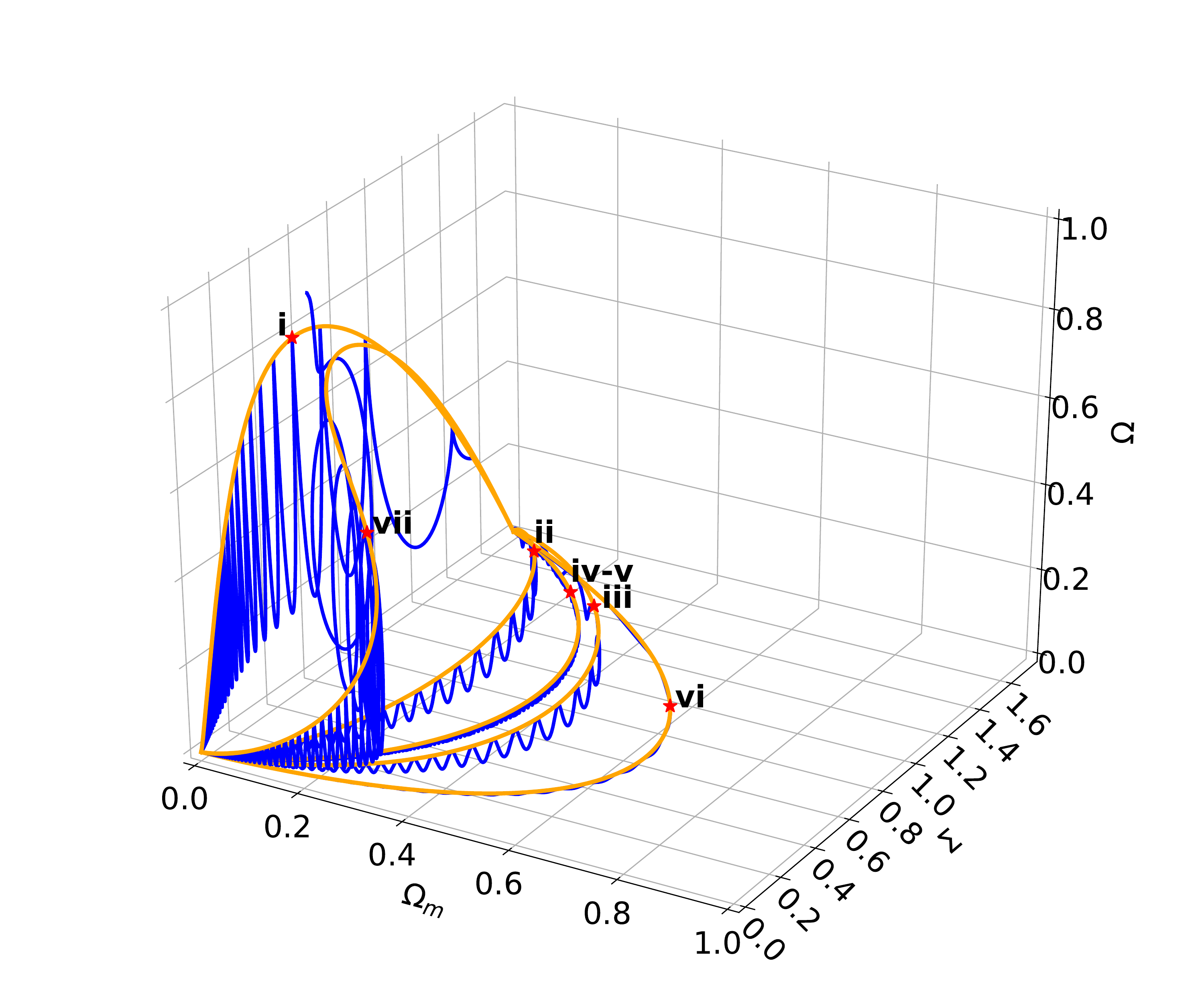}}
  \caption{Some solutions of the full system \eqref{YYEQ3.29YY} (blue) and time-averaged system \eqref{BIYEQ3.29Y} (orange) for a scalar field with generalized harmonic potential non-minimally coupled to matter in the Bianchi I metric when $\lambda = 0.1$, $f=0.3$ and $\gamma=\frac{2}{3}$. We have used for both systems the initial data sets presented in Table \ref{tab:BianchiI}.}
  \label{fig:BianchiINonminimallyBiff03}
\end{minipage}
\end{figure}

\begin{figure}[ht!]
\centering
\begin{minipage}{.48\textwidth}
  \centering
    \subfigure[\label{BianchiINonminimallyDustf033Dm} Projections in the space $(\Omega_{m},H,\Omega)$. The surface is given by the constraint $\Omega=1-\Omega_{m}$.]{\includegraphics[scale = 0.25]{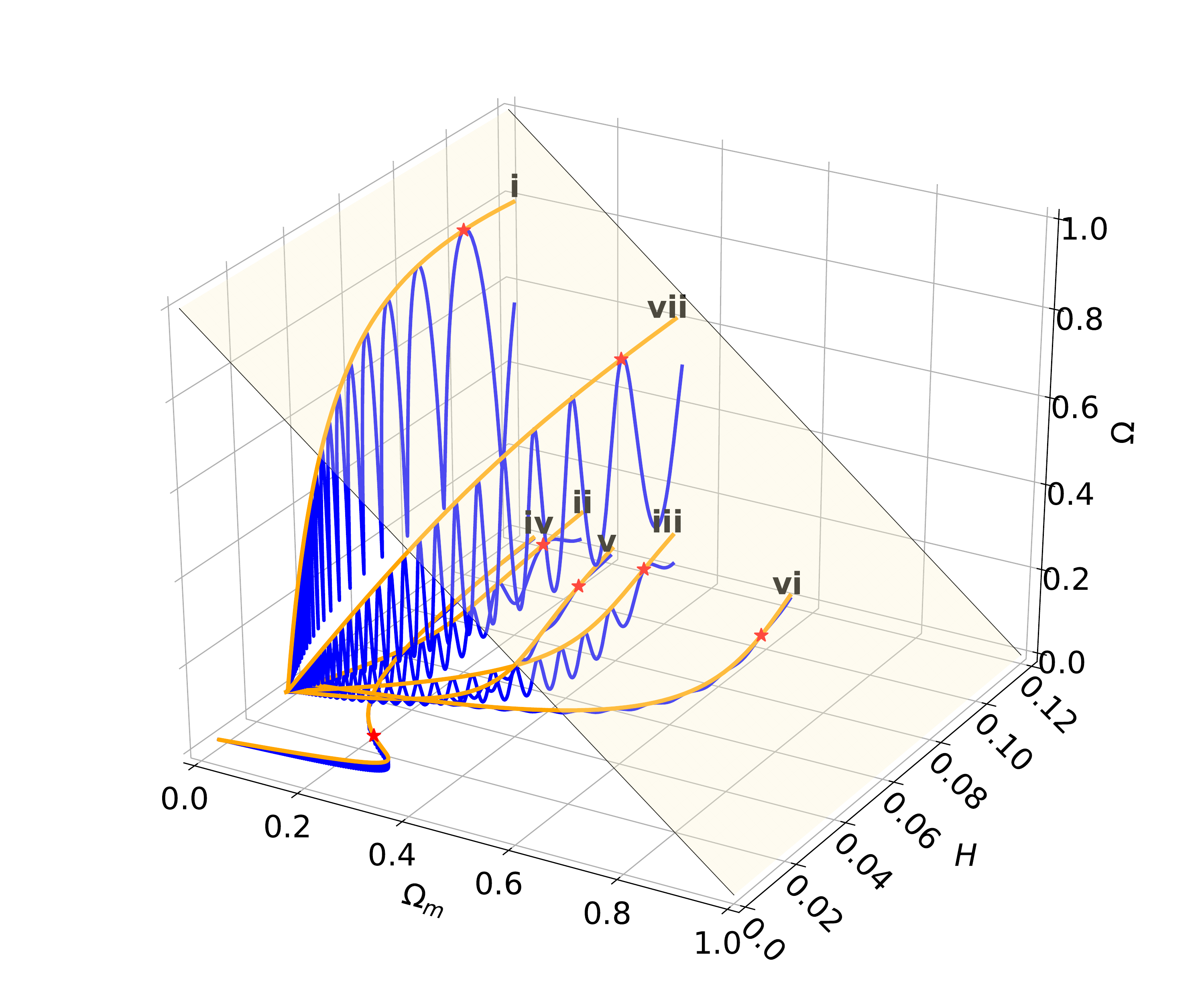}}
    \subfigure[\label{BianchiINonminimallyDustf033DS} Projections in the space $(\Sigma,H,\Omega)$. The surface is given by the constraint $\Omega=1-\Sigma^{2}/3$.]{\includegraphics[scale = 0.25]{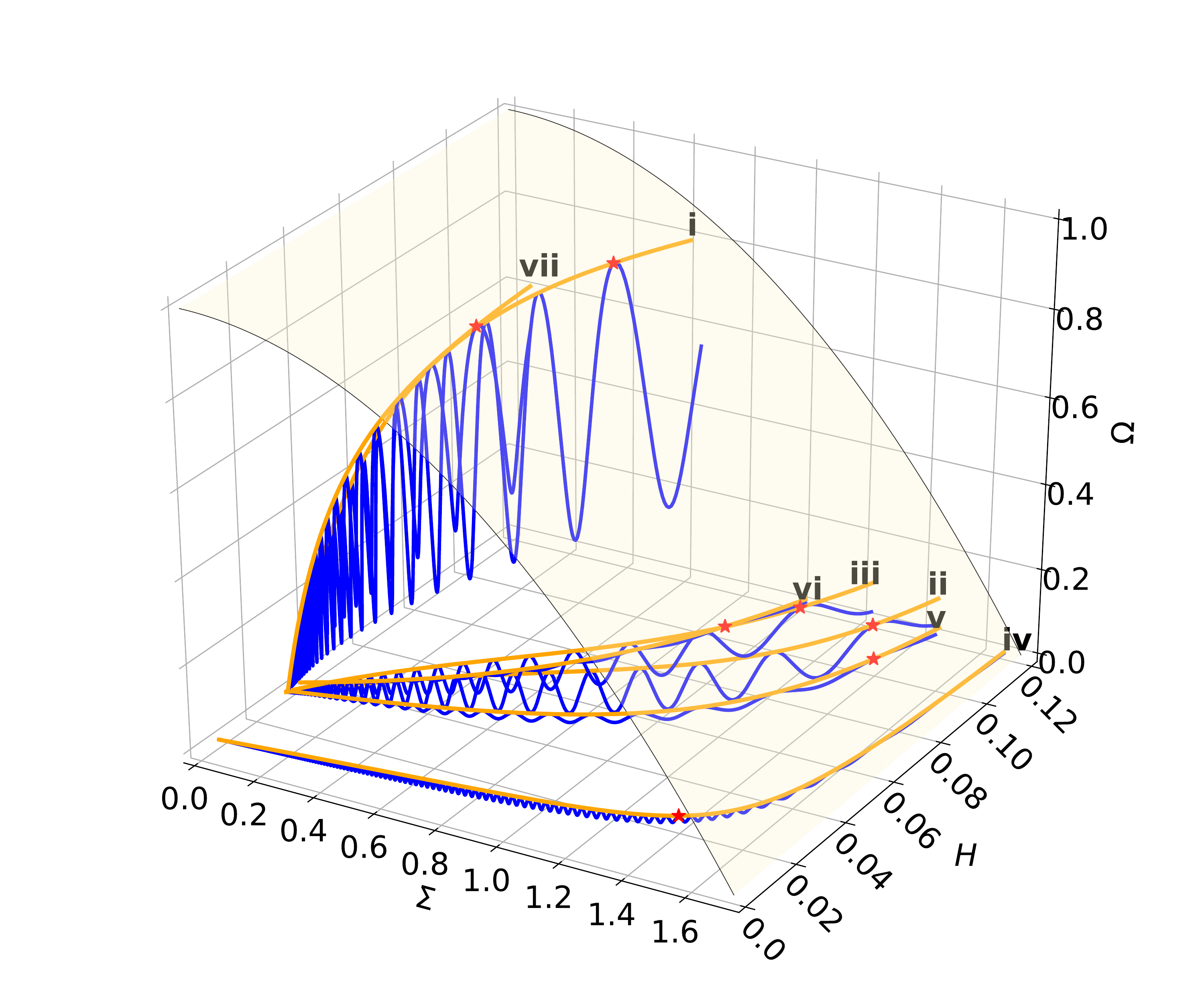}}
    \subfigure[\label{BianchiINonminimallyDustf033Dp} Projections in the space $(\Omega_{m},\Sigma,\Omega)$.]{\includegraphics[scale = 0.25]{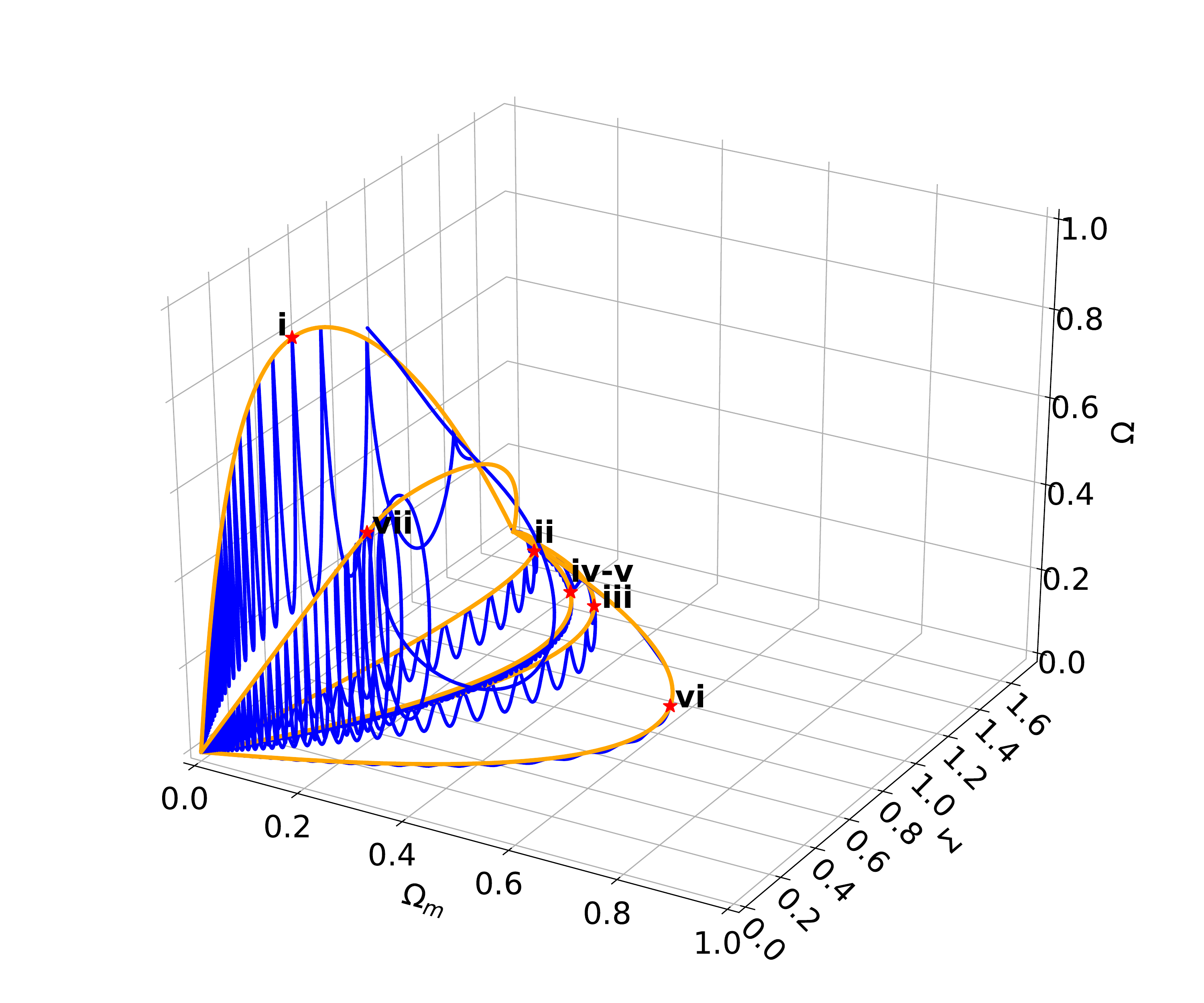}}
  \captionof{figure}{Some solutions of the full system \eqref{YYEQ3.29YY} (blue) and time-averaged system \eqref{BIYEQ3.29Y} (orange) for a scalar field with generalized harmonic potential non-minimally coupled to matter in the Bianchi I metric when $\lambda = 0.1$, $f=0.3$ and $\gamma=1$. We have used for both systems the initial data sets presented in Table \ref{tab:BianchiI}.}
  \label{fig:BianchiINonminimallyDustf03}
\end{minipage}%
\hspace{.02\textwidth}
\begin{minipage}{.48\textwidth}
  \centering
     \subfigure[\label{BianchiINonminimallyStifff033Dm} Projections in the space $(\Omega_{m},H,\Omega)$. The surface is given by the constraint $\Omega=1-\Omega_{m}$.]{\includegraphics[scale = 0.25]{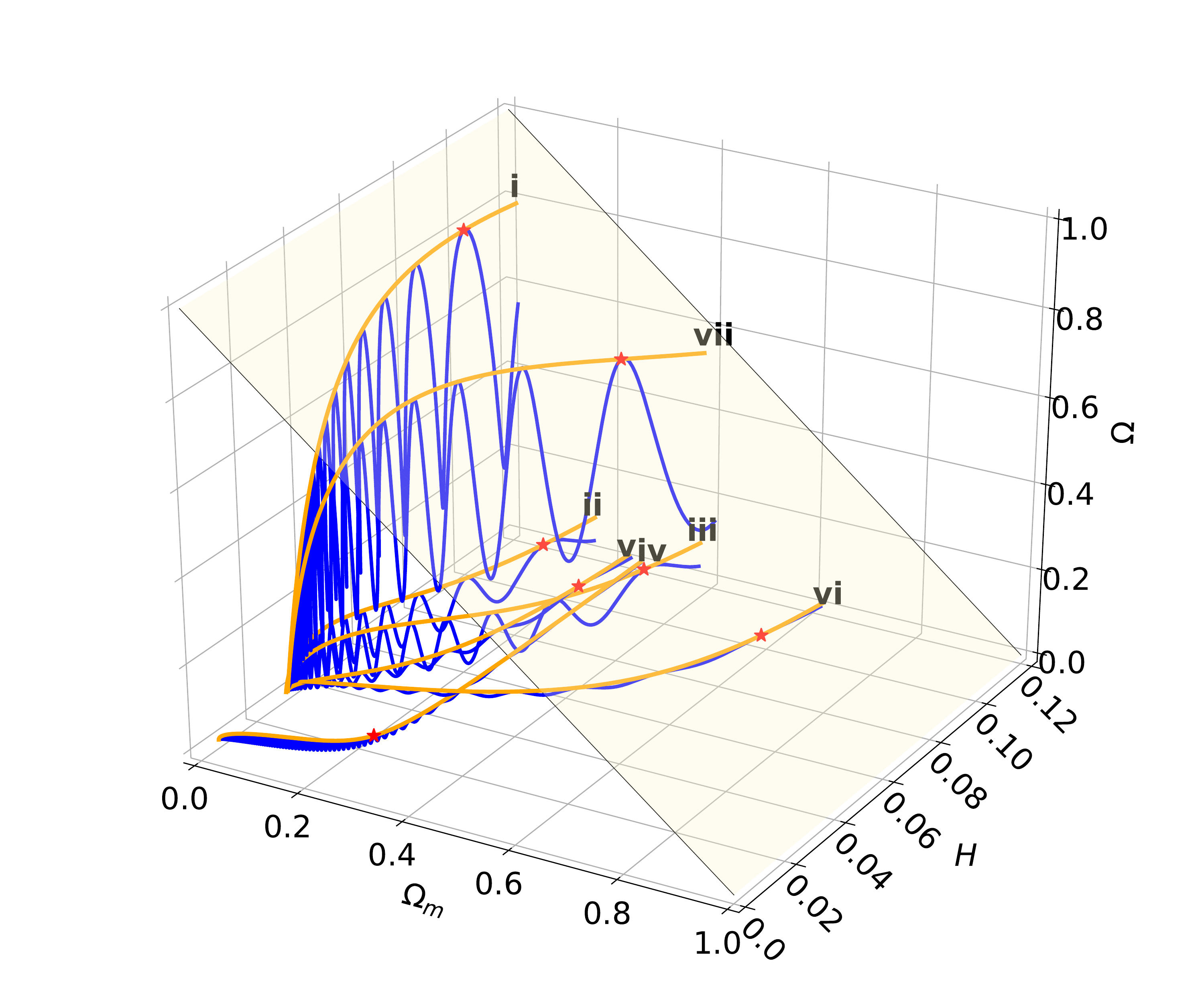}}
    \subfigure[\label{BianchiINonminimallyStifff033DS} Projections in the space $(\Sigma,H,\Omega)$. The surface is given by the constraint $\Omega=1-\Sigma^{2}/3$.]{\includegraphics[scale = 0.25]{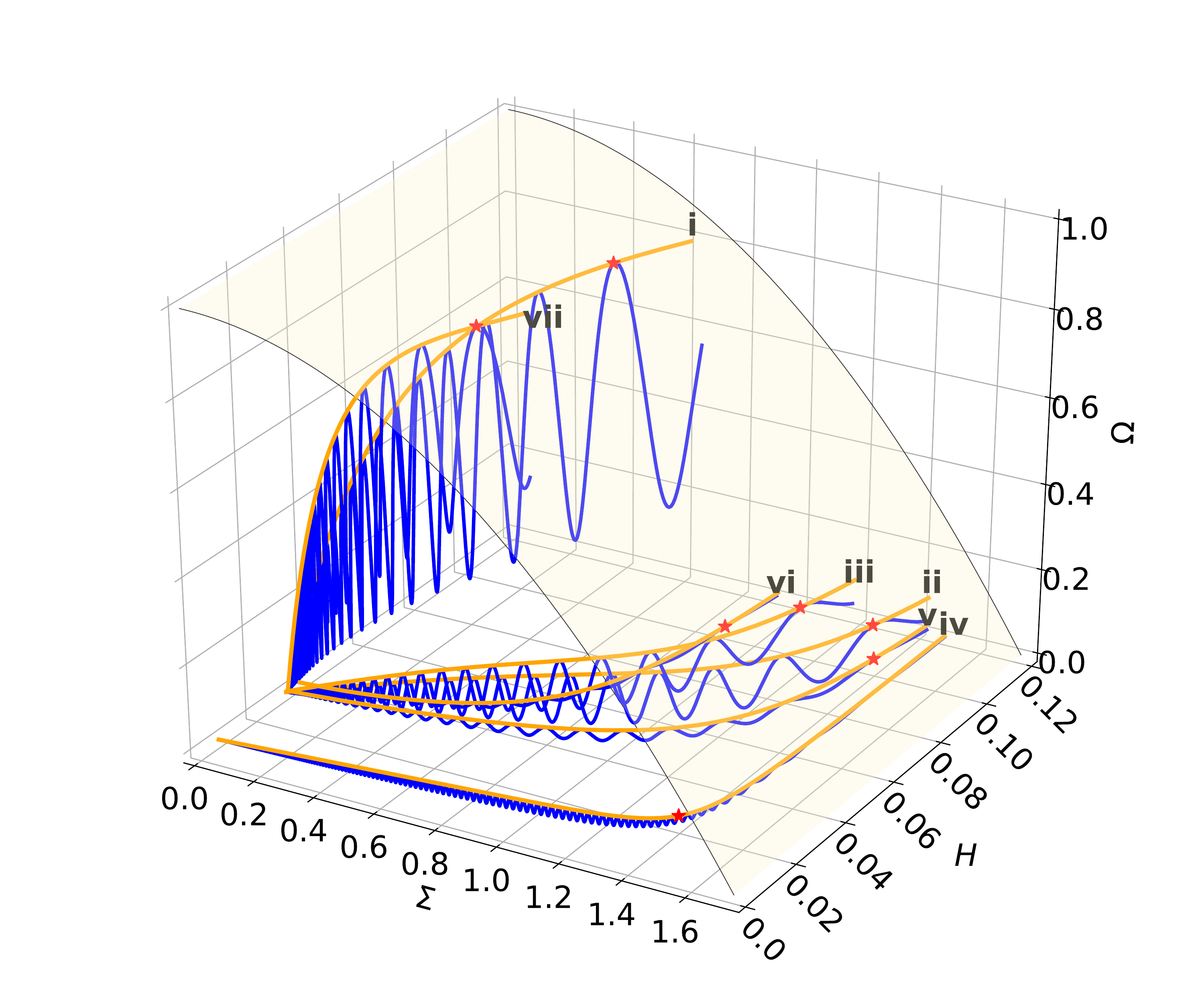}}
    \subfigure[\label{BianchiINonminimallyStifff033Dp} Projections in the space $(\Omega_{m},\Sigma,\Omega)$.]{\includegraphics[scale = 0.25]{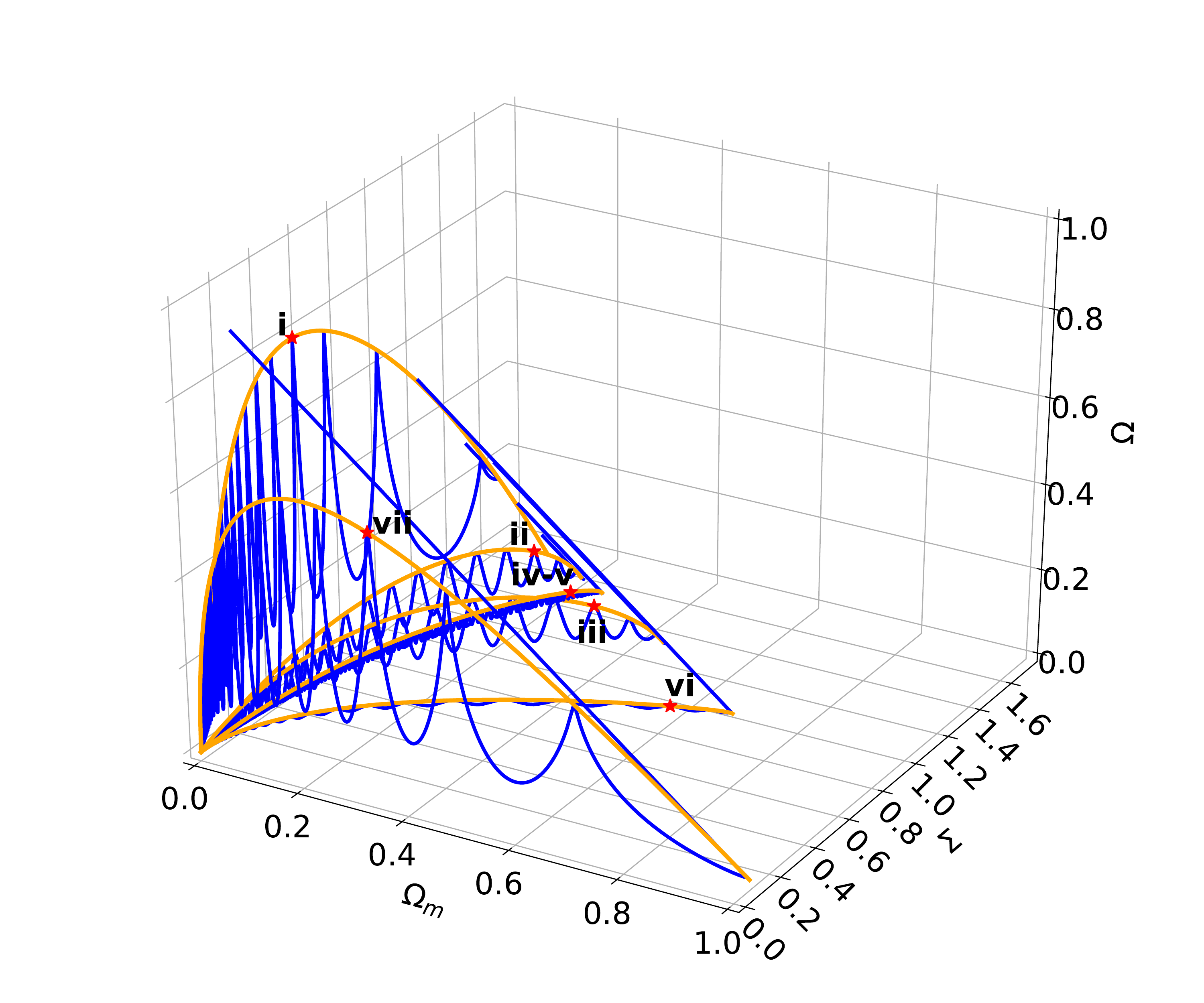}}
  \caption{Some solutions of the full system \eqref{YYEQ3.29YY} (blue) and time-averaged system \eqref{BIYEQ3.29Y} (orange) for a scalar field with generalized harmonic potential non-minimally coupled to matter in the Bianchi I metric when $\lambda = 0.1$, $f=0.3$ and $\gamma=2$. We have used for both systems the initial data sets presented in Table \ref{tab:BianchiI}.}
  \label{fig:BianchiINonminimallyStifff03}
\end{minipage}
\end{figure}

\begin{figure}[ht!]
\centering
\begin{minipage}{.48\textwidth}
  \centering
    \subfigure[\label{BianchiINonminimallyCCf093Dm} Projections in the space $(\Omega_{m},H,\Omega)$. The surface is given by the constraint $\Omega=1-\Omega_{m}$.]{\includegraphics[scale = 0.25]{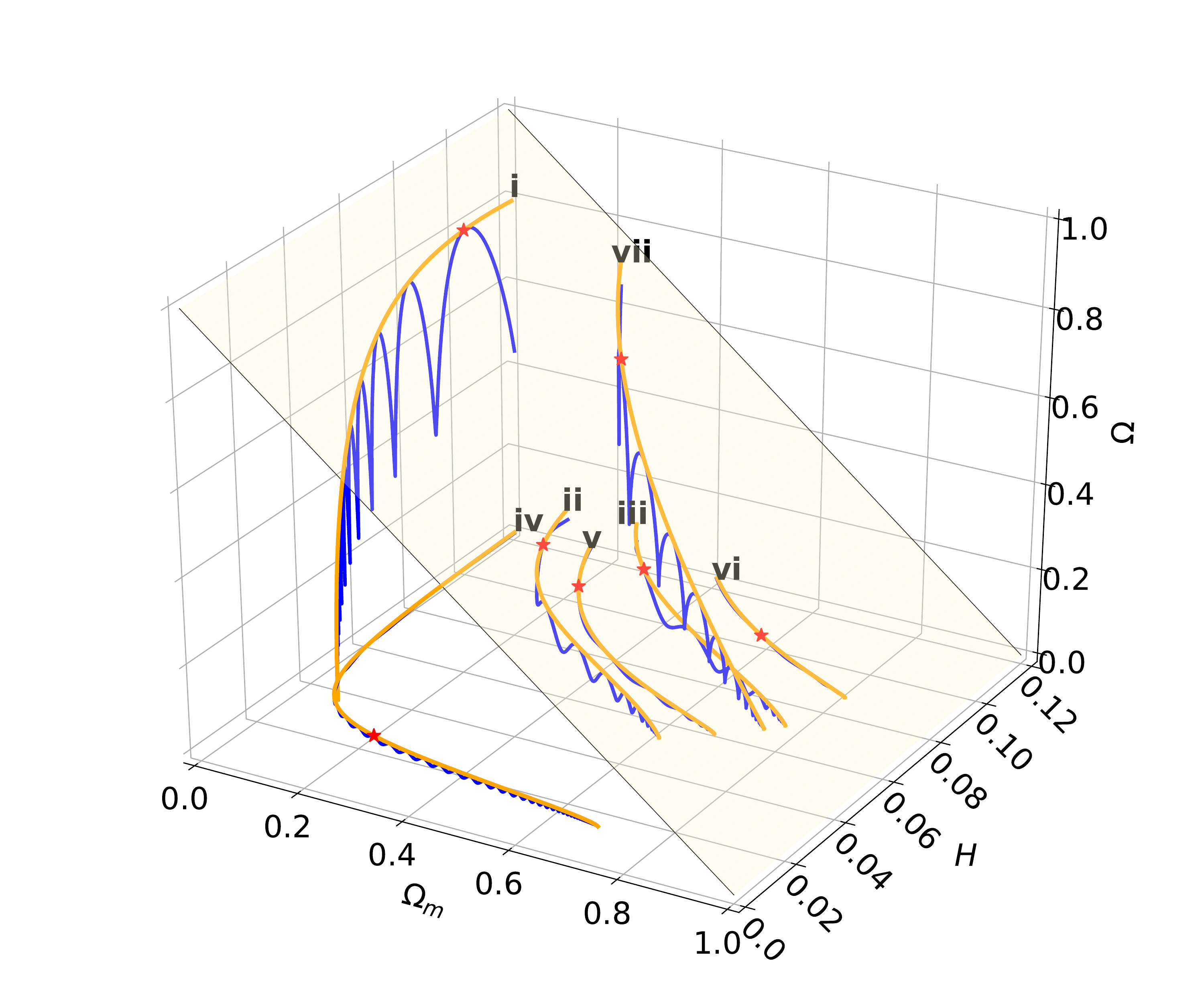}}
    \subfigure[\label{BianchiINonminimallyCCf093DS} Projections in the space $(\Sigma,H,\Omega)$. The surface is given by the constraint $\Omega=1-\Sigma^{2}/3$.]{\includegraphics[scale = 0.25]{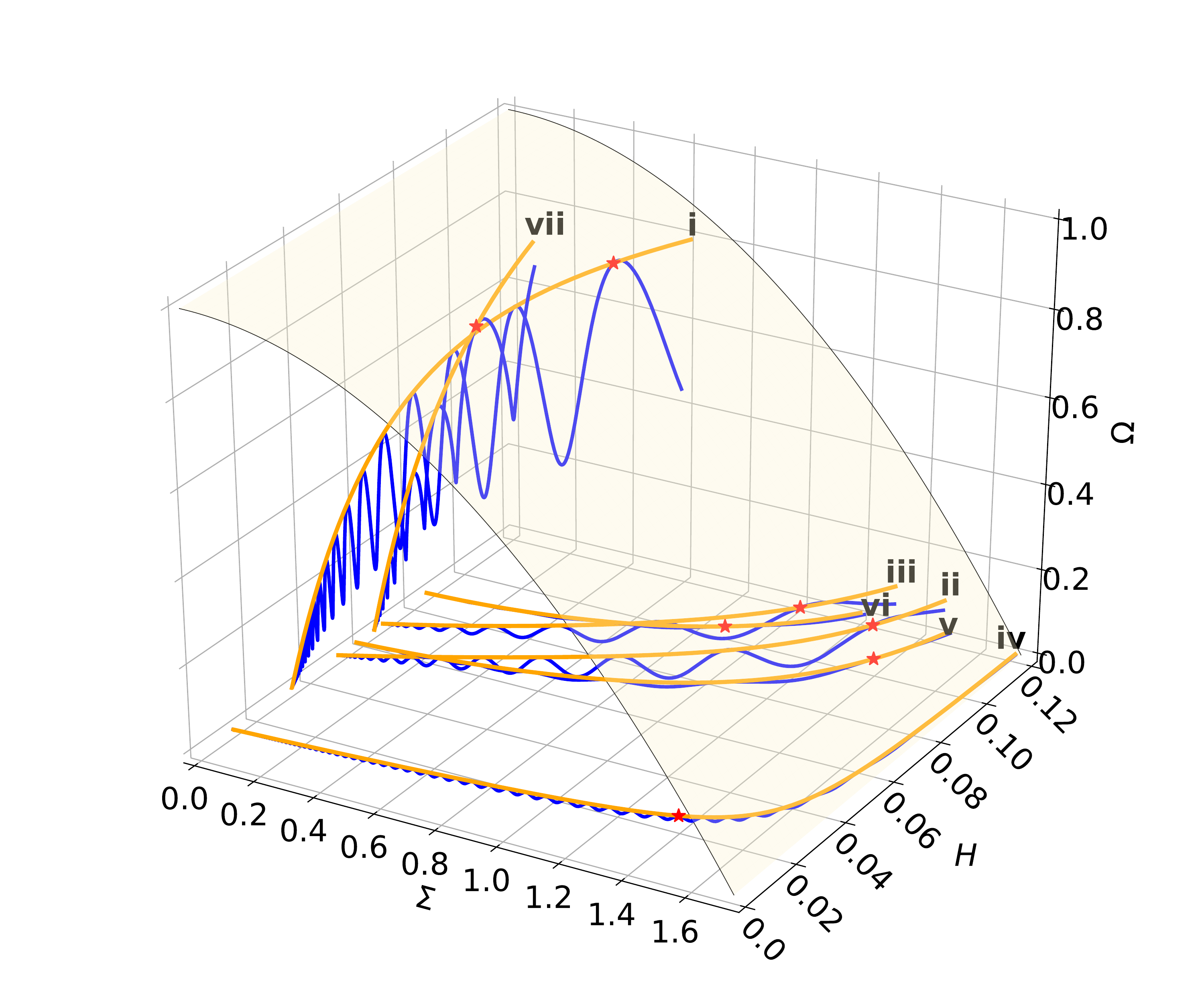}}
    \subfigure[\label{BianchiINonminimallyCCf093Dp} Projections in the space $(\Omega_{m},\Sigma,\Omega)$.]{\includegraphics[scale = 0.25]{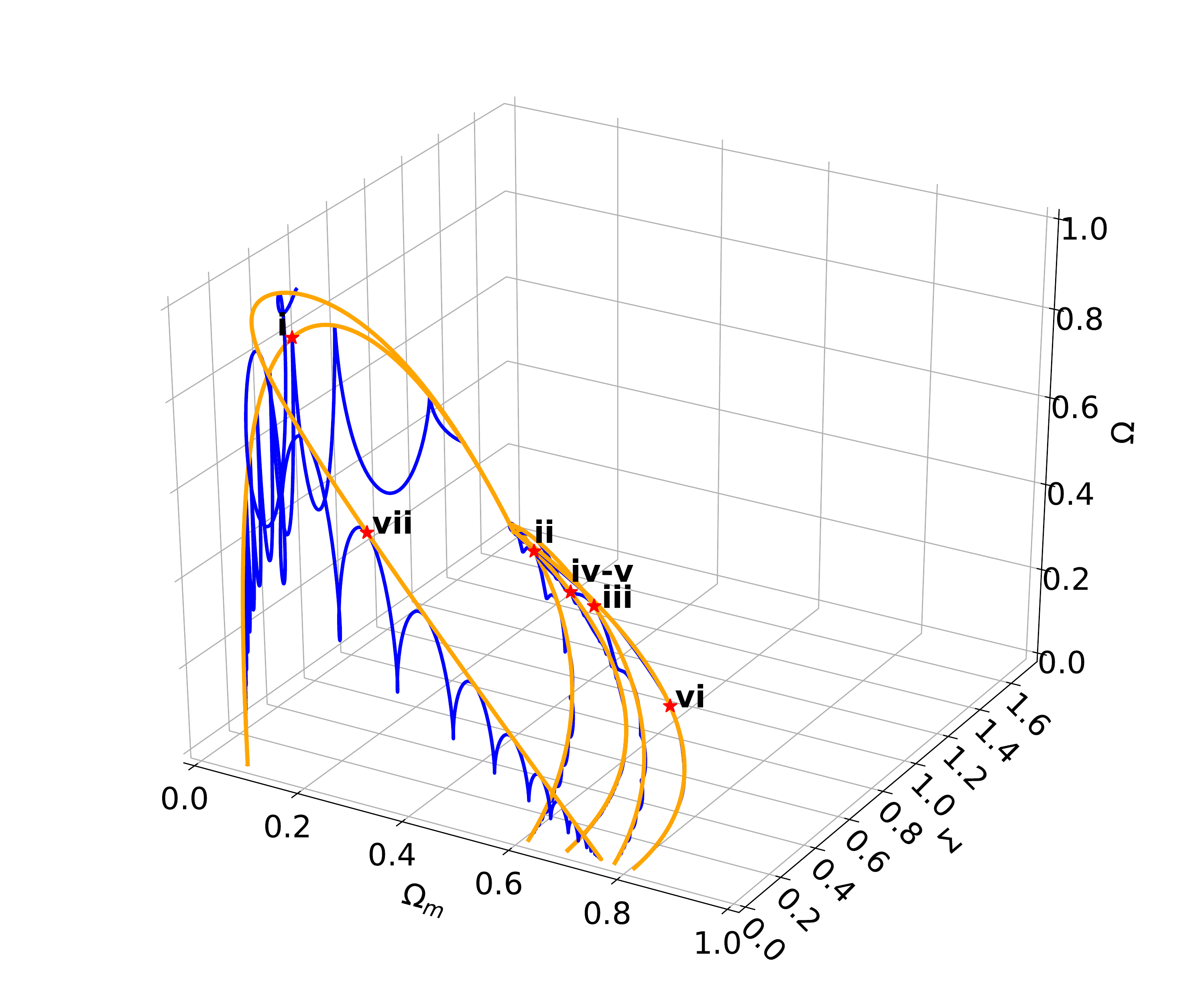}}
  \captionof{figure}{Some solutions of the full system \eqref{YYEQ3.29YY} (blue) and time-averaged system \eqref{BIYEQ3.29Y} (orange) for a scalar field with generalized harmonic potential non-minimally coupled to matter in the Bianchi I metric when $\lambda = 0.1$, $f=0.9$ and $\gamma=0$. We have used for both systems the initial data sets presented in Table \ref{tab:BianchiI}.}
  \label{fig:BianchiINonminimallyCCf09}
\end{minipage}%
\hspace{.02\textwidth}
\begin{minipage}{.48\textwidth}
  \centering
     \subfigure[\label{BianchiINonminimallyBiff093Dm} Projections in the space $(\Omega_{m},H,\Omega)$. The surface is given by the constraint $\Omega=1-\Omega_{m}$.]{\includegraphics[scale = 0.25]{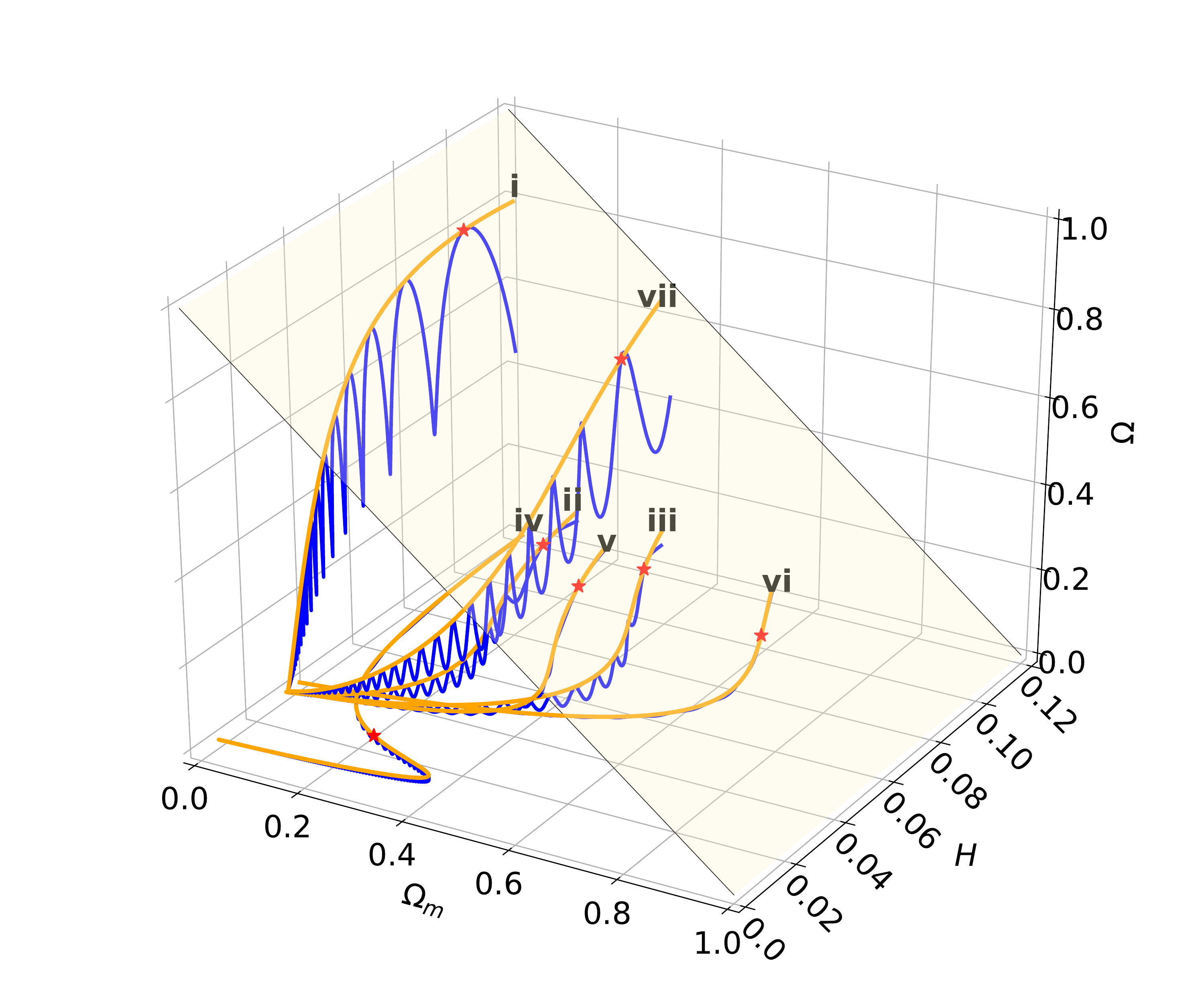}}
    \subfigure[\label{BianchiINonminimallyBiff093DS} Projections in the space $(\Sigma,H,\Omega)$. The surface is given by the constraint $\Omega=1-\Sigma^{2}/3$.]{\includegraphics[scale = 0.25]{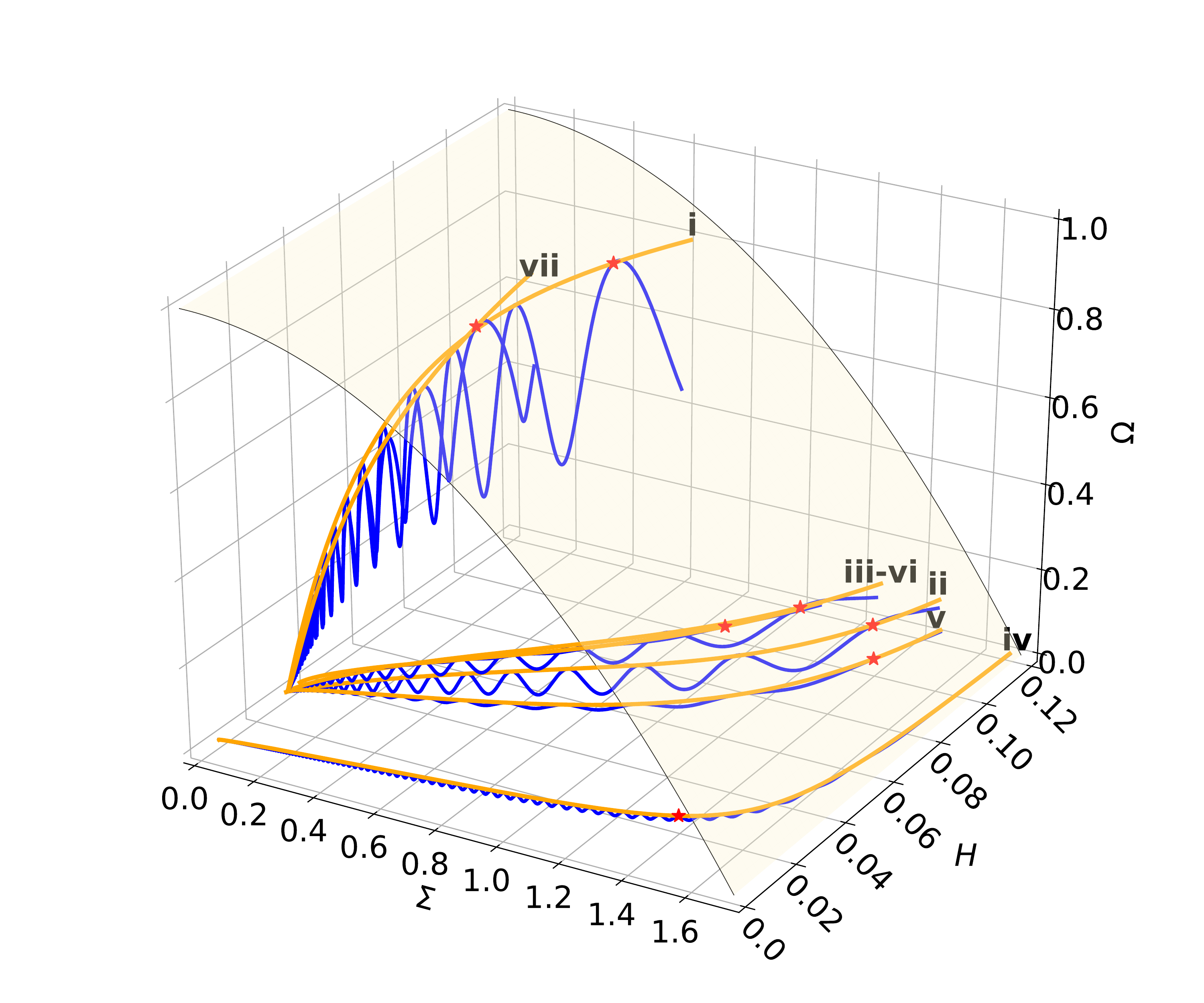}}
    \subfigure[\label{BianchiINonminimallyBiff093Dp} Projections in the space $(\Omega_{m},\Sigma,\Omega)$.]{\includegraphics[scale = 0.25]{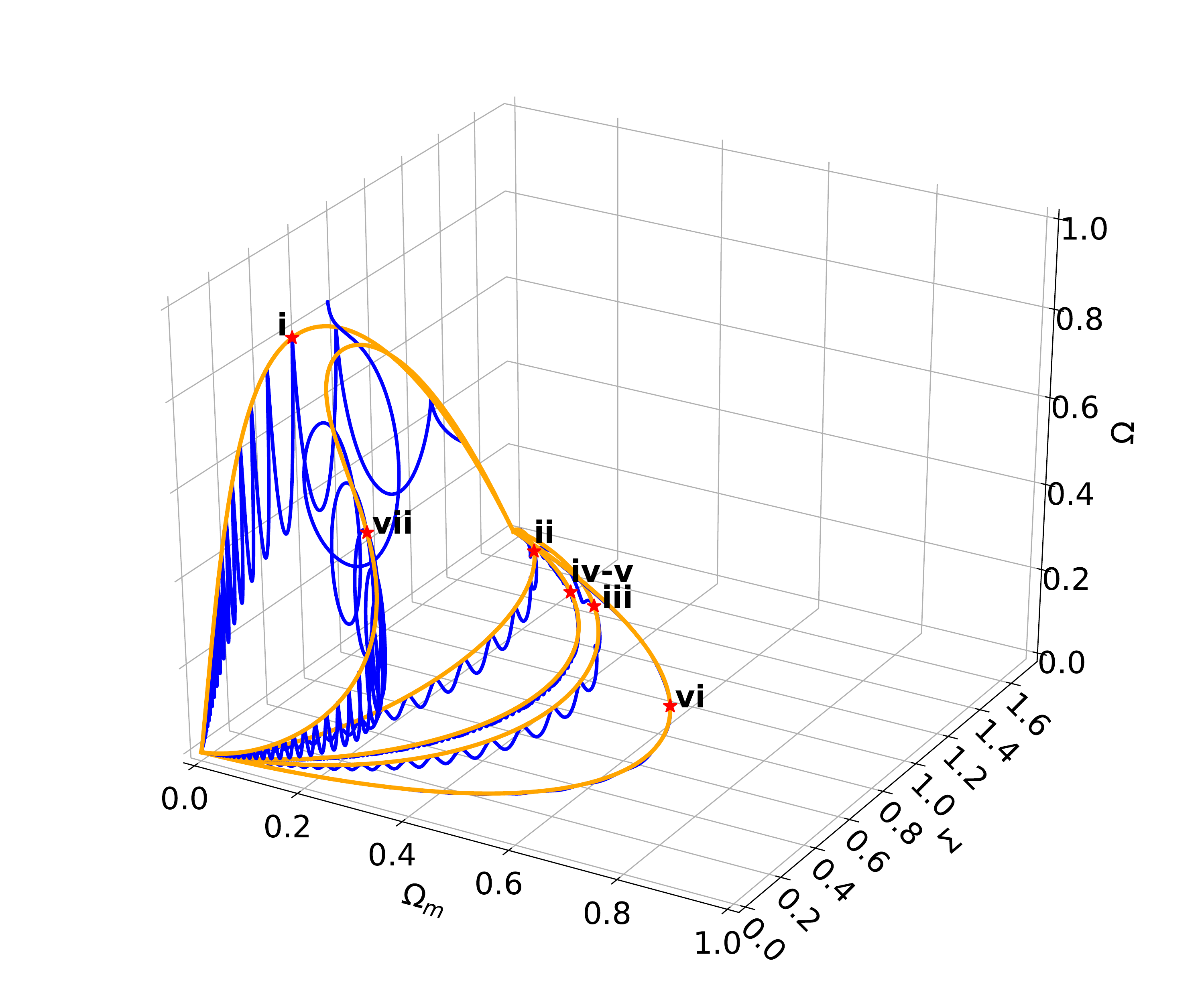}}
  \caption{Some solutions of the full system \eqref{YYEQ3.29YY} (blue) and time-averaged system \eqref{BIYEQ3.29Y} (orange) for a scalar field with generalized harmonic potential non-minimally coupled to matter in the Bianchi I metric when $\lambda = 0.1$, $f=0.9$ and $\gamma=\frac{2}{3}$. We have used for both systems the initial data sets presented in Table \ref{tab:BianchiI}.}
  \label{fig:BianchiINonminimallyBiff09}
\end{minipage}
\end{figure}

\begin{figure}[ht!]
\centering
\begin{minipage}{.48\textwidth}
  \centering
    \subfigure[\label{BianchiINonminimallyDustf093Dm} Projections in the space $(\Omega_{m},H,\Omega)$. The surface is given by the constraint $\Omega=1-\Omega_{m}$.]{\includegraphics[scale = 0.25]{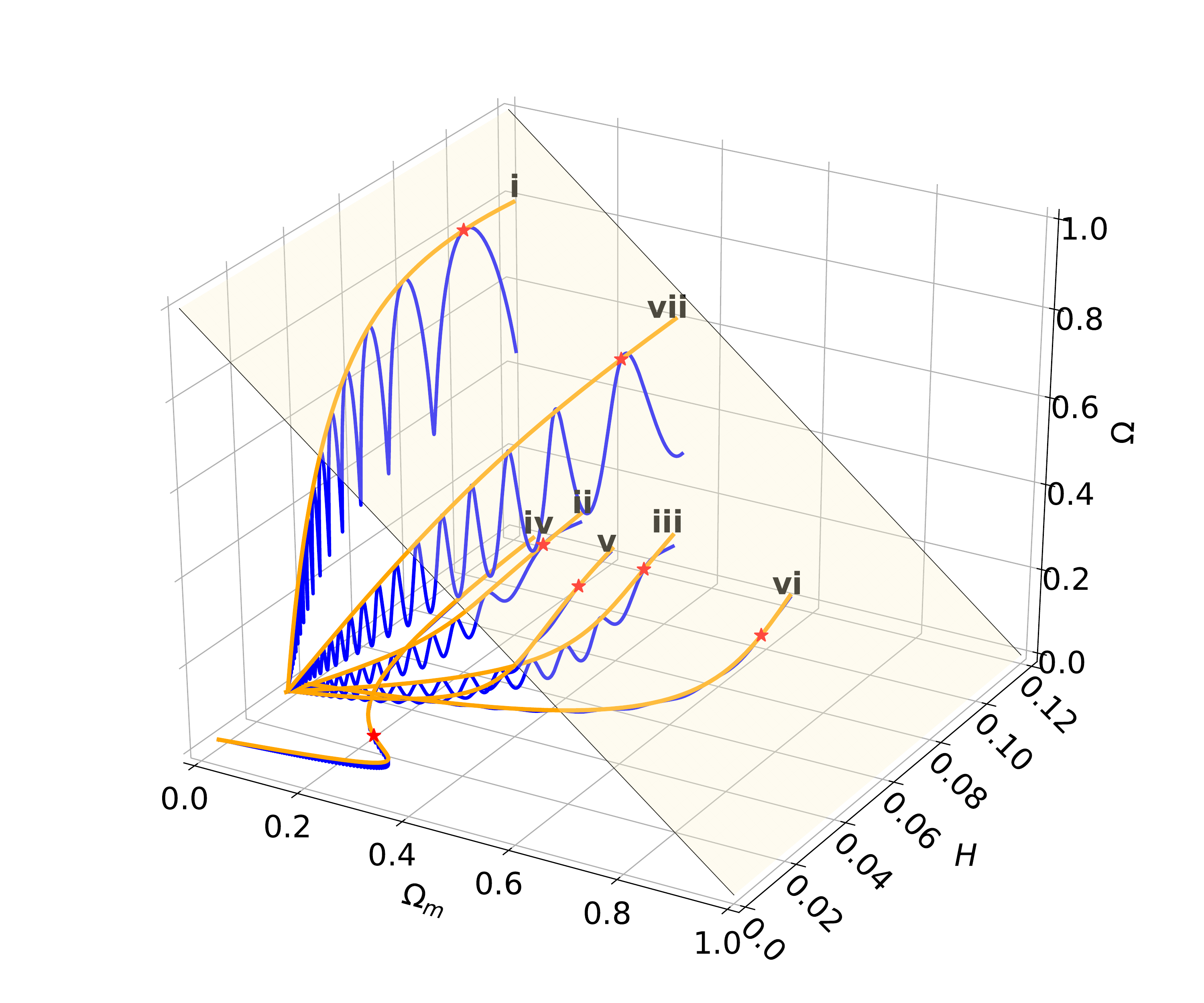}}
    \subfigure[\label{BianchiINonminimallyDustf093DS} Projections in the space $(\Sigma,H,\Omega)$. The surface is given by the constraint $\Omega=1-\Sigma^{2}/3$.]{\includegraphics[scale = 0.25]{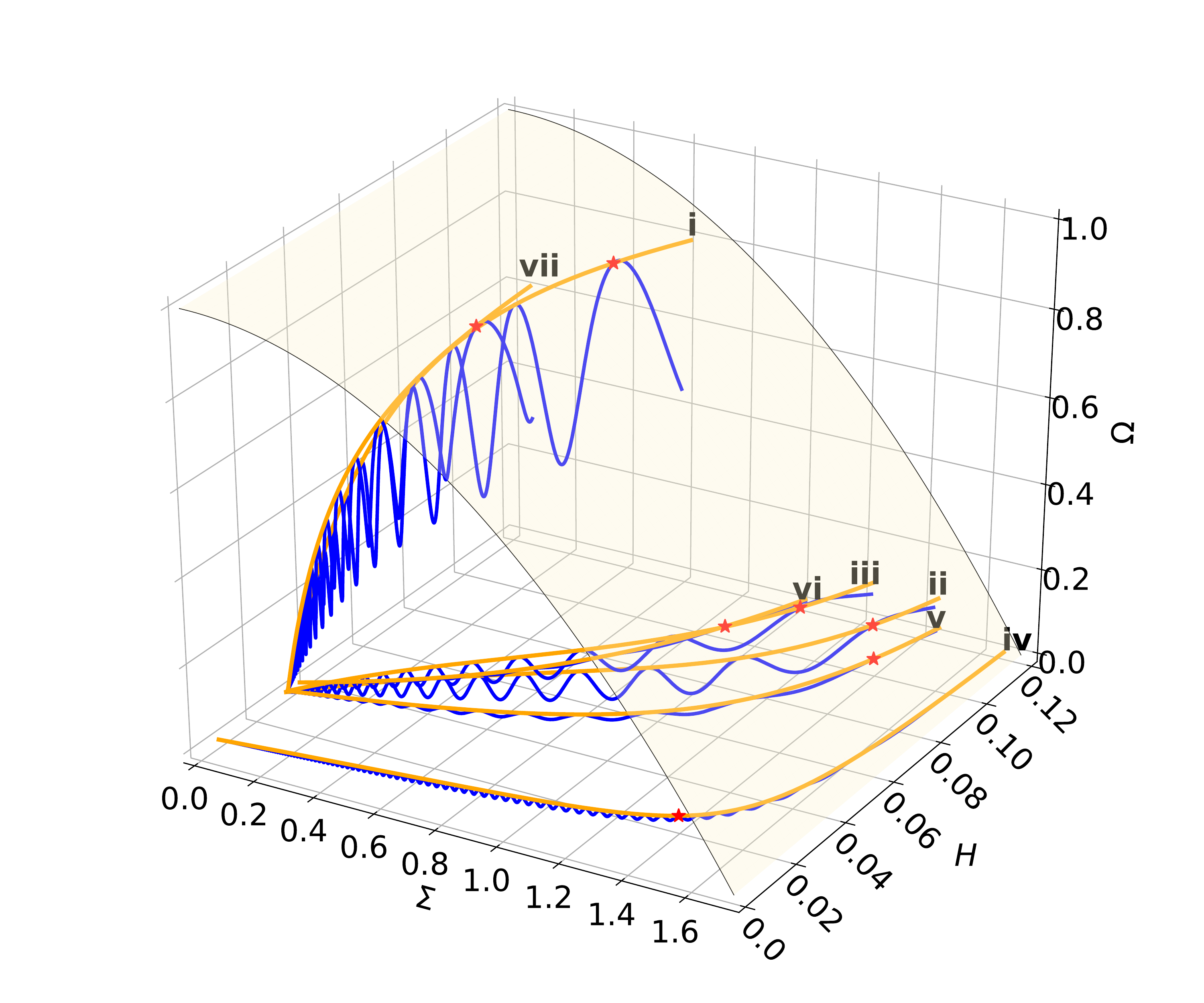}}
    \subfigure[\label{BianchiINonminimallyDustf093Dp} Projections in the space $(\Omega_{m},\Sigma,\Omega)$.]{\includegraphics[scale = 0.25]{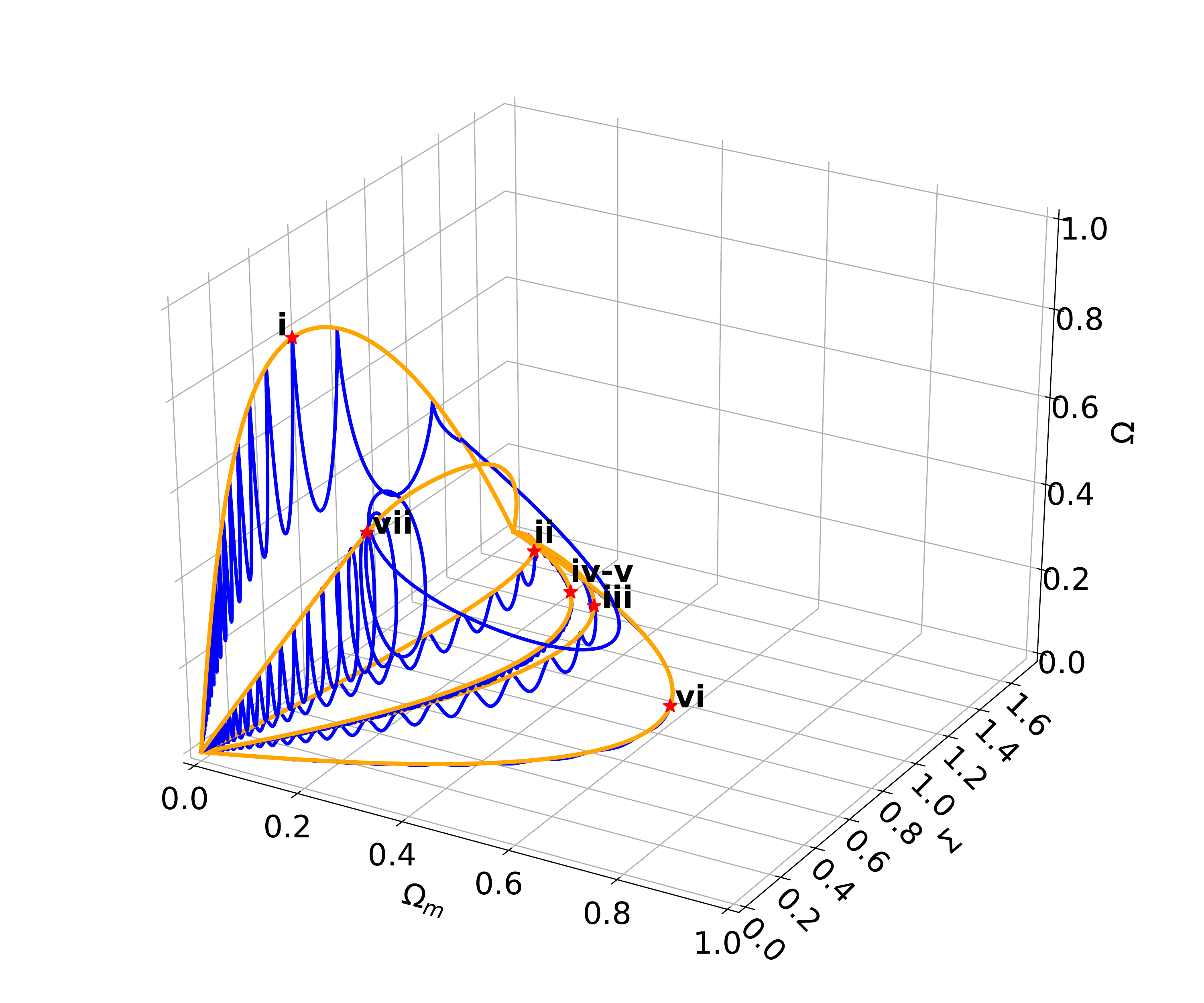}}
  \captionof{figure}{Some solutions of the full system \eqref{YYEQ3.29YY} (blue) and time-averaged system \eqref{BIYEQ3.29Y} (orange) for a scalar field with generalized harmonic potential non-minimally coupled to matter in the Bianchi I metric when $\lambda = 0.1$, $f=0.9$ and $\gamma=1$. We have used for both systems the initial data sets presented in Table \ref{tab:BianchiI}.}
  \label{fig:BianchiINonminimallyDustf09}
\end{minipage}%
\hspace{.02\textwidth}
\begin{minipage}{.48\textwidth}
  \centering
     \subfigure[\label{BianchiINonminimallyStifff093Dm} Projections in the space $(\Omega_{m},H,\Omega)$. The surface is given by the constraint $\Omega=1-\Omega_{m}$.]{\includegraphics[scale = 0.25]{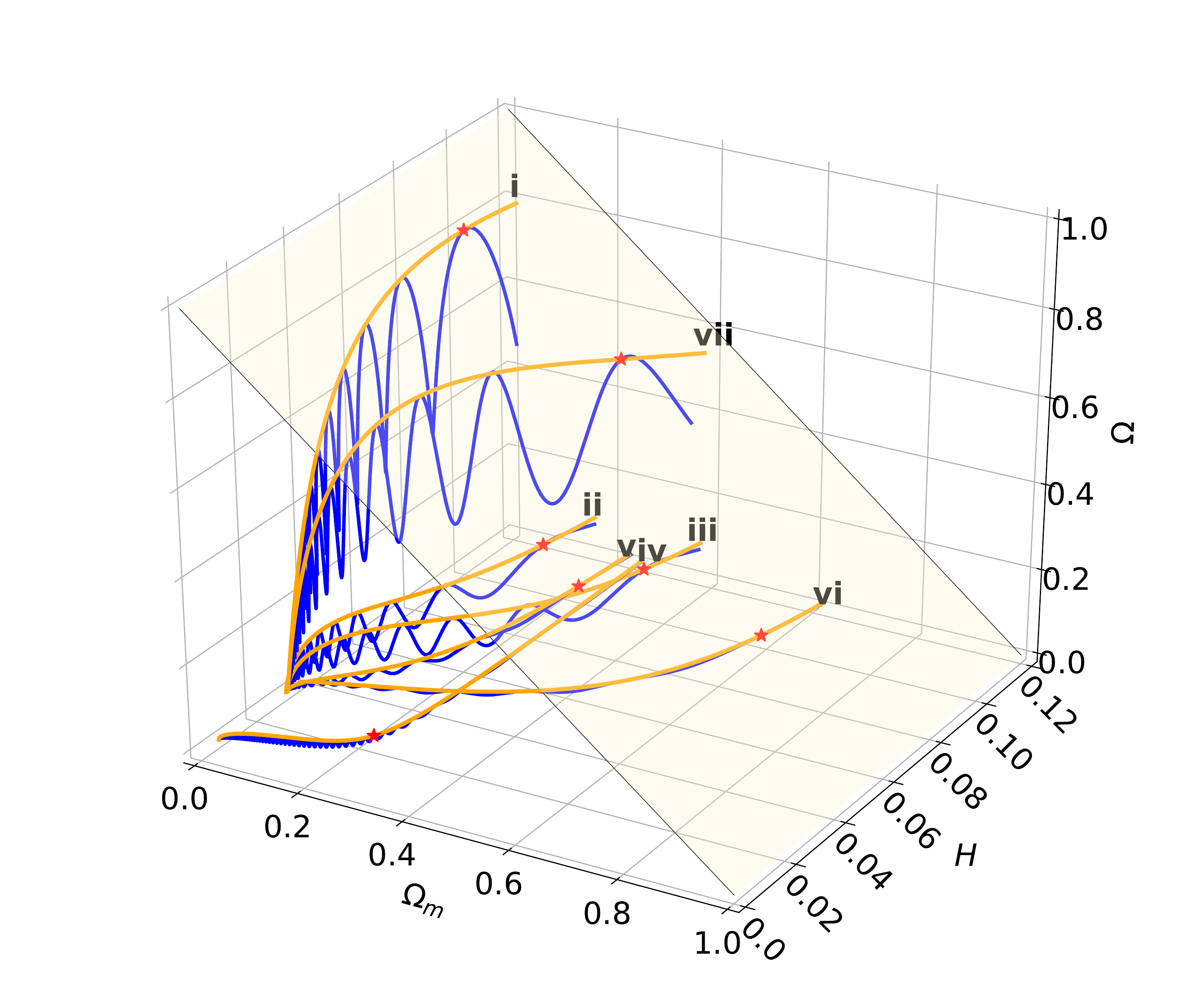}}
    \subfigure[\label{BianchiINonminimallyStifff093DS} Projections in the space $(\Sigma,H,\Omega)$. The surface is given by the constraint $\Omega=1-\Sigma^{2}/3$.]{\includegraphics[scale = 0.25]{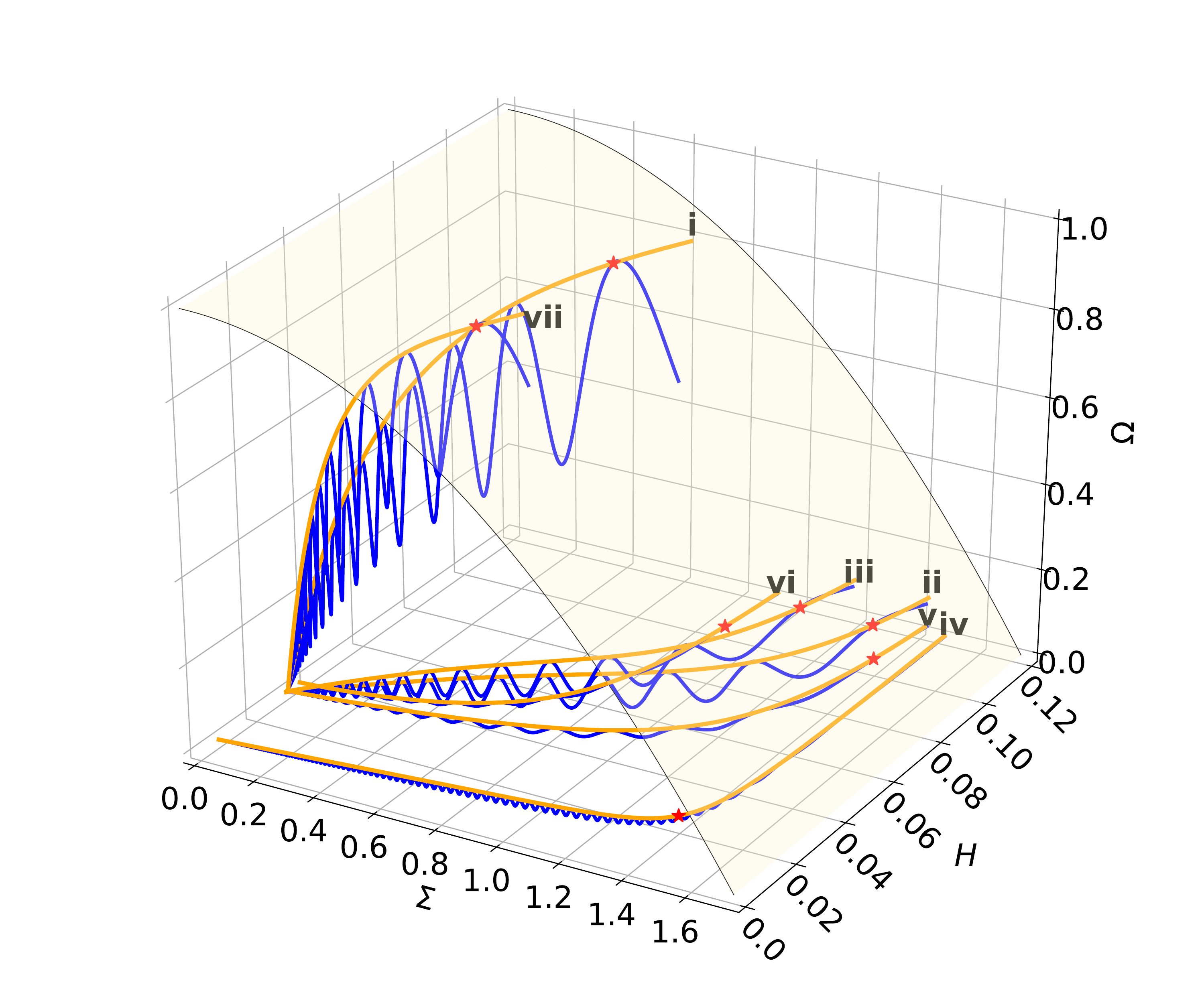}}
    \subfigure[\label{BianchiINonminimallyStifff093Dp} Projections in the space $(\Omega_{m},\Sigma,\Omega)$.]{\includegraphics[scale = 0.25]{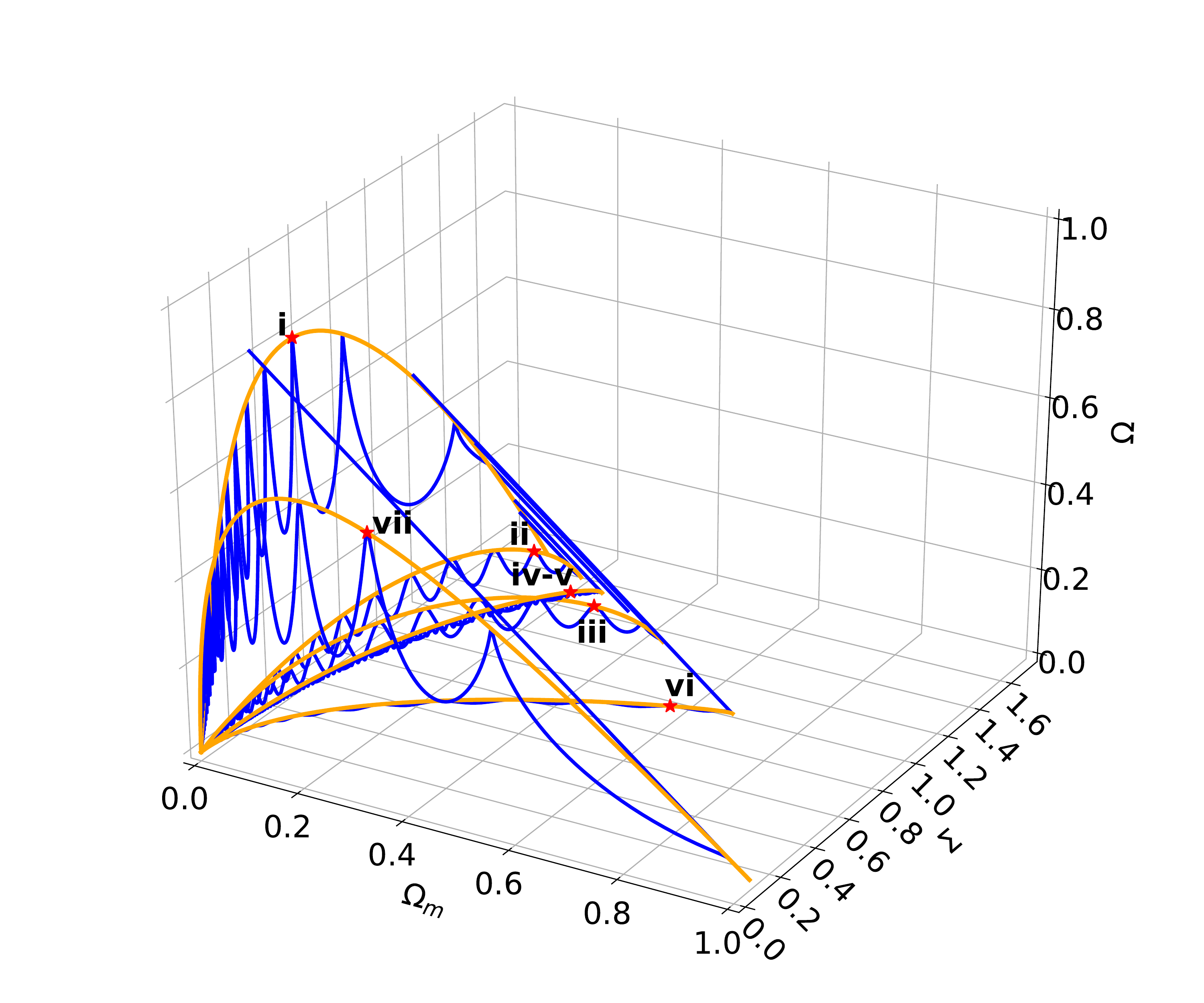}}
  \caption{Some solutions of the full system \eqref{YYEQ3.29YY} (blue) and time-averaged system \eqref{BIYEQ3.29Y} (orange) for a scalar field with generalized harmonic potential non-minimally coupled to matter in the Bianchi I metric when $\lambda = 0.1$, $f=0.9$ and $\gamma=2$. We have used for both systems the initial data sets presented in Table \ref{tab:BianchiI}.}
  \label{fig:BianchiINonminimallyStifff09}
\end{minipage}
\end{figure}

In Figures \ref{fig:BianchiINonminimallyCCf01}, \ref{fig:BianchiINonminimallyBiff01}, \ref{fig:BianchiINonminimallyDustf01}, \ref{fig:BianchiINonminimallyStifff01}, \ref{fig:BianchiINonminimallyCCf03}, \ref{fig:BianchiINonminimallyBiff03}, \ref{fig:BianchiINonminimallyDustf03}, \ref{fig:BianchiINonminimallyStifff03}, \ref{fig:BianchiINonminimallyCCf09}, \ref{fig:BianchiINonminimallyBiff09}, \ref{fig:BianchiINonminimallyDustf09} and \ref{fig:BianchiINonminimallyStifff09} we present the numerical results obtained from the integration of the full system \eqref{YYEQ3.29YY} (blue lines) and time-averaged system \eqref{BIYEQ3.29Y} (orange lines) for the non-minimally coupled case in the Bianchi I metric, using for both systems the seven initial data set presented in the Table \ref{tab:BianchiI}. Due to convergence problems, the integration range of the full system used in the $\gamma=2$ case was $-35\leq\tau\leq 3$.

In Figures \ref{fig:BianchiINonminimallyCCf01}, \ref{fig:BianchiINonminimallyCCf03} and \ref{fig:BianchiINonminimallyCCf09} we depict the results obtained for $\gamma=0$ when $f=0.1$, $0.3$ and $0.9$, respectively. Figures \ref{BianchiINonminimallyCCf013Dm}, \ref{BianchiINonminimallyCCf033Dm} and \ref{BianchiINonminimallyCCf093Dm} shows the projections in the space $(\Omega_{m},H,\Omega)$, Figures \ref{BianchiINonminimallyCCf013DS}, \ref{BianchiINonminimallyCCf033DS} and \ref{BianchiINonminimallyCCf093DS} shows the projections in the space $(\Sigma,H,\Omega)$, and Figures \ref{BianchiINonminimallyCCf013Dp}, \ref{BianchiINonminimallyCCf033Dp} and \ref{BianchiINonminimallyCCf093Dp} shows the projections in the space $(\Omega_{m},\Sigma,\Omega)$.

\begin{table}[ht!]
    \centering
      \begin{tabular}{ccccccc}
        \hline
        Sol. & $H(0)$ & $\Omega(0)$ & $\Omega_{m}(0)$ & $\Sigma(0)$ & $\varphi(0)$ & $t(0)$ \\
        \hline
        i & $0.1$ & $0.8$ & $0.01$ & $0.52$ & $0$ & $0$ \\
        ii & $0.1$ & $0.1$ & $0.16$ & $1.39$ & $0$ & $0$ \\
        iii & $0.1$ & $0.1$ & $0.36$ & $1.15$ & $0$ & $0$ \\
        iv & $0.02$ & $0.02$ & $0.2304$ & $1.3959$ & $0$ & $0$ \\
        v & $0.1$ & $0.02$ & $0.2304$ & $1.3959$ & $0$ & $0$ \\
        vi & $0.1$ & $0.01$ & $0.59$ & $0.9$ & $0$ & $0$ \\
        vii & $0.1$ & $0.584$ & $0.315$ & $0.055$ & $0$ & $0$ \\
        \hline
    \end{tabular}
      \caption{Seven initial data set for the simulations of the full and time averaged system for the Bianchi I metric in the non-minimal and minimal coupling cases. The initial conditions were chosen in order to fulfill the inequality $\Omega(0)+\Omega_{m}(0)+\Sigma(0)^{2}/3<1$.}
    \label{tab:BianchiI}
\end{table}

In Figures \ref{fig:BianchiINonminimallyBiff01}, \ref{fig:BianchiINonminimallyBiff03} and \ref{fig:BianchiINonminimallyBiff09} we depict the results obtained for $\gamma=\frac{2}{3}$ when $f=0.1$, $0.3$ and $0.9$, respectively. Figures \ref{BianchiINonminimallyBiff013Dm}, \ref{BianchiINonminimallyBiff033Dm} and \ref{BianchiINonminimallyBiff093Dm} shows the projections in the space $(\Omega_{m},H,\Omega)$, Figures \ref{BianchiINonminimallyBiff013DS}, \ref{BianchiINonminimallyBiff033DS} and \ref{BianchiINonminimallyBiff093DS} shows the projections in the space $(\Sigma,H,\Omega)$, and Figures \ref{BianchiINonminimallyBiff013Dp}, \ref{BianchiINonminimallyBiff033Dp} and \ref{BianchiINonminimallyBiff093Dp} shows the projections in the space $(\Omega_{m},\Sigma,\Omega)$.

In Figures \ref{fig:BianchiINonminimallyDustf01}, \ref{fig:BianchiINonminimallyDustf03} and \ref{fig:BianchiINonminimallyDustf09} we depict the results obtained for $\gamma=1$ when $f=0.1$, $0.3$ and $0.9$, respectively. Figures \ref{BianchiINonminimallyDustf013Dm}, \ref{BianchiINonminimallyDustf033Dm} and \ref{BianchiINonminimallyDustf093Dm} shows the projections in the space $(\Omega_{m},H,\Omega)$, Figures \ref{BianchiINonminimallyDustf013DS}, \ref{BianchiINonminimallyDustf033DS} and \ref{BianchiINonminimallyDustf093DS} shows the projections in the space $(\Sigma,H,\Omega)$, and Figures \ref{BianchiINonminimallyDustf013Dp}, \ref{BianchiINonminimallyDustf033Dp} and \ref{BianchiINonminimallyDustf093Dp} shows the projections in the space $(\Omega_{m},\Sigma,\Omega)$.

In Figures \ref{fig:BianchiINonminimallyStifff01}, \ref{fig:BianchiINonminimallyStifff03} and \ref{fig:BianchiINonminimallyStifff09} we depict the results obtained for $\gamma=2$ when $f=0.1$, $0.3$ and $0.9$, respectively. Figures \ref{BianchiINonminimallyStifff013Dm}, \ref{BianchiINonminimallyStifff033Dm} and \ref{BianchiINonminimallyStifff093Dm} shows the projections in the space $(\Omega_{m},H,\Omega)$, Figures \ref{BianchiINonminimallyStifff013DS}, \ref{BianchiINonminimallyStifff033DS} and \ref{BianchiINonminimallyStifff093DS} shows the projections in the space $(\Sigma,H,\Omega)$, and Figures \ref{BianchiINonminimallyStifff013Dp}, \ref{BianchiINonminimallyStifff033Dp} and \ref{BianchiINonminimallyStifff093Dp} shows the projections in the space $(\Omega_{m},\Sigma,\Omega)$.

These figures are evidence that the solutions of the full system (blue lines), obtained for a scalar field with generalized harmonic potential non-minimally coupled to matter in the Bianchi I metric, follow the track of the solutions of the averaged system (orange lines) when $H\to 0$ and, therefore, have the same asymptotic behavior. Furthermore, we can see that the amplitude of oscillations decreases when the value of $f$ increases.

\subsection{Scalar field  with generalized harmonic potential in vacuum}
\begin{figure}[ht!]
    \centering
    \subfigure[\label{Vacuumf01} Projections in the space $\left(H, \Omega\right)$ for $f=0.1$.]{\includegraphics[scale = 0.52]{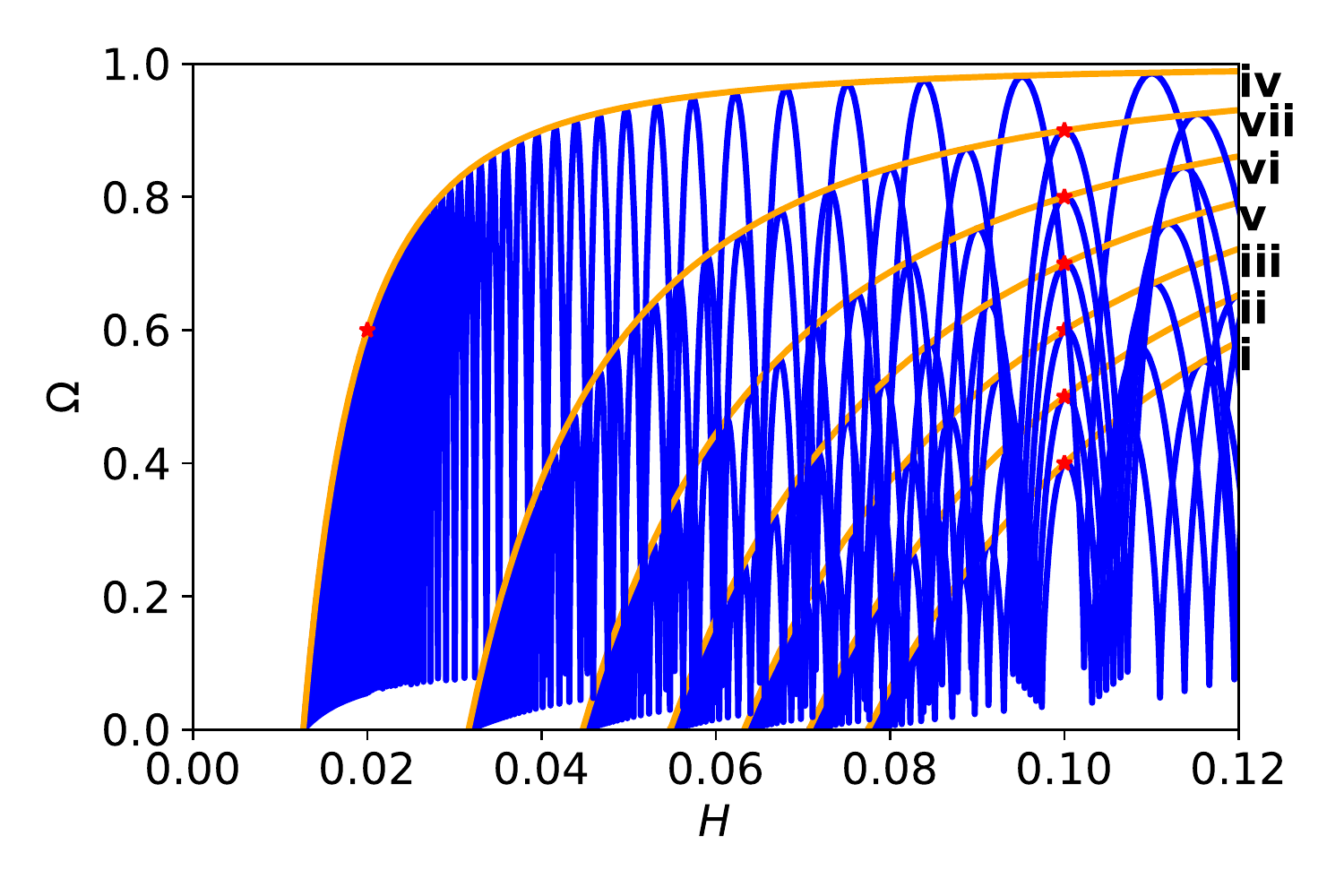}}
    \subfigure[\label{Vacuumf03} Projections in the space $\left(H, \Omega\right)$ for $f=0.3$.]{\includegraphics[scale = 0.52]{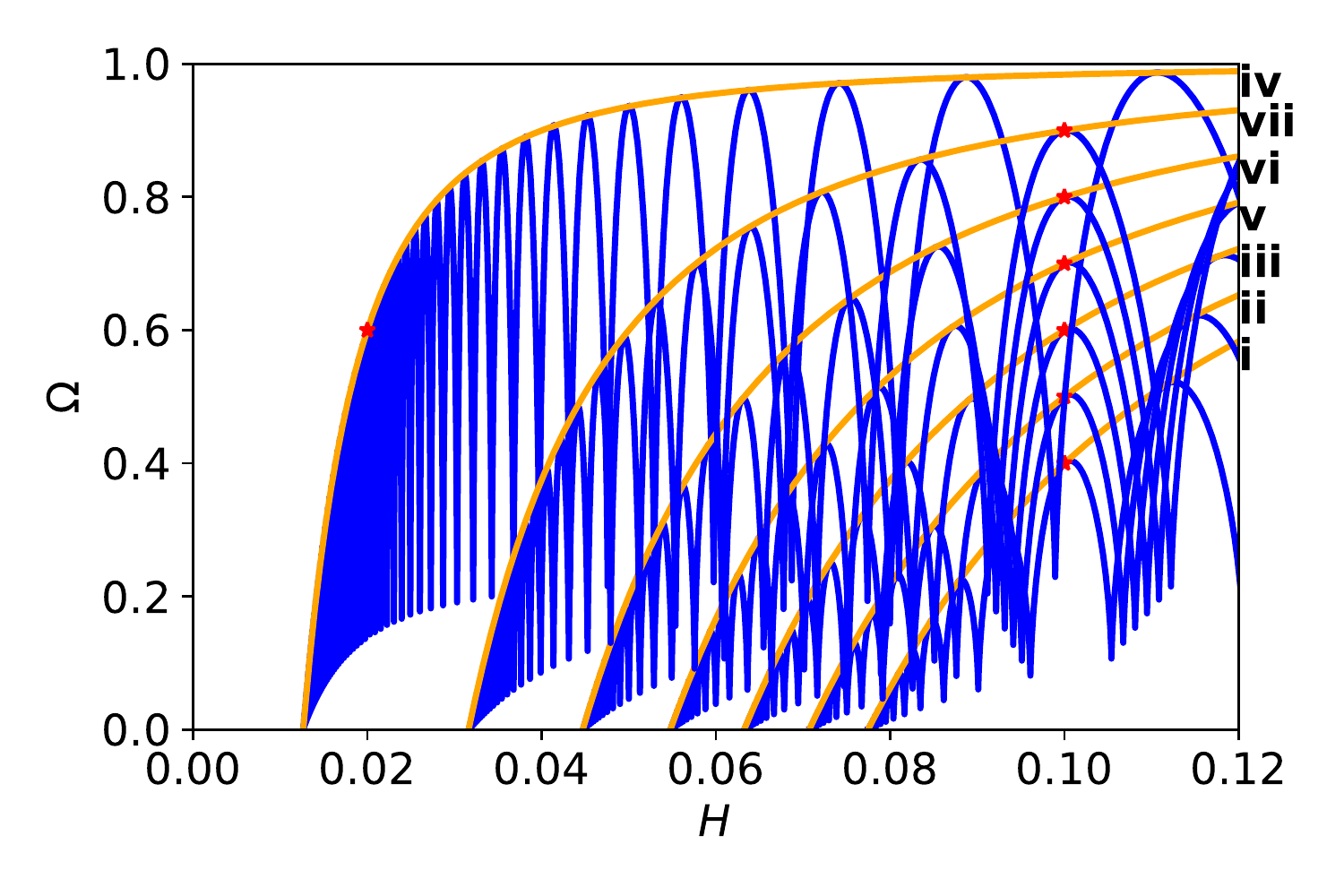}}
    \subfigure[\label{Vacuumf09} Projections in the space $\left(H, \Omega\right)$ for $f=0.9$.]{\includegraphics[scale = 0.52]{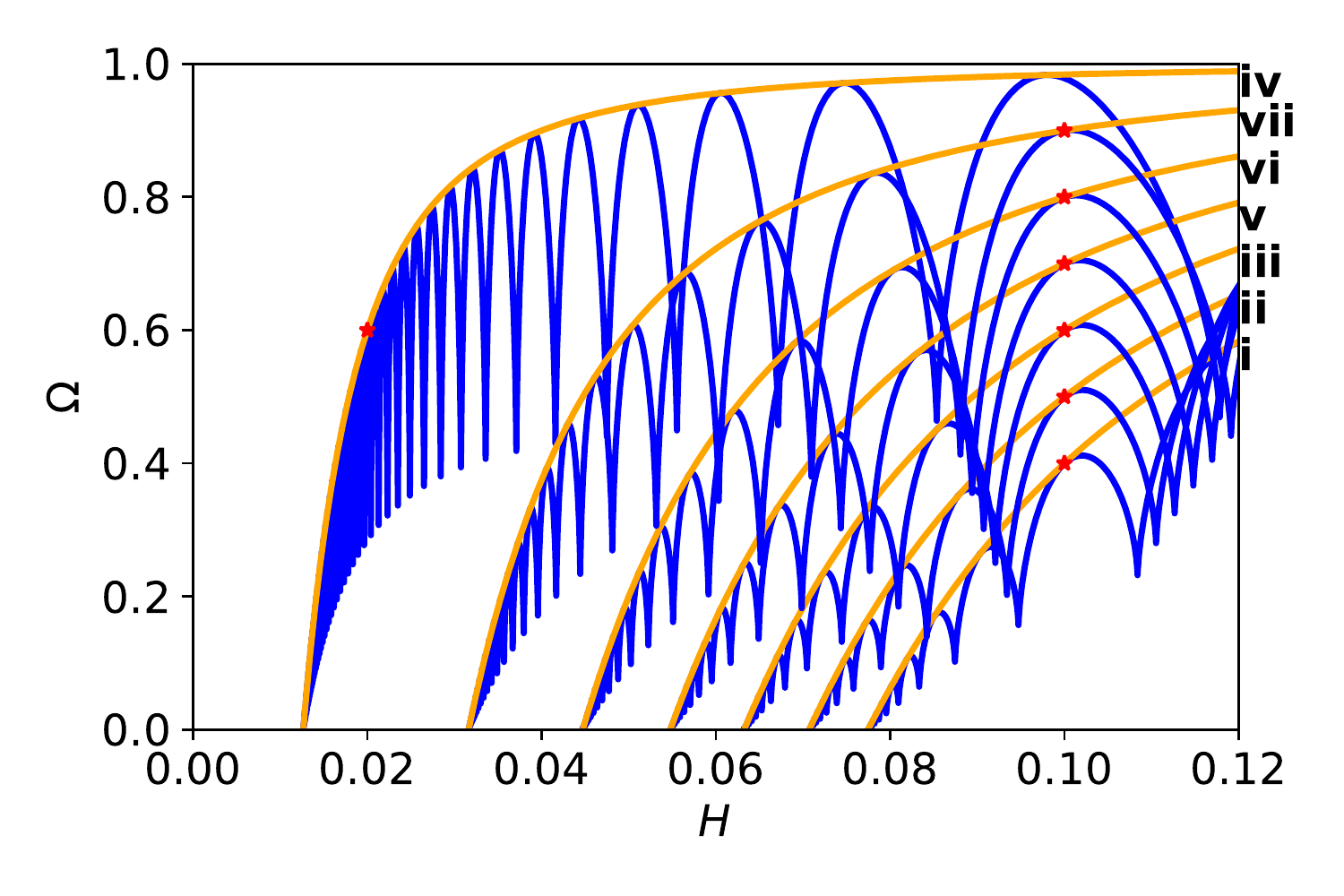}}
    \caption{Some solutions of the full system \eqref{AAsistemapertubado1a} (blue) and time-averaged system \eqref{Ypromedio2Y} (orange) for a scalar field with generalized harmonic potential in vacuum. We have used for both systems the seven initial data sets presented in Table \ref{tab:vacuum}.}
    \label{fig:Vacuum}
\end{figure}

In Figure \ref{fig:Vacuum} we present the numerical results obtained from the integration of the full system \eqref{AAsistemapertubado1a} (blue lines) and time-averaged system \eqref{Ypromedio2Y} (orange lines) for the vacuum case, using for both systemss the seven initial data set presented in the Table \ref{tab:vacuum}. 

In Figures \ref{Vacuumf01}, \ref{Vacuumf03} and \ref{Vacuumf09} we depict the results obtained for $f=0.1$, $0.3$ and $0.9$, respectively, in the $\left(H,\Omega\right)$ projection.

These figures are evidence that the solutions of the full system (blue lines), obtained for a scalar field with generalized harmonic potential in a vacuum, follow the track of the solutions of the averaged system (orange lines) when $H\to 0$ and, therefore, have the same asymptotic behavior. Furthermore, we can see that the amplitude of oscillations decreases when the value of $f$ increases.

It is important to mention that these figures confirm the result of Proposition \ref{Prop4}. Therefore, asymptotically we have a de Sitter solution with ``small'' $H \approx \frac{c_3}{2 \sqrt{3} \sqrt{c_2 f}}$, where $c_1$ and $c_2$ are integration constants that depends on the initial conditions.

\begin{table}[ht!]
    \centering
      \begin{tabular}{ccccccc}
        \hline
        Sol. & $H(0)$ & $\Omega(0)$ & $\varphi(0)$ & $t(0)$ \\
        \hline
        i & $0.1$ & $0.4$ & $0$ & $0$ \\
        ii & $0.1$ & $0.5$ & $0$ & $0$ \\
        iii & $0.1$ & $0.6$ & $0$ & $0$ \\
        iv & $0.02$ & $0.6$& $0$ & $0$ \\
        v & $0.1$ & $0.7$ & $0$ & $0$ \\
        vi & $0.1$ & $0.8$ & $0$ & $0$ \\
        vii & $0.1$ & $0.9$ & $0$ & $0$ \\
        \hline
    \end{tabular}
      \caption{Seven initial data set for the simulations of the full and time averaged system for the scalar field with generalized harmonic potential in vacuum. The initial conditions were chosen in order to fulfill the inequality $\Omega(0)<1$.}
    \label{tab:vacuum}
\end{table}

%%%%% FLRW nonminimally f=0.1 CC-Bif %%%%%
%%%%% FLRW nonminimally f=0.1 Dust-Stiff %%%%%
%%%%% FLRW nonminimally f=0.3 CC-Bif %%%%%
%%%%% FLRW nonminimally f=0.3 Dust-Stiff %%%%%
%%%%% FLRW nonminimally f=0.9 CC-Bif %%%%%
%%%%% FLRW nonminimally f=0.9 Dust-Stiff %%%%%

%%%%% Bianchi I nonminimally f=0.1 CC-Bif %%%%%
%%%%% Bianchi I nonminimally f=0.1 Dust-Stiff %%%%%
%%%%% Bianchi I nonminimally f=0.3 CC-Bif %%%%%
%%%%% Bianchi I nonminimally f=0.3 Dust-Stiff %%%%%
%%%%% Bianchi I nonminimally f=0.9 CC-Bif %%%%%
%%%%% Bianchi I nonminimally f=0.9 Dust-Stiff %%%%%

%%%%% Vacuum %%%%%

\section{Results and Conclusions}
\label{Sect:6}

This paper was devoted to the study of perturbation problems in scalar field  cosmologies in the FLRW metric with $k=-1, 0$, and Bianchi I metric in vacuum and with matter. In the last case, considering minimal and non-minimal couplings between matter and the scalar field. Qualitative techniques, asymptotic methods, and averaging theory were used to obtain relevant information about the solution's space of the aforementioned cosmologies. Variables that lead to regular equations in a bounded state space were chosen. This allows us to give a global description of the dynamics, in particular, the behavior in early and late times and the evolution in intermediate stages that may be of physical interest. Furthermore, differential equations, suitable for performing systematic numerical simulations, were derived. Averaged versions of original systems were constructed where the oscillations of the solutions are smoothed out. The analysis is then reduced to studying the late dynamics of a simpler averaged system where the oscillations entering the system without averaging can be controlled through the KG equation.

The tools of the averaging theory and the qualitative techniques of dynamical systems have been applied successfully in recent years in similar cosmological models, say in \cite {Leon:2021rcx,Leon:2021lct,Leon:2021hxc}. As the natural generalization of these models we considered spatially homogeneous and isotropic dark energy (scalar field) -matter interactive schemes. The relevant calculations depend on the shape of the potential and, in particular, are quite complicated for harmonic potentials. The result presented here shows that the oscillations arising due to harmonic functions can be ``averaged'', thus simplifying the problem. This approach is useful for describing the oscillations of the inflaton around the potential minimum during reheating after inflation in models like the $N$-field inflation model \cite{Dimopoulos:2005ac}. For non-zero $H$, this gives rise to time-dependent oscillatory dynamics. This is may be responsible for the production of particles through quantum field theory. Using some inverse transformations, one can find from the averaged version of the scalar field variables, the approximate temporal dependence of the original fields.
This approach is also suitable in the context of linear cosmological perturbations. In cosmological perturbation theory, cosmological perturbations at the linear level are governed by equations whose coefficients are made up of background quantities. Therefore, adequate knowledge of the background dynamics is necessary to perform further perturbation analysis.

To illustrate the relevance of these tools, we have discussed some basic examples of applications of perturbation techniques in section \ref{section_2} and section \ref{SECT:II}. Regarding the cosmological applications of these techniques (which are the core of the present research), there were obtained the following results.  In section \ref{section_5} some applications of perturbation and averaging methods in cosmology were presented. In particular, in section  \ref{SECT:4.5} it was studied a scalar field with generalized harmonic potential  \eqref{pot1} non-minimally coupled to matter with coupling \eqref{coupling}. Sections \ref{SECTION_4.5} and \ref{Sect:2.7.3} were devoted to the minimally coupled  and vacuum cases, respectively. The focus was to study the imprint of coupling function, as well as the influence of the metric on the dynamics of the averaged problem. As a first step towards generalization, we have considered an interaction with the background matter with strength of type $Q = \lambda/2\rho_m \dot\phi$  arising from the coupling function \eqref{coupling} within the interacting scheme \eqref{interacting-scheme}. Then we expect go increasing the degree of complexity until considering interaction models such as  $ Q = 3 \alpha H \rho_m, \; Q = 3 \beta H \rho_\phi $ y $ Q = 3 H (\alpha \rho_m + \beta \rho_\phi) $  \cite{Cardenas:2018nem,Lepe:2015qhq}.    
 When considering models with interaction like \eqref{interacting-scheme}, which have different physical implications, different results would be expected from the case without interaction. 
An interesting research path is to investigate the dynamics and asymptotic behavior of the solutions of the equations of the gravitational field for various interacting functions of the form $ Q = Q \left (H, \rho_ {m}, \rho_\phi \right) $.
It is worth noting that in the case of the scalar field with generalized harmonic potential minimally coupled to the matter model ($ \lambda = 0 $), for the FLRW metrics, the numerical results are very similar to their respective non-minimum coupling cases ($ \lambda \neq 0 $), and the same happens for the Bianchi I models. Note that the interaction appears in the equations explicitly in the form $ \lambda \sin (t- \varphi), \; \lambda \cos (t- \varphi) $, expressions that have zero average. Using averaging methods for periodic functions of a given period $T$, it can be concluded that, regardless of whether the scalar field is minimally or non-minimally coupled to the matter field, there is no difference in dynamics when performing the averaging process at least for interactions of the type $Q=\lambda/2\rho_m \dot{\phi}$. This indicates that the asymptotic results when $H\rightarrow 0$ are independent of this coupling function. Non averaged systems have quite different dynamics. 
There are several issues to be discussed within this line of research, but it is worth noting that the success in the implementation of mathematical techniques during this research allows an immediate implementation of these to the case with more general interaction terms, so that new results can be achieved as a continuation of this project.

\ack

The research of Genly Leon, Esteban González and Felipe Orlando Franz Silva was funded by Agencia Nacional de Investigaci\'on y Desarrollo - ANID 
through the program FONDECYT Iniciaci\'on grant no.
11180126. A.D. Millano was supported by Agencia Nacional de Investigación y Desarrollo - ANID-Subdirección de Capital Humano/Doctorado
Nacional/año 2020- folio 21200837. Esteban González was funded by Direcci\'on de Investigaci\'on y Postgrado at Universidad de Aconcagua. Additionally, this research is funded by Vicerrector\'ia de Investigaci\'on y Desarrollo Tecnol\'ogico at Universidad Católica del Norte.   Ellen de Los M. Fern\'andez Flores is acknowledged for proofreading this manuscript and improving the English.

\bigskip

  \appendix

\section{Proof of Proposition 1}
\label{App4}
Now is given the proof of proposition \ref{Prop4}.

Using Taylor series in a neighborhood of ${H}=0$ of \eqref{P({H})} the following holds: 
\begin{subequations}
\label{Sis2.205}
\begin{align}
&\dot {H}=-3 \Omega  {H} ^2 \cos ^2(t-\varphi ) + \mathcal{O}({H})^3, \label{expdotepsilon}\\
& \dot \Omega=-\frac{\Omega  \sin (2 (t-\varphi))}{f}+6 (\Omega -1) \Omega  {H}  \cos ^2(t-\varphi )  + \mathcal{O}({H})^2,\\
  &\dot \varphi =-\frac{\sin ^2(t-\varphi )}{f}-\frac{3}{2} {H}  \sin (2 (t-\varphi
   ))+ \mathcal{O}({H})^2.    
\end{align}
\end{subequations}
Then, as ${H} \rightarrow 0$, it follows the unperturbed problem:  
\begin{equation}
P(0): \left\{
\begin{array}{c}
\dot {H}=0\\
\dot{{\Omega}}=-\frac{\Omega  \sin (2 (t-\varphi))}{f}\\
\dot{{\varphi}} =-\frac{\sin ^2(t-\varphi)}{f}
    \end{array}\right.,
\end{equation}
whose solution is given by \eqref{solnoe}. $\qed$

\section*{Proof of Proposition 2}
\label{App5}
Now is given the proof of proposition \ref{Prop5}.

Continuing with the applications of the perturbation theory tools is proposed an expansion of kind \eqref{eqs_204}, where 
$\Omega_0(t)$ and $\varphi_0(t)$ are the solutions of the unperturbed problem  $P(0)$.
Applying the chain rule  and using the fact that  $\Omega_1\frac{d {H}}{dt}=\mathcal{O}({H}^2)$  according  to \eqref{expdotepsilon}, it follows: 
\begin{subequations}
\begin{align}
    &\frac{d \Omega}{d t}= \frac{d \Omega_0}{d t}+ {H} \frac{d \Omega_1}{d t} +\Omega_1 \frac{d {H}}{dt}+ \mathcal{O}({H}^2)=\frac{d \Omega_0}{d t}+ {H} \frac{d \Omega_1}{d t} + \mathcal{O}({H}^2),\\
    &\frac{d \varphi}{d t}= \frac{d \varphi_0}{d t}+ {H} \frac{d \varphi_1}{d t} +\varphi_1 \frac{d {H}}{dt}+ \mathcal{O}({H}^2)=\frac{d \varphi_0}{d t}+ {H} \frac{d \varphi_1}{d t} + \mathcal{O}({H}^2).
\end{align}
\end{subequations}
Hence, 
\begin{subequations}
\begin{align}
&{H} \frac{d \Omega_1}{d t}=  \frac{d{\Omega}}{dt}-\frac{d\Omega_0}{dt}\nonumber\\
&= 6 ((\Omega_0+{H} \Omega_1) -1)(\Omega_0+{H} \Omega_1){H}  \cos ^2(t-(\varphi_0+{H} \varphi_1) ) 
\nonumber \\
& -\frac{(\Omega_0+{H} \Omega_1)  \sin (2 (t-(\varphi_0+{H} \varphi_1) ))}{f}+\frac{\Omega_0 \sin^2  (({t-\varphi_0}))}{f} + \mathcal{O}({H}^2)\nonumber \\
&=\frac{{H}  \left(6 f (\Omega_0-1) \Omega_0 \cos ^2(t-\varphi_0)+2 \varphi_1 \Omega_0 \cos (2 (t-\varphi_0))-\Omega_1 \sin (2 (t-\varphi_0))\right)}{f}+\mathcal{O}\left({H} ^2\right),
\\
&{H} \frac{d \varphi_1}{d t}= \frac{d{\varphi}}{dt}-\frac{d\varphi_0}{dt}\nonumber \\
&=-\frac{3}{2} {H}  \sin (2 (t-(\varphi_0+{H} \varphi_1) )) -\frac{\sin ^2(t-(\varphi_0+{H} \varphi_1) )}{f}+ \frac{\sin ^2(t-\varphi_0)}{f}+ \mathcal{O}({H}^2)\nonumber \\
&=\frac{{H}  (2 \varphi_1-3 f) \sin (2 (t-\varphi_0))}{2 f} + \mathcal{O}({H}^2).
\end{align}
\end{subequations}
Therefore, to find analytically the functions $\Omega_1$ and $\varphi_1$, \begin{subequations}
\label{2solnoe}
\begin{align}
    &\frac{d \Omega_1}{d t}=\frac{\left(6 f (\Omega_0-1) \Omega_0 \cos ^2(t-\varphi_0)+2 \varphi_1 \Omega_0 \cos (2 (t-\varphi_0))-\Omega_1 \sin (2 (t-\varphi_0))\right)}{f},\\
    & \frac{d \varphi_1}{d t}=\frac{  (2 \varphi_1-3 f) \sin (2 (t-\varphi_0))}{2 f},
\end{align}
\end{subequations} 
 have to be solved with the substitution of $\Omega_0$ and $\varphi_0$ in \eqref{2solnoe}. Integrating for $\varphi_1$, it follows equation \eqref{eq_208}.  For $\Omega_1$ the following quadrature \eqref{eq_206} with $f(t)$ defined by 
\eqref{eq_207} is obtained. $\qed$

The next result is useful in the following proof.  

\begin{lemma}[\textbf{Gronwall's Lemma. Integral form}]
\label{Gronwall} Let be $\xi(t)$ a nonnegative function, summable over  $[0,T]$ which satisfies almost everywhere the integral inequality
        \begin{equation}
            \xi(t)\leq C_1 \int_0^t \xi(s)ds +C_2, \quad C_1, C_2\geq 0.
        \end{equation}
       Then 
        \begin{equation}
            \xi(t)\leq C_2 (1+C_1 t e^{C_1 t}),
        \end{equation}
        almost everywhere for $t$ in $0\leq t\leq T$.
       In particular, if    \begin{equation}
            \xi(t)\leq C_1 \int_0^t \xi(s)ds, \quad C_1\geq 0,
        \end{equation}
        almost everywhere for $t$ in $0\leq t\leq T$, then  \begin{equation}
           \eta \equiv 0,
       \end{equation}
        almost everywhere for $t$ in $0\leq t\leq T$.
\end{lemma}

\section{Proof of Proposition 3}
\label{App6}
Now is given the proof of proposition \ref{Prop_6}.

It is easy to see that the system  \eqref{Sis2.205} can be conveniently written as:   
\begin{subequations}
\label{Ysistemapertubado1aY}
\begin{align}
& \frac{d t}{d \eta}= {H}, \\ 
& \frac{d\Omega}{d \eta}=-\frac{{H}\Omega  \sin (2 (t-\varphi))}{f}+6 (\Omega -1) \Omega  {H}^2  \cos ^2(t-\varphi )  + \mathcal{O}({H})^3,\\
  &\frac{d \varphi}{d \eta} =-\frac{{H}\sin ^2(t-\varphi )}{f}-\frac{3}{2} {H}^2  \sin (2 (t-\varphi
   ))+ \mathcal{O}({H})^3.    
\end{align}
\end{subequations}
and the averaged problem is:
\begin{align}
& \frac{d t}{d \eta}= {H}, \quad \frac{d\overline{\Omega}}{d \eta}=-3 (1-\overline{\Omega}) \overline{\Omega}  {H}^2, \quad \frac{d\overline{\varphi}}{d \eta} =-\frac{H}{2 f}.
  \end{align}
Now, the expansion \eqref{YquasilinearY} is proposed. 
Next, it is proved that the equations for  ${t}_0, {\Omega}_0, {\varphi}_0$  have the same asymptotic that the averaged equations for  $\overline{t}, \overline{\Omega}, \overline{\varphi}$.

After some algebraic manipulations and recalling that   \begin{equation}
    \frac{d {H}}{d\eta}=-3 \Omega  {H} ^3 \cos ^2(t-\varphi ) = \mathcal{O}({H})^3,  
  \end{equation} it follows:  
  \begin{small}
\begin{subequations}
\begin{align}
      & \frac{d t_0}{d \eta}=H+H^2 \left(\frac{\alpha _{1\varphi_0}}{2 f}-\alpha _{1 t_0}\right)+\mathcal{O}\left({H} ^3\right),\\
      & \frac{d \Omega_0}{d \eta}=H^2 \Bigg\{\frac{\Omega_0 \left(\left(3 f^2 (\Omega_0-1)+2 f \eta +\eta \right) \cos (2 (t_0-\varphi_0))+3 f^2 (\Omega_0-1)+\eta
   \right)}{f^2} \nonumber \\
   & +\frac{\Omega_0 \alpha _{2 \varphi_0}}{2 f}-\frac{\alpha _2 \Omega_0 \sin (2 (t_0-\varphi_0))}{f}+\frac{2 \left(\alpha _3-\alpha _1\right)
   \Omega_0 \cos (2 (t_0-\varphi_0))}{f}-\Omega_0 \alpha _{2 t_0}\Bigg\}+\mathcal{O}\left({H} ^3\right),\\
   &   \frac{d \varphi_0}{d \eta}=-\frac{H}{2 f}+H^2 \left(\frac{\sin (2 (t_0-\varphi_0)) \left(-3 f^2-2 \alpha _1 f+2 \alpha _3 f+2 f \eta +\eta +\eta  \cos (2 (t_0-\varphi_0))\right)}{2 f^2}+\frac{\alpha _{3 \phi
   _0}}{2 f}-\alpha _{3 t_0}\right)+\mathcal{O}\left({H} ^3\right). 
\end{align}
\end{subequations}
\end{small}
Imposing the conditions
\begin{align}
    & \frac{\alpha _{1\varphi_0}}{2 f}-\alpha _{1 t_0}=0 \implies \alpha_1(t_0,\varphi_0)= c_1\left(\frac{t_0}{2 f}+\varphi_0\right), \\
    &\alpha _{2 \phi _0}-2 f \alpha _{2 t_0}-2 \alpha _2 \sin (2 (t_0-\varphi_0))=0 \implies \alpha_2(t_0,\varphi_0)= e^{\frac{\cos (2 (t_0-\varphi_0))}{2 f+1}} c_2\left(\frac{t_0}{2 f}+\varphi_0\right),
\end{align}
and assuming $\alpha_{3}= \alpha_1 + g(t_0,\varphi_0)$, the following equations are deduced:  \begin{small}
\begin{subequations}
\begin{align}
    & \frac{d t_0}{d \eta}={H} +\mathcal{O}\left({H} ^3\right),\\
    & \frac{d \Omega_0}{d \eta}=H^2 \Bigg\{\frac{\Omega_0 \cos (2 (t_0-\varphi_0)) \left(2 f g+3 f^2 (\Omega_0-1)+2 f \eta +\eta \right)}{f^2}+\Omega_0
   \left(\frac{\eta }{f^2}+3 \Omega_0-3\right)\Bigg\} +\mathcal{O}\left({H} ^3\right), \label{C.8.b}\\
   & \frac{d \varphi_0}{d \eta}=-\frac{H}{2 f}+H^2 \Bigg\{\frac{\eta  \sin (2 (t_0-\varphi_0)) (2 f+\cos (2 (t_0-\varphi_0))+1)}{2 f^2} \nonumber\\
   & +\frac{-2 f g_{t_0}-3 f \sin (2 (t_0-\varphi_0))+2 g \sin (2
   (t_0-\varphi_0))+g_{\phi _0}}{2 f}\Bigg\}+\mathcal{O}\left({H} ^3\right)  \label{C.8.c}.
\end{align}
\end{subequations}
\end{small}
The condition 
\begin{equation}
    -2 f g_{t_0}-3 f \sin (2 (t_0-\varphi_0))+2 g \sin (2
   (t_0-\varphi_0))+g_{\phi _0}=0 \implies g(t_0,\varphi_0)=\frac{3 f}{2}+e^{-\frac{\cos (2 (t_0-\varphi_0))}{2 f+1}} c_3\left(\frac{t_0}{2 f}+\varphi_0\right),
\end{equation}
leads to
\begin{small}
\begin{align}
  & \alpha_1(t_0,\varphi_0)= c_1\left(\frac{t_0}{2 f}+\varphi_0\right), \\
  & \alpha_2(t_0,\varphi_0)= e^{\frac{\cos (2 (t_0-\varphi_0))}{2 f+1}} c_2\left(\frac{t_0}{2 f}+\varphi_0\right),\\
  & \alpha_3(t_0,\varphi_0)= c_1\left(\frac{t_0}{2 f}+\varphi_0\right)+ \frac{3 f}{2}+e^{-\frac{\cos (2 (t_0-\varphi_0))}{2 f+1}} c_3\left(\frac{t_0}{2 f}+\varphi_0\right).
\end{align}
\end{small}
Equation \eqref{C.8.b}
becomes 
\begin{align}
   & \frac{d \Omega_0}{d \eta}=   H^2 \Bigg\{6 (\Omega_0-1) \Omega_0 \cos
   ^2(t_0-\varphi_0)+\frac{\eta  ((2 f+1) \Omega_0 \cos (2 (t_0-\varphi_0))+\Omega_0)}{f^2} \nonumber \\
   & + \Omega_0 \cos (2 (t_0-\varphi_0))  \left(3+\frac{2 e^{-\frac{\cos (2 (t_0-\varphi_0))}{2 f+1}} c_3\left(\frac{t_0}{2 f}+\varphi_0\right)}{f}\right)\Bigg\}. \label{C.12}
\end{align}
Equation \eqref{C.8.c}
becomes 
\begin{align}
   & \frac{d \varphi_0}{d \eta}=-\frac{H}{2 f} + 
\frac{\eta  H^2 \sin (2 (t_0-\varphi_0)) (f+ \cos^2(t_0-\varphi_0))}{f^2}.
\end{align}
From the equation  
  \begin{equation}
    \dot {H}=-3 \Omega  {H} ^2 \cos ^2(t-\varphi ) + \mathcal{O}({H})^3,  
  \end{equation} 
or its averaged version, it follows ${H}$ is a  monotonic decreasing function of  $t$ due to $0\leq \Omega, \overline{\Omega} \leq 1$. 
This allows to define recursively the sequences:
\begin{equation}
    \left\{\begin{array}{c}
       \eta_0=0   \\ \\
        {H}_0={H}(\eta_0) 
    \end{array}\right., \quad \left\{\begin{array}{c}
       {\eta_{n+1}}^2= {\eta_{n}}^2 +\frac{1}{{H}_n}   \\ \\
       {H}_{n+1}= {H}(\eta_{n+1})  
    \end{array}\right.,
\end{equation}
such that  $\lim_{n\rightarrow \infty}{H}_n=0$ and $\lim_{n\rightarrow \infty}\eta_n=\infty$.

Defining $\Delta \varphi(\eta)=\varphi_0(\eta)-\overline{\varphi}(\eta)$ and taking the same initial conditions at $\eta=\eta_n$,  $\varphi_0(\eta_n)=\overline{\varphi}(\eta_n)$, it follows: 
 \begin{align}
    &\displaystyle{|\Delta  \varphi(\eta)|=\Big{|}\int_{\eta_n}^{\eta}[\varphi_0'(s)-\overline{\varphi}'(s)]ds} \Big{|}\nonumber\\
    & = \Big{|}\int_{\eta_n}^{\eta} 2 s \Big[\underbrace{\frac{ H^2 \sin (2 (t_0-\varphi_0)) (  f+\cos^2 (t_0-\varphi_0))}{2 f^2}}_{|\cdot|\leq M_1 {H}_n^2}+\mathcal{O}\left({H} ^3\right)\Big] ds \Big{|} \nonumber\\
   &  \leq M_1 {H}_n^2  \Big{|}\int_{\eta_n}^{\eta} 2 s  ds \Big{|} + \mathcal{O}\left({H}_n ^3\right)\nonumber\\
   &\leq M_1 {H}_n^2 |\eta+\eta_n||\eta-\eta_n| + \mathcal{O}\left({H}_n ^3\right),
\end{align}
where $M_1$ is a constant, 
for all $\eta\geq \eta_{n}$. 
Then, for $\eta \in \left[\eta_n, \eta_{n+1}\right]$, it follows the inequality
\[|\Delta  \varphi(\eta)|\leq M_1 {H}_n.\] 
Finally, taking the limit as $n\rightarrow \infty$, it follows ${H}_n\rightarrow 0, \eta_n\rightarrow \infty$, then, it follows 
$\lim_{\eta\rightarrow \infty}|\Delta \varphi(\eta)|=0$. This means that $\varphi_0$ and $\overline{\varphi}$ have the same limit as $\eta\rightarrow \infty$.

Without losing generality, $c_3(\varphi_0)\equiv 0$ is chosen in \eqref{C.12}. Therefore, it follows 
\begin{small}
\begin{align}
    & \frac{d \Omega_0}{d \eta}={H} ^2 \Bigg\{-3 \Omega_0 (1-\Omega_0)+\frac{\eta  \Omega_0\left(1+ (2 f+1)  \cos (2 (t_0-\varphi_0))\right)}{f^2}+3 \Omega_0^2 \cos (2 (t_0-\varphi_{0}))\Bigg\} +\mathcal{O}\left({H} ^3\right).
\end{align}
\end{small}
Defining  $\Delta \Omega= \Omega_0 -\overline{\Omega}$,  it follows
\begin{small}
\begin{align}
    & \Delta\Omega'\left(s\right)=  -3 \Delta \Omega {H} ^2 (1-\overline{\Omega} -\Omega_0) + 2 {H} ^2\eta \Omega_{0} \left[   \frac{\left(1+ (2 f+1)  \cos (2 (t_0-\varphi_0))\right)}{2 f^2}\right]+3 {H} ^2\Omega_{0}^2  \cos (2 (t_0-\varphi_{0})).
\end{align}
\end{small}
Choosing the same initial conditions   $\Omega_0(\eta_n)=\overline{\Omega}(\eta_n)$ at $\eta=\eta_n$, it follows
\begin{small}
\begin{align}
    &\displaystyle{|\Delta \Omega(\eta)|=\Big{|}\int_{\eta_n}^{\eta}[\Omega_0'(s)-\overline{\Omega}'(s)]ds}\Big{|} \nonumber\\
    & = \Bigg{|}\bigints_{\eta_n}^{\eta}  \left[-3 \Delta \Omega  {H} ^2 \underbrace{(1-\overline{\Omega} -\Omega_0)}_{|\cdot|\leq 1}+ 2 {H} ^2  s \Omega_{0} \underbrace{\left[   \frac{\left(1+ (2 f+1)  \cos (2 (t_0-\varphi_0))\right)}{2 f^2}\right]}_{|\cdot|\leq M_2} \right. \nonumber \\
   & \left. +3 {H} ^2\Omega_{0}^2 \underbrace{\cos (2 (t_0-\varphi_{0}))}_{|\cdot|\leq 1} +\mathcal{O}\left({H}_n ^3\right)\right] ds \Bigg{|} \nonumber\\
   & \leq 3 {H}_n^2 \int_{\eta_n}^{\eta} \Big{|} \Delta \Omega(s) \Big{|} ds +  M M_2 {H}_n^2 |\eta + \eta_n||\eta - \eta_n|+ 3 M^2 {H}_n^2 |\eta - \eta_n|, \quad \eta \geq \eta_n,
\end{align}
\end{small}
where $M_2$ is a constant, and 
\[M=\max_{s\in[\eta_n, \eta]}  \Big{|} \Omega_0(s) \Big{|},\]
which exists due to the continuity of $\Omega_0$ on the compact set $[\eta_n, \eta]$. Applying Gronwall's Lemma \ref{Gronwall}, it follows:  
\begin{align}
    & |\Delta \Omega(\eta)|\leq \left[M M_2 {H}_n^2 |\eta + \eta_n||\eta - \eta_n|+ 3 M^2 {H}_n^2 |\eta - \eta_n|\right] \left[1+ {H}_n^2 \eta e^{{H}_n^2 \eta}\right]+\mathcal{O}({H}_n^3)\nonumber \\
    & = \left[M M_2 {H}_n^2 |\eta + \eta_n||\eta - \eta_n|+ 3 M^2 {H}_n^2 |\eta - \eta_n|\right] \left[1+ {H}_n^2 \eta \right]+\mathcal{O}({H}_n^3) \nonumber\\
    & =M M_2 {H}_n^2 |\eta + \eta_n||\eta - \eta_n|+ 3 M^2 {H}_n^2 |\eta - \eta_n| +\mathcal{O}({H}_n^3).
\end{align}
Then, for  $\eta \in \left[\eta_n, \eta_{n+1}\right]$   and for  $n$ large enough such that $|\eta+\eta_n|\geq 1$, it follows
\begin{align*}
    & M M_2|\eta + \eta_n||\eta - \eta_n|+ 3 M^2 |\eta - \eta_n|\leq \left(M M_2 +3 M^2\right)|\eta^2 - \eta_n^2|\\
    & \leq  \left(M M_2 +3 M^2\right)|\eta_{n+1}^2 - \eta_n^2|=\left(M M_2 +3 M^2\right) {H}_n^{-1}.
\end{align*} Therefore, it follows the inequality 
$|\Delta \Omega(\eta)|\leq K {H}_n$, for a positive constant  $K\geq \left(M M_2 +3 M^2\right)$. 
Finally, taking the limit as $n\rightarrow \infty$, it follows ${H}_n\rightarrow 0, \eta_n\rightarrow \infty$. Then, it follows 
$\lim_{\eta\rightarrow \infty}|\Delta \Omega(\eta)|=0$. This means that $\Omega_0$ and $\overline{\Omega}$ have the same limit as  $\eta\rightarrow \infty$. $\qed$


\bigskip



\begin{thebibliography}{999}


 


%\cite{Brans:1961sx}
\bibitem{Brans:1961sx}
C.~Brans and R.~H.~Dicke,
%``Mach's principle and a relativistic theory of gravitation,''
Phys. Rev. \textbf{124}, 925-935 (1961)
doi:10.1103/PhysRev.124.925
%3479 citations counted in INSPIRE as of 02 Nov 2020
	
	%\cite{Guth:1980zm}
\bibitem{Guth:1980zm} 
  A.~H.~Guth,
  %``The Inflationary Universe: A Possible Solution to the Horizon and Flatness Problems,''
  Phys.\ Rev.\ D {\bf 23}, 347 (1981)
  [Adv.\ Ser.\ Astrophys.\ Cosmol.\  {\bf 3}, 139 (1987)].
  doi:10.1103/PhysRevD.23.347
  %%CITATION = doi:10.1103/PhysRevD.23.347;%%
  %7588 citations counted in INSPIRE as of 19 Dec 2019
	
	%\cite{Horndeski:1974wa}
\bibitem{Horndeski:1974wa}
G.~W.~Horndeski,
%``Second-order scalar-tensor field equations in a four-dimensional space,''
Int. J. Theor. Phys. \textbf{10}, 363-384 (1974)
doi:10.1007/BF01807638
%1511 citations counted in INSPIRE as of 02 Nov 2020
	
	
%\cite{Copeland:1993jj}
\bibitem{Copeland:1993jj}
E.~J.~Copeland, E.~W.~Kolb, A.~R.~Liddle and J.~E.~Lidsey,
%``Reconstructing the inflation potential, in principle and in practice,''
Phys. Rev. D \textbf{48}, 2529-2547 (1993)
doi:10.1103/PhysRevD.48.2529
[arXiv:hep-ph/9303288 [hep-ph]].
%191 citations counted in INSPIRE as of 02 Nov 2020

%\cite{Lidsey:1995np}
\bibitem{Lidsey:1995np}
J.~E.~Lidsey, A.~R.~Liddle, E.~W.~Kolb, E.~J.~Copeland, T.~Barreiro and M.~Abney,
%``Reconstructing the inflation potential : An overview,''
Rev. Mod. Phys. \textbf{69}, 373-410 (1997)
doi:10.1103/RevModPhys.69.373
[arXiv:astro-ph/9508078 [astro-ph]].
%772 citations counted in INSPIRE as of 02 Nov 2020
	
%\cite{Ibanez:1995zs}
\bibitem{Ibanez:1995zs}
J.~Ibanez, R.~J.~van den Hoogen and A.~A.~Coley,
%``Isotropization of scalar field Bianchi models with an exponential potential,''
Phys. Rev. D \textbf{51}, 928-930 (1995)
doi:10.1103/PhysRevD.51.928
%52 citations counted in INSPIRE as of 02 Nov 2020
	
	%\cite{Copeland:1997et}
\bibitem{Copeland:1997et}
E.~J.~Copeland, A.~R.~Liddle and D.~Wands,
%``Exponential potentials and cosmological scaling solutions,''
Phys. Rev. D \textbf{57}, 4686-4690 (1998)
doi:10.1103/PhysRevD.57.4686
[arXiv:gr-qc/9711068 [gr-qc]].
%1052 citations counted in INSPIRE as of 02 Nov 2020
		
%\cite{Coley:1997nk}
\bibitem{Coley:1997nk}
A.~A.~Coley, J.~Ibanez and R.~J.~van den Hoogen,
%``Homogeneous scalar field cosmologies with an exponential potential,''
J. Math. Phys. \textbf{38}, 5256-5271 (1997)
doi:10.1063/1.532200
%88 citations counted in INSPIRE as of 02 Nov 2020


%\cite{Copeland:1998fz}
\bibitem{Copeland:1998fz}
E.~J.~Copeland, I.~J.~Grivell, E.~W.~Kolb and A.~R.~Liddle,
%``On the Reliability of Inflation Potential Reconstruction,''
Phys. Rev. D \textbf{58}, 043002 (1998)
doi:10.1103/PhysRevD.58.043002
[arXiv:astro-ph/9802209 [astro-ph]].
%27 citations counted in INSPIRE as of 02 Nov 2020


%\cite{Foster:1998sk}
\bibitem{Foster:1998sk}
S.~Foster,
%``Scalar field cosmologies and the initial space-time singularity,''
Class. Quant. Grav. \textbf{15}, 3485-3504 (1998)
doi:10.1088/0264-9381/15/11/014
[arXiv:gr-qc/9806098 [gr-qc]].
%64 citations counted in INSPIRE as of 11 Oct 2021

%\cite{Coley:1999mj}
\bibitem{Coley:1999mj}
A.~A.~Coley and R.~J.~van den Hoogen,
%``The Dynamics of multiscalar field cosmological models and assisted inflation,''
Phys. Rev. D \textbf{62}, 023517 (2000)
doi:10.1103/PhysRevD.62.023517
[arXiv:gr-qc/9911075 [gr-qc]].
%82 citations counted in INSPIRE as of 02 Nov 2020
	
	     
%\cite{vandenHoogen:1999qq}
\bibitem{vandenHoogen:1999qq}
R.~J.~van den Hoogen, A.~A.~Coley and D.~Wands,
%``Scaling solutions in Robertson-Walker space-times,''
Class. Quant. Grav. \textbf{16}, 1843-1851 (1999)
doi:10.1088/0264-9381/16/6/317
[arXiv:gr-qc/9901014 [gr-qc]].
%73 citations counted in INSPIRE as of 02 Nov 2020

%\cite{Albrecht:1999rm}
\bibitem{Albrecht:1999rm}
A.~Albrecht and C.~Skordis,
%``Phenomenology of a realistic accelerating universe using only Planck scale physics,''
Phys. Rev. Lett. \textbf{84}, 2076-2079 (2000)
doi:10.1103/PhysRevLett.84.2076
[arXiv:astro-ph/9908085 [astro-ph]].
%349 citations counted in INSPIRE as of 02 Nov 2020
		
%\cite{Coley:2000zw}
\bibitem{Coley:2000zw}
A.~Coley and M.~Goliath,
%``Selfsimilar spherically symmetric cosmological models with a perfect fluid and a scalar field,''
Class. Quant. Grav. \textbf{17}, 2557-2588 (2000)
doi:10.1088/0264-9381/17/13/309
[arXiv:gr-qc/0003080 [gr-qc]].
%19 citations counted in INSPIRE as of 02 Nov 2020
	
	
%\cite{Coley:2000yc}
\bibitem{Coley:2000yc}
A.~Coley and M.~Goliath,
%``Closed cosmologies with a perfect fluid and a scalar field,''
Phys. Rev. D \textbf{62}, 043526 (2000)
doi:10.1103/PhysRevD.62.043526
[arXiv:gr-qc/0004060 [gr-qc]].
%23 citations counted in INSPIRE as of 02 Nov 2020

	
%\cite{Coley:2003tf}
\bibitem{Coley:2003tf}
A.~Coley and Y.~J.~He,
%``Selfsimilar static spherically symmetric scalar field models,''
Gen. Rel. Grav. \textbf{35}, 707-749 (2003)
doi:10.1023/A:1022930418343
%4 citations counted in INSPIRE as of 02 Nov 2020


%\cite{Miritzis:2003ym}
\bibitem{Miritzis:2003ym}
J.~Miritzis,
%``Scalar field cosmologies with an arbitrary potential,''
Class. Quant. Grav. \textbf{20}, 2981-2990 (2003)
doi:10.1088/0264-9381/20/14/301
[arXiv:gr-qc/0303014 [gr-qc]].
%26 citations counted in INSPIRE as of 02 Nov 2020


%\cite{Rendall:2004ic}
\bibitem{Rendall:2004ic}
A.~D.~Rendall,
%``Accelerated cosmological expansion due to a scalar field whose potential has a positive lower bound,''
Class. Quant. Grav. \textbf{21}, 2445-2454 (2004)
doi:10.1088/0264-9381/21/9/018
[arXiv:gr-qc/0403070 [gr-qc]].
%44 citations counted in INSPIRE as of 11 Oct 2021


%\cite{Elizalde:2004mq}
\bibitem{Elizalde:2004mq}
E.~Elizalde, S.~Nojiri and S.~D.~Odintsov,
%``Late-time cosmology in (phantom) scalar-tensor theory: Dark energy and the cosmic speed-up,''
Phys. Rev. D \textbf{70}, 043539 (2004)
doi:10.1103/PhysRevD.70.043539
[arXiv:hep-th/0405034 [hep-th]].
%869 citations counted in INSPIRE as of 02 Nov 2020

%\cite{Capozziello:2005tf}
\bibitem{Capozziello:2005tf}
S.~Capozziello, S.~Nojiri and S.~D.~Odintsov,
%``Unified phantom cosmology: Inflation, dark energy and dark matter under the same standard,''
Phys. Lett. B \textbf{632}, 597-604 (2006)
doi:10.1016/j.physletb.2005.11.012
[arXiv:hep-th/0507182 [hep-th]].
%301 citations counted in INSPIRE as of 02 Nov 2020


%\cite{Curbelo:2005dh}
\bibitem{Curbelo:2005dh}
R.~Curbelo, T.~Gonzalez, G.~Leon and I.~Quiros,
%``Interacting phantom energy and avoidance of the big rip singularity,''
Class. Quant. Grav. \textbf{23}, 1585-1602 (2006)
doi:10.1088/0264-9381/23/5/010
[arXiv:astro-ph/0502141 [astro-ph]].
%75 citations counted in INSPIRE as of 02 Nov 2020

%\cite{Gonzalez:2005ie}
\bibitem{Gonzalez:2005ie}
T.~Gonzalez, G.~Leon and I.~Quiros,
%``Quintessence models of dark energy with non-minimal coupling,''
[arXiv:astro-ph/0502383 [astro-ph]].
%12 citations counted in INSPIRE as of 02 Nov 2020



%\cite{Miritzis:2005hg}
\bibitem{Miritzis:2005hg}
J.~Miritzis,
%``The Recollapse problem of closed FRW models in higher-order gravity theories,''
J. Math. Phys. \textbf{46}, 082502 (2005)
doi:10.1063/1.2009648
[arXiv:gr-qc/0505139 [gr-qc]].
%22 citations counted in INSPIRE as of 11 Oct 2021

%\cite{Rendall:2005if}
\bibitem{Rendall:2005if}
A.~D.~Rendall,
%``Intermediate inflation and the slow-roll approximation,''
Class. Quant. Grav. \textbf{22}, 1655-1666 (2005)
doi:10.1088/0264-9381/22/9/013
[arXiv:gr-qc/0501072 [gr-qc]].
%85 citations counted in INSPIRE as of 11 Oct 2021
	  
	  
%\cite{Rendall:2005fv}
\bibitem{Rendall:2005fv}
A.~D.~Rendall,
%``Dynamics of k-essence,''
Class. Quant. Grav. \textbf{23}, 1557-1570 (2006)
doi:10.1088/0264-9381/23/5/008
[arXiv:gr-qc/0511158 [gr-qc]].
%111 citations counted in INSPIRE as of 11 Oct 2021

%\cite{Gonzalez:2006cj}
\bibitem{Gonzalez:2006cj}
T.~Gonzalez, G.~Leon and I.~Quiros,
%``Dynamics of quintessence models of dark energy with exponential coupling to dark matter,''
Class. Quant. Grav. \textbf{23}, 3165-3179 (2006)
doi:10.1088/0264-9381/23/9/025
[arXiv:astro-ph/0702227 [astro-ph]].
%73 citations counted in INSPIRE as of 02 Nov 2020


	
%\cite{Rendall:2006cq}
\bibitem{Rendall:2006cq}
A.~D.~Rendall,
%``Late-time oscillatory behaviour for self-gravitating scalar fields,''
Class. Quant. Grav. \textbf{24}, 667-678 (2007)
doi:10.1088/0264-9381/24/3/010
[arXiv:gr-qc/0611088 [gr-qc]].
%27 citations counted in INSPIRE as of 02 Nov 2020


%\cite{Hertog:2006rr}
\bibitem{Hertog:2006rr}
T.~Hertog,
%``Towards a Novel no-hair Theorem for Black Holes,''
Phys. Rev. D \textbf{74}, 084008 (2006)
doi:10.1103/PhysRevD.74.084008
[arXiv:gr-qc/0608075 [gr-qc]].
%101 citations counted in INSPIRE as of 11 Oct 2021


%\cite{Gonzalez:2007ht}
\bibitem{Gonzalez:2007ht}
T.~Gonzalez and I.~Quiros,
%``Exact models with non-minimal interaction between dark matter and (either phantom or quintessence) dark energy,''
Class. Quant. Grav. \textbf{25}, 175019 (2008)
doi:10.1088/0264-9381/25/17/175019
[arXiv:0707.2089 [gr-qc]].
%35 citations counted in INSPIRE as of 02 Nov 2020

%\cite{Hrycyna:2007gd}
\bibitem{Hrycyna:2007gd}
O.~Hrycyna and M.~Szydlowski,
%``Extended Quintessence with non-minimally coupled phantom scalar field,''
Phys. Rev. D \textbf{76}, 123510 (2007)
doi:10.1103/PhysRevD.76.123510
[arXiv:0707.4471 [hep-th]].
%51 citations counted in INSPIRE as of 02 Nov 2020	

%\cite{Lazkoz:2007mx}
\bibitem{Lazkoz:2007mx}
R.~Lazkoz, G.~Leon and I.~Quiros,
%``Quintom cosmologies with arbitrary potentials,''
Phys. Lett. B \textbf{649}, 103-110 (2007)
doi:10.1016/j.physletb.2007.03.060
[arXiv:astro-ph/0701353 [astro-ph]].
%106 citations counted in INSPIRE as of 02 Nov 2020


%\cite{Elizalde:2008yf}
\bibitem{Elizalde:2008yf}
E.~Elizalde, S.~Nojiri, S.~D.~Odintsov, D.~Saez-Gomez and V.~Faraoni,
%``Reconstructing the universe history, from inflation to acceleration, with phantom and canonical scalar fields,''
Phys. Rev. D \textbf{77}, 106005 (2008)
doi:10.1103/PhysRevD.77.106005
[arXiv:0803.1311 [hep-th]].
%167 citations counted in INSPIRE as of 02 Nov 2020

\bibitem{Dania&Yunelsy} D. González Morales, Y. 
Nápoles Alvarez, Quintaesencia con acoplamiento no mínimo a la materia oscura desde la perspectiva de los sistemas dinámicos, Bachelor Thesis, Universidad Central Marta Abreu de Las Villas,  2008. 

		
%\cite{Giambo:2008ck}
\bibitem{Giambo:2008ck}
R.~Giambo, F.~Giannoni and G.~Magli,
%``The Dynamical behaviour of homogeneous scalar-field spacetimes with general self-interacting potentials,''
Gen. Rel. Grav. \textbf{41}, 21-30 (2009)
doi:10.1007/s10714-008-0647-z
[arXiv:0802.0157 [gr-qc]].
%12 citations counted in INSPIRE as of 02 Nov 2020

%\cite{Leon:2008de}
\bibitem{Leon:2008de}
G.~Leon,
%``On the Past Asymptotic Dynamics of Non-minimally Coupled Dark Energy,''
Class. Quant. Grav. \textbf{26}, 035008 (2009)
doi:10.1088/0264-9381/26/3/035008
[arXiv:0812.1013 [gr-qc]].
%47 citations counted in INSPIRE as of 11 Oct 2021


%\cite{Giambo:2009byn}
\bibitem{Giambo:2009byn}
R.~Giambo and J.~Miritzis,
%``Energy exchange for homogeneous and isotropic universes with a scalar field coupled to matter,''
Class. Quant. Grav. \textbf{27}, 095003 (2010)
doi:10.1088/0264-9381/27/9/095003
[arXiv:0908.3452 [gr-qc]].
%24 citations counted in INSPIRE as of 02 Nov 2020

%\cite{Leon:2009dt}

\bibitem{Leon:2009dt}
G.~Leon and E.~N.~Saridakis,
%``Phantom dark energy with varying-mass dark matter particles: acceleration and cosmic coincidence problem,''
Phys. Lett. B \textbf{693}, 1-10 (2010)
doi:10.1016/j.physletb.2010.08.016
[arXiv:0904.1577 [gr-qc]].
%48 citations counted in INSPIRE as of 02 Nov 2020

%\cite{Leon:2009rc}
\bibitem{Leon:2009rc}
G.~Leon and E.~N.~Saridakis,
%``Phase-space analysis of Horava-Lifshitz cosmology,''
JCAP \textbf{11}, 006 (2009)
doi:10.1088/1475-7516/2009/11/006
[arXiv:0909.3571 [hep-th]].
%120 citations counted in INSPIRE as of 02 Nov 2020


%\cite{Leon:2009ce}
\bibitem{Leon:2009ce}
G.~Leon, Y.~Leyva, E.~N.~Saridakis, O.~Martin and R.~Cardenas,
Falsifying Field-based Dark
Energy Models, in Dark Energy: Theories, Developments, and Implications, New York: Nova Science
Publishers, (2010).  arXiv:0912.0542 [gr-qc].
%20 citations counted in INSPIRE as of 02 Nov 2020
  
 %\cite{Leon:2010pu}
\bibitem{Leon:2010pu}
G.~Leon and E.~N.~Saridakis,
%``Dynamics of the anisotropic Kantowsky-Sachs geometries in $R^n$ gravity,''
Class. Quant. Grav. \textbf{28}, 065008 (2011)
doi:10.1088/0264-9381/28/6/065008
[arXiv:1007.3956 [gr-qc]].
%71 citations counted in INSPIRE as of 02 Nov 2020

%\cite{Basilakos:2011rx}
\bibitem{Basilakos:2011rx}
S.~Basilakos, M.~Tsamparlis and A.~Paliathanasis,
%``Using the Noether symmetry approach to probe the nature of dark energy,''
Phys. Rev. D \textbf{83}, 103512 (2011)
doi:10.1103/PhysRevD.83.103512
[arXiv:1104.2980 [astro-ph.CO]].
%98 citations counted in INSPIRE as of 02 Nov 2020


%\cite{Miritzis:2011zz}
\bibitem{Miritzis:2011zz}
J.~Miritzis,
%``FRW models in the conformal frame of f(R) gravity,''
J. Phys. Conf. Ser. \textbf{283}, 012024 (2011)
doi:10.1088/1742-6596/283/1/012024
%5 citations counted in INSPIRE as of 02 Nov 2020

%\cite{Xu:2012jf}
\bibitem{Xu:2012jf}
C.~Xu, E.~N.~Saridakis and G.~Leon,
%``Phase-Space analysis of Teleparallel Dark Energy,''
JCAP \textbf{07}, 005 (2012)
doi:10.1088/1475-7516/2012/07/005
[arXiv:1202.3781 [gr-qc]].
%169 citations counted in INSPIRE as of 02 Nov 2020


%\cite{Jamil:2012vb}
\bibitem{Jamil:2012vb}
M.~Jamil, D.~Momeni and R.~Myrzakulov,
%``Stability of a non-minimally conformally coupled scalar field in F(T) cosmology,''
Eur. Phys. J. C \textbf{72}, 2075 (2012)
doi:10.1140/epjc/s10052-012-2075-1
[arXiv:1208.0025 [gr-qc]].
%82 citations counted in INSPIRE as of 02 Nov 2020

%\cite{Leon:2012mt}
\bibitem{Leon:2012mt}
G.~Leon and E.~N.~Saridakis,
%``Dynamical analysis of generalized Galileon cosmology,''
JCAP \textbf{03}, 025 (2013)
doi:10.1088/1475-7516/2013/03/025
[arXiv:1211.3088 [astro-ph.CO]].
%95 citations counted in INSPIRE as of 02 Nov 2020


%\cite{Leon:2013qh}
\bibitem{Leon:2013qh}
G.~Leon, J.~Saavedra and E.~N.~Saridakis,
%``Cosmological behavior in extended nonlinear massive gravity,''
Class. Quant. Grav. \textbf{30}, 135001 (2013)
doi:10.1088/0264-9381/30/13/135001
[arXiv:1301.7419 [astro-ph.CO]].
%84 citations counted in INSPIRE as of 02 Nov 2020

%\cite{Skugoreva:2013ooa}
\bibitem{Skugoreva:2013ooa}
M.~A.~Skugoreva, S.~V.~Sushkov and A.~V.~Toporensky,
%``Cosmology with nonminimal kinetic coupling and a power-law potential,''
Phys. Rev. D \textbf{88}, 083539 (2013)
[erratum: Phys. Rev. D \textbf{88}, no.10, 109906 (2013)]
doi:10.1103/PhysRevD.88.083539
[arXiv:1306.5090 [gr-qc]].
%50 citations counted in INSPIRE as of 02 Nov 2020

%\cite{Fadragas:2013ina}
\bibitem{Fadragas:2013ina}
C.~R.~Fadragas, G.~Leon and E.~N.~Saridakis,
%``Dynamical analysis of anisotropic scalar-field cosmologies for a wide range of potentials,''
Class. Quant. Grav. \textbf{31}, 075018 (2014)
doi:10.1088/0264-9381/31/7/075018
[arXiv:1308.1658 [gr-qc]].
%54 citations counted in INSPIRE as of 02 Nov 2020

  
  
%\cite{Fadragas:2014mra}
\bibitem{Fadragas:2014mra}
C.~R.~Fadragas and G.~Leon,
%``Some remarks about non-minimally coupled scalar field models,''
Class. Quant. Grav. \textbf{31}, no.19, 195011 (2014)
doi:10.1088/0264-9381/31/19/195011
[arXiv:1405.2465 [gr-qc]].
%23 citations counted in INSPIRE as of 02 Nov 2020\textbf{}

%\cite{Kofinas:2014aka}
\bibitem{Kofinas:2014aka}
G.~Kofinas, G.~Leon and E.~N.~Saridakis,
%``Dynamical behavior in $f(T,T_G)$ cosmology,''
Class. Quant. Grav. \textbf{31}, 175011 (2014)
doi:10.1088/0264-9381/31/17/175011
[arXiv:1404.7100 [gr-qc]].
%105 citations counted in INSPIRE as of 02 Nov 2020



%\cite{Tzanni:2014eja}
\bibitem{Tzanni:2014eja}
K.~Tzanni and J.~Miritzis,
%``Coupled quintessence with double exponential potentials,''
Phys. Rev. D \textbf{89}, no.10, 103540 (2014)
doi:10.1103/PhysRevD.89.103540
[arXiv:1403.6618 [gr-qc]].
%15 citations counted in INSPIRE as of 02 Nov 2020


%\cite{Leon:2014yua}
\bibitem{Leon:2014yua}
G.~Leon and E.~N.~Saridakis,
%``Dynamical behavior in mimetic F(R) gravity,''
JCAP \textbf{04}, 031 (2015)
doi:10.1088/1475-7516/2015/04/031
[arXiv:1501.00488 [gr-qc]].
%97 citations counted in INSPIRE as of 02 Nov 2020



%\cite{Leon:2014bta}
\bibitem{Leon:2014bta}
G.~Le\'on Torres,
``Qualitative analysis and characterization of two cosmologies including scalar fields,''
 Phd Thesis, Universidad Central Marta Abreu de Las Villas,  2010. 
[arXiv:1412.5665 [gr-qc]].
%5 citations counted in INSPIRE as of 02 Nov 2020
	
	  

%\cite{Leon:2014rra}
\bibitem{Leon:2014rra}
G.~Leon and C.~R.~Fadragas,
``Cosmological dynamical systems: And Their Applications''. Saarbrücken: LAP
Lambert Academic Publishing, 2012,
[arXiv:1412.5701 [gr-qc]].
%32 citations counted in INSPIRE as of 02 Nov 2020

%\cite{Minazzoli:2014xua}
\bibitem{Minazzoli:2014xua}
O.~Minazzoli and A.~Hees,
%``Late-time cosmology of a scalar-tensor theory with a universal multiplicative coupling between the scalar field and the matter Lagrangian,''
Phys. Rev. D \textbf{90}, 023017 (2014)
doi:10.1103/PhysRevD.90.023017
[arXiv:1404.4266 [gr-qc]].
%36 citations counted in INSPIRE as of 02 Nov 2020

	

%\cite{Alho:2014fha}
\bibitem{Alho:2014fha}
A.~Alho and C.~Uggla,
%``Global dynamics and inflationary center manifold and slow-roll approximants,''
J. Math. Phys. \textbf{56}, no.1, 012502 (2015)
doi:10.1063/1.4906081
[arXiv:1406.0438 [gr-qc]].
%28 citations counted in INSPIRE as of 02 Nov 2020

%\cite{Paliathanasis:2014yfa}
\bibitem{Paliathanasis:2014yfa}
A.~Paliathanasis and M.~Tsamparlis,
%``Two scalar field cosmology: Conservation laws and exact solutions,''
Phys. Rev. D \textbf{90}, no.4, 043529 (2014)
doi:10.1103/PhysRevD.90.043529
[arXiv:1408.1798 [gr-qc]].
%48 citations counted in INSPIRE as of 02 Nov 2020

%\cite{DeArcia:2015ztd}
\bibitem{DeArcia:2015ztd}
R.~De Arcia, T.~Gonzalez, G.~Leon, U.~Nucamendi and I.~Quiros,
%``Cubic Derivative Interactions and Asymptotic Dynamics of the Galileon Vacuum,''
Class. Quant. Grav. \textbf{33}, no.12, 125036 (2016)
doi:10.1088/0264-9381/33/12/125036
[arXiv:1511.09125 [gr-qc]].
%15 citations counted in INSPIRE as of 02 Nov 2020



%\cite{Solomon:2015hja}
\bibitem{Solomon:2015hja}
A.~R.~Solomon,
%``Cosmology Beyond Einstein,''
doi:10.1007/978-3-319-46621-7
[arXiv:1508.06859 [gr-qc]].
%19 citations counted in INSPIRE as of 02 Nov 2020


%\cite{Harko:2015pma}
\bibitem{Harko:2015pma}
T.~Harko, F.~S.~N.~Lobo, J.~P.~Mimoso and D.~Pav\'on,
%``Gravitational induced particle production through a nonminimal curvature\textendash{}matter coupling,''
Eur. Phys. J. C \textbf{75}, 386 (2015)
doi:10.1140/epjc/s10052-015-3620-5
[arXiv:1508.02511 [gr-qc]].
%41 citations counted in INSPIRE as of 02 Nov 2020

%\cite{Paliathanasis:2015gga}
\bibitem{Paliathanasis:2015gga}
A.~Paliathanasis, M.~Tsamparlis, S.~Basilakos and J.~D.~Barrow,
%``Dynamical analysis in scalar field cosmology,''
Phys. Rev. D \textbf{91}, no.12, 123535 (2015)
doi:10.1103/PhysRevD.91.123535
[arXiv:1503.05750 [gr-qc]].
%60 citations counted in INSPIRE as of 02 Nov 2020


	%\cite{Leon:2015via}
\bibitem{Leon:2015via}
G.~Leon and E.~N.~Saridakis,
%``Cosmology in time asymmetric extensions of general relativity,''
JCAP \textbf{11}, 009 (2015)
doi:10.1088/1475-7516/2015/11/009
[arXiv:1504.07606 [gr-qc]].
%10 citations counted in INSPIRE as of 02 Nov 2020


%\cite{Matsumoto:2015hua}
\bibitem{Matsumoto:2015hua}
J.~Matsumoto and S.~V.~Sushkov,
%``Cosmology with nonminimal kinetic coupling and a Higgs-like potential,''
JCAP \textbf{11}, 047 (2015)
doi:10.1088/1475-7516/2015/11/047
[arXiv:1510.03264 [gr-qc]].
%16 citations counted in INSPIRE as of 02 Nov 2020

%\cite{Barrow:2016qkh}
\bibitem{Barrow:2016qkh}
J.~D.~Barrow and A.~Paliathanasis,
%``Observational Constraints on New Exact Inflationary Scalar-field Solutions,''
Phys. Rev. D \textbf{94}, no.8, 083518 (2016)
doi:10.1103/PhysRevD.94.083518
[arXiv:1609.01126 [gr-qc]].
%29 citations counted in INSPIRE as of 02 Nov 2020

%\cite{Barrow:2016wiy}
\bibitem{Barrow:2016wiy}
J.~D.~Barrow and A.~Paliathanasis,
%``Reconstructions of the dark-energy equation of state and the inflationary potential,''
Gen. Rel. Grav. \textbf{50}, no.7, 82 (2018)
doi:10.1007/s10714-018-2402-4
[arXiv:1611.06680 [gr-qc]].
%20 citations counted in INSPIRE as of 02 Nov 2020

%\cite{Cid:2017wtf}
\bibitem{Cid:2017wtf}
A.~Cid, F.~Izaurieta, G.~Leon, P.~Medina and D.~Narbona,
%``Non-minimally coupled scalar field cosmology with torsion,''
JCAP \textbf{04}, 041 (2018)
doi:10.1088/1475-7516/2018/04/041
[arXiv:1704.04563 [gr-qc]].
%17 citations counted in INSPIRE as of 02 Nov 2020

%\cite{Cruz:2017ecg}
\bibitem{Cruz:2017ecg}
M.~Cruz, A.~Ganguly, R.~Gannouji, G.~Leon and E.~N.~Saridakis,
%``Global structure of static spherically symmetric solutions surrounded by quintessence,''
Class. Quant. Grav. \textbf{34}, no.12, 125014 (2017)
doi:10.1088/1361-6382/aa70fc
[arXiv:1702.01754 [gr-qc]].
%11 citations counted in INSPIRE as of 02 Nov 2020


%\cite{Paliathanasis:2017ocj}
\bibitem{Paliathanasis:2017ocj}
A.~Paliathanasis,
%``Dust fluid component from Lie symmetries in Scalar field Cosmology,''
Mod. Phys. Lett. A \textbf{32}, no.37, 1750206 (2017)
doi:10.1142/S0217732317502066
[arXiv:1710.08666 [gr-qc]].
%9 citations counted in INSPIRE as of 02 Nov 2020

%\cite{Alhulaimi:2017ocb}
\bibitem{Alhulaimi:2017ocb}
B.~Alhulaimi, R.~J.~Van Den Hoogen and A.~A.~Coley,
%``Spatially homogeneous Einstein-Aether cosmological models: scalar fields with a generalized harmonic potential,''
JCAP \textbf{12}, 045 (2017)
doi:10.1088/1475-7516/2017/12/045
[arXiv:1707.08911 [gr-qc]].
%20 citations counted in INSPIRE as of 02 Nov 2020

%\cite{Dimakis:2017kwx}
\bibitem{Dimakis:2017kwx}
N.~Dimakis, A.~Giacomini, S.~Jamal, G.~Leon and A.~Paliathanasis,
%``Noether symmetries and stability of ideal gas solutions in Galileon cosmology,''
Phys. Rev. D \textbf{95}, no.6, 064031 (2017)
doi:10.1103/PhysRevD.95.064031
[arXiv:1702.01603 [gr-qc]].
%28 citations counted in INSPIRE as of 02 Nov 2020

%\cite{Giacomini:2017yuk}
\bibitem{Giacomini:2017yuk}
A.~Giacomini, S.~Jamal, G.~Leon, A.~Paliathanasis and J.~Saavedra,
%``Dynamical Analysis of an Integrable Cubic Galileon Cosmological Model,''
Phys. Rev. D \textbf{95}, no.12, 124060 (2017)
doi:10.1103/PhysRevD.95.124060
[arXiv:1703.05860 [gr-qc]].
%40 citations counted in INSPIRE as of 02 Nov 2020

%\cite{Karpathopoulos:2017arc}
\bibitem{Karpathopoulos:2017arc}
L.~Karpathopoulos, S.~Basilakos, G.~Leon, A.~Paliathanasis and M.~Tsamparlis,
%``Cartan symmetries and global dynamical systems analysis in a higher-order modified teleparallel theory,''
Gen. Rel. Grav. \textbf{50}, no.7, 79 (2018)
doi:10.1007/s10714-018-2400-6
[arXiv:1709.02197 [gr-qc]].
%22 citations counted in INSPIRE as of 02 Nov 2020

%\cite{Matsumoto:2017gnx}
\bibitem{Matsumoto:2017gnx}
J.~Matsumoto and S.~V.~Sushkov,
%``General dynamical properties of cosmological models with nonminimal kinetic coupling,''
JCAP \textbf{01}, 040 (2018)
doi:10.1088/1475-7516/2018/01/040
[arXiv:1703.04966 [gr-qc]].
%12 citations counted in INSPIRE as of 02 Nov 2020

%\cite{VanDenHoogen:2018anx}
\bibitem{VanDenHoogen:2018anx}
R.~J.~Van Den Hoogen, A.~A.~Coley, B.~Alhulaimi, S.~Mohandas, E.~Knighton and S.~O'Neil,
%``Kantowski-Sachs Einstein-Aether Scalar Field Cosmological Models,''
JCAP \textbf{11}, 017 (2018)
doi:10.1088/1475-7516/2018/11/017
[arXiv:1809.01458 [gr-qc]].
%19 citations counted in INSPIRE as of 02 Nov 2020



%\cite{Leon:2018lnd}
\bibitem{Leon:2018lnd}
G.~Leon, A.~Paliathanasis and J.~L.~Morales-Mart\'\i{}nez,
%``The past and future dynamics of quintom dark energy models,''
Eur. Phys. J. C \textbf{78}, no.9, 753 (2018)
doi:10.1140/epjc/s10052-018-6225-y
[arXiv:1808.05634 [gr-qc]].
%23 citations counted in INSPIRE as of 02 Nov 2020

%\cite{Leon:2018skk}
\bibitem{Leon:2018skk}
G.~Leon, A.~Paliathanasis and L.~Velazquez Abab,
%``Stability of a modified Jordan-Brans-Dicke theory in the dilatonic frame,''
Gen. Rel. Grav. \textbf{52}, 71 (2020)
doi:10.1007/s10714-020-02718-7
[arXiv:1812.03830 [physics.gen-ph]].
%6 citations counted in INSPIRE as of 02 Nov 2020

%\cite{DeArcia:2018pjp}
\bibitem{DeArcia:2018pjp}
R.~De Arcia, T.~Gonzalez, F.~A.~Horta-Rangel, G.~Leon, U.~Nucamendi and I.~Quiros,
%``Dynamical systems analysis of the cubic galileon beyond the exponential potential and the cosmological analogue of the vDVZ discontinuity,''
Class. Quant. Grav. \textbf{35}, no.14, 145001 (2018)
doi:10.1088/1361-6382/aac6a5
[arXiv:1801.02269 [gr-qc]].
%10 citations counted in INSPIRE as of 02 Nov 2020

%\cite{Tsamparlis:2018nyo}
\bibitem{Tsamparlis:2018nyo}
M.~Tsamparlis and A.~Paliathanasis,
%``Symmetries of Differential Equations in Cosmology,''
Symmetry \textbf{10}, no.7, 233 (2018)
doi:10.3390/sym10070233
[arXiv:1806.05888 [gr-qc]].
%23 citations counted in INSPIRE as of 02 Nov 2020

%\cite{Paliathanasis:2018vru}
\bibitem{Paliathanasis:2018vru}
A.~Paliathanasis, G.~Leon and S.~Pan,
%``Exact Solutions in Chiral Cosmology,''
Gen. Rel. Grav. \textbf{51}, no.9, 106 (2019)
doi:10.1007/s10714-019-2594-2
[arXiv:1811.10038 [gr-qc]].
%18 citations counted in INSPIRE as of 02 Nov 2020 


%\cite{Barrow:2018zav}
\bibitem{Barrow:2018zav}
J.~D.~Barrow and A.~Paliathanasis,
%``Szekeres Universes with Homogeneous Scalar Fields,''
Eur. Phys. J. C \textbf{78}, no.9, 767 (2018)
doi:10.1140/epjc/s10052-018-6245-7
[arXiv:1808.00173 [gr-qc]].
%10 citations counted in INSPIRE as of 02 Nov 2020

	
%\cite{Basilakos:2019dof}
\bibitem{Basilakos:2019dof}
S.~Basilakos, G.~Leon, G.~Papagiannopoulos and E.~N.~Saridakis,
%``Dynamical system analysis at background and perturbation levels: Quintessence in severe disadvantage comparing to $\Lambda$CDM,''
Phys. Rev. D \textbf{100}, no.4, 043524 (2019)
doi:10.1103/PhysRevD.100.043524
[arXiv:1904.01563 [gr-qc]].
%11 citations counted in INSPIRE as of 02 Nov 2020



%\cite{Leon:2019mbo}
\bibitem{Leon:2019mbo}
G.~Leon and A.~Paliathanasis,
%``Extended phase-space analysis of the Ho\v{r}ava\textendash{}Lifshitz cosmology,''
Eur. Phys. J. C \textbf{79}, no.9, 746 (2019)
doi:10.1140/epjc/s10052-019-7236-z
[arXiv:1902.09961 [gr-qc]].
%14 citations counted in INSPIRE as of 02 Nov 2020

%\cite{Paliathanasis:2019qch}
\bibitem{Paliathanasis:2019qch}
A.~Paliathanasis and G.~Leon,
%``Cosmological solutions in Ho\textbackslash{}v\{r\}ava-Lifshitz gravity,''
doi:10.1515/zna-2020-0003
[arXiv:1903.10821 [gr-qc]].
%5 citations counted in INSPIRE as of 02 Nov 2020


%\cite{Leon:2019jnu}
\bibitem{Leon:2019jnu}
G.~Leon, A.~Coley and A.~Paliathanasis,
%``Static spherically symmetric Einstein-\ae{}ther models II: Integrability and the modified Tolman\textendash{}Oppenheimer\textendash{}Volkoff approach,''
Annals Phys. \textbf{412}, 168002 (2020)
doi:10.1016/j.aop.2019.168002
[arXiv:1906.05749 [gr-qc]].
%19 citations counted in INSPIRE as of 02 Nov 2020
	
%\cite{Paliathanasis:2019pcl}
\bibitem{Paliathanasis:2019pcl}
A.~Paliathanasis, G.~Papagiannopoulos, S.~Basilakos and J.~D.~Barrow,
%``Dynamics of Einstein\textendash{}Aether scalar field cosmology,''
Eur. Phys. J. C \textbf{79}, no.8, 723 (2019)
doi:10.1140/epjc/s10052-019-7229-y
[arXiv:1906.03872 [gr-qc]].
%16 citations counted in INSPIRE as of 02 Nov 2020

%\cite{Quiros:2019ktw}
\bibitem{Quiros:2019ktw}
I.~Quiros,
%``Selected topics in scalar\textendash{}tensor theories and beyond,''
Int. J. Mod. Phys. D \textbf{28}, no.07, 1930012 (2019)
doi:10.1142/S021827181930012X
[arXiv:1901.08690 [gr-qc]].
%33 citations counted in INSPIRE as of 02 Nov 2020


%\cite{Shahalam:2019jgs}
\bibitem{Shahalam:2019jgs}
M.~Shahalam, R.~Myrzakulov and M.~Y.~Khlopov,
%``Late time evolution of a nonminimally coupled scalar field system,''
Gen. Rel. Grav. \textbf{51}, no.9, 125 (2019)
doi:10.1007/s10714-019-2610-6
[arXiv:1905.06856 [gr-qc]].
%5 citations counted in INSPIRE as of 02 Nov 2020


%\cite{Nojiri:2019riz}
\bibitem{Nojiri:2019riz}
S.~Nojiri, S.~D.~Odintsov and V.~K.~Oikonomou,
%``$F(R)$ Gravity with an Axion-like Particle: Dynamics, Gravity Waves, Late and Early-time Phenomenology,''
Annals Phys. \textbf{418}, 168186 (2020)
doi:10.1016/j.aop.2020.168186
[arXiv:1907.01625 [gr-qc]].
%14 citations counted in INSPIRE as of 02 Nov 2020

%\cite{Humieja:2019ywy}
\bibitem{Humieja:2019ywy}
F.~Humieja and M.~Szyd\l{}owski,
%``Bifurcations in Ratra\textendash{}Peebles quintessence models and their extensions,''
Eur. Phys. J. C \textbf{79}, no.9, 794 (2019)
doi:10.1140/epjc/s10052-019-7299-x
[arXiv:1901.06578 [gr-qc]].
%7 citations counted in INSPIRE as of 02 Nov 2020


















	







%\cite{Giambo:2019ymx}
\bibitem{Giambo:2019ymx}
R.~Giamb\`o, J.~Miritzis and A.~Pezzola,
%``Late time evolution of negatively curved FLRW models,''
Eur. Phys. J. Plus \textbf{135}, no.4, 367 (2020)
doi:10.1140/epjp/s13360-020-00370-3
[arXiv:1905.01742 [gr-qc]].
%8 citations counted in INSPIRE as of 02 Nov 2020







\bibitem{Coddington55} E. A. Coddington y Levinson, N.
``{Theory of Ordinary Differential Equations}'', New York,
MacGraw-Hill, (1955).



\bibitem{Hale69} J. K. Hale, ``{Ordinary Differential
Equations}'', New York, Wiley (1969).

\bibitem{AP}
D. K. Arrowsmith y C. M. Place, ``{An introduction to dynamical systems}'',
 Cambridge University Press, Cambridge, England, (1990).
 
 
\bibitem{wiggins}
S. Wiggins. ``Introduction to Applied Nonlinear dynamical systems
and Chaos''. Springer (2003).

\bibitem{perko} L. Perko, ``{Differential equations and dynamical
systems, third edition}'' (Springer-Verlag, New York, 2001).
pp 272-273 \& pp 281-282.



\bibitem{160}  V.I. Arnold, ``{Ordinary differential equations}''.
\newblock{ Cambridge: M.I.T. Press.}, 1973. 

\bibitem{Hirsch}
M.~W. Hirsch and S.~Smale.
``{Differential equations, dynamical systems, and linear algebra}''.
\newblock New York: Academic Press (1974).

\bibitem{165} J. Hale.
``{Ordinary differential equations}''.
\newblock Malabar, Florida: Robert E. Krieger Publishing Co., Inc. (1980).



 \bibitem{LaSalle} Lasalle, J. P.,  J. Diff. Eq., \textbf{4}, pp. 57-65, 1968.

\bibitem{aulbach}
B. Aulbach, ``Continuous and Discrete Dynamics near Manifolds
of Equilibria'' (Lecture Notes in Mathematics No. 1058, Springer,
1984).


\bibitem{TWE}
R. Tavakol, ``{Introduction to dynamical systems}'', ch 4. Part
one, pp. 84--98,
 Cambridge University Press, Cambridge, England, (1997).


\bibitem{coleybook}  A.A. Coley, 2003,
{``{Dynamical systems and cosmology}''} (Kluwer Academic,
Dordrecht: ISBN 1-4020-1403-1).



%\cite{Coley:1999uh}
\bibitem{Coley:1999uh}
A.~A.~Coley,
%``Dynamical systems in cosmology,''
[arXiv:gr-qc/9910074 [gr-qc]].
%71 citations counted in INSPIRE as of 02 Nov 2020

  
 \bibitem{bassemah} Bassemah Alhulaimi (2017), Einstein-Aether Cosmological Scalar Field Models (Phd Thesis, Dalhousie University).


%\cite{LeBlanc:1994qm}
\bibitem{LeBlanc:1994qm}
V.~G.~LeBlanc, D.~Kerr and J.~Wainwright,
%``Asymptotic states of magnetic Bianchi V2I0 cosmologies,''
Class. Quant. Grav. \textbf{12}, 513-541 (1995)
doi:10.1088/0264-9381/12/2/020
%58 citations counted in INSPIRE as of 02 Nov 2020


%\cite{Heinzle:2009zb}
\bibitem{Heinzle:2009zb}
J.~M.~Heinzle and C.~Uggla,
%``Monotonic functions: Why they exist and how to find them,''
Class. Quant. Grav. \textbf{27}, 015009 (2010)
doi:10.1088/0264-9381/27/1/015009
[arXiv:0907.0653 [gr-qc]].
%6 citations counted in INSPIRE as of 02 Nov 2020

			
%\cite{Alho:2015cza}
\bibitem{Alho:2015cza}
A.~Alho, J.~Hell and C.~Uggla,
%``Global dynamics and asymptotics for monomial scalar field potentials and perfect fluids,''
Class. Quant. Grav. \textbf{32}, no.14, 145005 (2015)
doi:10.1088/0264-9381/32/14/145005
[arXiv:1503.06994 [gr-qc]].
%24 citations counted in INSPIRE as of 02 Nov 2020


%\cite{Alho:2019pku}
\bibitem{Alho:2019pku}
A.~Alho, V.~Bessa and F.~C.~Mena,
%``Global dynamics of Yang\textendash{}Mills field and perfect-fluid Robertson\textendash{}Walker cosmologies,''
J. Math. Phys. \textbf{61} (2020) no.3, 032502
doi:10.1063/1.5139879
[arXiv:1910.04678 [gr-qc]].
%7 citations counted in INSPIRE as of 09 Aug 2021

	
%\cite{Paliathanasis:2015cza}
\bibitem{Paliathanasis:2015cza}
A.~Paliathanasis, S.~Pan and S.~Pramanik,
%``Scalar field cosmology modified by the Generalized Uncertainty Principle,''
Class. Quant. Grav. \textbf{32}, no.24, 245006 (2015)
doi:10.1088/0264-9381/32/24/245006
[arXiv:1508.06543 [gr-qc]].
%20 citations counted in INSPIRE as of 02 Nov 2020
	
	
	
%\cite{Leon:2020ovw}
\bibitem{Leon:2020ovw}
G.~Leon and F.~O.~F.~Silva,
%``Generalized scalar field cosmologies: a global dynamical systems formulation,''
Class. Quant. Grav. \textbf{38} (2021) no.1, 015004
doi:10.1088/1361-6382/abc095
[arXiv:2007.11990 [gr-qc]].
%7 citations counted in INSPIRE as of 03 Nov 2021


 %\cite{Llibre:2012zz}
\bibitem{Llibre:2012zz}
J.~Llibre and C.~Vidal,
%``Periodic orbits and non-integrability in a cosmological scalar field,''
J. Math. Phys. \textbf{53}, 012702 (2012)
doi:10.1063/1.3675493
%2 citations counted in INSPIRE as of 02 Nov 2020

\bibitem{Zhuravlev_2021}  V. M. Zhuravlev,  and S. V. Chervon,  
%Method of multiple scales in scalar field cosmology 
Journal of Physics: Conference Series, IOP Publishing, \textbf{2021}, 2081, 012037


%\cite{Fajman:2020yjb}
\bibitem{Fajman:2020yjb}
D.~Fajman, G.~Hei\ss{}el and M.~Maliborski,
%``On the oscillations and future asymptotics of locally rotationally symmetric Bianchi type III cosmologies with a massive scalar field,''
Class. Quant. Grav. \textbf{37} (2020) no.13, 135009
doi:10.1088/1361-6382/ab8c97
[arXiv:2001.00252 [gr-qc]].
%9 citations counted in INSPIRE as of 19 Aug 2021

 %\cite{Fajman:2021cli}
\bibitem{Fajman:2021cli}
D.~Fajman, G.~Hei\ss{}el and J.~W.~Jang,
%``Averaging with a time-dependent perturbation parameter,''
Class. Quant. Grav. \textbf{38} (2021) no.8, 085005
doi:10.1088/1361-6382/abe883
%4 citations counted in INSPIRE as of 19 Aug 2021

%\cite{Leon:2021lct}
\bibitem{Leon:2021lct}
G.~Leon, E.~Gonz\'alez, S.~Lepe, C.~Michea and A.~D.~Millano,
%``Averaging generalized scalar field cosmologies I: locally rotationally symmetric Bianchi III and open Friedmann\textendash{}Lema\^\i{}tre\textendash{}Robertson\textendash{}Walker models,''
Eur. Phys. J. C \textbf{81} (2021) no.5, 414
doi:10.1140/epjc/s10052-021-09185-7
[arXiv:2102.05465 [gr-qc]].
%3 citations counted in INSPIRE as of 09 Aug 2021


%\cite{Leon:2021rcx}
\bibitem{Leon:2021rcx}
G.~Leon, S.~Cuellar, E.~Gonzalez, S.~Lepe, C.~Michea and A.~D.~Millano,
%``Averaging generalized scalar field cosmologies II: locally rotationally symmetric Bianchi I and flat Friedmann\textendash{}Lema\^\i{}tre\textendash{}Robertson\textendash{}Walker models,''
Eur. Phys. J. C \textbf{81} (2021) no.6, 489
doi:10.1140/epjc/s10052-021-09230-5
[arXiv:2102.05495 [gr-qc]].
%3 citations counted in INSPIRE as of 09 Aug 2021

%\cite{Leon:2021hxc}
\bibitem{Leon:2021hxc}
G.~Leon, E.~Gonz\'alez, S.~Lepe, C.~Michea and A.~D.~Millano,
%``Averaging generalized scalar-field cosmologies III: Kantowski\textendash{}Sachs and closed Friedmann\textendash{}Lema\^\i{}tre\textendash{}Robertson\textendash{}Walker models,''
Eur. Phys. J. C \textbf{81} (2021) no.10, 867
doi:10.1140/epjc/s10052-021-09580-0
[arXiv:2102.05551 [gr-qc]].
%3 citations counted in INSPIRE as of 14 Jan 2022

	




%\cite{Kaloper:1997sh}
\bibitem{Kaloper:1997sh}
N.~Kaloper and K.~A.~Olive,
%``Singularities in scalar tensor cosmologies,''
Phys. Rev. D \textbf{57}, 811-822 (1998)
doi:10.1103/PhysRevD.57.811
[arXiv:hep-th/9708008 [hep-th]].
%48 citations counted in INSPIRE as of 11 Oct 2021


%\cite{Coley:2003mj}
\bibitem{Coley:2003mj}
A.~A.~Coley,
%``Dynamical systems and cosmology,''
Astrophys. Space Sci. Libr. \textbf{291} (2003)
doi:10.1007/978-94-017-0327-7
%68 citations counted in INSPIRE as of 02 Nov 2020

%\cite{Kolda:1998wq}
\bibitem{Kolda:1998wq}
C.~F.~Kolda and D.~H.~Lyth,
%``Quintessential difficulties,''
Phys. Lett. B \textbf{458} (1999), 197-201
doi:10.1016/S0370-2693(99)00657-7
[arXiv:hep-ph/9811375 [hep-ph]].
%181 citations counted in INSPIRE as of 23 Nov 2021

%\cite{Sahni:2002kh}
\bibitem{Sahni:2002kh}
V.~Sahni,
%``The Cosmological constant problem and quintessence,''
Class. Quant. Grav. \textbf{19}, 3435-3448 (2002)
doi:10.1088/0264-9381/19/13/304
[arXiv:astro-ph/0202076 [astro-ph]].
%310 citations counted in INSPIRE as of 11 Oct 2021

%\cite{Padmanabhan:2002ji}
\bibitem{Padmanabhan:2002ji}
T.~Padmanabhan,
%``Cosmological constant: The Weight of the vacuum,''
Phys. Rept. \textbf{380}, 235-320 (2003)
doi:10.1016/S0370-1573(03)00120-0
[arXiv:hep-th/0212290 [hep-th]].
%2690 citations counted in INSPIRE as of 02 Nov 2021


%\cite{Gasperini:2007zz}
\bibitem{Gasperini:2007zz}
M.~Gasperini,
``Elements of string cosmology,'' ISBN: 9780511332296 (eBook), 9780521187985 (Print), 9780521868754 (Print). Cambridge University Press , Cambridge, 2007. 
%41 citations counted in INSPIRE as of 11 Oct 2021



%\cite{Fujii:2003pa}
\bibitem{Fujii:2003pa}
Y.~Fujii and K.~Maeda,
``The scalar-tensor theory of gravitation,''
doi:10.1017/CBO9780511535093
%199 citations counted in INSPIRE as of 11 Oct 2021


%\cite{Faraoni:2004pi}
\bibitem{Faraoni:2004pi}
V.~Faraoni,
``Cosmology in scalar tensor gravity,''
doi:10.1007/978-1-4020-1989-0
%179 citations counted in INSPIRE as of 28 Oct 2021



%\cite{Capozziello:2007ec}
\bibitem{Capozziello:2007ec}
S.~Capozziello and M.~Francaviglia,
%``Extended Theories of Gravity and their Cosmological and Astrophysical Applications,''
Gen. Rel. Grav. \textbf{40}, 357-420 (2008)
doi:10.1007/s10714-007-0551-y
[arXiv:0706.1146 [astro-ph]].
%704 citations counted in INSPIRE as of 01 Nov 2021

%\cite{Waterhouse:2006wv}
\bibitem{Waterhouse:2006wv}
T.~P.~Waterhouse,
%``An Introduction to Chameleon Gravity,''
[arXiv:astro-ph/0611816 [astro-ph]].
%55 citations counted in INSPIRE as of 11 Oct 2021

%\cite{Amendola:1999er}
\bibitem{Amendola:1999er}
L.~Amendola,
%``Coupled quintessence,''
Phys. Rev. D \textbf{62}, 043511 (2000)
doi:10.1103/PhysRevD.62.043511
[arXiv:astro-ph/9908023 [astro-ph]].
%1450 citations counted in INSPIRE as of 15 Oct 2021




%\cite{Tocchini-Valentini:2001wmi}
\bibitem{Tocchini-Valentini:2001wmi}
D.~Tocchini-Valentini and L.~Amendola,
%``Stationary dark energy with a baryon dominated era: Solving the coincidence problem with a linear coupling,''
Phys. Rev. D \textbf{65}, 063508 (2002)
doi:10.1103/PhysRevD.65.063508
[arXiv:astro-ph/0108143 [astro-ph]].
%145 citations counted in INSPIRE as of 11 Oct 2021

%\cite{Billyard:2000bh}
\bibitem{Billyard:2000bh}
A.~P.~Billyard and A.~A.~Coley,
%``Interactions in scalar field cosmology,''
Phys. Rev. D \textbf{61}, 083503 (2000)
doi:10.1103/PhysRevD.61.083503
[arXiv:astro-ph/9908224 [astro-ph]].
%188 citations counted in INSPIRE as of 11 Oct 2021


%\cite{Carloni:2007eu}
\bibitem{Carloni:2007eu}
S.~Carloni, S.~Capozziello, J.~A.~Leach and P.~K.~S.~Dunsby,
%``Cosmological dynamics of scalar-tensor gravity,''
Class. Quant. Grav. \textbf{25}, 035008 (2008)
doi:10.1088/0264-9381/25/3/035008
[arXiv:gr-qc/0701009 [gr-qc]].
%94 citations counted in INSPIRE as of 11 Oct 2021


	


%\cite{Tsujikawa:2008uc}
\bibitem{Tsujikawa:2008uc}
S.~Tsujikawa, K.~Uddin, S.~Mizuno, R.~Tavakol and J.~Yokoyama,
%``Constraints on scalar-tensor models of dark energy from observational and local gravity tests,''
Phys. Rev. D \textbf{77}, 103009 (2008)
doi:10.1103/PhysRevD.77.103009
[arXiv:0803.1106 [astro-ph]].
%129 citations counted in INSPIRE as of 11 Oct 2021



%\cite{Boehmer:2008av}
\bibitem{Boehmer:2008av}
C.~G.~Boehmer, G.~Caldera-Cabral, R.~Lazkoz and R.~Maartens,
%``Dynamics of dark energy with a coupling to dark matter,''
Phys. Rev. D \textbf{78}, 023505 (2008)
doi:10.1103/PhysRevD.78.023505
[arXiv:0801.1565 [gr-qc]].
%274 citations counted in INSPIRE as of 11 Oct 2021


%\cite{Chimento:2003iea}
\bibitem{Chimento:2003iea}
L.~P.~Chimento, A.~S.~Jakubi, D.~Pavon and W.~Zimdahl,
%``Interacting quintessence solution to the coincidence problem,''
Phys. Rev. D \textbf{67}, 083513 (2003)
doi:10.1103/PhysRevD.67.083513
[arXiv:astro-ph/0303145 [astro-ph]].
%487 citations counted in INSPIRE as of 03 Nov 2021

 
%\cite{Leon:2020pfy}
\bibitem{Leon:2020pfy}
G.~Leon and F.~O.~F.~Silva,
%``Generalized scalar field cosmologies: theorems on asymptotic behavior,''
Class. Quant. Grav. \textbf{37} (2020) no.24, 245005
doi:10.1088/1361-6382/abbd5a
[arXiv:2007.11140 [gr-qc]].
%7 citations counted in INSPIRE as of 03 Nov 2021



%\cite{Sharma:2018vnv}
\bibitem{Sharma:2018vnv}
M.~Sharma, M.~Shahalam, Q.~Wu and A.~Wang,
%``Preinflationary dynamics in loop quantum cosmology: Monodromy Potential,''
JCAP \textbf{11}, 003 (2018)
doi:10.1088/1475-7516/2018/11/003
[arXiv:1808.05134 [gr-qc]].
%10 citations counted in INSPIRE as of 02 Nov 2020


\bibitem{Verhulst} Ferdinand Verhulst, (2000) ``Methods and Applications of Singular Perturbations: Boundary Layers and Multiple Timescale Dynamics'' (Springer-Verlag New York, ISBN 978-0-387-22966-9)   \href{}{https://doi.org/10.1007/0-387-28313-7}.



\bibitem{SandersEtAl2010}
Sanders J., Verhulst  F. , Murdock, J.,  Averaging Methods in Nonlinear Dynamical Systems (Applied Mathematical Sciences vol 59, Springer Science + Business Media, LLC, 2010) 2nd edn (Berlin). 
		
\bibitem{dumortier} F. Dumortier  and R. Roussarie (1995) ``Canard cycles and center manifolds'', (Memoirs of the American Mathematical Society, 577). 

  
\bibitem{fenichel} N. Fenichel (1979) ``Geometric singular perturbation theory for ordinary differential equations''. Journal of Differential Equations {\bf{31}}, 53-98. 


\bibitem{Fusco} G. Fusco and J.K. Hale, Journal of Dynamics and Differential Equations 1, 75 (1988).

\bibitem{Berglund} N. Berglund and B. Gentz, ``Noise-Induced Phenomena in Slow-Fast Dynamical Systems'', Series: Probability and Applications, Springer-Verlag: London, (2006).

 \bibitem{holmes} 
M. H. Holmes (2013) ``Introduction to Perturbation Methods'', (Springer Science+Business Media New York, ISBN 978-1-4614-5477-9) \href{}{https://doi.org/10.1007/978-1-4614-5477-9}.

\bibitem{Kevorkian1} Jirair  Kevorkian, 
J.D. Cole (1981) ``Perturbation Methods in Applied Mathematics'' (Applied Mathematical Sciences Series,  Volume 34, Springer-Verlag New York
eBook ISBN 978-1-4757-4213-8 \href{}{https://doi.org/10.1007/978-1-4612-3968-0}.



%\cite{Kamenshchik:2013dga}
\bibitem{Kamenshchik:2013dga}
A.~Y.~Kamenshchik, E.~O.~Pozdeeva, A.~Tronconi, G.~Venturi and S.~Y.~Vernov,
%``Integrable cosmological models with non-minimally coupled scalar fields,''
Class. Quant. Grav. \textbf{31}, 105003 (2014)
doi:10.1088/0264-9381/31/10/105003
[arXiv:1312.3540 [hep-th]].
%51 citations counted in INSPIRE as of 02 Nov 2020
	
%\cite{Andrianov:2011fg}
\bibitem{Andrianov:2011fg}
A.~A.~Andrianov, F.~Cannata and A.~Y.~Kamenshchik,
%``General solution of scalar field cosmology with a (piecewise) exponential potential,''
JCAP \textbf{10}, 004 (2011)
doi:10.1088/1475-7516/2011/10/004
[arXiv:1105.4515 [gr-qc]].
%38 citations counted in INSPIRE as of 02 Nov 2020
  
%\cite{Cid:2015pja}
\bibitem{Cid:2015pja}
A.~Cid, G.~Leon and Y.~Leyva,
%``Intermediate accelerated solutions as generic late-time attractors in a modified Jordan-Brans-Dicke theory,''
JCAP \textbf{02}, 027 (2016)
doi:10.1088/1475-7516/2016/02/027
[arXiv:1506.00186 [gr-qc]].
%16 citations counted in INSPIRE as of 02 Nov 2020











%\cite{Chakraborty:2021vcr}
\bibitem{Chakraborty:2021vcr}
S.~Chakraborty, E.~Gonz\'alez, G.~Leon and B.~Wang,
%``Time-averaging axion-like interacting scalar fields models,''
Eur. Phys. J. C \textbf{81} (2021) no.11, 1039
doi:10.1140/epjc/s10052-021-09802-5
[arXiv:2107.04651 [gr-qc]].
%1 citations counted in INSPIRE as of 09 Dec 2021


%\cite{Trodden:2004st}
\bibitem{Trodden:2004st}
M.~Trodden and S.~M.~Carroll,
``TASI lectures: Introduction to cosmology,''
[arXiv:astro-ph/0401547 [astro-ph]].
%124 citations counted in INSPIRE as of 29 Oct 2021





%%\cite{Chimento:1995da}
\bibitem{Chimento:1995da}
L.~P.~Chimento and A.~S.~Jakubi,
%``Scalar field cosmologies with perfect fluid in Robertson-Walker metric,''
Int. J. Mod. Phys. D \textbf{5}, 71-84 (1996)
doi:10.1142/S0218271896000084
[arXiv:gr-qc/9506015 [gr-qc]].
%64 citations counted in INSPIRE as of 02 Nov 2020

\bibitem{Wainwrightellis1997} 
 J. Wainwright and  G. F. R. Ellis, Eds. Dynamical Systems in Cosmology. Cambridge Univ. Press, Cambridge, 1997
 
 
 %\cite{Dimopoulos:2005ac}
\bibitem{Dimopoulos:2005ac}
S.~Dimopoulos, S.~Kachru, J.~McGreevy and J.~G.~Wacker,
%``N-flation,''
JCAP \textbf{08} (2008), 003
doi:10.1088/1475-7516/2008/08/003
[arXiv:hep-th/0507205 [hep-th]].
%580 citations counted in INSPIRE as of 14 Jan 2022

%\cite{Cardenas:2018nem}
\bibitem{Cardenas:2018nem}
V.~H.~C\'ardenas, D.~Grand\'on and S.~Lepe,
%``Dark energy and Dark matter interaction in light of the second law of thermodynamics,''
Eur. Phys. J. C \textbf{79} (2019) no.4, 357
doi:10.1140/epjc/s10052-019-6887-0
[arXiv:1812.03540 [astro-ph.CO]].
%17 citations counted in INSPIRE as of 14 Jan 2022

%\cite{Lepe:2015qhq}
\bibitem{Lepe:2015qhq}
S.~Lepe and F.~Pe\~na,
%``Interacting cosmic fluids and phase transitions under a holographic modeling for dark energy,''
Eur. Phys. J. C \textbf{76} (2016) no.9, 507
doi:10.1140/epjc/s10052-016-4347-7
[arXiv:1511.07186 [gr-qc]].
%7 citations counted in INSPIRE as of 14 Jan 2022

\end{thebibliography}
\end{document}